\documentclass[11pt,12pt,twoside,onecolumn,acm,bind]{hepthesis}
\usepackage{thesis,color}
\usepackage[T1]{fontenc}                                                        
\usepackage{times}
\usepackage{pbsi}
\usepackage{amsmath,amssymb,bbm,indentfirst,multirow}
\usepackage{fancyhdr,tocbibind,rotating,subfigure,ccaption,caption,wrapfig}
\usepackage[comma,square,sort&compress]{natbib}
\usepackage[colorlinks=true,linkcolor=blue,citecolor=magenta,bookmarks=true]{hyperref}

\graphicspath{{./graphics/}}

\providecommand{\chapabsbegin}{\begin{center}\begin{minipage}{0.9\linewidth} \it \small}
\providecommand{\chapabsend}{\end{minipage}\end{center}}

\providecommand{\putfigurelong}[4]{\begin{figure}\includegraphics{#1}\caption[#4]{\label{figure:#3}#2}\end{figure}}
\providecommand{\putfigureflags}[4]{\begin{figure}[#4]\includegraphics{#1}\caption{\label{figure:#3}#2}\end{figure}}
\providecommand{\putfigureflagslong}[5]{\begin{figure}[#4]\includegraphics{#1}\caption[#5]{\label{figure:#3}#2}\end{figure}}

\providecommand{\putfigurescaleyoudecidelong}[5]{\begin{figure}\includegraphics[scale=#4]{#1}\caption[#5]{\label{figure:#3}#2}\end{figure}}
\providecommand{\putfigureflagsscale}[5]{\begin{figure}[#4]\includegraphics[scale=#5]{#1}\caption{\label{figure:#3}#2}\end{figure}}
\providecommand{\putfigureflagsscalelong}[6]{\begin{figure}[#4]\includegraphics[scale=#5]{#1}\caption[#6]{\label{figure:#3}#2}\end{figure}}

\providecommand{\putfigurerightherelong}[4]{\begin{figure}[!htb]\includegraphics{#1}\caption[#4]{\label{figure:#3}#2}\end{figure}}

\providecommand{\refeq}[1]{(\ref{#1})}
\providecommand{\re}[1]{\refeq{eq:#1}}
\providecommand{\eqrefeq}[1]{Eq.~(\ref{eq:#1})}
\providecommand{\ere}[1]{\eqrefeq{#1}}
\providecommand{\esre}[1]{Eqs.~\re{#1}}
\providecommand{\reffig}[1]{\ref{figure:#1}}
\providecommand{\rf}[1]{\reffig{#1}}
\providecommand{\figreffig}[1]{Fig.~\ref{figure:#1}}
\providecommand{\frf}[1]{\figreffig{#1}}
\providecommand{\reftable}[1]{\ref{table:#1}}
\providecommand{\tablereftable}[1]{Table~\ref{table:#1}}
\providecommand{\trt}[1]{\tablereftable{#1}}
\providecommand{\refchap}[1]{\ref{chapter:#1}}

\providecommand{\refsec}[1]{\ref{section:#1}}
\providecommand{\secrefsec}[1]{Sec.~\ref{section:#1}}
\providecommand{\srs}[1]{\secrefsec{#1}}
\providecommand{\rs}[1]{\ref{section:#1}}
\providecommand{\refap}[1]{\ref{ap:#1}}
\providecommand{\aprefap}[1]{Appendix~\ref{ap:#1}}
\providecommand{\ara}[1]{\aprefap{#1}}
\providecommand{\lsbra}{\left[}
\providecommand{\rsbra}{\right]}
\providecommand{\lcbra}{\left\{}
\providecommand{\rcbra}{\right\}}
\providecommand{\lnbra}{\left(}
\providecommand{\rnbra}{\right)}
\providecommand{\linv}{\left . }
\providecommand{\rinv}{\right . }
\providecommand{\n}{\nonumber \\ }
\providecommand{\nn}{\nonumber}
\providecommand{\symmetrydeltasymbol}[2]{\Delta^{(#1)}_{#2}}
\providecommand{\symmetrydelta}[4]{\symmetrydeltasymbol{#1}{#2,#3}( #4 )}
\providecommand{\symmetrydeltalg}[3]{\symmetrydeltasymbol{#1}{#2}( #3 )}
\providecommand{\symmetrydeltana}[3]{\symmetrydeltasymbol{#1}{#2,#3}}
\providecommand{\symmetrydeltasymboltopalign}[2]{\Delta^{(#1)^{\phantom{\prime}}}_{#2}}
\providecommand{\symmetrydeltatopalign}[4]{\symmetrydeltasymboltopalign{#1}{#2,#3}( #4 )}
\providecommand{\average}[1]{\langle #1 \rangle}
\providecommand{\averageref}[1]{\overline{#1}}
\providecommand{\orderparameter}[3]{\average{\symmetrydeltasymbol{#1}{#2,#3}}}
\providecommand{\orderparameterref}[3]{\averageref{\symmetrydeltasymbol{#1}{#2,#3}}}
\providecommand{\rvec}{\vec{\mathbf{r}}}
\providecommand{\be}{\begin{equation}}
\providecommand{\ee}{\end{equation}}
\providecommand{\bea}{\begin{eqnarray}}
\providecommand{\eea}{\end{eqnarray}}
\providecommand{\befi}{\begin{figure}}
\providecommand{\enfi}{\end{figure}}
\providecommand{\eqdot}{\, \textrm{.}}
\providecommand{\eqcolon}{\, \textrm{,}}
\providecommand{\orientation}{\hat{\boldm{\Omega}}}
\providecommand{\orientationin}{{\hat{\boldm{\Omega}}}^{\prime}}
\providecommand{\relativeorientation}{{\orientation}^{\prime -1}\orientation}
\providecommand{\cltrace}{\textnormal{Tr}}
\providecommand{\grandpotentialsymbol}{\mathbf{\Omega}}
\providecommand{\grandpotentialf}[1]{\grandpotentialsymbol\lsbra #1 \rsbra}

\providecommand{\densitysymbol}{\varrho}
\providecommand{\densityf}[1]{\densitysymbol(#1)}
\providecommand{\densityff}[2]{\densitysymbol_{#2}(#1)}
\providecommand{\ensembleavg}[1]{\langle #1 \rangle }
\providecommand{\microdensityf}[1]{\hat{\densitysymbol} \lnbra #1 \rnbra}
\providecommand{\intrinsicfesymbol}{{\cal{F}}}
\providecommand{\intrinsicfef}[1]{\intrinsicfesymbol\lsbra #1 \rsbra}
\providecommand{\idealfef}[1]{\intrinsicfesymbol_{id}\lsbra #1 \rsbra}
\providecommand{\excessivefef}[1]{\intrinsicfesymbol_{ex}\lsbra #1 \rsbra}
\providecommand{\excessivefesymbol}{\intrinsicfesymbol_{ex}}
\providecommand{\excessfef}[1]{\intrinsicfesymbol_{ex}\lsbra #1 \rsbra}
\providecommand{\excessfesymbol}{\intrinsicfesymbol_{ex}}
\providecommand{\fderivate}[3]{\frac{\delta#3}{\delta #1 \lnbra #2 \rnbra}}
\providecommand{\fderivatede}[2]{\frac{\delta #1}{\delta \densitysymbol \lnbra #2 \rnbra}}
\providecommand{\fdederivatede}[3]{\frac{\delta^{2} #1}{\delta \densitysymbol \lnbra #2 \rnbra\delta \densitysymbol \lnbra #3 \rnbra}}
\providecommand{\firstdcf}[2]{c_{1}( #1,\lsbra #2 \rsbra )}
\providecommand{\seconddcf}[3]{c_{2}( #1,#2,\lsbra #3 \rsbra )}
\providecommand{\fderivatepsi}[2]{\frac{\delta #1}{\delta \psi \lnbra #2 \rnbra}}
\providecommand{\opdfsymbol}{P}
\providecommand{\opdff}[1]{\opdfsymbol(#1)}

\providecommand{\opdfff}[2]{\opdfsymbol_{#2}(#1)}
\providecommand{\opdffin}[1]{\opdfsymbol_{#1}(\orientation)}
\providecommand{\opdfffin}[2]{\opdfsymbol_{#2}(#1)}
\providecommand{\opdf}{\opdff{\orientation}}

\providecommand{\opdfs}{\opdff{\orientation,z}}
\providecommand{\meyerfsymbol}{c}
\providecommand{\meyerf}[2]{\meyerfsymbol(#1,#2)}
\providecommand{\meyerfunc}{\meyerf{\coord{1}}{\coord{2}}}

\providecommand{\dimensionless}[1]{#1^{*}}
\providecommand{\enbx}{\lambda_{\epsilon}}
\providecommand{\shbx}{\lambda_{\sigma}}
\providecommand{\gbsete}[1]{$(#1)_{\epsilon}$}
\providecommand{\gbsets}[1]{$(#1)_{\sigma}$}
\providecommand{\sx}{\sigma_{x}}
\providecommand{\sy}{\sigma_{y}}
\providecommand{\sz}{\sigma_{z}}
\providecommand{\ex}{\epsilon_{x}}
\providecommand{\ey}{\epsilon_{y}}
\providecommand{\ez}{\epsilon_{z}}
\providecommand{\studiedrho}{$\dimensionless{\rho}=0.18$}
\providecommand{\sdplane}{$(\enbx^{sd},\shbx^{sd})$}
\providecommand{\sigmaset}{$\{\sigma_{i}\}$}
\providecommand{\epsilonset}{$\{\epsilon_{i}\}$}
\providecommand{\felastic}{{\textnormal{F}}_{elastic}}
\providecommand{\sbentcore}{bent-core }
\providecommand{\sbbentcore}{Bent-core }
\providecommand{\sopdf}{one-particle distribution function }
\providecommand{\sbopdf}{One-particle distribution function }
\providecommand{\sopdfns}{one-particle distribution function}
\providecommand{\sopdfs}{one-particle distribution functions }

\providecommand{\sdcf}{direct correlation function }

\providecommand{\sdft}{Density Functional Theory }
\providecommand{\sldft}{{Local Density Functional Theory}}

\providecommand{\sceq}{self--consistent equation}

\providecommand{\stbxn}{thermotropic biaxial nematic}

\providecommand{\coordsymbol}{x}
\providecommand{\coord}[1]{\coordsymbol_{#1}}
\providecommand{\coords}[2]{\boldv{r}_{#1},\orientation_{#2}}
\providecommand{\unitvv}[2]{{\mathbf{\hat{#1}}}_{#2}}
\providecommand{\unitv}[1]{{{\mathbf{\hat{#1}}}}}
\providecommand{\boldv}[1]{\vec{\mathbf{#1}}}
\providecommand{\oper}[1]{{\mathbf{\hat{#1}}}}
\providecommand{\ssts}{side-to-side}
\providecommand{\sftf}{face-to-face}
\providecommand{\sete}{end-to-end}
\providecommand{\ep}{\epsilon}
\providecommand{\pref}{P_{ref}}

\providecommand{\bcorf}[2]{p_{#1}(#2)}
\providecommand{\bcorr}[1]{p_{#1}}
\providecommand{\kernelsymbol}{K}

\providecommand{\kernelbif}[2]{\kernelsymbol_{2}(#1,[ #2 ])}
\providecommand{\kernelbifn}[3]{\kernelsymbol_{#3}(#1,[ #2 ])}

\providecommand{\innerpro}[2]{(\,#1,#2\,)}
\providecommand{\clmn}[3]{c^{(#1)}_{#2,#3}}
\providecommand{\almn}[2]{a^{(#1)}_{#2}}
\providecommand{\slmn}[3]{s^{(#1)}_{#2,#3}}
\providecommand{\olmn}[2]{\omega^{(#1)}_{#2}}
\providecommand{\bifmat}[1]{\oper{\boldm{\omega}}^{(#1)}}
\providecommand{\labv}[1]{{\unitv{l}}_{#1}}
\providecommand{\labx}{{\unitv{l}}_{1}}
\providecommand{\laby}{{\unitv{l}}_{2}}
\providecommand{\labz}{{\unitv{l}}_{3}}
\providecommand{\bodv}[1]{{\unitv{b}}_{#1}}
\providecommand{\bodx}{{\unitv{b}}_{1}}
\providecommand{\body}{{\unitv{b}}_{2}}
\providecommand{\bodz}{{\unitv{b}}_{3}}
\providecommand{\wignerd}[4]{D^{(#1)}_{#2,#3}( #4 )}
\providecommand{\wignerdna}[3]{D^{(#1)}_{#2,#3}}
\providecommand{\wignerdnaconjugate}[3]{D^{(#1)\,*}_{#2,#3}}
\providecommand{\wignerdnaconjugatetopalign}[3]{D^{(#1)^{\phantom{\prime}}*}_{#2,#3}}
\providecommand{\irrten}[2]{{\bf{T}}^{\,#1}_{#2}}
\providecommand{\irrtensymm}[3]{\linv\widetilde{{\bf{T}}}^{\,#1}_{\,#2}\right|_{#3}}
\providecommand{\symmg}[1]{{\cal{#1}}}
\providecommand{\abs}[1]{\left| #1 \right| }
\providecommand{\deltasnorm}{\frac{8\,\pi^{2}}{5}}
\providecommand{\deltasnorml}[1]{\frac{8\,\pi^{2}}{2\,#1+1}}
\providecommand{\dwcoeff}[2]{N^{#1,#2}_{\Delta}}
\providecommand{\deltasnormalization}[1]{N^{\,#1}_{\Delta\Delta}}
\providecommand{\clebschgordan}[5]{C\lnbra #1,#2,#3;#4,#5 \rnbra}
\providecommand{\smecticbasef}[3]{S^{\,(#1)}_{\,#2,#3}}
\providecommand{\smecticbaseftopalign}[3]{S^{\,(#1)^{\phantom{\prime}}}_{\,#2,#3}}
\newcommand{\boldm}[1]{\mbox{\boldmath{$#1$}}}
\providecommand{\dg}[1]{#1^{\circ}}
\providecommand{\dtwh}{$D_{2h}$ }
\providecommand{\dinfh}{$D_{\infty h}$ }
\providecommand{\dinfhns}{$D_{\infty h}$}

\providecommand{\iso}{$Iso$}
\providecommand{\nrod}{$\textnormal{N}_{U+}$}
\providecommand{\ndisc}{$\textnormal{N}_{U-}$}
\providecommand{\ndisk}{\ndisc}
\providecommand{\nun}{$\textnormal{N}_{U}$}
\providecommand{\nbx}{$\textnormal{N}_{B}$}
\providecommand{\sbx}{$Sm\textnormal{A}_{B}$}
\providecommand{\sun}{$Sm\textnormal{A}_{U}$}
\providecommand{\sma}{$Sm\textnormal{A}$}
\providecommand{\director}{\hat{{\mathbf{n}}}}
\providecommand{\ital}[1]{{\it #1}}
\providecommand{\nl}{\newline }
\providecommand{\nlin}{\newline\indent }

\title{Density Functional Theory of Model Systems with the Biaxial Nematic Phase}                   
\author{Piotr Grzybowski}
\begin{document}

 \begin{frontmatter}
\titlepage[~]%
{A dissertation submitted to the Jagiellonian University in Krak\'ow \linebreak for the degree of Doctor of Philosophy}

\frontquote{Gaudium in Litteris.}{}

\begin{acknowledgements}
 
\indent There are many people to whom I owe much. Without a great
help from my supervisor, Professor Lech Longa, this thesis would have
never come into existence. In first place sincere thanks go to him.
I always received an incredible amount of support from my Parents,
I am forever grateful to them. Warm thanks go to my fiancée Ana,
who has always stood by me. I am grateful to my Brother for reading (and correcting)
the manuscript. I would also like to express my appreciation
to Dr Michael Ciesla for many useful and interesting discussions.
\nlin During my years as Ph.D. student, I could always count on a warm
welcome and words of encouragement from lovely ladies at the editorial board 
of the Acta Physica Polonica B, especially Mrs Maria Czyz
and Mrs Krystyna Stankiewicz. Last, but certainly not least, I am grateful
to all the great people at the Institute of Physics and Mathematics
at Jagiellonian University; I always felt like a member of a great family,
they are in my heart forever. Finally, I could not forget my first
physics teacher, Mrs Teresa Kownacka, I would also like to thank her.

\end{acknowledgements}

\dedication{To my Grandmother}

\begin{abstract}

\indent Present work is a theoretical study on the stability of
the thermotropic biaxial nematic liquid crystal phase in model systems.
Its main aim is to present the phase diagrams of spatially uniform
liquid mesophases and to identify the molecular parameters
that influence the stability of the biaxial nematic.
The diagrams are obtained by means of the \sldft\ in
low density approximation, and the relation between the molecular
parameters of the models and macroscopic properties of the system
close to the transition point are obtained by means of bifurcation
analysis. We consider three model systems; the so-called $L=2$ model
(the lowest coupling model of the orientational part of pair potential),
the biaxial Gay-Berne interaction, and the \sbentcore system. For the
second one, we also briefly investigate the elastic constants and comment on
the smectic phases. In every case we take into account rigid molecules.
\nlin Firstly, we study a generalized version of Straley model, which should be
considered as the simplest one giving rise to the biaxial nematic phase;
it is to the biaxial nematic what Maier-Saupe model is to the uniaxial nematic.
Its analysis allows to construct a generic, exact in mean field
bifurcation diagram as function of potential parameters. By considering the
symmetries of the $L=2$ expansion of pair interaction, valid also for 
pair direct correlation function, we find the duality transformation which
connects the states at different temperatures. The so-called self-dual points,
i.e., the points in space of potential parameters left invariant under the
action of the duality transformation, coincide with Landau points. 
The following Landau region is found as well as the tricritical points.
Despite its simplicity, the model exhibits a rich behaviour. Many systems
can be accurately approximated using this approach, and located on
the presented diagram. 
\nlin The second model studied is a Gay-Berne potential generalized
to the soft ellipsoids of three different axes. It showed
the biaxial nematic in Monte Carlo study, and therefore it gives us
the chance to confront our results with simulations and study the
interaction parameter space in more detail. We also compare the
bifurcation diagram following from \sldft\ with transition points
acquired by the minimisation of the Helmholtz free energy. In comparison
with the simulations, the approach used overestimates
the transition temperatures. We study the interaction further and locate the 
Landau points (also called self-dual) induced by the increase of both the
shape and energy biaxiality. In the former case, our results indicate that
the introduction of attractive forces slightly shifts the position of
the self-dual point from its location for hard biaxial ellipsoids.
We find that the direct isotropic -- biaxial nematic transition occurs
at Landau points, and show how the concurring biaxialities
influence their position. The results indicate that for this model 
self-dual region predicted for hard, biaxial ellipsoids by the so-called
square root rule retains its significance, and provides qualitatively correct
estimations of Landau points positions.
\nlin Analysis of the \sbentcore system is aimed at studying the
shape induced effects and the role of the dipole-dipole interaction. The
model molecules are built using two and three Gay-Berne interacting,
prolate, soft ellipsoids of uniaxial and biaxial symmetry. We study the
influence of dipole-dipole interaction, in case of two and three uniaxial arms,
by introducing a transverse dipole moment lying along the $C_{2}$ symmetry axis,
in the plane of an angle between the arms (opening angle). For non-polar case
of two uniaxial arms, we find that by changing the opening angle we can obtain
a Landau point at $\dg{107}$, in agreement with previous studies for
hard \sbentcore molecules. Surprisingly, for this model the inclusion of 
attractive forces does not influence its position. Once the arms deviate from
the uniaxial symmetry and become biaxial ellipsoids
interacting via biaxial Gay-Berne potential,
the self-dual point evolves to the line of Landau points ranging in the
opening angle between $\dg{121}$ and $\dg{128}$. In both cases, the
opening angles at which self-dual points occur, differ
from the experimental estimates of $\dg{140}$. However, a study on
system of \sbentcore molecules with opening angle of $\dg{90}$ was published
recently, and those results remain in qualitative agreement with our analysis of the
model of three uniaxial Gay-Berne parts where we find the bifurcation diagram 
with the isolated Landau point at the angle between the arms equal
to $\dg{89}$. The introduction of weak dipoles (relative to the strength of
the Gay-Berne energy) leads to the shifting of the Landau point towards lower
angles. For three uniaxial parts model, moderate dipole magnitudes
provide the diagram with a line of self-dual points, range of which in
opening angle increases with dipole strength. The maximum length
of the interval of the opening angle for
which a direct isotropic -- biaxial nematic transition occurs is $\dg{23}$
between $\dg{63}$ and $\dg{86}$. For the strongest dipoles studied, that
line begins to shrink and shift towards higher angles. In \sbentcore models,
again, the Landau points are the only places where
isotropic -- biaxial nematic transition takes place,
the novel feature is the appearance of the self-dual line.
\nlin Since the elastic constants are important quantities in the theory and
applications of\linebreak liquid mesophases, in final chapter we present the
preliminary studies on temperature dependence of a set of bulk biaxial elastic constants 
for the biaxial Gay-Berne interaction in $L=2$ model. Our results
are in agreement with previous findings for the lattice biaxial model; we recover
the splay-bend degeneracy, and find that the constants associated with one
of three directors in the biaxial nematic are always negative. Also, in the
rod-like molecular regime the relative values of the uniaxial elastic constants
are higher than biaxial ones. This behaviour changes in vicinity
of the Landau point, where the constants associated with one of the
biaxial directors become dominant.
\nlin Finally, we make a comment on the smectic-A phases. Although the thesis is
devoted to the spatially uniform liquids, the competition between the
smectic order and biaxial nematic should be considered since the
former may gain stability earlier and prevent the formation of the
spatially uniform biaxial ordering. In order to address this issue, and as an
example, we limit ourselves to the orthogonal smectic-A phases. We present the behaviour of a complete
set of leading order parameters for the biaxial Gay-Berne potential, based on
the \sceq\ for equilibrium \sopdf in $L=2$ model, including uniaxial
and biaxial smectic-A states. The temperature dependence of order parameters is
presented for two sets of interaction parameters. One of them, namely
the one where molecules are strongest attracted to their sides (the lateral
forces are strongest), was the only case for which Monte Carlo simulations
discovered a stable biaxial nematic. We find that in this case the temperature
range of the lower symmetric nematic state is significantly wider than
for the model of strong \sftf\ attractions. For fixed
density and with increasing temperature, biaxial smectic is found to melt to
biaxial nematic which in turn transforms to uniaxial nematic and for
higher temperatures the system undergoes a phase transition to isotropic liquid.
This phase sequence is in agreement with simulations results. In final section,
we extend the approach used for spatially uniform states to include the
orthogonal smectic phases of uniaxial and biaxial symmetry,
and we present the additional bifurcation equations.

\end{abstract}

\tableofcontents

\end{frontmatter}
 
 \begin{mainmatter}
 \chapter{Introduction}
 \label{chapter:intro}
{ \chapabsbegin A hundred and twenty years ago it became clear that between 
the usual fluid and solid crystal there exists an intermediate phase of matter.
Substance in the new phase remained liquid, however, system exhibited
extraordinary anisotropic optical properties, which could be explained
by microscopic studies. It was found that in the new states a degree of
order was formed on the molecular level. Apparently the molecules no
longer possessed random orientations, as in the normal liquids, but were
on average oriented along one, system-wide direction.
A certain subclass of those intermediate phases is a subject of the
present work for which this chapter serves as an introduction. We will begin
with a brief description of the historical background and review of the research
conduced so far, then continue to introduce the matter of the thesis,
its aim, and the employed methods. \chapabsend }
\section{Discovery of ordered liquids}
 The first indications of the new phase of matter are due to W. Heintz who
in 1850 studied stearin. He found that the substance apparently possessed two
melting points, and that was a new phenomenon altogether. With the increase of
temperature, it firstly became cloudy, then opaque, and eventually turned into
a clear liquid. Later, in 1854, Rudolf Virchow described a soft, floating
substance present in the human body, protecting the nerve fibres, which
he called \ital{myelin} \cite{virchowhist} (see \figreffig{myelin}).
By the work of Carl Mettenheimer three years later this substance was found
to show birefringence. Those were the first indications of the new phenomena.
Today the systems studied by Virchow and Mettenheimer would be called
\ital{lyotropic}, since concentration of water
is a driving force for the phase transitions. In contrast, the materials
for which this role is played by temperature would be called 
\ital{thermotropic}\footnote{The distinction is a bit more subtle, since in lyotropic materials the temperature is also important, but definitely for thermotropic phases the concentration of water is not the issue.}. Surprisingly, it took
over thirty years to realize the meaning of those findings. It is the
year 1888 that is recognised as the year of the discovery.
At that time Friedrich Reinitzer synthesized cholesteryl benzoate 
($C_{34}H_{50}O_{2}$, see Fig. \reffig{cholesterylbenzoate}) and
observed with his microscope iridescent colours as well as similar temperature
driven behaviour as Heintz did 38 years earlier. Intrigued by the phenomenon
he consulted Hofrath von Zepharovich, and by his advice wrote a letter to 
Otto Lehmann on 14th of March, 1888 \cite{historyliquidcrystals}. Otto Lehmann,
born a year before Virchow's discoveries, at the time
of receiving the letter was a young associate professor, although
already a well known figure in the field of phase transitions
phenomena. He just finished his ''Molekularphysik'', an impressive work of
two volumes enclosing more than 1500 pages with 620 figures and
over 2000 citations. Friedrich Reinitzer, born 1857, at the time was a botanist.
His conscientious observations \cite{reinitzerhist} and correct conclusions
were one of the main factors that helped to realize the meaning of 
the discovery. He was the right man in the right time and place, and a lot
of credit for the early discoveries is rightfully attributed to him.
On March 14th 1888 he wrote:
\begin{quotation}
The substance has two melting points, if it can be expressed in such a manner. 
At $\dg{145.5}$C it melts to a cloudy, but fully liquid melt which at 
$\dg{178.5}$C suddenly becomes completely clear. On cooling a violet and 
blue colour phenomenon appears, which then quickly disappears leaving the 
substance cloudy but still liquid. On further cooling the violet and blue
colouration appears again and immediately afterwards the substance 
solidifies to a white, crystalline mass.
\end{quotation}
\putfigureflags{part1/chapter1/myelin.eps}{Drawing of myelin as shown by O. Lehmann, after \cite{mclchistory}.}{myelin}{h}
Those intermediate states, ordered and yet liquid, at first were called 
by Lehmann ''Fliessende Kristalle'' that is ''flowing crystals'', which
later evolved to ''liquid crystals'' or {\it meso}phases from a Greek word
''meso'' meaning ''middle'' or ''intermediate''. It was the cooperation of 
Reinitzer and Lehmann that marked the first chapter of the liquid crystals
history, which is recognized to began on 14th March 1888,
the day the letter was sent which opened a new field of physics.
\nlin Today there is a great amount of known liquid crystalline structures,
only a brief description of those would consume many pages, if not a bundle
of volumes. The mesogenic substances play an important role in our world.
They are not only essential for a widely available and most useful
LC displays, but also, as was mentioned, were found early on in 
living organisms; besides myelin, the living cell membranes manifest the behaviour 
of the lyotropic liquid crystals. The overall experimental issues,
technological applications and interesting challenges in theoretical 
description make those materials an important part of modern physics.
\nlin The reasons for the emerging of liquid crystalline ordering are
one of the fundamental problems that have been addressed many times.
There exists a magnificent variety of substances that
manifest liquid crystalline behaviour, and yet many more can be synthesized,
and there are numerous factors that can be responsible for mesophase formation.
At least in most cases some anisotropy of shape of molecules is a required
condition. Another is the anisotropy of the intermolecular forces. Those
two basic issues can be set in the centre of the problem (together with
others of which we say nothing).
\nlin One of the interesting problems is the question of the
conditions which are required for the stabilization of a liquid crystal 
structure, like temperature, density, pressure, or external fields.
We can assume that we have at our disposal the microscopic
quantities like molecular shape and characterisations of forces, and 
we want to find the macroscopic, equilibrium properties of the system.
The perfect way to establish the relation between those is the statistical
description. This work is an example of such approach.
\putfigureflagsscale{part1/chapter1/cholesteryl_benzoate.eps}{Cholesteryl benzoate. The first synthesized liquid crystalline substance.}{cholesterylbenzoate}{b}{0.6}
\putfigureflagslong{part1/chapter1/isotropicnematic.eps}{Snapshot of molecular structure of isotropic phase (on the left) and uniaxial nematic (on the right).}{isotropicnematic}{t}{Snapshot of molecular structure of isotropic phase and uniaxial nematic.}
\section{Nematic order}
\putfigureflagslong{part1/chapter1/rodsdiscs.eps}{Snapshot of molecular structure of disc-like (on the left) and rod-like nematic (on the right). Director $\bf{\hat{n}}$ is marked.}{rodsdiscs}{b}{Snapshot of molecular structure of disc-like and rod-like nematic.}
 Most of the mesogenic substances consist of molecules that do not 
possess any definite shape. However, in a given liquid crystal state
the molecules are rotating much faster then they are moving through 
the sample, i.e., the time scale of the rotation is few orders of
magnitude different from the time scale in which a measurement takes
place. In effect the molecule acts as an object of certain
averaged shape. Therefore we can approximate the complicated structure
of the compound by a simpler object of a given symmetry and think
of the molecules as if they possessed a defined shape. In particular, many
liquid crystals are built from molecules of average shape resembling
a rod, which can be approximated by a prolate ellipsoid of revolution.
In the system of elongated, rod-like molecules in high temperature or low density
regime we will observe the usual, disordered fluid which is traditionally
called an isotropic phase (\iso). By lowering the temperature, or increasing
the density, it can evolve to a more ordered structure, possibly a 
liquid crystalline phase. The simplest of those is called a
\ital{nematic} phase, from a Greek word ''nemato'' meaning ''thread''.
The complete disorder of isotropic liquid in the nematic state is replaced
by a tendency of long molecular axes to on average align along given direction,
called a \ital{director}. The long range correlation of orientational
degrees of freedom appears while the full translational symmetry is upheld.
Because there is only one system wide, distinguished direction, the state
is referred to as \ital{uniaxial} nematic (\nun). Snapshots of the system
exhibiting the isotropic and nematic states are depicted in 
\figreffig{isotropicnematic}.
This state is additionally characterized by the invariance with respect to the rotation around the director, and
reflection in the perpendicular plane, and therefore it is said to be of
\dinfh symmetry, since this group contains those transformations. 
In present work a phase or molecule is considered to be uniaxial if it
is left invariant under the operation of \dinfh symmetry group. 
\nlin The model molecular uniaxial symmetry can be realised in another way.
Some mesogens consist of molecules that can be well approximated by
oblate ellipsoids of revolution, resembling a disc. The uniaxial nematic phase
in a system of those objects is characterized by order of shorter axes.
It is called a disc-like or oblate nematic (\ndisk), in opposition to 
formerly described rod-like or prolate nematic (\nrod). Both phases are 
shown in \figreffig{rodsdiscs}. Naturally, these
states may be realized in a systems of molecules with more complex shapes,
inclusion of which may bring a possible new behaviour of the system.
\nlin We can consider a deviation from the \dinfh symmetry and take into account
the model molecules possessing the shape of ellipsoid with three different
axes. If we choose one of them to be significantly longer, relative 
to the remaining two, we can obtain the \nrod\ resulting from the 
ordering of the longer axes. While the system is in uniaxial nematic phase,
we can wonder if it is possible to introduce the correlations of shorter
axes, while maintaining a lack of long-range order of molecular centres
of masses. In that way we would acquire a second director in the plane 
perpendicular to the uniaxial one, and third perpendicular to those two.
The symmetry of this state consists of reflections in three perpendicular
planes and is denoted by \dtwh. This phase is called a \ital{biaxial} nematic
(\nbx) and is represented in \figreffig{biaxialnematic}. As of now,
it is a ''hot subject'' in the soft matter field from both theoretical
and experimental point of view. On the experimental side, it is still
unclear why some materials, especially \sbentcore and tetrapode systems,
exhibit spatially uniform biaxiality while the others do not,
or what molecular parameters are required to stabilize the biaxial nematic.
In the light of recent discoveries, it seems that we are getting closer to 
obtaining the answer to these issues, and once they are resolved, 
the technological applications will be most promising. In this respect the theory can be closely related
to experiment, since it can point out the way which should be undertaken.
\nbx\ is one of the two simplest mesophases, yet the introduction of
two directors in addition to the uniaxial one brings a rich variety 
of new behaviour. Present work is devoted to that state. In the following
sections we present a description of the research conduced so far
in context of the biaxial nematic phase.
\section{Biaxial nematics}
\subsection{First indications of stable biaxial nematic phase}
\begin{figure}[b]
 \centering
 \subfigure[Picture of mixture of rods and discs. If the two uniaxial phases
{\nrod} and {\ndisc} mix two directors $\hat{\mathbf{n}}_{1}$ and
$\hat{\mathbf{n}}_{2}$ can be perpendicular.]{\label{figure:mixturescheme} \includegraphics{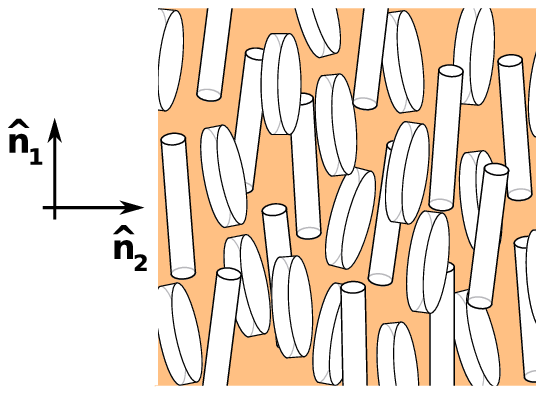}}
 \subfigure[One of the phase diagrams presented in \cite{albenmix}. Effective temperature $\textnormal{T}^{*}$ as in \cite{albenmix} versus discs concentration for elongation of prolate molecules equal to $5$ and width of oblate ones $3$.]{\label{figure:albenmixdiag} \includegraphics{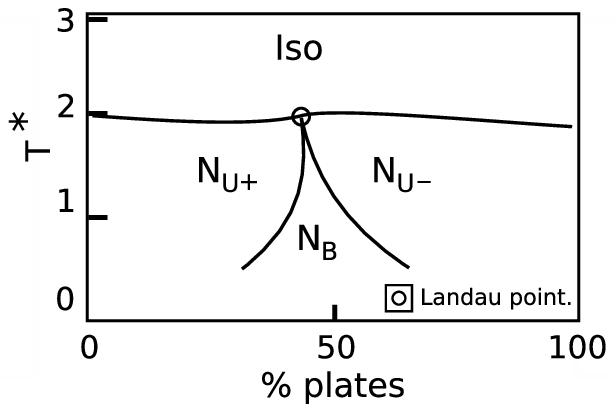}}
 \caption[Schematic picture of mixture and phase diagram from Alben's work \cite{albenmix}.]{\label{figure:mixturesschemeanddiag} Schematic picture of rods and
discs giving rise to biaxial nematic phase, and example of phase diagram for
model mixture of rods and discs (on the right).}
\end{figure}

 The lower symmetric nematic phase was predicted for the first time
by Freiser \cite{freiser} in 1970. Essentially he generalized
the Maier--Saupe model \cite{maiersaupe} to non-cylindrical molecules.
In the mean field approach, he found the first order transition from isotropic
liquid to the uniaxial nematic phase to be followed by a second order 
transition to the biaxial nematic. 
\nlin One year later another approach to the biaxial phase was presented.
The Landau \cite{landau} description of phase transitions was extended by
de Gennes and shown \cite{landaudegennes} to produce a stable biaxial nematic.  \putfigurerightherelong{part1/chapter1/biaxialnematic.ps}{Snapshot of molecular structure of biaxial nematic (on the right) with two directors marked as arrows, the third one is perpendicular to the surface of the picture. On the left an example of the biaxial molecule is shown and three perpendicular planes; reflections in those constitute the \dtwh symmetry.}{biaxialnematic}{Snapshot of molecular structure of \nbx\ and \dtwh -- symmetric shape.}
\nlin In 1973, Alben proposed a different method of stabilisation of the
biaxial ordering \cite{albenmix}. It was based on an assumption that in
a system of molecules of oblate and prolate shape the two types of nematic
ordering \ndisk\ and \nrod\ can form a mixture, i.e., the system would not
exhibit any areas rich in rods and with low concentration of disks or
vice-versa. Furthermore, the two directors should then be perpendicular and in
effect biaxiality could emerge (see \figreffig{mixturescheme}). Alben studied
the lattice model of hard molecules, where only steric interactions are
taken into account. The biaxial phase was found to be stable, and at a certain
concentration of discs even a direct {\iso} -- {\nbx} transition took place.
An example of the diagram from that study is presented in \figreffig{albenmixdiag}. \nlin A year after Alben's work, another study was presented;
following from the success of Maier--Saupe model for isotropic -- uniaxial
nematic transition Straley proposed a generalization of the orientational part
of their potential to the case of objects of \dtwh symmetry \cite{straley}. 
The new two-point interaction had four parameters, which were related to the 
dimensions of rectangular blocks by requiring the consistency with 
excluded volume present in Onsager's theory. The potential parametrized 
in this way was then used to construct a mean field theory. The resulting
phase diagram revealed a first order {\iso} -- {\nun} transition which in
lower temperatures was followed by a second order {\nun} -- {\nbx} transition.
Also it exhibited a direct transition between isotropic liquid and 
biaxial nematic at self-dual point, as can bee seen in \figreffig{straleydiagram}. In this work Straley also for the first time presented a complete set of
four molecular order parameters needed to describe uniaxial and biaxial
nematics. \putfigurerightherelong{part1/chapter1/straleydiagram.eps}{Phase
diagram adapted from \cite{straley}, breadth B as function of dimensionless temperature t.  Length and width of rectangular blocks are set to $10$ and $1$ respectively. {\iso} - {\nbx} transition occurs at self-dual point $\textnormal{B}=\sqrt{10}$.}{straleydiagram}{Phase diagram by Straley for $L=2$ model adapted from \cite{straley}.}

 Most of the above theoretical studies in comparison to real systems
can be considered huge simplifications. Not only many microscopic
aspects of liquid crystal phenomena, like intermolecular forces, were neglected
or crudely approximated, but also some structures, like smectics, were
not included. Nevertheless, those models provided a stable biaxial nematic
and predicted no particular difficulties in the stabilisation.
Since the fundamental problems in the formation of biaxial ordering,
had they existed, should have emerged in those studies, it came as a surprise
that the biaxial nematic was much more difficult to observe than other 
mesophases. It took more than ten years from the first prediction to the first
experimental realisation of \nbx. In 1980, Yu and Saupe
published a report on the spatially uniform biaxial ordering
in lyotropic ternary system of potassium laurate-1-decanol-$\textnormal{D}_{2}\textnormal{O}$ \cite{yulyo}.
By means of microscopic studies and deuteron resonance (NMR), two uniaxial
nematic states denoted $N_{L}$(\ndisc) and $N_{C}$(\nrod), and the biaxial phase were found. 
Phase diagram from that study is presented in \figreffig{yusaupediagram}. For 
$\textnormal{D}_{2}\textnormal{O}$ concentration above $68$ wt. \% the
system exhibits only the $N_{L}$ phase, characterized by micelles of disc--like
shape. When $\textnormal{D}_{2}\textnormal{O}$ concentration is lowered
to $67.8$ wt. \% the phase $N_{C}$ forms, with rod--like micelles. Between those
two states, the biaxial nematic phase occurs which on heating and on cooling
is transformed into the $N_{L}$ state. \putfigurerightherelong{part1/chapter1/yusaupediagram.eps}{Phase diagram of potassium laurate-1-decanol-$\textnormal{D}_{2}\textnormal{O}$ as presented in \cite{yulyo}. Uniaxial nematic $N_{L}$ at a certain concentration of $\textnormal{D}_{2}\textnormal{O}$ transforms to the biaxial nematic phase \nbx.}{yusaupediagram}{Phase diagram of potassium laurate-1-decanol-$\textnormal{D}_{2}\textnormal{O}$ by Yu and Saupe \cite{yulyo}.}
\subsection{Further theoretical studies}
\subsubsection{Mixtures}
\begin{figure}[!htb] \centering
 \subfigure[Structure of shape amphiphilic compound from \cite{datemix}.]{\label{figure:shapeamphicompound} \includegraphics[scale=0.8]{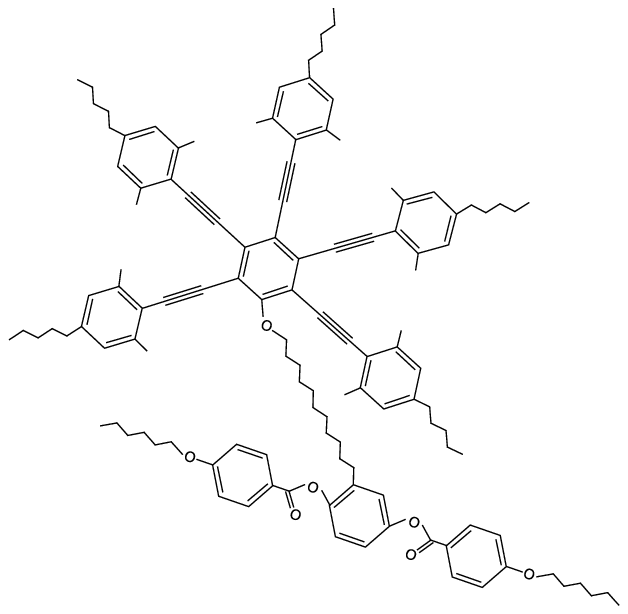}}
 \subfigure[Schematic picture of the connection of a rod to a disc.]{\label{figure:shapeamphischeme} \includegraphics{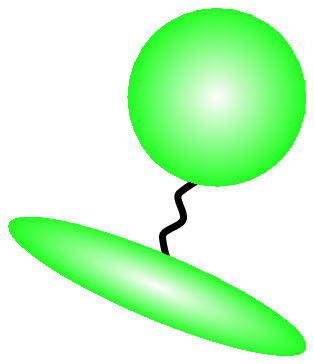}}
 \caption{\label{figure:shapeamphiphilic} The amphiphilic molecule.}
\end{figure}

 Study by Alben work rendered the binary mixtures of rod--like and
disc--like molecules as a possible candidate for the realization of a stable
biaxial nematic phase in thermotropic systems \cite{albenmix}. In the early 80s,
this matter was re-examined. Models of mixtures of hard molecules of equal
volume were shown to produce a stable biaxial phase \cite{rabinmix,stroobantsmix}.
The effects of excluded volume which were studied in Onsager's approximation lead to a first order
transition from isotropic to uniaxial nematic phase, and to a second
order {\nun} -- {\nbx} transition \cite{rabinmix}. This work also
confirmed that point of a direct \iso\ -- \nbx\ transition
can be found. A similar study on the extension of Onsager theory showed as well
a stable biaxial nematic phase at a certain concentration of molecules
and fraction of discs and rods \cite{stroobantsmix}. The inclusion
of both long and short range interactions in a mean field approach
using a type of Van der Waals theory also gave \nbx \cite{chenmix}.
In this study the system was divided into cells of
a given volume; within a cell molecules adopted three
discrete orientations, and interacted via a short--range repulsive potential,
while the long range attractive forces acted between the cells.
\nlin The predictions of the above studies \cite{rabinmix,stroobantsmix,chenmix}, as well as of Alben \cite{albenmix},
are correct, provided the two uniaxial nematic phases of discs and rods mix.
It is not obvious however, if the system does not undergo a demixing transition
before the formulation of the biaxial nematic phase. The configuration where states 
{\ndisk} and {\nrod} are separated, i.e., in the system appear regions rich
in molecules of one type, may gain stability instead of {\nbx}.
This issue was addressed by Palffy-Muhoray and de Bruyn \cite{palffymix2}.
They applied previously developed \cite{palffymix0}, generalized Maier -- Saupe
model in mean field approximation to binary mixture of rods and discs
of equal volume and found it phase separates when the arithmetic mean rule was
used for interactions between different types of molecules \cite{palffymix2}.
Similar model was developed to study the effects of external fields and
induced biaxiality \cite{palffymix1}. Again, the system exhibited demixing
and the biaxial nematic phase lost stability at the expense of two separated
uniaxial nematics. 
\nlin Studies mentioned above turned the attention to the question of the 
required conditions for biaxial phase to become stable against mixture
separation. For the Palffy-Muhoray and de Bruyn model \cite{palffymix2} it
turned out that in order to stabilise the biaxial nematic phase against demixing
it was enough to remove the constraint of the arithmetic mean for the inter-type
potential strengths. It was also noted that the demixing phenomena was related
to the interaction strengths between unlike molecules \cite{sharmamix}.
This issue was partly addressed in \cite{vanakarasmix1}
where Onsager-like theory (variational cluster approximation) was used
to identify the molecular size, elongation, and interaction strength ratio
between different kinds of molecules that make biaxial nematic phase possible.
There it was suggested that the increase in rod-disc interactions
can be of importance and should be taken into account. Later, on grounds of
Onsager theory, which was generalized in order to include selective 
attractions between rods and discs, it was demonstrated that these interactions
could stabilize the \nbx\ against demixing \cite{vanakarasmix}.
Similar results were presented for a mixture of biaxial, \dtwh - symmetric
molecules in \cite{goetzmix}, where within mean field approximation it was
proven that with high enough biaxiality and interaction strength {\nbx} was
stable. Changing the relative potential strengths between the unlike molecules
brings a degree of asymmetry to the mixture. In terms of Onsager approximation
the asymmetric systems of oblate and prolate molecules can be studied by
changing the ratio between excluded volumes of rods and discs.
Such an attempt was made and several demixing scenarios were demonstrated,
and at a certain excluded volume ratio the biaxial nematic was proven
to be stable \cite{wensinkmix}.
\nlin In Onsager model for {\iso} -- {\nun} transition the orientational
entropy, which is maximal in the isotropic phase, competes with the entropy
of packing. At high enough density the latter can ''win'' and the global minimum
of the free energy can be associated with the orientationally ordered state.
In the case of binary mixtures another term is added, which corresponds to the
entropy of mixing, which is maximal when the two phases are mixed. 
The competition of these three terms may lead to stabilisation of the biaxial nematic phase. 
However, when the two components do not mix it is possible for the entropy
of mixing to ''loose'' and for the system to separate into two uniaxial nematic
phases. It was believed that it can happen mainly due
to attractive van der Waals interactions, but there is another possibility of
phase separation. It was shown that a system of two kinds of hard spheres can
undergo a demixing transition \cite{JCP.52.1670} (for the exact formulation of
the problem of miscibility for mixtures of spheres see also
\cite{JCP.41.133}). In the case of rod-disc mixtures the demixing
is also driven by steric interactions.
It was proven using Onsager theory that only at elongation of rods
equal to $15$ and discs equal to $1/15$ stable biaxial nematic phase is possible
in a narrow region in number density \cite{JPF.4.1763}. That inspired further
research on the issues of phase separation. The Monte Carlo simulations
study of hard discs of elongation $1/15$ and $1/20$ and hard rods with
elongations $15$ and $20$ \cite{campmix} confirmed the findings
from \cite{JPF.4.1763}. \nbx\ was found to be stable
in a region of molecular elongations and fractions of discs and rods
significantly limited by areas of separated, coexisting uniaxial nematic phases.
The ultimate solution to the demixing problem was proposed 
in 2003 \cite{datemix}. In that study a so--called shape amphiphiles
were proposed, built by covalently linking the rod-like molecule
to the disc-like one. The resulting compound possessed the molecules of 
high average biaxiality, but so far even that approach failed to provide
the biaxial nematic. A picture of the shape amphiphile is shown
in \figreffig{shapeamphiphilic}.
\nlin The binary mixtures of rod-like and disc-like molecules were
believed to be good candidates for the discovery of a stable biaxial
nematic phase. However, after the intensive studies it was found that
most probably the system undergoes a demixing transition into two coexisting
uniaxial phases {\nrod} and {\ndisc}, before the transition to
the biaxial nematic takes place. Even if {\nbx} can be stabilized for
certain molecular elongations and/or intermolecular potential
strengths as suggested in \cite{sharmamix,vanakarasmix,JPF.4.1763,campmix},
the region of stability against demixing has to be very narrow. In practice
this prevents the observation of the thermotropic biaxial nematic in binary
mixtures. Also, on the experimental level it is extremely hard to realize
the suggested model where the molecules of different types are more attracted
to each other than to their own kind. It is so because the oblate nematogens
are significantly different than their prolate counterparts, and separated
uniaxial phases are favoured over the biaxial mixture. 
\nlin In the mixtures the interaction between rods and discs
is the source of biaxiality. Alternatively it can be introduced by taking 
into account a system consisting of molecules of the same type with a degree 
of shape biaxiality and interacting via a biaxial potential. In the subsequent
sections we take a look at the approaches that follow this path.
\subsubsection{\dtwh - symmetric molecules}
 Straley proved that the simplest deviation from the uniaxial molecular symmetry, i.e., inclusion of \dtwh - symmetric molecules (see \figreffig{biaxialnematic}), leads to a stable biaxial nematic \cite{straley}. Furthermore, he showed
that at a certain shape anisotropy a direct transition from isotropic liquid
to \nbx\ can be observed at an isolated point. This so-called Landau point
(also known as bicritical, or self-dual) marks the place where
the three nematic phases \ndisc, \nrod, \nbx, and isotropic state ''meet''.
An example of such place on the phase diagram can be seen
in \figreffig{straleydiagram}. At that point the biaxial nematic is
most stable, in the sense that there takes place a direct second order
\iso\ -- \nbx\ transition. It significantly changes the topology
of a phase diagram, therefore it is of interest. In 1989, Mulder published
a paper on isotropic symmetry breaking bifurcations, where a condition for this
point was derived \cite{mulder}. Then it was tested for Straley model 
and hard spheroplatelet using Onsager approximation. Further studies on
spheroplatelets and systems of biaxial ellipsoids confirmed
the condition for the Landau point \cite{holystponiewierski} where
Density Functional Theory and the so-called smoothed (or weighted) density approximation \cite{PRL.61.2461,PRA.39.2742} were used.
In that approach the free energy part that includes the contribution
from intermolecular forces is taken as the free energy of isotropic fluid
calculated at ''smoothed density'', which is related to the actual density
by a formula involving the pair excluded volume. Landau points were
identified for a range of molecular dimensions, confirming the earlier findings.
For hard ellipsoids the Landau point was found to occur for molecules
with axes fulfilling square root rule: $\sigma_{x}=\sqrt{\sigma_{y}\,\sigma_{z}}$ (for definitions of $\sigma_{x},\sigma_{y},\sigma_{z}$ see \figreffig{biaxialnematic}) which was called a self-dual geometry.
Furthermore, the transition density at this geometry was found to be
significantly lower than for close-packing, which suggests that the biaxial phase
realisation is possible. These findings were confirmed by Monte Carlo
simulations of hard ellipsoids with full translational and orientational 
freedom \cite{allenhard}. The biaxial nematic was found to be most stable
near the self-dual molecular geometry, in agreement with previous
studies, also the existence of Landau point was confirmed, and
{\iso}, {\nrod}, {\ndisk}, and {\nbx} states were identified. In 1997,
the simulations of \dtwh - symmetric hard ellipsoids presented in 
\cite{allenhard} were studied in more detail \cite{JCP.106.6681}. 
Again the position of Landau point was confirmed to occur at the self-dual 
geometry. The isotropic, two uniaxial nematics, and biaxial nematic were found. 
Additionally it was found that the first order {\iso} -- {\nun} transition
with increasing molecular shape biaxiality becomes weakened; even for
small deviations from uniaxial symmetry the discontinuity associated with
first order transitions is considerably reduced, in agreement with earlier
indications \cite{AccountsOfChemicalResearch.13.290}.
\nlin The above studies concerned mainly the hard core interactions,
which are practical, yet are known only to accurately reproduce the 
quantitative properties of the phase diagram. A more complete study,
including the attractive dispersion forces, on the existence of the biaxial 
nematic was presented in the year 2000 \cite{bz2k}. Berardi and Zannoni showed
that Monte Carlo simulations of a system of biaxial molecules interacting
via a biaxial version of the Gay-Berne potential (developed 5 years earlier 
\cite{bzdevelopmentofpotential}) give a stable biaxial nematic. 
The authors studied a single molecular geometry in the rod-like region,
with a number of interaction strength parameters sets.
Surprisingly, \nbx\ was found only when the lateral attractive
forces for biaxial ellipsoids were dominant, that is, the configuration
where the molecules are facing their sides was preferred.
\nlin The research on biaxial molecules confirmed the mean field predictions of
Freiser \cite{freiser} and Straley \cite{straley}. Firstly, both
Density Functional Theory and Monte Carlo simulations predicted that purely
repulsive, steric forces between molecules of \dtwh symmetry give rise to
the stable biaxial nematic phase. Furthermore, it was proven that for certain
molecular dimensions there exists a Landau point, where system undergoes
a second order transition from isotropic phase directly to the biaxial nematic.
All the studies confirmed that this is an isolated self-dual point; so far
there are no microscopic phase diagrams with a line of
{\iso} -- {\nbx} transitions, although Landau theory can predict
such behaviour \cite{allenderlongaldg}.
\nlin The theoretical approaches predicted the existence of stable
biaxial nematic in the system of rigid molecules of \dtwh symmetry.
There exists a class of mesogens behaviour of which due to their molecular
structure cannot be accurately modelled by systems of such objects. 
Those materials, known as \sbentcore systems, exhibited a variety of new phases,
and also proved to be of importance in matter of thermotropic biaxial nematic.
In the following section we review some of the properties of these systems.
\subsubsection{Bent-core systems}
\begin{figure}[t]
 \centering
 \subfigure[Diagram predicted by mean field analysis of a lattice model of V-shaped molecules as presented in \cite{luckhursttsf}.]{\label{figure:bentmfdiagram} \includegraphics{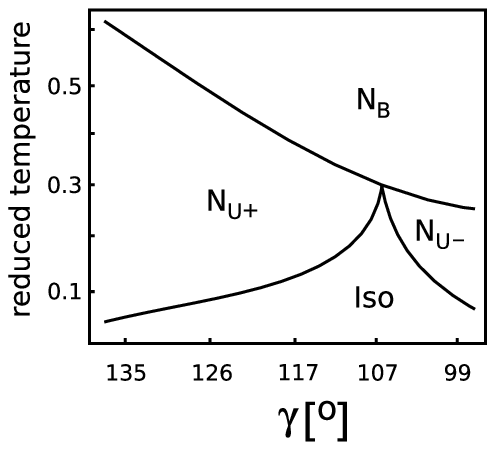}}
 \subfigure[Model \sbentcore molecule with opening angle $\gamma$. Dashed line represents $C_{2}$ symmetry axis.]{\label{figure:modelbanana} \includegraphics{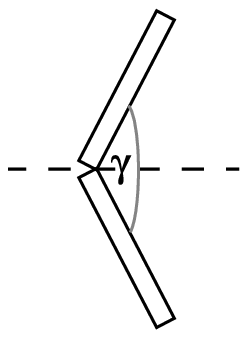}}
 \subfigure[Bifurcation diagram for \sbentcore molecules composed from two hard spherocylinders from \cite{teixeira}.]{\label{figure:bentharddiagram} \includegraphics{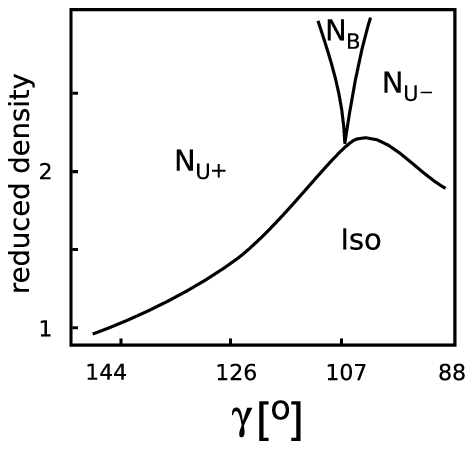}}
\caption[Mean field and bifurcation diagrams for \sbentcore molecules from \cite{luckhursttsf,teixeira}.]{\label{figure:bentdiagrams} Predictions of phase diagram for model \sbentcore molecules with bend angle $\gamma$.}
\end{figure}

 In the so-called \ital{\sbentcore}systems, the molecular structure resembles
that of a boomerang, or a banana. They behave like entities of $C_{2}$ symmetry
(see \figreffig{modelbanana}), and were found to exhibit the biaxial
nematic phase. The first theoretical study aimed at understanding of the
biaxial ordering in \sbentcore systems was based on a model of 
two, hard, rod-like spherocylinders joined at the ends at a given bend angle (also called opening angle).
Onsager theory was applied together with bifurcation analysis, and
the phase diagram in plane of reduced density and bend angle was
presented. It showed the isotropic phase, two uniaxial, and biaxial nematic.
At the angle of $\dg{107}$ the Landau point was found at the value of reduced
density accessible in experiment \cite{teixeira}. Similar results followed
from the mean field analysis of a lattice model composed of V-shaped
molecules \cite{luckhursttsf}; both diagrams are shown
in \figreffig{bentdiagrams}. The issue of a possible biaxial nematic
and phase sequence in \sbentcore systems was also widely addressed by 
simulations. In 1999, a Monte Carlo study of a system of \sbentcore molecules
composed of two hard, oblate spherocylinders was presented \cite{JCP.111.9871}. 
The elongation of the resulting banana-like molecule for the bend angle
of $\dg{180}$ was chosen to be 4:1. A slight deviation from the rod-like shape
resulted in appearance of uniaxial nematic for the opening angle between $\dg{160}$ and
$\dg{170}$ in higher densities followed by smectic-A phase. Smaller angles between arms
destabilized the nematic phase and system crystallized to a biaxial solid.
For angles of $\dg{120}$ and $\dg{110}$ molecules formed interlocking pairs and 
thus orientational ordering was not seen. Also the case of elongation 18:1
(in the rod-like limit) was studied for bend angle of $\dg{108}$;
the simulations in the biaxial nematic proved it to be mechanically stable
against isotropic liquid. No other sign of the spatially uniform biaxial phase
was found. The inclusion of Gay-Berne interacting parts as model arms of the
\sbentcore molecules also failed to produce the elusive phase.
For the model of two Gay-Berne arms each elongated in proportion of 3:1 
Monte Carlo simulations for bend angle of $\dg{140}$ showed that
isotropic phase was followed by uniaxial nematic and then by smectic-A
\cite{memmer}. For the opening angle of $\dg{170}$ the isotropic liquid lost
stability at the expense of smectic-A, and no uniaxial nematic was seen;
it appeared for lower angles of $\dg{160}$ at a single point \cite{johnston2},
where also tilted smectic-B phase was seen. No alignment of
the steric dipole axis was found \cite{johnston2}.
\nlin The issue of polar structures, which may follow from the steric
and electric dipoles present in the \sbentcore molecules, was addressed by
molecular dynamic simulations for a two Gay-Berne molecule \sbentcore model
\cite{johnston} where the electric transverse dipole was introduced in the
plane of the opening angle. The dipole--dipole interactions stabilized the
smectic-A and B phases at the expense of the uniaxial nematic. The influence
of the location of the dipole on phase sequence was also studied for a model
of the banana constructed from three Gay-Berne molecules by Monte Carlo
simulations \cite{orlandi}. When the dipoles were localized on the arms of
the molecule, the uniaxial nematic was observed. For a central dipole
along $C_{2}$ symmetry axis, the isotropic phase lost stability
to smectic structures with lowering of the temperature.
\nlin None of the above simulation studies detected any sign of the
biaxial nematic. Some proof of biaxial ordering was presented in
the atomistic simulations study \cite{atomistic}, however, the degree
of order was low. Only a simple lattice Lebwohl-Lasher model
\cite{mcasymmetric,mcflexible,mcbatesdipoles} exhibited a
stable biaxial nematic in Monte Carlo studies. It was found that in the
system of V-shaped molecules, the increase in asymmetry of the arms shifts
the Landau point to lower angles \cite{mcasymmetric}. Similar behaviour
was observed when molecular flexibility was introduced by allowing
the bend angle to vary \cite{mcflexible}. The dipole-dipole interactions
were also studied, and were found to lead to a new topology of a phase diagram,
namely the line of Landau points in the range of opening angles
at a certain dipole strength was observed \cite{mcbatesdipoles}.
\nlin Despite the predictions of the mean field and Onsager approaches
for the \sbentcore mole-\linebreak cules \cite{luckhursttsf,teixeira},
the simulations \cite{JCP.111.9871,memmer,johnston2,orlandi}
were unable (with the above exceptions \cite{atomistic,mcasymmetric,mcflexible,mcbatesdipoles}) to find a stable biaxial nematic phase. Biaxiality was
exhibited only in crystalline structures, and in the smectic phases. 
Usually the latter dominated the phase diagrams obtained in the simulations.
Those approaches also indicated that the polar long-range nematic order
due to the presence of the steric and/or electric dipoles is
less stable than non-polar nematic and smectic structures. 
\subsection{Unsuccessful experimental attempts until year 2003}
\begin{figure}[t]
 \centering
 \subfigure[Biaxial ellipsoid motif considered in \cite{chandrasekharnotbx1,chandrasekharnotbx2}.]{\label{figure:attemptedshape1} \includegraphics[scale=0.5]{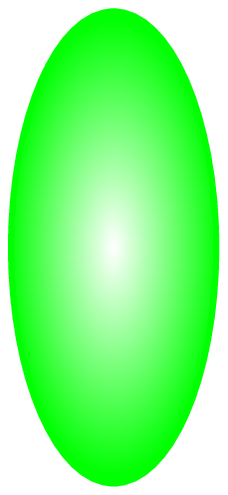}}
 \subfigure[''Dumbbell'' - like shape studied in \cite{praefckenotbx,chandrasekharnotbx3}]{\label{figure:attemptedshape2} \includegraphics[scale=0.5]{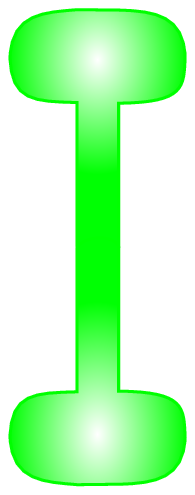}}
 \subfigure[''Pipette'' proposed in \cite{bruce}]{\label{figure:attemptedshape3} \includegraphics[scale=0.5]{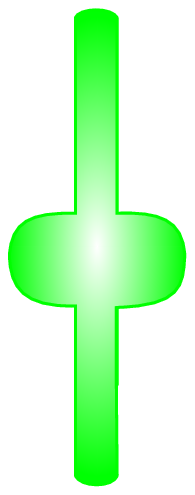}}
 \subfigure[''Lollipop'' motif from \cite{malthetenotbx}]{\label{figure:attemptedshape4} \includegraphics[scale=0.5]{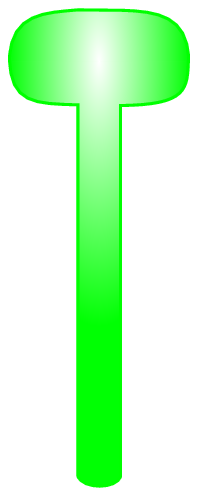}}
 \caption[Molecular motifs of unsuccessful attempts at stabilization of \nbx.]{\label{figure:shapesofunsuccessfulmeso}Schematic representations of molecular motifs tested in unsuccessful attempts of finding the biaxial phase \cite{praefckenewsonbxmol}.}
\end{figure}

 The predictions of Freiser and Straley of the early 70s and later theoretical
and simulation studies suggested that the biaxial nematic can be stabilized in a 
system of molecules possessing ''sufficiently'' biaxial shape. 
That inspired a series of experimental studies (following the discovery 
of Yu and Saupe in 1980 \cite{yulyo}) focused on the synthesis
of compounds with molecules that possessed a structure which would
bring enough biaxiality to the system to make the observation of the
thermotropic biaxial nematic phase possible. \figreffig{shapesofunsuccessfulmeso} shows the approximate molecular shapes that were advanced as good
candidates for the realisation of {\nbx} phase \cite{praefckenewsonbxmol}.
All of those were tested experimentally and apart from one, which was found
to be uniaxial \cite{bruce} (\figreffig{attemptedshape3}), were claimed to form the \stbxn\ phase \cite{chandrasekharnotbx1,chandrasekharnotbx2,malthetenotbx,praefckenotbx,chandrasekharnotbx3}, mainly on the basis of the optical 
observations of textures. They were all reinvestigated by
means of ${}^{2}\textnormal{H}$ NMR and the nematic phases were found
to be actually uniaxial or the phase exhibited a too small degree of
biaxiality to consider it to be truly biaxial.
\nlin The attempt to stabilise the thermotropic biaxial nematic based on 
the molecular design of low-mass compounds failed, but the search continued.
Finally, after over thirty years of struggle, at the beginning of year 2004
came the long-expected discovery. Next section is devoted to a
brief description of experiments showing stable biaxial nematic phase in
thermotropic materials. \putfigureflagslong{part1/chapter1/sideonattachment.eps}{Side-on attachment of mesogen to a polymer chain \cite{severingpoly}, giving rise to the biaxial nematic phase.}{sideonattachment}{b}{Side-on attachment of mesogen to a polymer chain \cite{severingpoly}.}
\subsection{The discoveries after 2003}
\begin{figure}
 \centering
 \subfigure[ ]{\label{figure:bentcoremoleculeacharya}\includegraphics[scale=0.9]{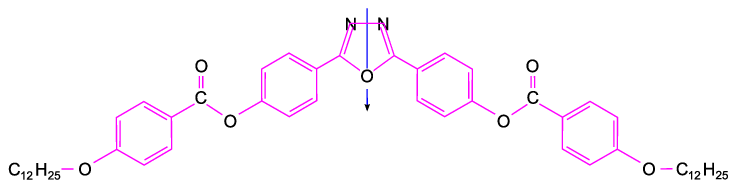}}
 \subfigure[ ]{\label{figure:bentcoremoleculelehmann}\includegraphics[scale=0.9]{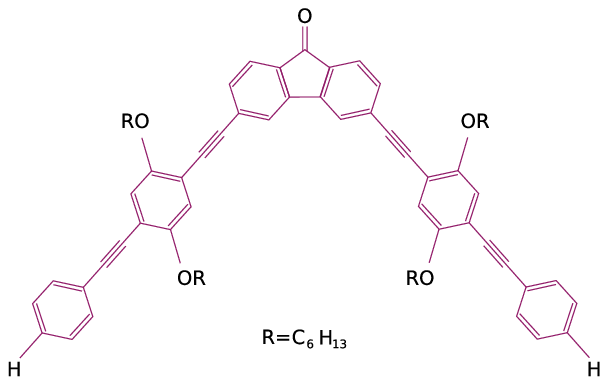}}
 \caption[Structure of two \sbentcore compounds exhibiting \nbx\ (\cite{acharya},\cite{mlehmann}).]{\label{figure:bananasstructural}Structure of two \sbentcore compounds; \subref{figure:bentcoremoleculeacharya} \sbbentcore molecules studied in \cite{acharya}, opening angle in biaxial nematic phase was estimated to be $140^{\circ}$. \subref{figure:bentcoremoleculelehmann} \sbbentcore molecule with bend angle of $90^{\circ}$ (after \cite{mlehmann}).}
\end{figure}

 The first convincing report on the thermotropic biaxial nematic phase was
presented by Severing and Saalw\"achter in March, 2004 (the paper arrived within the editors four moths earlier) \cite{severingpoly}. The authors studied a
system consisting of molecules composed of rod-like mesogen attached to
a polymer (see \figreffig{sideonattachment}). That connection of
the low-mass nematogenic compound to the side-chain polyatomic structure
resulted in the biaxial nematic phase of the attached mesogens.
The discovery was confirmed by ${}^{2}\textnormal{H}$ NMR.
\nlin A few weeks later, in April of 2004, two independent reports were
published, first one by the Kent group (arrived within editors in July of 2003)
\cite{madsen}, second in October \cite{acharya} (being a reprint and a confirmation of earlier, preliminary study \cite{acharyapanarama}), both concerned the
experimental investigation of the thermotropic biaxial nematic phase
in systems of low-mass \sbentcore molecules of similar structure,
see \figreffig{bananasstructural}. As can be seen, each molecule has a 
strong electric and steric dipole moment along $C_{2}$ axis, and two
flexible ''tails'' connected to the ends of arms. The findings of
the biaxial order in the x-ray diffraction \cite{acharya} were confirmed by the
${}^{2}\textnormal{H}$ NMR experiments \cite{madsen}, and the bend angle in {\nbx} 
phase was estimated to be about $\dg{140}$. A year later another report
of a stable biaxial nematic order for a different \sbentcore compound 
was presented \cite{prasad}. \nbx\ was observed using 
x-ray diffraction and optical studies following {\nrod} phase,
and when the temperature was lowered further, a sequence of three smectic phases
(smectic-C, X and Y) was detected. Very recently the \sbentcore mesogens
with the opening angle of $\dg{90}$ were synthesised and showed to produce both the
biaxial and uniaxial nematic phases \cite{mlehmann}.
All these results for banana-shaped molecules were surprising; however, earlier
it had been indicated that \sbentcore systems
can exhibit the elusive phase \cite{earlybentcorebx}.
\nlin The third realisation of the {\stbxn} phase came from systems of molecules
which in their structure link the properties of two \sbentcore mesogens;
the so-called tetrapodes. These compounds consist of four mesogenic molecules connected
to a rigid siloxane core through siloxane spacers. In result the system forms
the molecules resembling on the average a flat platelet (\figreffig{tetrapodecompound}). In 2004, Merkel, Kocot, and co-workers reported that materials
composed of such molecules form a stable \stbxn\ \cite{merkeltetra1}.
The findings were based on the infra-red absorbance measurements, which
allowed for order parameters to be calculated. It was shown
that the system undergoes a second order transition from the
uniaxial nematic to the biaxial nematic phase. A significant
degree of biaxial ordering was determined, the results were confirmed
by conoscopy and texture observations, and the temperature dependence
of order parameters was found to be in agreement with mean field
predictions \cite{degennesprost}. The phase behaviour of similar systems
was later studied by means of ${}^{2}\textnormal{H}$ NMR and \nbx\ was also found \cite{figuerinhastetra}.

\begin{figure}[t]
 \centering
 \subfigure[Side-on attachment of mesogen to polymer.]{\label{figure:motifpolymer} \includegraphics{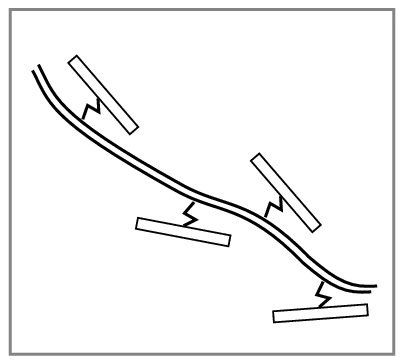}}
 \subfigure[Model banana-like molecule.]{\label{figure:motifbanana} \includegraphics{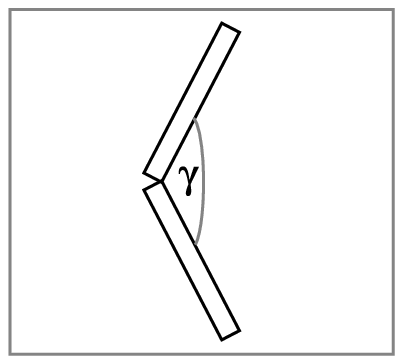}}
 \subfigure[Schematic representation of tetrapode (side view).]{\label{figure:motiftetrapode} \includegraphics{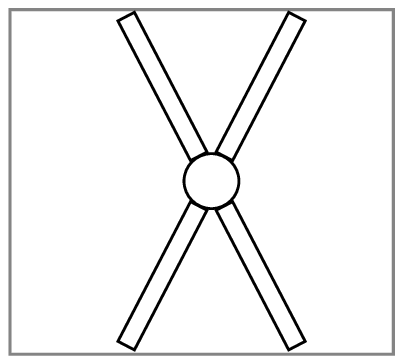}}
\caption[Schematic structure of compounds giving \nbx.]{\label{figure:motifsbiaxial} Schematic structure of molecular motifs of the compounds that gave rise to the thermotropic biaxial nematic.}
\end{figure}

 The discovery of the biaxial nematic phase in systems of banana-like 
mesogens came as a surprise mainly because these compounds of low molar
mass where known to produce mostly smectic phases, the behaviour which had been
recovered in some simulations. Neither they were a newcomers to the
world of liquid crystal physics; the first \sbentcore mesogenic molecule
was synthesised by Vorlander in 1929 \cite{vorlander} and the banana-like material were a subject of intensive studies since then. 
Those revealed a great variety of polar and chiral smectic
structures, many of them seen for the first time thanks to the \sbentcore
mesogens \cite{takezoe}. Since one of the threats for the biaxial nematic ordering
is for a system to stabilise the smectic phase before the biaxial nematic,
the materials best known for producing smectics would be the last place
to look for a stable thermotropic biaxial nematic. However, it proved to be 
otherwise, and now it is clear that the \sbentcore and tetrapode systems
are the only low-mass compounds where the {\stbxn} was found to be stable
so far. Therefore they are of interest.
\putfigureflagslong{part1/chapter1/tetrapodecompound.eps}{Structure of the tetrapode compound for which the biaxial nematic phase was found as shown in \cite{merkeltetra1}, and its schematic shape.}{tetrapodecompound}{t}{Structure of the tetrapode compound giving stable \nbx \cite{merkeltetra1}.}
\nlin As we have mentioned, the molecules in uniaxial nematic phase
are rotating relatively fast about ''the longer'' axis, which essentially
causes the constituent of the rod-like nematogen to
behave on average as uniaxial objects. The above examples of tetrapodes,
\sbentcore and polymeric systems prove that effective hindering of that 
rotation may lead to the stable \stbxn. On the other hand, more than 
thirty years of research resulted in four classes of substances 
(including the lyotropic system) giving rise to \nbx. They seem
very different from each other (see \figreffig{motifsbiaxial}) and from earlier
proposals (see \figreffig{shapesofunsuccessfulmeso}).
Thus it is clear that the biaxial nematic order is not easy to induce. However,
there probably exist certain factors that make the elusive phase more likely to
occur. It is not straightforward to deduce them just by looking at
\figreffig{motifsbiaxial} but we can investigate the models
of biaxial nematic phase and try to localize some of those factors.
This is the main goal of the present thesis, as will be described in the
following section.
\section{Purpose of the thesis}
\indent Present work is devoted to the investigation of some of the factors 
that influence the stability of the biaxial nematic phase. As we have seen,
there are many unresolved issues concerning stabilisation of this
liquid crystal state. In the case of banana-like molecules the theory dictates
that {\nbx} is most stable, in comparison to \nun\ and \iso, for the bend angle
near $\dg{107}$, which differs
significantly from the experimentally estimated value of $\dg{140}$.
Also the influence of electrical dipoles, present in these systems, on
biaxial nematic ordering is unclear. For \dtwh -- symmetric molecules
the hard interactions provided the stable \nbx, however,
for the more complex but also more realistic biaxial Gay-Berne model
only one Monte Carlo study was presented, and biaxial nematic was stable 
only for one set of potential parameters. We are going to address those
and other issues of stabilisation of \nbx\ phase in a more systematic way,
namely by taking into account models of {\sbentcore} molecules and studying
the biaxial Gay-Berne interaction. We aim at finding approximate phase diagram
and identifying molecular factors responsible for appearance of the
biaxial nematic order.  Much effort is devoted to the investigation of Landau points\footnote{In present work the Landau points coincide with self-dual points obtained from duality transformation of $L=2$ model, so the both names refer to the same physical entity.}, since there the action of microscopic model parameters on
macroscopic biaxial nematic ordering is strongest, and its stability is
enhanced. These points are indicated as the best places to look
for a stable \nbx. 
\nlin Up to date there are only few theoretical treatments that allow to
predict a phase diagram of liquid mesophases from first
principles. One of them is the Local Density Functional Theory (DFT),
which is known to give results of reasonable agreement with simulations and
experiment for a variety of liquid crystalline models.
Yet to our knowledge it has not been employed in a systematic way to the
systems with biaxial nematic phase. It is of interest to see how that
theory works in this case. In majority of cases studied, we have decided to use this approach in low-density 
limit (also called second order virial approximation), and analyse the
equation for equilibrium by means of bifurcations to obtain behaviour
of a system close to transition point. Both methods
will be described in the following chapter. It is a well known fact that
this approach overestimates transition temperatures,
therefore we will not aim at the exact values of critical parameters,
but rather try to identify possible topologies of the phase
diagrams\footnote{Most of the diagrams will be presented at bifurcation point, therefore will be approximated, possible phase diagrams.}.
\nlin Firstly, we consider an effective model with three
coupling constants based on Straley's \cite{straley} generalization
of the orientational part of the Maier-Saupe potential \cite{maiersaupe},
and present the exact bifurcation diagram in mean field, 
including spatially uniform phases. It serves as an introduction of some
general concepts concerning studies on nematic order, in particular the case
when the coupling constants are interpreted as coefficients of expansion of
the pair direct correlation function used later. The derived diagram
shows Landau (self-dual) and tricritical points. Next, the mentioned earlier
biaxial potential of Berardi, Fava, and Zannoni \cite{bzdevelopmentofpotential}
is studied using the low-density approximation. This model is the only
soft interaction involving translational degrees of freedom for which the
biaxial nematic phase was discovered in simulations \cite{bz2k}. Therefore it
presents a possibility of comparing our results with those obtained from
Monte Carlo and then investigating the potential parameter space further.
We begin with calculation of the bifurcation diagram for parameters sets
studied in the simulations, and also compare the method of
minimisation of the Helmholtz free energy with bifurcation for
uniaxial -- biaxial nematic transition.
Then we address the issue of the Landau point, and locate it by varying
the shape and energy biaxiality. Finally, we study the \sbentcore models.
In that case we pursue the shape related effects as well as the influence
of dipole--dipole interactions, in search for the decisive factors that
stabilise the biaxial ordering. We consider the models of Gay-Berne interacting
parts by choosing the arms of \sbentcore molecules to be soft, prolate ellipsoids.
We study the cases of two and three uniaxial arms with elongation 5:1,
and also take into account two biaxial ellipsoids as the model arms. Then, we turn our attention to
the elastic issues and study the bulk elastic free energy density in absence
of chiral order by calculating the set of elastic constants in
the biaxial nematic phase for the biaxial Gay-Berne interaction.
We show the temperature dependence of the constants in prolate molecular regime
and at Landau point. Finally, we present the bifurcation equations
with the inclusion of the transitions involving orthogonal smectic phases
of uniaxial and biaxial symmetry. We also briefly address the issue
of apparent importance of lateral interactions in the stabilization
of \nbx\ \cite{bz2k}, by presenting temperature dependence of order parameters
including biaxial and uniaxial smectic-A phases for two sets of
potential parameters, calculated for biaxial Gay-Berne in $L=2$ model.
We show how the temperature range of biaxial nematic, limited by
biaxial smectic-A, gets wider for the case of strongest lateral forces.
\nlin This work is aimed at the investigation of the way transition point
to the biaxial nematic phase is affected by shape of molecules and parameters
of the intermolecular forces strengths\footnote{In the models used in the present work, both the molecular shape dimensions and forces strength parameters are all incorporated into pair interaction potential.}.
These microscopic quantities are related to the macroscopic properties
of the system by \sldft, and the transition temperatures and densities
are approximated by means of bifurcation analysis of the \sceq\ for
equilibrium \sopdfns. In that way we obtain a possible phase diagram
parametrized by the constants entering pair potential, which can be used as a
guidelines of the role different microscopic quantities play in
the stabilisation of biaxial nematic.
\nlin The thesis is organized as follows. Firstly we present the general
introduction to the \sldft, which is particularly useful in the studies
of phase transitions. Next we describe a general bifurcation scheme for
finding the point where a new, lower symmetric solution branches of
the reference one of higher symmetry, together with methods of finding
the Landau and tricritical points. In subsequent chapters the three models
describing the biaxial ordering are studied. Firstly, the model of
three coupling constants in a general expansion of
\dtwh -- symmetric pair interaction is investigated in mean field approximation.
Then, we take into account the biaxial Gay--Berne potential. We start by
comparing the DFT results with Monte Carlo simulations and with minimisation of
free energy, and continue to investigate the interaction parameter space
further; the roles of shape and energy biaxiality are studied while
acting separately and simultaneously, and the behaviour of Landau point
is considered. Next, we take into account the \sbentcore molecules models,
and pursuit the issues of shape and dipole-dipole interaction. In concluding
chapters we briefly address the issues of elastic constants for
biaxial Gay-Berne model, and comment on uniaxial and biaxial smectic-A phases,
including the calculation of the order parameters in $L=2$ model, and
the derivation of bifurcation equations for additional transitions.

 \chapter{Local Density Functional Theory}
 \label{chapter:DFT}

{ \chapabsbegin
 The \sldft\ is a tool that presents a way for determining the 
equilibrium properties of a system from first principles. Once applied 
to liquid crystal models, it allows to determine the phase
diagram. In the present work, using this method, we obtain
the densities and temperatures for phase transitions in
the systems with stable isotropic, uniaxial and biaxial nematic
phases. Current chapter is devoted to a systematic approach
to the \sldft. Also a particular form of the theory used
in this thesis is introduced. \chapabsend }

\section{General formulation for one--component systems}
In this chapter we introduce the theory that is used as a tool in the analysis
of models with the biaxial nematic phase. We consider the liquid crystal
mesophases by postulating pair intermolecular potential
and seek the phase diagram in density-temperature plane. In order to do this,
we need to establish a connection between the microscopic
parameters associated with molecular dimensions and interaction strengths,
present in a model, and macroscopic system properties.
We are interested in the behaviour of the system at equilibrium
in the vicinity of phase transition point. Local Density Functional Theory (DFT)
provides a way of systematic introduction of intermolecular forces.
It can be used to determine the conditions for equilibrium, and to establish
the connection between microscopic and macroscopic parameters. Furthermore,
it allows to derive exact formulas for the point where a new structure emerges
from a given equilibrium state.
\nlin In this section we present the main concepts of DFT 
and derivation of the most important formulas, following
\cite{hk1964,mermin,evans}\footnote{It is probably worth noting that a certain level of generality is maintained here. We believe it will stress out the flexibility and tractability of the approach.}. The results were published by us
in \cite{longa}. Although the approach can be easily modified to include
molecular flexibility or to model polyatomic systems, presently we restrict
ourselves to the case of rigid molecules, orientation of which in euclidean
space can be described by supplying the right-handed, orthonormal tripod of vectors associated
with a given molecule, and the position by the vector linking to the molecular
centre of mass.
\nlin Let's consider a system of rigid, identical, biaxial molecules in
grand canonical ensemble, where volume V, temperature $T$, and
chemical potential $\mu$ are fixed. State of each molecule is described with
respect to the global laboratory frame by the position $\vec{\mathbf{r}}_{i}$
of centre of mass, and orientation $\orientation_{i}$ of reference frame
associated with the molecule with respect to the laboratory system,
and parametrized, e.g., by three Euler angles $\alpha_{i},\beta_{i},\gamma_{i}$ \cite{altmann,rose}. In the following we
will use the notation $\coord{i}\equiv(\vec{\mathbf{r}}_{i},\orientation_{i})$
and $\int d\coord{i}\equiv \int d\vec{\mathbf{r}}_{i} \int^{2\pi}_{0} d\alpha_{i} \int^{1}_{-1} d\cos(\beta_{i}) \int^{2\pi}_{0} d\gamma_{i}$ to represent
the degrees of freedom of the molecules, and available positional-orientational
phase space. The system considered can be described by the Hamiltonian of
the following form:
\begin{equation}
 \label{eq:systemhamiltonian}
 H_{N}=T_{N}+U_{N}+V_{N} \, ,
\end{equation}
where $T_{N}$, $U_{N}$, $V_{N}$ are the kinetic energy, the potential energy,
and the external field interaction, respectively. They read:
\begin{align}
\begin{split}
 \label{eq:energiesdefs}
 T_{N} = & \sum_{i=1}^{N} \frac{\vec{\mathbf{p}}^{2}_{i}}{2\,m}+\frac{1}{2} \sum_{i=1}^{N} \lnbra I_{1}\,\omega^{2}_{1,i}+I_{2}\,\omega^{2}_{2,i}+I_{3}\,\omega^{2}_{3,i} \rnbra \, , \\
 U_{N} = & \, U(\coord{1}, \dots , \coord{N}) \, , \\
 V_{N} = & \sum_{i=1}^{N} V_{ext}(\coord{i}) \, ,
\end{split}
\end{align}
where $N$ stands for the number of molecules, $\vec{\mathbf{p}}_{i}$ for 
the momenta, $m$ for mass, $\omega_{1,i},\omega_{2,i},\omega_{3,i}$ denote
the angular velocities, $I_{1},I_{2},I_{3}$ are the principal moments of
inertia, $V_{ext}(\coord{i})$ is the external potential,
and $U(\coord{1}, \dots , \coord{N})$ is the potential energy.
As we can see we are not making any assumptions
concerning $U$, especially we do not yet introduce two-body terms.
The class of systems considered includes all systems where the kinetic energy
can be integrated out. Therefore, we will neglect the momentum variables,
since the corresponding contribution does not affect the phase diagram.
\nlin Each microstate is realized with a certain probability; it is described by
the distribution\linebreak function $f(\coord{1}, \dots , \coord{N})$, which
is the probability density of a configuration where $N$ molecules are in
a given state, taken between $\lsbra \coord{1}, \dots , \coord{N} \rsbra$ 
and $\lsbra \coord{1}+d\coord{1}, \dots , \coord{N}+d\coord{N} \rsbra$,
therefore called N-particle distribution function. Following Evans \cite{evans}
we consider a functional $\grandpotentialsymbol$ of $f$ (for clarity we will drop the arguments), defined as
\begin{equation}
 \label{eq:grandpotential}
 \grandpotentialf{f}=\cltrace\lsbra\lnbra H_{N}-\mu N+\beta^{-1}\ln f \rnbra f \rsbra \, \textrm{,}
\end{equation}
where $\beta^{-1}=k_{B}T$, with Boltzmann constant $k_{B}$.
We used the following notation for a ''classical trace'':
\begin{equation}
 \cltrace{(A)}\equiv \sum_{N=0}^{\infty}\frac{1}{h^{3N}N!}\int d\coord{1} \dots d\coord{N}A(\coord{1}, \dots, \coord{N}) \, , \nonumber
\end{equation}
where $h$ is the Planck constant. Clearly, the functional $\grandpotentialf{f}$ for an equilibrium distribution $f_{eq}$ reduces to the grand potential
\begin{equation}
 \label{eq:grandomega}
 \grandpotentialsymbol \equiv \grandpotentialf{f_{eq}}=\beta^{-1}\ln\Xi \, ,
\end{equation}
where the grand partition function 
\begin{equation}
 \Xi = \cltrace \lcbra \exp\lsbra -\beta \lnbra H_{N}-\mu N \rnbra \rsbra \rcbra \, ,
\end{equation}
and
\begin{equation}
 f_{eq} = \Xi^{-1} \exp\lsbra -\beta \lnbra H_{N}-\mu N \rnbra \rsbra \, .
\end{equation}
As usual, $f_{eq}$ is normalized with respect to the classical trace:
$\cltrace(f_{eq})=1$, and the ensemble averages are calculated with the
help of $f_{eq}$, namely $\average{A}\equiv\cltrace(f_{eq}A)$.
\nlin Let's consider an arbitrary distribution $f_{1}$ of unit trace
($\cltrace(f_{1})=1$). The functional $\grandpotentialf{f_{1}}$ can be written as:
\begin{equation}
 \label{eq:grandfurther}
\begin{split}
 \grandpotentialf{f_{1}} = & \, \cltrace\lcbra \lsbra -\beta^{-1}\lnbra \ln{\Xi}+\ln{f_{eq}} \rnbra +\beta^{-1}\ln{f_{1}} \rsbra f_{1} \rcbra \\
 = & \, \grandpotentialf{f_{eq}} +\beta\cltrace\lnbra f_{1}\ln{f_{1}}-f_{1}\ln{f_{eq}} \rnbra \\
 = & \, \grandpotentialf{f_{eq}} +\beta\cltrace\lnbra f_{1}\ln{ \frac{f_{1}}{f_{eq}} } \rnbra \, .
\end{split}
\end{equation}
Since $\ln(f_{eq}/f_{1}) \leq f_{eq}/f_{1} - 1$, where 
equality holds for $f_{1}=f_{eq}$, we can see that the last term is
positive-definite. It represents the well-known relative entropy $S_{rel}$
of Kullback and Leibler, introduced in 1951 \cite{kullbackleibler}.
From definition $S_{rel}[f_{1}/f_{2}]=-k_{B} \cltrace \lsbra f_{1} \ln\lnbra  f_{1}/f_{2} \rnbra \rsbra$, and always $S_{rel}[f_{1}/f_{2}] \leq 0$.
So we can conclude that for any normalized $f \neq f_{eq}$ we have
\begin{equation}
 \label{eq:variational1}
 \grandpotentialf{f} > \grandpotentialf{f_{eq}} \, \textrm{.}
\end{equation}
It means that there exists a variational principle, which states
that the global minimum of $\grandpotentialf{f}$ is attained when $f$
coincides with equilibrium distribution $f_{eq}$.
Another key quantity is a functional $\intrinsicfef{f}$, which for $f=f_{eq}$
plays a role of the Helmholtz free energy of the system in the absence
of external field. It is defined as follows:
\begin{equation}
 \label{eq:intrinsicfe}
 \intrinsicfef{f}=\cltrace \lsbra \lnbra T_{N}+U_{N}+\beta^{-1}\ln{f} \rnbra f \rsbra \, ,
\end{equation}
and is traditionally called the \ital{intrinsic} Helmholtz free energy
functional, while the total Helmholtz free energy reads $F=\cltrace \lnbra f_{eq}V_{N} \rnbra + \intrinsicfef{f_{eq}}$.
\nlin The knowledge of the N-particle distribution gives a complete
information about the behaviour of a system. Fortunately, in the description of
equilibrium properties we do not need to know all details of $f_{eq}$.
Actually, it turns out that the equilibrium behaviour can be determined
by noticing that both the grand potential $\grandpotentialsymbol$
and the Helmholtz free energy $\intrinsicfesymbol$ are functionals 
of only one-particle distribution $\densityf{\coordsymbol}$, since $f$
is also a functional of $\densityf{\coordsymbol}$ \cite{evans}. Even more,
it can be proven that both $\grandpotentialsymbol$ and $\intrinsicfesymbol$
are unique, although unknown explicitly, functionals
of $\densityf{\coordsymbol}$ \cite{evans}. Using this general theorem
we can proceed to construct approximation schemes for $\grandpotentialsymbol$
and $\intrinsicfesymbol$, we will now describe the most common ones.
\nlin We start by introducing the \sopdf $\densityf{\coordsymbol}$, which 
in equilibrium is denoted by $\densityff{\coordsymbol}{eq}$ and defined with
the help of the microscopic density operator
\begin{equation}
 \label{eq:microdensity}
 \microdensityf{\coordsymbol}=\sum_{i=1}^{N} \delta \lnbra \coordsymbol-\coord{i} \rnbra \, \textrm{,}
\end{equation}
as an equilibrium average of the above, over the grand canonical ensemble:
\begin{equation}
 \label{eq:eqdensityf}
 \densityff{\coordsymbol}{eq}=\ensembleavg{\microdensityf{\coordsymbol}}\equiv\cltrace{\lsbra f_{eq}\microdensityf{\coordsymbol} \rsbra} \, .
\end{equation}
$\densityff{\coordsymbol}{eq}$ is normalized to the average number of
particles $\average{N}$,
\begin{equation}
 \label{eq:normalizationdensityf}
 \int d\coordsymbol\, \densityff{\coordsymbol}{eq} = \average{N} \, , \,\,\,\, \densityff{\coordsymbol}{eq}=\average{N}f_{eq}(\coordsymbol) \, .
\end{equation}
Now we rewrite \refeq{eq:grandpotential} with the help
of \refeq{eq:intrinsicfe}, and obtain \cite{mermin,evans}:
\begin{equation}
 \label{eq:grandpotentialrho}
 \grandpotentialf{\densitysymbol}=\intrinsicfef{\densitysymbol}-\mu\int d\coordsymbol\, \densityf{\coordsymbol}+\int d\coordsymbol\, \densityf{\coordsymbol} V_{ext}\lnbra \coordsymbol \rnbra \, \textrm{,}
\end{equation}
which for equilibrium \sopdf reduces to the grand canonical potential:
\begin{equation}
 \label{eq:omegatogrand}
\begin{split}
 \grandpotentialf{\densitysymbol_{eq}} = & \intrinsicfef{\densitysymbol_{eq}}+\int d\coordsymbol\, \cltrace{\lsbra f_{eq}\microdensityf{\coordsymbol}\rsbra}V_{ext}\lnbra \coordsymbol \rnbra -\mu\int d\coordsymbol\, \cltrace{\lsbra f_{eq}\microdensityf{\coordsymbol}\rsbra} \\
 = & \cltrace\lsbra \int d\coordsymbol \, \microdensityf{\coordsymbol}V_{ext}+T_{N}+U_{N}+\beta^{-1}\ln{f_{eq}}-\mu\int d\coordsymbol \, \microdensityf{\coordsymbol} \rsbra f_{eq} \\
 = & \cltrace\lnbra H_{N}-\mu N+\beta^{-1}\ln{f_{eq}} \rnbra f_{eq}\equiv\grandpotentialsymbol \, .
\end{split}
\end{equation}
$\intrinsicfef{\densitysymbol}$ and $\grandpotentialf{\densitysymbol}$ 
are related by the Legendre transformation, since
$\grandpotentialf{\densitysymbol}$ can also be treated as a functional of $\psi=\mu-V_{ext}$, 
\begin{equation}
 \label{eq:grandinpsi}
 \intrinsicfef{\densitysymbol}=\grandpotentialf{\psi}+\int d\coordsymbol \, \psi \lnbra \coordsymbol \rnbra \densityf{\coordsymbol} \, \textrm{.}
\end{equation}
\indent Since in the first place we are interested in equilibrium properties
of a system, we need to determine the condition which can be used
to find $\densityff{\coordsymbol}{eq}$. Keeping in mind that $\grandpotentialf{\densitysymbol}$ is a unique functional of \sopdfns, the
variational principle \refeq{eq:variational1} leads to the
following (necessary) condition:
\begin{equation}
 \label{eq:variational}
 \left. \fderivatede{\grandpotentialf{\densitysymbol}}{\coordsymbol} \right|_{\densitysymbol=\densitysymbol_{eq}}=0 \eqcolon
\end{equation}
which according to \re{grandpotentialrho} can also be written as
\begin{equation}
 \label{eq:variationalforf}
 \fderivatede{\,}{\coord{1}}\lcbra \intrinsicfef{\densitysymbol}+\int d\coord{2} \lsbra V_{ext}\lnbra \coord{2} \rnbra -\mu \rsbra \densityf{\coord{2}} \rcbra=0 \, .
\end{equation}
\eqrefeq{variational} is a consequence of the Hohenberg-Kohn-Mermin
theorem \cite{hk1964,mermin} stating that the minimum of the
functional $\grandpotentialf{\densitysymbol}$ is attained when 
$\densitysymbol$ coincides with equilibrium \sopdfns. In other words,
$\densitysymbol_{eq}$ can be found by minimizing \refeq{eq:grandpotentialrho},
for which the equation \refeq{eq:variational} is the necessary condition.
As we can see, the minimum of $\grandpotentialf{\densitysymbol}$
can be found as a minimum of $\intrinsicfef{\densitysymbol}$ with the condition
of normalization of $\densityf{\coordsymbol}$, where the chemical potential
plays the role of the Lagrange multiplier. So we turn
our attention to the functional representing the intrinsic part of
the Helmholtz free energy. We can immediately divide it into two parts:
$\idealfef{\densitysymbol}$ describing the ideal gas, and the \ital{excess}
part, $\excessivefef{\densitysymbol}$ that includes the contribution 
from intermolecular forces;
\begin{equation}
 \label{eq:idealandexcesive}
 \intrinsicfef{\densitysymbol}=\idealfef{\densitysymbol}+\excessivefef{\densitysymbol} \, \textrm{.}
\end{equation}
The ideal part is given by
\begin{equation}
 \label{eq:idealpartdef}
 \idealfef{\densitysymbol}=\beta^{-1}\int d\coordsymbol \densityf{\coordsymbol} \lcbra \ln{\lsbra \Lambda \densityf{\coordsymbol} \rsbra}-1\rcbra \eqcolon
\end{equation}
where $\Lambda=\sqrt{h^{12}\beta^{6}/(2\,\pi)^{6}m^{3}I_{1}I_{2}I_{3}}$
is the constant resulting from the integration over momenta. Using the
above we can employ the variational principle expressed
by \eqrefeq{variationalforf} to finally get
\begin{equation}
 \label{eq:selfconsistent}
 \ln{\lcbra \Lambda \densityf{\coordsymbol} \rcbra}=-\beta \fderivatede{\excessivefef{\densitysymbol}}{\coordsymbol}+\beta \lcbra \mu-V_{ext}\lnbra \coordsymbol \rnbra \rcbra \, \textrm{.}
\end{equation}
Since the yet unknown $\excessivefef{\densitysymbol}$
depends on the \sopdfns, the above formula becomes the non-linear, \sceq\ 
for $\densityf{\coordsymbol}$. It reads
\begin{equation}
 \label{eq:selfconsistent2}
 \densityf{\coordsymbol}=\Lambda^{-1}e^{\beta\mu}\exp\lcbra -\beta \lsbra \fderivatede{\excessivefef{\densitysymbol}}{\coordsymbol} + V_{ext}\lnbra \coordsymbol \rnbra \rsbra \rcbra \, .
\end{equation}
For given $\delta\excessivefef{\densitysymbol}/\delta\densityf{\coordsymbol}$,
the above equation can be solved for $\densityf{\coordsymbol}$, e.g., in an
iterative manner. The solutions coincide with minima, maxima, or saddle points
of the grand canonical potential functional; the equilibrium distribution is
the one that minimises $\grandpotentialf{\densitysymbol}$. For $V_{ext}=0$ and
for $U_{N}$ invariant under a global translation and rotation,
\eqrefeq{selfconsistent2} always possesses a trivial isotropic
solution $\densityf{\coordsymbol}=const$.
The fact that $\delta\excessivefef{\densitysymbol}/\delta\densityf{\coordsymbol}$
depends on $\densitysymbol$ in a complex way makes
the existence of many non--trivial, non--isotropic solutions of 
\eqrefeq{selfconsistent2} possible.
\nlin In order to seek the non-trivial solutions of the equation \refeq{eq:selfconsistent2},
we need to turn our attention to the excess part of the Helmholtz 
free energy, $\excessivefef{\densitysymbol}$. It is of course impossible
to carry out the calculations without some approximations. A standard
method to approximate $\excessivefef{\densitysymbol}$ is to perform a functional Taylor expansion about some reference state\footnote{Here we do not yet make any specific assumptions as to the form of the reference state or its relation to the other states that may became stationary for different temperature and/or density.} described by $\densityff{\coordsymbol}{ref}$. The expansion reads
\begin{equation}
 \label{eq:taylorexp}
\begin{split}
 \excessivefef{\densitysymbol} = \, & \excessivefef{\densitysymbol_{ref}}+\int 
 d\coord{1}
 \left. \fderivatede{\excessivefef{\densitysymbol}}{\coord{1}} \right|_{\densitysymbol=\densitysymbol_{ref}} \hspace{-10mm} \lcbra \densityf{\coord{1}}-\densityff{\coord{1}}{ref} \rcbra + \\
 & \, \frac{1}{2!}\int d\coord{1} d\coord{2} \left. \fdederivatede{\excessivefef{\densitysymbol}}{\coord{1}}{\coord{2}} \right|_{\densitysymbol=\densitysymbol_{ref}} \hspace{-10mm} \lcbra \densityf{\coord{1}}-\densityff{\coord{1}}{ref} \rcbra \lcbra \densityf{\coord{2}}-\densityff{\coord{2}}{ref} \rcbra + \dots \eqdot 
\end{split}
\end{equation}
The above expansion should be treated with caution, not only because its 
existence can be questioned, but also because it requires a 
detailed knowledge of the reference state, determination of which in general 
case can be a great challenge in itself, even with some approximation for
the derivatives of the Helmholtz free energy. Here we restrict ourselves to 
$\excessivefesymbol$ which is analytical in $\densityf{\coordsymbol}$,
and so the Taylor expansion \refeq{eq:taylorexp}
is assumed to exists and converge. This approach apparently works quite well in
determining the phase diagrams, as can be verified by making a comparison
with computer simulations (see, e.g., \cite{jozefowicz}).
\nlin The idea of treating the free energy as an expansion
in $\densityf{\coordsymbol}$ is similar in essence to the Landau--de Gennes
approach \cite{landaudegennes}. Actually \eqrefeq{taylorexp} gives the
thermodynamical foundation for this theory. Also from the above expansion one
can derive exact formulas for, e.g., the Oseen-Zocher-Frank elastic constants,
assuming $\densityff{\coordsymbol}{ref}$ describes an undeformed state,
as well as get an accurate description of the freezing transition (Ramakrishnan-Yussouff theory \cite{ramakrishnanyussouff}). 
\nlin The expansion \refeq{eq:taylorexp} is traditionally rewritten with the
help of a set of the so-called direct correlation functions $c_{n}$,
for which $\excessivefesymbol$ is by definition a generating functional.
More specifically, they are introduced as functional derivatives 
of $\excessivefef{\densitysymbol}$ with respect to $\densityf{\coordsymbol}$:
\begin{equation}
 \label{eq:dcfs}
\begin{split}
 \firstdcf{\coord{1}}{\densitysymbol_{ref}} \equiv & \, -\beta\linv\fderivatede{\excessivefef{\densitysymbol}}{\coord{1}}\right|_{\densitysymbol=\densitysymbol_{ref}} \eqcolon \\
 \seconddcf{\coord{1}}{\coord{2}}{\densitysymbol_{ref}} \equiv & \, -\beta\linv\fdederivatede{\excessivefef{\densitysymbol}}{\coord{1}}{\coord{2}}\right|_{\densitysymbol=\densitysymbol_{ref}} \eqcolon \\
 & \vdots \\
 c_{n}\lnbra \coord{1},\,\dots\,,\coord{n},\lsbra\densitysymbol\rsbra \rnbra = \linv\fderivate{\densitysymbol}{\coord{n}}{c_{n-1}}\right|_{\densitysymbol=\densitysymbol_{ref}} \equiv & \, -\beta\linv\frac{\delta^{n}\excessivefef{\densitysymbol}}{\delta \densitysymbol \lnbra \coord{1} \rnbra \dots \delta \densitysymbol \lnbra \coord{n} \rnbra}\right|_{\densitysymbol=\densitysymbol_{ref}} \, .
\end{split}
\end{equation}
Interestingly, each $c_{n}$ is related through integral equations to the set of ordinary
distribution functions $\lcbra \densityf{\coord{1}},\densitysymbol_{2}(\coord{1},\coord{2}),\dots,\densitysymbol_{n}(\coord{1},\dots,\coord{n}) \rcbra$.
For example, the first function of the set \refeq{eq:dcfs} can be connected to $\densityf{\coord{1}}$ 
as (see Appendix \ref{ap:firstdcf}):
\begin{equation}
 \label{eq:firstdcf}
 c_{1}\lnbra \coord{1},\lsbra\densitysymbol\rsbra \rnbra = \lcbra\mu-V_{ext}\lnbra \coord{1} \rnbra \rcbra-\beta^{-1}\ln{\lcbra\Lambda\densityf{\coord{1}} \rcbra} \, ,
\end{equation}
and $c_{2}$ can be related to the pair correlation
function $h_{2}(\coord{1},\coord{2})=\densitysymbol_{2}(\coord{1},\coord{2})/\densityf{\coord{1}}\densityf{\coord{2}}-1$, by the Ornstein--Zernike relation:
\begin{equation}
 \label{eq:orsteinzernike}
 h_{2}(\coord{1},\coord{2})=\seconddcf{\coord{1}}{\coord{2}}{\densitysymbol}+\int d\coord{3} \, \seconddcf{\coord{1}}{\coord{3}}{\densitysymbol} \densityf{\coord{3}} h_{2}(\coord{3},\coord{2}) \, .
\end{equation}
Now \refeq{eq:taylorexp} can be written as\footnote{We have assumed the equality of the chemical potentials of the reference and actual states.}:
\begin{equation}
 \label{eq:taylorexpwithdcfs}
\begin{split}
 \excessivefef{\densitysymbol} = & \, \excessivefef{\densitysymbol_{ref}} - \sum^{\infty}_{n=1} \int d\coord{1}\dots d\coord{n} c_{n}(\coord{1},\dots,\coord{n}) \delta \densityf{\coord{1}}\dots\delta \densityf{\coord{n}}=\\
 & \, \excessivefef{\densitysymbol_{ref}}
 +\beta^{-1}\int d\coord{1} \, \delta \densityf{\coord{1}} \ln\densityff{\coord{1}}{ref} - \\
 & \, \frac{1}{2}\int d\coord{1}d\coord{2} \, \delta \densityf{\coord{1}} \delta \densityf{\coord{2}} \seconddcf{\coord{1}}{\coord{2}}{\densitysymbol_{ref}} - \dots \, ,
\end{split}
\end{equation}
where $\delta \densityf{\coord{i}} \equiv \densityf{\coord{i}}-\densityff{\coord{i}}{ref}$. With the definitions \re{dcfs} and for $V_{ext}=0$, the {\sceq} \refeq{eq:selfconsistent} can now be rewritten in a compact form as
\begin{equation}
 \label{eq:selfconsistentequniform}
 \densityf{\coordsymbol}=Z^{-1}_{\densitysymbol}\exp\lcbra \, \firstdcf{\coordsymbol}{\densitysymbol} \rcbra \, ,
\end{equation}
where $Z_{\densitysymbol}=\int d\coordsymbol \exp\lcbra\firstdcf{\coordsymbol}{\densitysymbol}\rcbra/\average{N}$ assures the normalization \refeq{eq:normalizationdensityf}. We can conclude that the equilibrium \sopdf is determined by the effective
one-particle ''potential'': $-\beta^{-1}\firstdcf{\coordsymbol}{\densitysymbol}$, in a way specified by \eqrefeq{selfconsistent2}. In what follows we
set $V_{ext}=0$, for this work is devoted to the liquids in the absence
of external fields.

\section{Bifurcation analysis. Exact results.}
 \label{section:bif}
\indent Information about the behaviour of a system close to a point of
phase transition is stored in  the \sceq\ \refeq{eq:selfconsistentequniform}.
Due to the structure of this equation, we can locate the points where a
new solution branches off the reference one. This, by construction, takes place
at a critical point or close to a real phase transition if the transition is
of first order. We can also find sectors in parameter space where character of 
phase transition changes from continuous to first order -- the so-called
tricritical points \cite{longatri}. In current section, we present a derivation
of bifurcation equations, which are used to obtain the phase behaviour of
biaxial nematics in subsequent part of this study. We also show the condition
for bifurcation point to be a tricritical point.
\nlin As mentioned before, the solutions of the {\sceq} \refeq{eq:selfconsistentequniform} describe the local extrema and saddle points of the
intrinsic Helmholtz free energy $\intrinsicfef{\densitysymbol}$. The
stable state is associated with the solution in the class of local minima
corresponding to the global minimum of $\intrinsicfef{\densitysymbol}$.
Remaining local minima are usually associated with metastable states, while
the maxima are not realized as macroscopic states of a system. At
sufficiently low density (or high temperature) $\intrinsicfef{\densitysymbol}$
has only one minimum corresponding to the isotropic distribution $\densityf{\coordsymbol}=const$, which describes the unordered, isotropic phase. We expect
that when the density is increased (or the temperature lowered) the system
undergoes a phase transition, which means that a new solution 
of \eqrefeq{selfconsistentequniform} and a new local minimum of
$\intrinsicfef{\densitysymbol}$ emerges. In cases studied here
the symmetry group of high temperature phase contains all elements of
the low temperature state. Therefore we say that the former has higher
and the latter lower symmetry, and the relation between them is
of group-subgroup type. The point where the higher symmetric solution loses
the property of being the minimum of the free energy and becomes unstable
with respect to the lower symmetric one which branches off is called
a bifurcation point. For the first order transitions, it corresponds to
a spinodal point, thus the bifurcating solution describes a metastable state.
In practice, for the first order {\iso} -- {\nun} transition, it means that
at the bifurcation density the minimum associated with the isotropic state
changes to a local maximum of the free energy, and the only local minima left
correspond to the uniaxial nematic phases. For continuous phase transitions, the
bifurcation point corresponds to the critical point, and the
lower symmetric solution becomes immediately stable. Since present study deals
with the weakly first order and second order transitions, the
bifurcation analysis provides good estimates for the transition points.
\nlin The choice of the class of phase transitions where the states are connected
by group -- subgroup relation and analysis of the solutions of the {\sceq}
that branch off the reference (higher symmetric) one constitute the so--called
\ital{symmetry breaking bifurcation analysis}. For the first time it was applied to liquid
crystals by Kayser and Ravech\'e for Onsager model of hard, long rods
\cite{kayserraveche}, and later was generalized by Mulder \cite{mulder}.
It is a numerically tractable method that allows to acquire some insight into
the behaviour of the system close to phase transition.
For the transitions studied in the present work it is a good choice,
yet in general case it does not exhaust all the possible scenarios
for the phase sequence. It does not tell which of the bifurcating
states will be preferred. However, we can determine
the symmetry of bifurcating state and character of the bifurcation.
Present section is devoted to the description of this approach.
\nlin We begin with finding the exact equations for points where a given
solution branches off the reference one, then continue to present the exact
bifurcation equations for the spatially uniform phases and their specific form
for \iso\ -- \nun, \iso\ -- \nbx, and \nun\ -- \nbx\ transitions.
Next, we make a remark on a possibility
of Landau points, and in the concluding section present formulas for
the tricritical point. We follow the scheme presented by Mulder for
the isotropic reference state \cite{mulder} and later extended by us
to the case of reference state of arbitrary symmetry \cite{longa}.

\subsection{General bifurcation equations}
\label{section:generalpropertiesofsc}
\begin{figure}[h]
 \centering
 \subfigure[First order phase transition.]{\label{figure:bifscenariofirstorder} \includegraphics{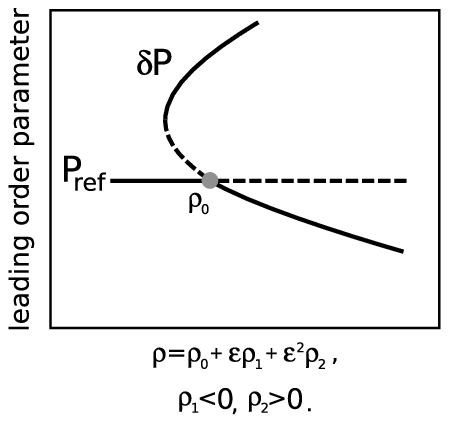}}
 \subfigure[First order phase transition in vicinity of a tricritical point.]{\label{figure:bifscenariofirstclosetotri} \includegraphics{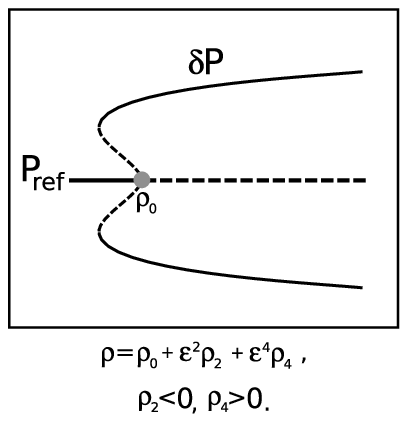}}
 \subfigure[Continuous phase transition.]{\label{figure:bifscenariosecondorder} \includegraphics{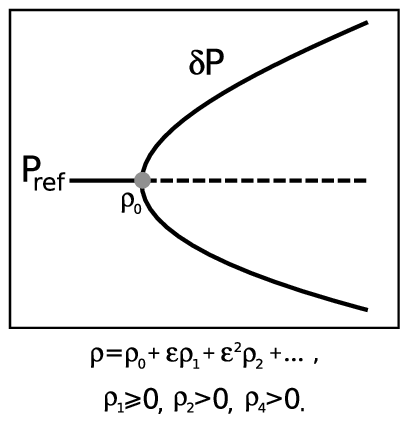}}
 \caption[Generic bifurcation diagrams.]{\label{figure:bifurcationscenarios} Generic bifurcation diagrams. Schematic representation of behaviour of leading order parameter near bifurcation point $\rho_{0}$. Path leading from \subref{figure:bifscenariofirstclosetotri} to \subref{figure:bifscenariosecondorder} describes a change of character of phase transition as observed at tricritical point. In the reference state $\pref$ the order parameter vanishes.}
\end{figure}

 Let's start by noting that a straightforward consequence of
the expansion \refeq{eq:taylorexpwithdcfs} is that the {\sceq} \refeq{eq:selfconsistentequniform} can be written as:
\begin{equation}
\label{eq:selfconsistentdeltap}
 \delta\opdff{\coordsymbol}=Z_{ref}\opdfff{\coordsymbol}{ref}Z^{-1}_{s}\exp\lcbra \sum^{\infty}_{n=1}\rho^{n}K_{n+1}\lnbra \coordsymbol,\lsbra\delta\opdfsymbol\rsbra \rnbra \rcbra-\opdfff{\coordsymbol}{ref} \, ,
\end{equation}
where we have introduced average density $\rho\equiv \average{N}/V$ (with $V$ standing for volume);
\begin{equation}
 \densityf{\coordsymbol}=\rho\opdff{\coordsymbol} \, , \,\,\,\, \frac{1}{V}\int d\coordsymbol \, \opdff{\coordsymbol}=1 \, ,
\end{equation}
and where $\delta\opdff{\coordsymbol} \equiv \opdfff{\coordsymbol}{s}-\opdfff{\coordsymbol}{ref}$ denotes a small deviation of the bifurcating $\opdfff{\coordsymbol}{s}$ from the reference state $\opdfff{\coordsymbol}{ref}$, and where
\begin{equation}
\label{eq:kerndef}
 K_{n+1}\lnbra \coordsymbol,\lsbra\delta\opdfsymbol\rsbra \rnbra=\frac{1}{n!}\int c_{n+1}( \coordsymbol,\coord{1},\dots,\coord{n},\lsbra\delta\opdfsymbol\rsbra) \prod^{n}_{i=1} \delta \opdff{\coord{i}} d\coord{i} \, ,
\end{equation}
with the normalization constant
\begin{equation}
 Z_{s}=\frac{Z_{ref}}{V}\int d\coordsymbol \, \opdfff{\coordsymbol}{ref}\exp\lcbra \sum^{\infty}_{n=1}\rho^{n}K_{n+1}\lnbra \coordsymbol,\lsbra\opdfsymbol\rsbra \rnbra \rcbra \eqdot \nonumber
\end{equation}
The normalization of $\opdfff{\coordsymbol}{s}$ and $\opdfff{\coordsymbol}{ref}$
implies that $\int d\coordsymbol \, \delta\opdff{\coordsymbol}=0$.
By construction, $\delta\opdff{\coordsymbol}=0$ satisfies
the equation \refeq{eq:selfconsistentdeltap} giving the \sceq\ for
$\pref$ equivalent to \eqrefeq{selfconsistentequniform}. It is natural to seek
the solutions $\delta\opdff{\coordsymbol} \neq 0$ that branch off from the
trivial one $\delta\opdff{\coordsymbol}=0$ since they represent possible
equilibrium states accessible from $\pref$ state through a phase transition.
As described in \cite{mulder} and \cite{longa}, these can be found by performing
the following expansions in powers of some
arbitrary control parameter $\epsilon$:
\begin{equation}
\label{eq:expansionsclosetobif}
\begin{split}
 \delta \opdfsymbol(\coordsymbol) & = \ep\,\bcorf{1}{\coordsymbol}+\ep^{2}\,\bcorf{2}{\coordsymbol}+\,\dots \, , \\
 \rho      & = \rho_{0}+\ep\,\rho_{1}+\ep^{2}\,\rho_{2}+\,\dots \, .
\end{split}
\end{equation}
Now inserting the above into \eqrefeq{selfconsistentdeltap}, and comparing
the terms of equal order in $\epsilon$ we obtain the formulas
fulfilled by $\bcorr{n}$ at the bifurcation point. They read
\begin{equation}
\label{eq:bifeqsgeneralcase}
\begin{split}
 \bcorf{1}{\coordsymbol} & = \rho_{0} \, \pref(\coordsymbol)\lcbra \kernelbif{\coordsymbol}{\bcorr{1}} - \averageref{\kernelbif{\coordsymbol}{\bcorr{1}}} \rcbra \, , \\
 \bcorf{2}{\coordsymbol} & = \rho_{0} \, \pref(\coordsymbol)\lcbra \kernelbif{\coordsymbol}{\bcorr{2}} - \averageref{\kernelbif{\coordsymbol}{\bcorr{2}}} \rcbra \\
  & + \rho_{1} \, \pref(\coordsymbol) \lcbra \kernelbif{\coordsymbol}{\bcorr{1}} - \averageref{\kernelbif{\coordsymbol}{\bcorr{1}}} \rcbra \\
  & + \rho_{0}^{2} \, \pref(\coordsymbol) \bigg( \! \linv \averageref{\kernelbif{\coordsymbol}{\bcorr{1}}} \lcbra \averageref{\kernelbif{\coordsymbol}{\bcorr{1}}}-\kernelbif{\coordsymbol}{\bcorr{1}} \rcbra \right. \\
  & + \left. \frac{1}{2}\lcbra \kernelbif{\coordsymbol}{\bcorr{1}}^{2} - \averageref{\kernelbif{\coordsymbol}{\bcorr{1}}^{2}} \rcbra + \kernelbifn{\coordsymbol}{\bcorr{1}}{3}-\averageref{\kernelbifn{\coordsymbol}{\bcorr{1}}{3}} \rnbra \, , \\
 \bcorf{3}{\coordsymbol} & = \, \dots \, ,
\end{split}
\end{equation}
where all the averages in the above expressions are calculated with respect
to the reference state: $\averageref{A}\equiv\int d\coordsymbol A(\coordsymbol) \pref(\coordsymbol)$. Also, due to the normalization of \sopdfns,
we have $\int d\orientation \, \bcorf{n}{\orientation}=0$ and we can impose $1/V\int d\coordsymbol\,\bcorf{n}{\coordsymbol}\bcorf{1}{\coordsymbol}=\delta_{n,1}$\footnote{It follows from the fact that \esre{bifeqsgeneralcase} are also fulfilled by $\bcorr{n}^{\,\prime}=\bcorr{n}+\alpha_{n}\bcorr{1}$ (it expresses the freedom of monotonic reparametrization of $\ep$ in \esre{expansionsclosetobif}), where $\alpha_{n}$ are arbitrary parameters. By setting $1/V\int d\coordsymbol\,\bcorf{n}{\coordsymbol}\bcorf{1}{\coordsymbol}=\delta_{n,1}$ we fix $\alpha_{n}=0$.}. Equations \re{bifeqsgeneralcase} can be simplified further by 
recalling that the symmetry of the bifurcating state is lower than that
of $\pref$ and that they remain in the group--subgroup relation, that implies
$\averageref{\kernelbifn{\coordsymbol}{\bcorr{1}}{n}}=0$ and $\int d\coordsymbol\,\pref(\coordsymbol)\bcorf{1}{\coordsymbol}=0$. After substituting
to the above, we obtain a set of hierarchical equations for
$\bcorf{1}{\coordsymbol}$, $\bcorf{2}{\coordsymbol}$, $\dots$:
\begin{align}
 \bcorf{1}{\coordsymbol} = &\, \rho_{0} \, \pref(\coordsymbol) \kernelbif{\coordsymbol}{\bcorr{1}} \, , \label{eq:bifurcationeq} \\
 \bcorf{2}{\coordsymbol} = &\, \rho_{0} \, \pref(\coordsymbol) \kernelbif{\coordsymbol}{\bcorr{2}} + \rho_{1} \, \pref(\coordsymbol) \kernelbif{\coordsymbol}{\bcorr{1}} \n
 & + \frac{1}{2}\rho_{0}^{2} \, \pref(\coordsymbol) \lcbra \kernelbif{\coordsymbol}{\bcorr{1}}^{2} - \averageref{\kernelbif{\coordsymbol}{\bcorr{1}}^{2}} + 2\kernelbifn{\coordsymbol}{\bcorr{1}}{3}\rcbra \, ,  \label{eq:bifeqsecondorder} \\
 \dots \, . \;\;\;\;\;\; & \nonumber
\end{align}
The first of the above is commonly known as the bifurcation equation,
describing the branching point to a new solution $\bcorf{1}{\coordsymbol}$.
Thus in the limit of small $\epsilon$, the low-symmetry state is
$\opdfff{\coordsymbol}{s}=\pref(\coordsymbol)+\epsilon\bcorf{1}{\coordsymbol}$.
The remaining equations determine the character of the bifurcation
(first order vs continuous), and also allow to obtain a condition where this
character changes (tricritical or, generally, multicritical point). Interestingly, as we can see, once we
know $\pref$ (either we have postulated it or found it by solving the \sceq),
in order to determine the bifurcation point, it is enough to know
the second direct correlation function. Higher order direct correlations
contribute only to the subsequent equations, also in the hierarchical
manner (in the first order equation only $c_{2}$ is present, in second order
additionally $c_{3}$ appears, and so forth).
\nlin Possible scenarios for bifurcation are schematically depicted in
\figreffig{bifurcationscenarios}.
\nlin In the next section we show examples of applications of the formulas
\re{selfconsistentdeltap}-\re{bifeqsecondorder} for spatially
uniform phases, and then continue to derive the bifurcation equations for
the systems with isotropic, uniaxial nematic, and biaxial nematic phases.
\subsection{Bifurcation equations in case of spatially uniform states}
\label{section:bifeqsspatiallyuniform}
\indent The equations presented above are exact as long as we do not make
any approximations for the direct correlation functions and the reference state.
We can still make some general assumptions as to the form of the
pair direct correlation and \sopdf and acquire the form of the bifurcation equations that is
easier to use. In case of the uniaxial nematic reference state, in $c_{2}$
we disregard the terms that depend on the director orientation. This assumption
is consistent with low-density approximation used later, and is also equivalent
of $c_{2}$ describing polydomain sample where the director dependence is
averaged out $c_{2}(\coord{1},\coord{2})\equiv\frac{1}{4\,\pi}\int d^{2}\unitv{n} \, c_{2}(\coord{1},\coord{2},\unitv{n})$. Also we assume that $c_{2}$ is
invariant under the global rotation and the particle interchange operation,
i.e., $c_{2}(\coord{1},\coord{2})=c_{2}(\coord{2},\coord{1})$, which imply
that it depends on the relative orientations and positions of molecules.
In present work we consider spatially uniform structures, namely isotropic,
uniaxial nematic, and biaxial nematic phases. Thus, we take into account the
bifurcations from \iso\ to \nun, from \iso\ to \nbx, and from \nun\ to \nbx.
For those transitions \sopdf depends only on orientational degrees of freedom,
i.e., $\opdff{\coordsymbol}=\opdf$, with $\int d\orientation\, \opdf=1$, $\pref(\coordsymbol)=\pref(\orientation)$, also $\bcorr{i}(\coordsymbol)=\bcorr{i}(\orientation)$, and pair direct correlation function $c_{2}(\coord{1},\coord{2})=c_{2}(\orientation_{1},\orientation_{2},\boldv{r}_{1},\boldv{r}_{2})=c_{2}(\orientation^{-1}_{1}\orientation_{2},\boldv{r}_{1}-\boldv{r}_{2})$, where $\orientation^{-1}_{1}\orientation_{2}$ stands for relative orientation. With those assumptions we can
present more specific and simpler form of \ere{bifurcationeq}.
\nlin Close to bifurcation we can assume that $\bcorr{1}(\orientation)$ can be
expressed as
\begin{equation}
 \label{eq:isotosomep1expanded}
 \bcorr{1}(\orientation)=\sum_{l,m,n} \almn{l}{m,n} \symmetrydelta{l}{m}{n}{\orientation} \, ,
\end{equation}
where $\almn{l}{m,n}$ are constant coefficients, and $\symmetrydelta{l}{m}{n}{\orientation}$ stand for real, orthogonal,
symmetrized, linear combinations of Wigner matrices $\wignerd{l}{m}{n}{\orientation}$ \cite{rose,lindner}. As described in \ara{deltas}, $\symmetrydelta{l}{m}{n}{\orientation}$ can be defined for arbitrary symmetry of phase and
molecule, those correspond to the first and second of lower indices,
respectively. In the above equation the summation goes over the physical
and relevant for a given model ranges of indices; always $-l \leq m,n \leq l$,
and, e.g., in case of \dtwh symmetry all indices are positive and even.
The coefficients $\almn{l}{m,n}$ are unknown parameters which can be determined
by means of bifurcation equations. Presently we proceed to find the
bifurcation point $\rho_{0}$. The \ital{symmetry adapted} functions $\symmetrydelta{l}{m}{n}{\orientation}$ form an
orthogonal base in the space of real functions, so we have the advantage
of following orthogonality relations (for derivation see \srs{basefunctions}):
\begin{align}
 \int\,d\orientation\,\symmetrydeltatopalign{l}{m}{n^{\phantom{\prime}}}{\orientation}\,\symmetrydelta{l^{\prime}}{m^{\prime}}{n^{\prime}}{\orientation} = & \, \deltasnormalization{l}\,\delta_{l,l^{\prime}}\,\delta_{m,m^{\prime}}\,\delta_{n,n^{\prime}}\, ,\,\,\deltasnormalization{l}\equiv\frac{8\,\pi^2}{2\,l+1} \, , \label{eq:orthogonalityofdelta} \\
 \int\,d\orientationin\,\symmetrydeltatopalign{l}{m}{n^{\phantom{\prime}}}{\relativeorientation}\,\symmetrydelta{l^{\prime}}{m^{\prime}}{n^{\prime}}{\orientationin} = & \, \deltasnormalization{l}\dwcoeff{m}{n^{\prime}}\,\symmetrydelta{l}{m^{\prime}}{n}{\orientation}\sum_{\sigma,\sigma^{\prime}\in\lcbra -1,1 \rcbra}\delta_{\sigma m,\sigma^{\prime} n^{\prime}} \, \delta_{l,l^{\prime}} \, , \label{eq:orthogonalityofdeltarelative}
\end{align}
where $\dwcoeff{m}{n} \equiv \, \lnbra \sqrt{2}/2 \rnbra^{2+\delta_{m,0}+\delta_{n,0}}$.
Using the above we can construct the equations for $\almn{l}{m,n}$ fulfilled
at bifurcation point; by inserting \re{isotosomep1expanded}
to \re{bifurcationeq} and casting it on $\symmetrydelta{l}{m}{n}{\orientation}$
we arrive at
\begin{equation}
 \label{eq:bifurcationeqexpanded}
 \almn{l}{m,n}=\lnbra\deltasnormalization{l}\rnbra^{-1}\rho_{0}\sum_{k,p,q} \almn{k}{p,q} \int d\orientation d\orientationin \, \pref(\orientation) c_{2}(\relativeorientation) \symmetrydelta{k}{p}{q}{\orientationin} \symmetrydelta{l}{m}{n}{\orientation} \, ,
\end{equation}
where $c_{2}(\relativeorientation)=\int d^{3}\vec{\mathbf{r}} c_{2}(\relativeorientation,\vec{\mathbf{r}})$.
\nlin If we fix $l=\tilde{l}$ in \ere{isotosomep1expanded}, i.e., if we choose one
subspace of infinitely numerous set of subspaces labelled by the
squared angular momentum index $l$, we can cast the above equation as
\begin{equation}
 \label{eq:bifurcationeqcompiled}
 \almn{\tilde{l}}{\{A\}}=\lnbra\deltasnormalization{\tilde{l}}\rnbra^{-1}\rho_{0} \sum_{B} \olmn{\tilde{l}}{\{A\},\{B\}} \almn{\tilde{l}}{\{B\}} \, ,
\end{equation}
where capital letters stand for two indices; $\{ A \} \equiv m,n$, and
the summation over $A$ means the summation over $m$ and $n$ in the manner described above, and where the elements of a bifurcation matrix $\oper{\boldm{\omega}}^{(\tilde{l})}$ read 
\begin{equation}
 \label{eq:omegafrombifeqcompiled}
 \olmn{\tilde{l}}{\{A\},\{B\}}=\int d\orientation d\orientationin \, \pref(\orientation) c_{2}(\relativeorientation) \symmetrydeltalg{\tilde{l}}{\{A\}}{\orientation} \symmetrydeltalg{\tilde{l}}{\{B\}}{\orientationin} \, .
\end{equation}
\ere{bifurcationeqcompiled} is the eigenequation with
eigenvalue $(\deltasnormalization{\tilde{l}}/\rho_{0})$. The
parameter $\rho_{0}$ depends only on the elements of
matrix $\oper{\boldm{\omega}}^{(\tilde{l})}$ and
can be calculated as one of the roots of
characteristic polynomial, namely from the solutions of the following equation:
\begin{equation}
 \label{eq:eigenequation}
 \det \lsbra \oper{\boldm{\omega}}^{(\tilde{l})} - \deltasnormalization{\tilde{l}} \rho^{-1}_{0} \hat{\mathbbm{1}} \rsbra=0 \, ,
\end{equation}
where $\hat{\mathbbm{1}}$ is the unit matrix.
\nlin We need to decide which of the eigenvalues following from the
above equation correspond to the bifurcation point. If we associate $\rho_{0}$
with the density, then in the absence of coupling, when $\rho_{0}=0$, we can
only expect the isotropic phase. With the increase of density an assumed phase
transition takes place, and in higher densities the bifurcation occurs.
Therefore for non--zero $\rho_{0}$ the relevant point where the
reference solution loses the property of being stable is described by
minimal value of $\rho_{0}$, thus the bifurcation point is chosen from the set
of solutions of the \ere{eigenequation} as the one corresponding to
the lowest $\rho_{0}$.
\nlin In general case, in \ere{omegafrombifeqcompiled} terms with higher
angular momentum index than $\tilde{l}$ can be present, but only
due to the presence of $\pref(\orientation)$. As can be seen
from \ere{orthogonalityofdeltarelative}, the integral over $\orientationin$
in \ere{omegafrombifeqcompiled} will leave only terms proportional
to $\symmetrydelta{\tilde{l}}{a}{b}{\orientation}$ (which is a consequence of the global rotational invariance of $c_{2}(\relativeorientation)$).
It means that the leading contribution to the bifurcation from the
pair direct correlation function will come from the part of $c_{2}(\relativeorientation) \sim \symmetrydelta{\tilde{l}}{a}{b}{\relativeorientation}$,
which justifies the use of the expansion of $c_{2}$ in base of $\symmetrydelta{l}{a}{b}{\orientation}$ functions. Only the presence of terms
with $\symmetrydelta{k}{d}{e}{\orientation}$ with non-zero $k$
in $\pref(\orientation)$ can generate in effect terms with higher 
angular momentum index
in $\bifmat{\tilde{l}}$ (see, e.g. \ere{threedeltasortho}).
In the special case when $\pref(\orientation)=const$, the representations
numbered by $\tilde{l}$ decouple and $\bifmat{\tilde{l}}$ contains only 
terms of order $\tilde{l}$. It is studied in the following section.
\subsection{Results for \dtwh -- symmetric model}
 Presently we evaluate the above equations further and obtain the specific
form of Eqs.~\re{bifurcationeqcompiled}-\re{omegafrombifeqcompiled} for the case
of \dtwh -- symmetric molecules and phase. It means that the
symmetry adapted base functions $\symmetrydelta{l}{m}{n}{\orientation}$
are defined for even and positive indices, and the summations go
over the relevant values of $l \geq 0$ and $0 \leq m,n \leq l$, they read
\begin{equation}
 \symmetrydelta{l}{m}{n}{\orientation} = \dwcoeff{m}{n}\,\sum_{\sigma,\sigma^{'}\in\lcbra -1,1 \rcbra} \wignerd{l}{\sigma\,m}{\sigma^{'}\,n}{\orientation} \, ,
\end{equation}
where $\dwcoeff{m}{n} \equiv \, \lnbra \sqrt{2}/2 \rnbra^{2+\delta_{m,0}+\delta_{n,0}}$, and where $\symmetrydelta{0}{0}{0}{\orientation}=1$. We also assume
that for nematics we can truncate the expansion \re{isotosomep1expanded}
at $l=2$, i.e., that the leading contribution to bifurcation
for spatially uniform phases \iso, \nun, and \nbx\ comes from
coefficients $\almn{2}{m,n}$. It is consistent with the symmetry of these
states and the choice of leading order parameters.
\nlin We start with the special case when the reference state is
isotropic which means that
\begin{equation}
 \pref(\orientation)=\frac{1}{8\,\pi^{2}} \, .
\end{equation}
Then, the bifurcation matrix $\oper{\boldm{\omega}}^{(\tilde{l})}$ \re{omegafrombifeqcompiled} contains only terms with given $\tilde{l}$, and
the representations labelled by $\tilde{l}$s
bifurcate independently (it is straightforward to check that this only
happens when $\pref(\orientation)$ corresponds to isotropic state).
\nlin Since, as we have mentioned above, the integral $\int d\orientationin \, c_{2}(\relativeorientation) \symmetrydeltalg{\tilde{l}}{m,n}{\orientationin}$
in \re{omegafrombifeqcompiled} leaves in $\oper{\boldm{\omega}}^{(\tilde{l})}$
only the terms with $\tilde{l}$, in order to find $\rho_{0}$, we can postulate
the expansion of $c_{2}(\relativeorientation)$ in the symmetry adapted base:
\begin{equation}
 \label{eq:expansionofc2l2}
 c_{2}(\relativeorientation) = & \, \sum_{l,m,n} \clmn{l}{m}{n} \symmetrydelta{l}{m}{n}{\relativeorientation} \, ,
\end{equation} 
where
\begin{equation}
\label{eq:clmncoeffs}
\begin{split}
 \clmn{l}{m}{n} \equiv & \, \frac{2\,l+1}{8\pi^2} \int d(\relativeorientation)\, c_{2}(\relativeorientation) \, \symmetrydelta{l}{m}{n}{\relativeorientation} \, , \\
 c_{2}(\relativeorientation) \equiv & \int d^{3}\vec{\mathbf{r}}\, c_{2}(\relativeorientation,\vec{\mathbf{r}}) \, ,
\end{split}
\end{equation}
and where due to the particle interchange symmetry we have $\clmn{l}{m}{n}=\clmn{l}{n}{m}$. Now, using orthogonality of the base functions \re{orthogonalityofdelta}-\re{orthogonalityofdeltarelative}, and inserting the expansion of pair direct correlation function \re{expansionofc2l2}, \ere{bifurcationeqcompiled} can be expressed in a simple form with $\tilde{l}=2$, namely
\begin{equation}
 \label{eq:isotosomebifmsimple}
 \almn{2}{m,n}=\frac{\rho_{0}}{5}\sum_{q\in\{0,2\}} \clmn{2}{q}{n} \almn{2}{m,q} \, .
\end{equation}
Above equation is a $4x4$ eigenproblem. We recall that the first of
lower indices of the base functions corresponds to the symmetry of phase.
In this sense $\{\almn{2}{0,0},\almn{2}{0,2}\}$ are associated with
uniaxial and $\{\almn{2}{2,0},\almn{2}{2,2}\}$ with biaxial state. Another 
consequence of \re{orthogonalityofdeltarelative} is the fact that those
two sets bifurcate independently, and in effect we get a $2x2$ eigenproblem.
It is so because $c_{2}(\relativeorientation)$ expressed by \re{expansionofc2l2}
gives the splitting of the space into subspaces labelled
by angular momentum index $l$ and phase related index $m$ through the following
term in \re{omegafrombifeqcompiled}: $\int d\orientationin\, c_{2}(\relativeorientation)\symmetrydelta{l}{m}{n}{\orientationin}$ $\sim\sum_{k}\symmetrydelta{l}{m}{k}{\orientation}$. So, for isotropic reference state, not only the
representations numbered by angular momentum index bifurcate independently, but
also the phases of different symmetry branch off in separate subspaces.
Furthermore, it is clear that \ere{isotosomebifmsimple} is the same for
uniaxial and biaxial $\almn{2}{m,n}$. It can be easily solved, and we
arrive at the following expression for bifurcation point:
\begin{equation}
 \label{eq:exactbifisoldl2}
 \rho_{0} = \frac{10}{\clmn{2}{0}{0}+\clmn{2}{2}{2}-\sqrt{4 \lnbra \clmn{2}{0}{2} \rnbra^{2} + \left( \clmn{2}{0}{0} - \clmn{2}{2}{2}\right)^{2}}} \, .
\end{equation}
This equation describes the location of bifurcation point for transitions where
the high-\linebreak symmetry phase is the isotropic liquid and
lower-symmetric state
possesses spontaneously broken orientational symmetry $O(3)$ to
uniaxial \dinfh or biaxial \dtwh symmetry group. As we can see, the bifurcation point is fully determined by
coefficients of the expansion of pair direct correlation function in the
subspace of angular momentum index $\tilde{l}=2$, and due to \ere{orthogonalityofdeltarelative}
depends on all of the coefficients $\clmn{2}{m}{n}$.
Above expression is equivalent to Eq.~(16) from \cite{straley}.

 Situation changes qualitatively when the reference state is taken to be
a stable uniaxial nematic phase. In this case we need equilibrium,
uniaxial $\pref$; to determine it we solve the \sceq\ \re{selfconsistentequniform}. The \sopdf is taken to be of the following form 
\begin{equation}
 \label{eq:opdfinops}
 \opdf=\sum_{l,m,n} \frac{2\,l+1}{8\pi^{2}}\,\orderparameter{l}{m}{n}\symmetrydelta{l}{m}{n}{\orientation} \, . 
\end{equation}
where the coefficients $\orderparameter{l}{m}{n}$ are calculated as
the averages over the ensemble, and identified with order parameters
(for the explanation of this model see \ara{opdf}). Using the
orthogonality relations \re{orthogonalityofdelta} we can find that
\begin{equation}
 \label{eq:orderparamsset}
 \orderparameter{l}{m}{n} = \int d\orientation\,\opdf\symmetrydelta{l}{m}{n}{\orientation} \, .
\end{equation}
\indent From the above we choose the set with $l=2$ as the leading order parameters, which is 
consistent with the symmetry of states in consideration. As we have mentioned,
the first of the lower indices in $\symmetrydelta{l}{m}{n}{\orientation}$
is identified with the symmetry of a phase, while the remaining one with
a molecule. Along this convention $\orderparameter{2}{0}{0}$ and $\orderparameter{2}{0}{2}$ are associated with uniaxial nematic, and $\orderparameter{2}{2}{0}$ and $\orderparameter{2}{2}{2}$ with biaxial nematic. They all vanish in the 
isotropic phase. With apropriate choice of the moleuclar and lablatory axes,
in \nun\ $\orderparameter{2}{0}{0} \neq 0$ while
$\orderparameter{2}{2}{0}=\orderparameter{2}{2}{2}=0$,
for \nbx\ $\orderparameter{2}{0}{0} \neq 0$, $\orderparameter{2}{2}{2} \neq 0$.\putfigureflagslong{part1/misc/bodylab.eps}{Two orthogonal, right-handed tripods of unit vectors corresponding to the frames associated with laboratory ($\{\labx,\laby,\labz\}$) and molecular (body) frame ($\{\bodx,\body,\bodz\}$).}{bodylab}{b}{Molecular and laboratory frames of reference.}The interpretation is clearer once we explicitly write the set of symmetry adapted functions $\symmetrydelta{l}{m}{n}{\orientation}$ for $l=2$; the uniaxial functions read
\begin{equation}
 \label{eq:deltasl2setun}
\begin{split}
 \symmetrydelta{2}{0}{0}{\orientation} = &  P_{2}(cos(\beta)) = \, \frac{1}{4}+\frac{3}{4}\,cos\lnbra2\,\beta\rnbra=-\frac{1}{2}+\frac{3}{2} \lnbra\bodz\cdot\labz \rnbra^{2} \, , \\
 \symmetrydelta{2}{0}{2}{\orientation} = & \, \frac{\sqrt{3}}{2}cos\lnbra2\,\gamma\rnbra sin\lnbra \beta \rnbra^2=\frac{\sqrt{3}}{2} \lcbra \lnbra \bodx\cdot\labz \rnbra^{2} - \lnbra \body\cdot\labz \rnbra^{2} \rcbra \, ,
\end{split}
\end{equation}
where $P_{2}(x)=\frac{1}{2}\lnbra 3\,x^{2}-1\rnbra$
is the second Legendre polynomial, and the biaxial functions are
\begin{equation}
 \label{eq:deltasl2setbx}
 \begin{split}
 \symmetrydelta{2}{2}{0}{\orientation} = & \, \frac{\sqrt{3}}{2}cos\lnbra2\,\alpha\rnbra sin\lnbra \beta \rnbra^2=\frac{\sqrt{3}}{2} \lcbra \lnbra \bodz\cdot\labx \rnbra^{2} - \lnbra \bodz\cdot\laby \rnbra^{2} \rcbra \, , \\
 \symmetrydelta{2}{2}{2}{\orientation} = & \, \frac{1}{4}\lsbra 3+cos\lnbra 2\,\beta \rnbra \rsbra\,cos\lnbra 2\,\alpha\rnbra\, cos\lnbra 2\,\gamma\rnbra -cos\lnbra \beta \rnbra sin\lnbra 2\,\alpha \rnbra sin\lnbra 2\, \gamma \rnbra=\\
 & \, \lnbra\bodx\cdot\labx \rnbra^{2} + \lnbra\body\cdot\laby \rnbra^{2} - \frac{1}{2}\lnbra\bodz\cdot\labz \rnbra^{2} - \frac{1}{2} \, ,
\end{split}
\end{equation}
where $\lcbra \labx,\laby,\labz \rcbra$ and $\lcbra \bodx,\body,\bodz \rcbra$
stand for right-handed, orthogonal tripods associated with laboratory (director)
and molecular (body) frame, respectively (see \frf{bodylab}). Now it is clear that $\orderparameter{2}{0}{0}$ measures the degree of order of chosen molecular axis with respect
to the laboratory-fixed direction, while $\orderparameter{2}{2}{2}$ reflects
the ordering of complete tripod of vectors of molecular and laboratory frames.
It is straightforward to check that $-\frac{1}{2}\leq\orderparameter{2}{0}{0} \leq 1$, and $-1\leq\orderparameter{2}{2}{2} \leq 1$. We can also see
that when we choose $\labz$ as the director, then the oblate nematic order 
\ndisc\ with respect to molecular axis $\bodz$ entails in ideal case
$\orderparameter{2}{0}{0}=-\frac{1}{2}$ and $\orderparameter{2}{0}{2}$
can be non-zero, while in ideal ordering of $\bodz$ along $\labz$ we
have $\orderparameter{2}{0}{0}=1$, $\orderparameter{2}{0}{2}=0$. In these
cases $\orderparameter{2}{2}{0}=\orderparameter{2}{2}{2}=0$.
On the other hand when $\labx$ is parallel to $\bodx$, $\laby$ to $\body$,
and $\labz$ to $\bodz$, i.e., we have ideally aligned \nbx, then
$\orderparameter{2}{2}{2}=\orderparameter{2}{0}{0}=1$
and remaining averages vanish. We can construct linear combinations
of $\symmetrydelta{2}{m}{n}{\orientation}$ that satisfy the uniaxial symmetry
restrictions with respect to other laboratory axes, e.g., 
states $a (\symmetrydeltana{2}{0}{0}+\sqrt{3}\symmetrydeltana{2}{2}{0})+b (\symmetrydeltana{2}{0}{2}+\sqrt{3}\symmetrydeltana{2}{2}{2})$ and 
$a (\symmetrydeltana{2}{0}{0}-\sqrt{3}\symmetrydeltana{2}{2}{0})+b (\symmetrydeltana{2}{0}{2}-\sqrt{3}\symmetrydeltana{2}{2}{2})$ describe \nun\ phases uniaxial
about $\laby$ and $\labx$, respectively.
$\orderparameter{2}{0}{0}$ and $\orderparameter{2}{2}{2}$ are referred to
as leading, dominant order parameters. The remaining two,
as we can see from \re{deltasl2setun}-\re{deltasl2setbx}, 
$\orderparameter{2}{0}{2}$ is sensitive to the orientation of the
plane determined by two vectors associated with the molecule with respect to
the laboratory chosen direction, while $\orderparameter{2}{2}{0}$ depends
on the orientation of the plane determined by two vectors in the
laboratory frame in comparison to the chosen molecular axis.
\nlin In order to further work out the bifurcation equation \re{bifurcationeq} 
for the \nun\ -- \nbx\ transition, we need a stable reference phase.
Thinking in terms of the expansion \re{opdfinops}, we need to know the 
order parameters $\orderparameter{l}{0}{m}$ for even $l$ and for given temperature
and density. Since now the reference state has non-trivial structure, we cannot
expect that the bifurcation point obtained from \re{bifurcationeqcompiled} will
depend only on order parameters with lowest possible angular momentum index $l=2$.
\nlin The reference phase can be determined with the help of
the \sceq\ \re{selfconsistentequniform}. The equation can now be cast in
the following form:
\begin{equation}
 \label{eq:scuniformcase}
 \opdf=Z^{-1}\exp\lsbra \rho \int d\orientationin \, c_{2}(\relativeorientation) \opdff{\orientationin} \rsbra \, ,
\end{equation}
where $Z=\int d\orientation \exp\lsbra \rho \int d\orientationin \, c_{2}(\relativeorientation) \opdff{\orientationin} \rsbra$ ensures the normalization.
Once we insert the expansion of $c_{2}(\relativeorientation)$ \re{expansionofc2l2} up to angular momentum index $l=2$ to the above, we can write explicitly the equations for
order parameters using the definition
of $\orderparameter{l}{m}{n}$ \ere{orderparamsset}, they read
\begin{equation}
 \label{eq:opsinl2sceq}
 \orderparameter{l}{m}{n} = Z^{-1} \int d\orientation \, \symmetrydelta{l}{m}{n}{\orientation} \exp \lcbra \rho \lsbra c^{(2)}_{2,u}(\orientation)+c^{(2)}_{2,b}(\orientation) \rsbra \rcbra \, ,
\end{equation}
where
\begin{equation}
\begin{split}
 c^{(2)}_{2,u}(\orientation) = & \, \sum\limits_{m,n \in \{0,2\}}\clmn{2}{m}{n}\orderparameter{2}{0}{m}\symmetrydelta{2}{0}{n}{\orientation} \, , \\
 c^{(2)}_{2,b}(\orientation) = & \, \sum\limits_{m,n \in \{0,2\}}\clmn{2}{m}{n}\orderparameter{2}{2}{m}\symmetrydelta{2}{2}{n}{\orientation} \, .
\end{split}
\end{equation}
Setting $\orderparameter{2}{2}{0}=\orderparameter{2}{2}{2}=0$, i.e., $c^{(2)}_{2,b}(\orientation)=0$, we can solve the self-consistent equations for
order parameters \re{opsinl2sceq} interactively, provided
we know the expansion coefficients $\clmn{2}{m}{n}$, and then use the
order parameters to build the uniaxial reference \sopdf
\begin{equation}
 \label{eq:prefinun}
 \pref(\orientation)=Z^{-1}_{ref} \exp \lsbra \rho \, c^{(2)}_{2,u}(\orientation) \rsbra \, ,
\end{equation}
where $Z_{ref}=\int d\orientation \, \exp \lsbra \rho \, c^{(2)}_{2,u}(\orientation) \rsbra$.
\nlin Having determined the reference state, using \esre{opsinl2sceq}-\re{prefinun}, we can continue to find the
equation for bifurcation point $\rho_{0}$. The eigenproblem \re{bifurcationeqcompiled} now involves a $2x2$ matrix 
\begin{equation}
 \label{eq:unbxmatrixomega}
\begin{split}
 \olmn{2}{\{a,b\},\{p,q\}} = & \int d\orientation d\orientationin \, \pref(\orientation) c_{2}(\relativeorientation) \symmetrydelta{2}{a}{b}{\orientation} \symmetrydelta{2}{p}{q}{\orientationin} \, ,
\end{split}
\end{equation}
where $\{a,b\}\in\{2,0\},\{2,2\}$, and $\{p,q\}\in\{2,0\},\{2,2\}$. 
Using the above with the expansion of $c_{2}(\relativeorientation)$ \re{expansionofc2l2}, orthogonality of base functions \re{orthogonalityofdelta}-\re{orthogonalityofdeltarelative}, and substituting the reference phase \re{prefinun},
we can cast the eigenequation \re{bifurcationeqcompiled} as
\begin{equation}
 \label{eq:unbxeigenequationmatrices}
\lnbra \begin{array}{c}
 \almn{2}{2,0} \\
 \almn{2}{2,2} 
\end{array} \rnbra = \rho_{0}
\lnbra \begin{array}{cc}
 A_{0} \clmn{2}{0}{0}+B_{0} \clmn{2}{0}{2} & B_{0} \clmn{2}{2}{2}+A_{0} \clmn{2}{0}{2} \\
 A_{2} \clmn{2}{0}{0}+B_{2} \clmn{2}{0}{2} & B_{2} \clmn{2}{2}{2}+A_{2} \clmn{2}{0}{2}
\end{array} \rnbra
\lnbra \begin{array}{c}
 \almn{2}{2,0} \\
 \almn{2}{2,2} 
\end{array} \rnbra \, ,
\end{equation}
where
\begin{align}
 & A_{2}\equiv\frac{2}{7} \orderparameterref{2}{0}{2}+\frac{1}{14} \sqrt{\frac{3}{5}} \orderparameterref{4}{0}{2}=B_{0} \, , \nn \\
 & B_{2}\equiv\frac{1}{5}+\frac{2}{7} \orderparameterref{2}{0}{0}+\frac{1}{70} \orderparameterref{4}{0}{0}+\frac{ \orderparameterref{4}{0}{4}}{2\sqrt{35}} \, , \nn \\
 & A_{0}\equiv\frac{1}{5}-\frac{2}{7} \orderparameterref{2}{0}{0}+\frac{3}{35} \orderparameterref{4}{0}{0} \, , \nn
\end{align}
and where the averages are calculated in the reference, uniaxial nematic state.
The bifurcation point can be calculated from the above eigenequation
using \re{eigenequation}, it reads
\begin{equation}
 \label{eq:bifunbx}
 \rho_{0}=\frac{2}{A_{0}\clmn{2}{0}{0}+2\,B_{0}\clmn{2}{0}{2}+B_{2}\clmn{2}{2}{2}+\sqrt{4\lnbra B_{0} \clmn{2}{0}{0}+B_{2}\clmn{2}{0}{2} \rnbra\lnbra A_{0}\clmn{2}{0}{2}+B_{0}\clmn{2}{2}{2} \rnbra+\lnbra A_{0}\clmn{2}{0}{0}-B_{2}\clmn{2}{2}{2} \rnbra^{2}}} \, .
\end{equation}

 Equations \re{exactbifisoldl2} and \re{bifunbx} describe the bifurcation 
scenarios for the system with three spatially uniform phases: isotropic,
uniaxial and biaxial nematic. The symmetry considerations of those structures
indicate that the leading order parameters are associated with the
symmetry adapted base functions with angular momentum index $l$ equal $2$.
That leads to the truncation of the expansion of the
bifurcating state $\bcorf{1}{\orientation}$ in \re{isotosomep1expanded}
after $l=2$ and implies that the leading contribution to the
bifurcation matrix $\oper{\boldm{\omega}}^{(2)}$ \re{omegafrombifeqcompiled}
from the pair direct correlation function comes from the
coefficients $\clmn{2}{m}{n}$. When the isotropic phase is taken as
the reference, those quantities determine the bifurcation point completely.
In case of \nun\ -- \nbx\ transition, $\rho_{0}$ additionally
depends on the uniaxial order parameters $\averageref{\symmetrydeltana{l}{0}{m}}$, which due to the non-isotropic
structure of the reference state come into the bifurcation equation 
up to $l=4$.
\nlin We would like to note that both equations \re{exactbifisoldl2}
and \re{bifunbx} are non-linear relations fulfilled at bifurcation point for
density $\rho=\rho_{0}$ and temperature $t$ since the coefficients $\clmn{2}{m}{n}$
and order parameters $\averageref{\symmetrydeltana{l}{0}{m}}$ in general depend
on $t$ and $\rho$. The description of the method used in the determination
of $\clmn{2}{m}{n}$ and solutions of Eqs.~\re{exactbifisoldl2},
\re{opsinl2sceq}, and \re{bifunbx} are presented in \ara{numericaldetails}.
\nlin Considering the transitions from isotropic phase and using only
the bifurcation equation \re{bifurcationeq}, we cannot say what
state is described by $\bcorf{1}{\orientation}$ because the
equations for the bifurcation point of biaxial and uniaxial states 
are the same \re{exactbifisoldl2}. In order to obtain the symmetry of the
bifurcating phase, we need to consult the second order equation \re{bifeqsecondorder}. This approach will also lead to the Landau points, as is described in the following section. 
\subsection{Exact equations for Landau point}
\label{section:exactlandaueqs}
\indent The Landau point, as we have indicated earlier, is the point
on the phase diagram where four phases: isotropic, uniaxial rod-like,
uniaxial disc-like, and biaxial nematic meet. Thus, in order to locate it,
we calculate the bifurcating state,
find its symmetry using \eqrefeq{bifeqsecondorder}, and determine the places
where the two types of uniaxial nematic ordering, \nrod\ and \ndisc,
become undistinguishable \cite{mulder}. The other method is based on the
analysis of the duality transformation of the part of $c_{2}$ 
expansion \ere{expansionofc2l2} with angular index equal $2$. The
thermodynamic states left invariant under this operation (self-dual) coincide
with Landau points \cite{longa}. Both methods lead to the same conditions.
We start with description of the former approach.
\nlin As was shown by Mulder \cite{mulder}, in order to acquire some information
about the structure of solution that branches of the reference state,
we should calculate the eigenvectors of operator $\kernelsymbol_{2}$
in \refeq{eq:bifurcationeq}--\refeq{eq:bifeqsecondorder}. Let's consider
transitions from isotropic phase. The bifurcation equation states that
\begin{equation}
 \label{eq:bifeqfromiso}
 \bcorf{1}{\orientation}=\frac{1}{8\pi^{2}}\rho_{0}\kernelbif{\orientation}{\bcorr{1}} \, .
\end{equation}
We mentioned before, that the above has the form of an eigenproblem with
$\kernelsymbol_{2}$ being a linear operator acting in the space of 
real functions in which the orthogonal symmetry adapted set \refeq{eq:symmetryadapteddeltas} was chosen as a base. In that space we can define an inner product
\begin{equation}
 \label{eq:innerproductdef}
 \innerpro{f_{1}}{f_{2}}\equiv\int d\orientation \, f_{1}(\orientation) f_{2}(\orientation) \, ,
\end{equation}
with respect to which the operator $\kernelsymbol_{2}$ \ere{kerndef}
is Hermitian:
\begin{equation}
 \label{eq:ceffhermitian}
 \innerpro{\kernelbif{\orientation}{f_{1}}}{f_{2}}=\innerpro{f_{1}}{\kernelbif{\orientation}{f_{2}}} \, .
\end{equation}
We also know that the eigenvalues $\rho_{0}$ calculated from
\ere{exactbifisoldl2} are the same for the sets of biaxial and uniaxial
coefficients $\almn{2}{m,n}$ (\ere{isotosomep1expanded}), which comes as a direct consequence of \re{orthogonalityofdeltarelative} and the trivial (isotropic) structure of the reference state.
\nlin Now following \cite{mulder} we turn to look for states that
branch off the isotropic (reference) phase. Eigenvectors associated
with the eigenvalue \refeq{eq:exactbifisoldl2} are
\begin{equation}
 \label{eq:eigenchi}
\begin{split}
 \chi_{0}=e_{0}\,\symmetrydelta{2}{0}{0}{\orientation} + e_{2}\,\symmetrydelta{2}{0}{2}{\orientation} \, , \\
 \chi_{2}=e_{0}\,\symmetrydelta{2}{2}{0}{\orientation} + e_{2}\,\symmetrydelta{2}{2}{2}{\orientation} \, .
\end{split}
\end{equation}
The $e_{n}$ coefficients are chosen such that $\chi_{n}$ are normalized with
respect to the inner product \refeq{eq:innerproductdef}:
\begin{equation}
 \label{eq:eigenvnormalization}
 \innerpro{\chi_{m}}{\chi_{n}}=\deltasnorm\delta_{m,n}\, .
\end{equation}
They read
\begin{equation}
 \label{eq:e0e2defs}
\begin{split}
 e_{0} = & \, \frac{\clmn{2}{0}{2}}{\sqrt{\lnbra\clmn{2}{0}{2}\rnbra^{2}+\tau^{2}}} \, , \\
 e_{2} = & \, \frac{\tau}{\sqrt{\lnbra\clmn{2}{0}{2}\rnbra^{2}+\tau^{2}}} \, , \\
\end{split}
\end{equation}
where
\begin{equation}
 \tau=\frac{1}{2}\lsbra \clmn{2}{0}{2}-\clmn{2}{2}{2}+\sqrt{\lnbra \clmn{2}{0}{0}-\clmn{2}{2}{2} \rnbra^{2} +4\,\clmn{2}{2}{2}} \rsbra \, .
\end{equation}
In the symmetry adapted base \re{deltasl2setun}-\re{deltasl2setbx}, a general solution of \eqrefeq{bifeqfromiso}
is given by:
\begin{equation}
 \label{eq:generalsolutiontobifeq}
 \bcorf{1}{\orientation}=\alpha_{0}\chi_{0}+\alpha_{2}\chi_{2} \, .
\end{equation}
We can add a normalization condition for $\bcorr{1}$
by setting $\alpha^{2}_{0}+\alpha^{2}_{2}=1$ and calculate the
coefficients $\alpha_{n}$ using $\innerpro{\bcorr{2}}{\pref}=0$
and $\innerpro{\bcorr{2}}{\bcorr{1}}=0$.
With second order bifurcation equation \refeq{eq:bifeqsecondorder}
and \re{ceffhermitian} we get
\begin{equation}
 \rho_{1}\innerpro{\chi_{n}}{\kernelbif{\orientation}{\bcorr{1}}}+\frac{1}{2}\rho_{0}^{2}\innerpro{\chi_{n}}{\kernelbif{\orientation}{\bcorr{1}}^{2}}=0\, .
\end{equation}
Combining Eqs.~\refeq{eq:generalsolutiontobifeq} and \re{eigenvnormalization}
with the bifurcation equation \refeq{eq:bifeqfromiso} we can further
simplify the above to obtain
\begin{equation}
\begin{split}
 & \frac{1}{5} \rho_{1} \alpha_{0} + \frac{1}{2} \rho_{0} \lsbra \alpha^{2}_{0}\innerpro{\chi_{0}}{\chi^{2}_{0}}+\alpha^{2}_{2}\innerpro{\chi_{0}}{\chi^{2}_{2}}+2\alpha_{0}\alpha_{2}\innerpro{\chi_{2}}{\chi^{2}_{0}} \rsbra=0 \, , \\
 & \frac{1}{5} \rho_{1} \alpha_{2} + \frac{1}{2} \rho_{0} \lsbra \alpha^{2}_{0} \innerpro{\chi_{2}}{\chi^{2}_{0}} + \alpha^{2}_{2} \innerpro{\chi_{2}}{\chi^{2}_{2}} + 2\alpha_{0}\alpha_{2}\innerpro{\chi_{0}}{\chi^{2}_{2}} \rsbra=0 \, .
\end{split}
\end{equation}
The products of the eigenvectors $\chi_{n}$ can be calculated; in those
equations each of them contains three symmetry--adapted functions. 
Using \refeq{eq:threedeltasortho} we arrive at the following conditions:
\begin{equation}
 \label{eq:eqforp2coeffs}
\begin{split}
 & \rho_{1} \alpha_{0} + \frac{1}{7}8\pi^{2}\rho_{0}\,\xi\lnbra \alpha^{2}_{0}-\alpha^{2}_{2} \rnbra=0  \, , \\
 & \rho_{1} \alpha_{2} - \frac{2}{7}8\pi^{2}\rho_{0}\,\xi\,\alpha_{0}\,\alpha_{2}=0  \, , \\
\end{split}
\end{equation}
where $\xi \equiv e_{0}\lnbra e^{2}_{0}-3\,e^{2}_{1} \rnbra$.
\nlin Now we see that for $\xi \ne 0$ the allowed values of $(\alpha_{0},\alpha_{2})$ are: $(\alpha_{0},\alpha_{2})=(\pm 1,0)$, $(\alpha_{0},\alpha_{2})=(\pm 1/2,\pm\sqrt{3}/2)$, and $(\alpha_{0},\alpha_{2})=(\pm 1/2,\mp\sqrt{3}/2)$. The 
corresponding solutions for $\bcorr{1}$, \ere{generalsolutiontobifeq}, are 
\begin{equation}
 \label{eq:p1solutions}
\begin{split}
 \bcorr{1}^{(3)}= & \, \pm\chi_{0} \, , \\
 \bcorr{1}^{(2)}= & \, \pm\frac{1}{2}\chi_{0}\pm\frac{\sqrt{3}}{2}\chi_{2} \, , \\
 \bcorr{1}^{(1)}= & \, \pm\frac{1}{2}\chi_{0}\mp\frac{\sqrt{3}}{2}\chi_{2} \, .
\end{split}
\end{equation}
All of the above are of uniaxial symmetry. It can be easily seen once we use
the representation of $\symmetrydelta{l}{m}{n}{\orientation}$ in directional cosines between the laboratory and body associated tripods $\lcbra \labx,\laby,\labz \rcbra$ and $\lcbra \bodx,\body,\bodz \rcbra$ as in \re{deltasl2setun}-\re{deltasl2setbx}. $\bcorr{1}^{(3)}$ does not change
under rotations about axis $\labz$, while the $\bcorr{1}^{(2)}$ and $\bcorr{1}^{(1)}$ about $\laby$ and $\labx$, respectively. The appearance of the three solutions
is a natural consequence of the freedom of renumbering of the axes, i.e., 
the permutation symmetry.
\nlin The choice of positive solutions in \refeq{eq:p1solutions} for 
$\xi>0$ entails the ordering of $\bodz$ axes along $\labz$ (see \re{deltasl2setun}-\re{deltasl2setbx}), thus marking the behaviour of molecules in
rod-like uniaxial nematic phase, while the case
of $\xi<0$, where negative solutions are used, stands for disk-like behaviour.
The one distinguished case is $\xi=0$, which marks the point where
the difference between oblate and prolate nematic phases ceases to exist.
In that place a second order (since $\xi=0\Rightarrow\lambda_{1}=0$, as can be seen from \re{eqforp2coeffs}) transition to lower symmetric, biaxial structure
is possible. The necessary conditions can be cast in the following form:
\begin{align}
 \clmn{2}{0}{2}=0\, , \textnormal{ for } \clmn{2}{0}{0}-\clmn{2}{2}{2}>0 \, , \\
 \clmn{2}{2}{2}=\clmn{2}{0}{0}-\frac{2}{\sqrt{3}}\abs{\clmn{2}{0}{2}} \, . \label{eq:landaupointcondition}
\end{align}
The above is a non-linear equation for Landau point, relating temperature
and model specific parameters. It proves that the location of Landau point can
be found once we know the coefficients $\clmn{2}{m}{n}$ \cite{mulder,mulderhard}. If $\clmn{2}{0}{2} \neq 0$ and $\clmn{2}{0}{0} \neq 0$ we can rewrite the above
as: $\clmn{2}{2}{2}/\clmn{2}{0}{0}=1-2/\sqrt{3}\abs{\clmn{2}{0}{2}}/\clmn{2}{0}{0}$. This
defines a line in $(\clmn{2}{0}{2}/\clmn{2}{0}{0},\clmn{2}{2}{2}/\clmn{2}{0}{0})$ space
that marks a boundary between the disc-like (when 
$\clmn{2}{2}{2}/\clmn{2}{0}{0}>1-2/\sqrt{3}\abs{\clmn{2}{0}{2}}/\clmn{2}{0}{0}$) and rod-like
(when $\clmn{2}{2}{2}/\clmn{2}{0}{0}<1-2/\sqrt{3}\abs{\clmn{2}{0}{2}}/\clmn{2}{0}{0}$) states.
The above analysis can be generalized further, and the existence of
Landau points can be proven \cite{mulderhard}.
\nlin There exists another way of determining Landau points following from
general considerations on the contribution of the terms with
angular momentum index $l=2$ in $c_{2}(\relativeorientation)$
in the expansion \re{expansionofc2l2}. Let's write it down
explicitly, up to the term with $l=2$
\begin{equation}
 \label{eq:c2pexpandeddeltas}
\begin{split}
 c_{2}(\relativeorientation) = \, & c_{0,0}\symmetrydelta{2}{0}{0}{\relativeorientation}+c_{0,2}\symmetrydelta{2}{0}{2}{\relativeorientation} \\
 + & c_{2,0}\symmetrydelta{2}{2}{0}{\relativeorientation}+c_{2,2}\symmetrydelta{2}{2}{2}{\relativeorientation} \, ,
\end{split}
\end{equation}
where for clarity we have dropped the upper index ($c_{m,n} \equiv \clmn{2}{m}{n}$), and we keep in mind that it is an approximation, since we have truncated
the expansion; however, we already know that it gives the
leading contribution to the bifurcation. As we have mentioned, the particle interchange symmetry requires $c_{2,0}=c_{0,2}$, which can be explicitly seen once we have used the Cartesian representation for $\symmetrydelta{2}{m}{n}{\orientation}$ functions (see \re{deltasl2setun}-\re{deltasl2setbx} and also \aprefap{deltas}):
\begin{equation}
 \label{eq:cartesianc2pexpanded}
\begin{split}
 c_{2}(\lcbra\labv{i},\bodv{i}\rcbra)=-\abs{c_{0,0}} & \lsbra \lnbra \widetilde{c}_{2}-\sqrt{3}\widetilde{c}_{0} \rnbra \lnbra \bodx\cdot\labx \rnbra^{2} + 
 \lnbra \widetilde{c}_{2}+\sqrt{3}\widetilde{c}_{0} \rnbra \lnbra \body\cdot\laby \rnbra^{2} \right . \\
  & \linv + \lnbra \frac{3}{2}sgn(c_{0,0})-\frac{1}{2}\widetilde{c}_{2} \rnbra \lnbra \bodz\cdot\labz \rnbra^{2} - \frac{sgn\lnbra c_{0,0} \rnbra + \widetilde{c}_{2}}{2} \rsbra \, ,
\end{split}
\end{equation}
where $\widetilde{c}_{0}\equiv c_{0,2}/\abs{c_{0,0}}$ and 
$\widetilde{c}_{2}\equiv c_{2,2}/\abs{c_{0,0}}$.
\nlin The thermodynamic properties of the system in our approach are determined
by the \sceq\ \re{scuniformcase}, we recall that the pair direct correlation function contributes to this equation via term $\rho c_{2}(\relativeorientation)$, namely
\begin{equation}
 \label{eq:l2scuniform}
 \opdf=Z^{-1}\exp\lsbra \widetilde{\rho} \int d\orientationin \, c_{r}(\relativeorientation) \opdff{\orientationin} \rsbra \, , 
\end{equation}
where $\widetilde{\rho}\equiv\rho \, |c_{0,0}|$ and a reduced $c_{2}$ is denoted $c_{r}(\relativeorientation) \equiv c_{2}(\relativeorientation)/|c_{0,0}|$. In order to
obtain \re{cartesianc2pexpanded} and in effect the above equation,
we needed to number the axes associated with laboratory and molecular frames.
We can always renumber(permute) the axes and require that the phase diagram 
does not change. It means that $\widetilde{\rho}^{\,\,'} \, c^{\,\,'}_{r}(\relativeorientation)=\widetilde{\rho} \, c_{r}(\relativeorientation)$, where the former stands for the $\widetilde{\rho}\,c_{r}(\relativeorientation)$ with renumbered axes. In particular, if we change the ''2'' and ''3''
axes of the molecules, and replace $c_{0,0}$, $\widetilde{c}_{0}$
and $\widetilde{c}_{2}$ with $c^{'}_{0,0}$, $\widetilde{c}^{\,\,'}_{0}$
and $\widetilde{c}^{\,\,'}_{2}$, where
\begin{equation}
 \label{eq:dualitytransformationc2}
\begin{split}
 sgn(c^{'}_{0,0}) |c^{'}_{0,0}|= & \, \frac{1}{4}|c_{0,0}|\lsbra sgn(c_{0,0})+2\sqrt{3}+3\,\widetilde{c}_{2} \rsbra \, , \\
 sgn(c^{'}_{0,0})\widetilde{c}^{\,\,'}_{0} = & \, \frac{6\,\widetilde{c}^{\,2}_{0}+\sqrt{3}\,\widetilde{c}_{0}\lsbra 3\,sgn(c_{0,0})-\widetilde{c}_{2} \rsbra+3\lsbra sgn(c_{0,0})-\widetilde{c}_{2} \rsbra \widetilde{c}_{2}}{\lsbra sgn(c_{0,0})+2\sqrt{3}\,\widetilde{c}_{0}+3\,\widetilde{c}_{2} \rsbra\lnbra 3\,\widetilde{c}_{0}+\sqrt{3}\,\widetilde{c}_{2} \rnbra} \, , \\
 sgn(c^{'}_{0,0})\widetilde{c}^{\,\,'}_{2} = & \, \frac{-6\,\widetilde{c}_{0}+\sqrt{3}\lsbra 3\,sgn(c_{0,0})+\widetilde{c}_{2} \rsbra}{6\,\widetilde{c}_{0}+\sqrt{3}\lsbra sgn(c_{0,0})+3\,\widetilde{c}_{2} \rsbra} \, ,
\end{split}
\end{equation}
then the \sceq\ will give the same phase behaviour of the system,
whether we use $\rho^{'}_{0}$, $c^{'}_{0,0}$, $\widetilde{c}^{\,\,'}_{0}$, $\widetilde{c}^{\,\,'}_{2}$ or $\rho_{0}$, $c_{0,0}$, $\widetilde{c}_{0}$, $\widetilde{c}_{2}$, because $\widetilde{\rho}^{\,\,'} \, c^{\,\,'}_{r}(\relativeorientation)=\widetilde{\rho} \, c_{r}(\relativeorientation)$. We have obtained a non-trivial
\ital{duality transformation} which connects the states of different density
and temperature. The interesting points are the self-dual ones, i.e., those left unchanged by the above duality transformation, they fulfil $\widetilde{c}^{\,\,'}_{0}=\widetilde{c}_{0}$ and $\widetilde{c}^{\,\,'}_{2}=\widetilde{c}_{2}$. Those are the point $(\widetilde{c}_{0},\widetilde{c}_{2})=(-sgn(c_{0,0})/\sqrt{3},-sgn(c_{0,0}))$ and the line
\begin{equation}
 \label{eq:selfdualregions}
 \widetilde{c}_{2} = & \, sgn\lnbra c_{0,0} \rnbra-2 \widetilde{c}_{0}/\sqrt{3} \, .
\end{equation}
This condition is the same line as the one obtained previously from
bifurcation analysis and expressed by \eqrefeq{landaupointcondition}.
It marks the cross-over between the states of prolate
and oblate symmetry \cite{longa,straley,ams}. In the similar
way we can work the other duality transformations \cite{longa}.
\nlin The Landau points can be found by both the analysis of the bifurcating 
state, and the study of the regions invariant under the duality
transformation. In the first method the Landau points are expressed by means
of eigenvectors of the operator related to the
pair direct correlation function, without employing any 
additional expansions. On the other hand the analysis of the self-dual
points is purely geometrical and requires nothing more but the $L=2$ 
model of $c_{2}$ and reasonable requirement of invariance under the permutation
of the axes of the molecule-fixed orthogonal tripod of vectors. 
\nlin The above analysis allowed to determine the symmetry of state bifurcating
from isotropic phase and the equation for Landau point. It is also possible
to find the condition for the point where the character of bifurcation changes,
it is presented in the next section.
\subsection{Equations for tricritical point}
\label{section:tricritical}
 As can be seen from \figreffig{bifurcationscenarios}, we can locate the points
where the character of the bifurcation changes and the corresponding
first order phase transition becomes continuous. This behaviour is observed
at tricritical point. The required condition
is $\rho_{1}=\rho_{2}=0$ \cite{longa} which marks the path
between schemes \figreffig{bifscenariofirstclosetotri}
and \figreffig{bifscenariosecondorder}. Detailed calculations assuming 
the expansions \re{c2pexpandeddeltas} and \re{opdfinops} give
\begin{equation}
\label{eq:tricriticaleqc2}
\begin{split}
& 3\overline{\eta_2^2}\,^2 - \overline{\eta_2^4} + 3 \frac{1}{\rho^{*\,2}} \left[
\left( \overline{\xi_0^2}- \overline{\xi_0}\,^2 - \frac{1}{\rho^{*}}
sgn(c_{0,0})\right) e^2 - 2\,(\overline{\eta_0}\,\overline{\xi_0} -
\overline{\eta_0 \xi_0} + \frac{1}{\rho^{*}} \widetilde{c}_0)\,e f \right. \\
 -& \left.
 (\overline{\eta_0}\,^2 - \overline{\eta_0^2} + \frac{1}{\rho^{*}} \widetilde{c}_2) f\,^2\right] +
 6 \,\overline{\eta_2^2}\, d \left( 2\, \overline{\eta_2\xi_2} +
 d \,\overline{\xi_2^2} \right) +
 d \left[ - 4 \,\overline{\eta_2^3
 \xi_2} \rinv \\ 
 & \linv + d \left( 12 \,\overline{\eta_2 \xi_2}\,^2 - 24 \,
 \overline{\eta_2^2 \xi_2^2} (3\, \overline{\eta_2 \xi_2}\, \overline{\xi_2^2}
 - \overline{\eta_2 \xi_2^3}\,)\, d  + \left( 3\,\overline{\xi_2^2}\,^2
 - \overline{\xi_2^4}\,\right) d^2 \right) \right]=0 \, ,
\end{split}
\end{equation}
where
\begin{equation}
\label{eq:triwheredefs}
\begin{split}
  d = & \, \frac{\overline{\eta_2 \xi_2} - \rho^{*\,-1} \widetilde{c}_0}{\rho^{*\,-1} sgn(c_{00})
  - \overline{\xi_2^2}} \, , \\
  e = & \, g\left[ g_3(\,lnZ_{2,3}\, \rho^{*\,-1}-\widetilde{c}_0) - g_2(\,lnZ_{3,3}\, 
  \rho^{*\,-1}-\widetilde{c}_2)\right] \, , \\
  f = & \, g\left[ -g_3(\,lnZ_{2,2} \,\rho^{*\,-1}-sgn(c_{0,0}) +
         g_2(\, lnZ_{2,3} \, \rho^{*\,-1}-\widetilde{c}_0) \right] \, , \\
  g^{-1} = & \, \rho^{*} Z_0 \left[ \,( \,lnZ_{2,3}\, \rho^{*\,-1} - \widetilde{c}_0)^2
  - (\,lnZ_{2,2}\, \rho^{*\,-1} - sgn(c_{0,0}) (\,lnZ_{3,3}\, \rho^{*\,-1} - \widetilde{c}_2)
  \right] \, , \\
  g_m = & \, lnZ_{m}\, Z_{0,0} - Z_{0,0,m} +
     d\, ( 2\, lnZ_{m}\, Z_{0,1}- 2\, Z_{0,1,m}
   +  d \,(\, lnZ_{m}\, Z_{1,1} - Z_{1,1,m})) \, ,
\end{split}
\end{equation}
and where all of the averages are calculated in the reference,
uniaxial nematic phase, $\rho^{*}$ stands for bifurcation point for the case
of \nun\ -- \nbx\ transition, $Z$ is from \ere{scuniformcase} and $Z_{m,n}$
are derivatives of $Z$ with respect to $x_0 \equiv
\average{\Delta^{2}_{2,2}}$, $x_1 \equiv
\average{\Delta^{2}_{2,0}}$, $x_2 \equiv
\average{\Delta^{2}_{0,0}}$, $x_3 \equiv
\average{\Delta^{2}_{0,2}}$, and also $\xi_m = sgn(c_{0,0})\average{\Delta^{2}_{m,0}} + \widetilde{c}_0
\average{\Delta^{2}_{m,2}}$,
$\eta_m = \widetilde{c}_0\average{\Delta^{2}_{m,0}} + \widetilde{c}_2
\average{\Delta^{2}_{m,2}}$.
\nlin The formulas \re{tricriticaleqc2}-\re{triwheredefs} become more
complicated when in the expansion of $c_{2}$ \re{expansionofc2l2} terms
with angular momentum index $l>2$ are included and when the
higher order direct correlation functions are taken into account,
therefore, the above equations are approximate expressions.
\section{Summary}
\indent We have presented a general formulation of the so--called \sldft.
It is a well known theoretical tool widely used for simple liquids
and in the crystallization phenomena, applied with great success to
the inhomogeneous fluids and interfaces, as well as in studies of
the elasticity of liquid and solid crystals. It was also used
in the description of mesophase diagrams and yielded reasonable results
in comparison with simulations.
\nlin In the present work we are only interested
in the behaviour of the system close to the transition point. So in essence
we are looking for the points where the number or structure of local minima of 
$\intrinsicfef{\densitysymbol}$ changes. One way of addressing
this issue is to postulate a model for the direct correlation functions and
use the variational principle to locate numerically the extrema of the
intrinsic Helmholtz free energy. A first step towards this goal is to study the
{\sceq} \refeq{eq:selfconsistent2} in pursuit of the points where from the
stable reference state a new solution bifurcates. This approach allows
to establish a connection between the microscopic quantities such as
molecular dimensions, interaction parameters, etc., and global,
macroscopic properties of the system, such as the type and degree
of order of the bifurcating state.
\nlin We derived exact bifurcation equations for arbitrary
reference state $\pref$. The point of bifurcation was fully determined 
by the pair direct correlation function $c_{2}$ and \sopdf\ $\pref$,
while its character (first order vs multicritical) depended on higher 
direct correlations. Then, we presented a specific form of the equations
in spatially uniform regime for arbitrary symmetry of phase and molecules
and given $c_{2}$ and $\pref$. The bifurcation equation was cast in the form
of eigenproblem for bifurcation matrix $\oper{\boldm{\omega}}^{(l)}$
of dimension not exceeding $(2\,l+1)x(2\,l+1)$,
where $l$ stands for the angular momentum index of bifurcating state.
In particular, we showed that $c_{2}$ contributes to the bifurcation point
via coefficients of the expansion in symmetry adapted base with 
the same angular momentum index as the bifurcating representation.
We also commented on the splitting of the space of solutions of
bifurcation equations induced by $c_{2}$, which in conjunction with
with trivial structure of $\pref$ leads to the same bifurcation
equations for isotropic reference state.
\nlin Then we turned to the case of molecules and phase of \dtwh symmetry.
We chose the order parameters accordingly to the symmetry of the nematic states,
and work out the bifurcation equations. For transitions where higher symmetric
phase is taken to be isotropic, the bifurcation point was fully determined by
the coefficients $\{\clmn{l}{m}{n}\}$ of the expansion of
pair direct correlation function with angular momentum index $l=2$.
The eigenvectors associated with biaxial and uniaxial phase gave the
eigenequations with the same bifurcation matrix, leaving the symmetry
of the bifurcating state to be determined by the
second order bifurcation equation. For the uniaxial -- biaxial nematic
bifurcation the structure of the equation was similar, but it contained 
explicitly the order parameters calculated in the uniaxial(reference) state
for angular momentum index $2 \leq l \leq 4$. In concluding sections we
presented the method of finding Landau and tricritical points. The
analysis of the duality transformation of the
expansion of pair direct correlation function up to $l=2$
showed the existence of self-dual regions, which coincided with Landau points,
as was proven by calculation of the bifurcating state symmetry. Finally, we
gave the condition for the tricritical point expressed in terms of $\{\clmn{2}{m}{n}\}$. In this case the choice of the
subspace with $l=2$ in the expansion of $c_{2}$, and the use of only this
element from the set of direct correlation functions was an approximation,
since to this equation also terms with higher angular index, as well as $c_{n}$
for $n>2$ contribute.
\nlin As was mentioned before, the real point of the first order transition lies
in slightly lower densities (or higher temperatures) than the bifurcation point,
therefore in this case, bifurcation diagrams are only an approximation for the
true phase sequence. The true equilibrium state can be derived by the analysis
of the minima of the free energy. This procedure, although numerically more
complex, can give a definite answer to the question of the state that gains
stability. \nlin In this work, as far as the transitions from the isotropic phase are
concerned, we can expect only two scenarios: to either enter uniaxial or
biaxial nematic, and when the reference phase is taken to be uniaxial nematic
only one: to stabilize the biaxial nematic phase. Those transitions are
weakly first or second order, and while other liquid crystal mesophases are
neglected, the bifurcation analysis is a good method for estimation
of the potential possibility that phase of given symmetry could be
realized in practice. In more complex cases a care should be taken, for 
the bifurcations do not necessary provide the correct physical results.
Especially it should be noted once more, that certain transitions are not
immediately accessible from the bifurcation analysis; only the transitions
between states the symmetries of which can be connected by the group -- subgroup
relation can be straight away determined in this approach.
\nlin Present work is based on the analysis of
the {\sceq} in the manner described above, that is, we focus our attention
on the determination of transition densities and temperatures by analysing 
the equation for equilibrium \sopdfns, while assuming a given
pair direct correlation function, which will be in fact modelled
with the help of the mean field and virial expansion (as described in the following chapter).
It is also possible to address the issue of determination of thermodynamic
properties of the system by calculating direct correlation functions as, e.g.,
in \cite{percusyevick}.
\nlin In the following chapters we turn to the determination
of the behaviour of the model systems in vicinity of the phase transition
mainly by finding the solutions of the bifurcation equations \refeq{eq:exactbifisoldl2} and \refeq{eq:bifunbx}, and also by analysing the tricritical condition \re{tricriticaleqc2} and minimising the free energy.
To perform the actual calculations we postulate
a model pair direct correlation function $c_{2}$, which is equivalent to making
certain approximations for the excess Helmholtz free energy $\excessfesymbol$ \re{taylorexp}
and the reference state in the Taylor expansion \refeq{eq:taylorexpwithdcfs}.
In the next chapter we start with the description of these additional assumptions,
and then continue to study transitions to biaxial nematic for
model pair potentials.

 \chapter{Model calculations}
 \label{chapter:results}

{ \chapabsbegin In the previous chapter we introduced the
\sldft\ and bifurcation analysis with a special emphasis on
the presentation of exact results. We developed the conditions for
tricritical and Landau points following from duality transformation
as well as from the analysis of the symmetry of bifurcating state,
and derived bifurcation equations for transitions involving
spatially uniform phases. In this chapter the solutions for
those are presented in the form of phase diagrams at
bifurcation point, for three models of the biaxial nematic. \chapabsend }
\section{Introduction}
 We start with the description of numerically tractable approximation
of the excess\linebreak Helmholtz free energy carried out through modelling the 
direct correlation functions, which is used in calculations.
Then, we consider the mean field of a simple two-point interaction as an
introduction of the basic ideas, as well as an example of
the generic \dtwh -- symmetric model giving rise to the biaxial nematic phase,
Landau, and tricritical points. In this case we illustrate the use of the
duality transformation using Eqs.~\refeq{eq:dualitytransformationc2}
and formula \re{tricriticaleqc2}.
In following sections we consider the biaxial Gay-Berne potential and a model
of the \sbentcore molecule using second order virial approximation for the
excess Helmholtz free energy \refeq{eq:taylorexp}. In the first case we aim at
determining the topologies of the phase diagrams as function of the
potential parameters; we focus on the competition between molecular dimensions
and various interaction forces in the stabilisation of the biaxial nematic
phase and their influence on the Landau point. Then, we study models of
\sbentcore molecules, where we seek the main factors that make {\nbx} stable
in those systems. We use Gay-Berne interacting segments to model
the molecules, and include dipole-dipole interaction. Most of the diagrams
presented in this chapter follow from the numerical solution of the
bifurcation equations \refeq{eq:exactbifisoldl2} and \refeq{eq:bifunbx},
where $\rho_{0}$ is identified with the bifurcation average number density,
and temperature dependence is stored in coefficients $\clmn{2}{m}{n}$ and
order parameters $\averageref{\symmetrydeltana{l}{0}{m}}$.
The equations are solved in iterative manner, with $\clmn{2}{m}{n}$ calculated
numerically (see \aprefap{numericaldetails}).
\section{Models of direct correlation functions studied}
\label{section:modelsofdcfs}
\begin{figure}[t]
 \centering
 \subfigure[Example of pair intermolecular potential.]{\label{figure:intermolecularpotentialexample} \includegraphics{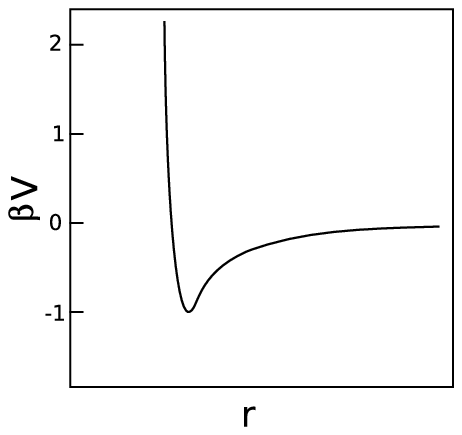}}
 \subfigure[Mean field approximation.]{\label{figure:mfexampleofc2} \includegraphics{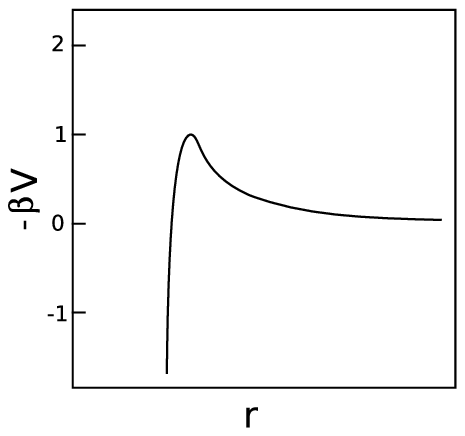}}
 \subfigure[Low-density approximation.]{\label{figure:ldappofc2} \includegraphics{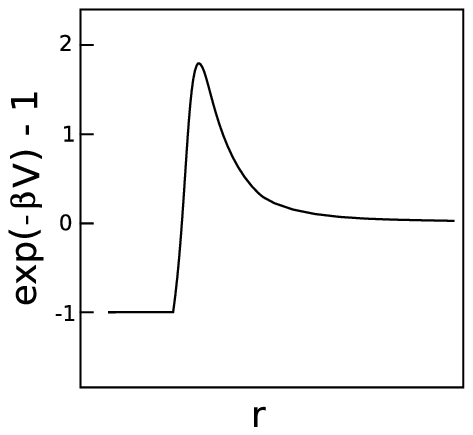}}
\caption[Popular approximations for pair direct correlation function $c_{2}$.]{\label{figure:potentialmfld} Two popular approximations for the pair direct correlation function $c_{2}$ (Figs.~\subref{figure:mfexampleofc2} and \subref{figure:ldappofc2}) for an example of model two-point potential $V$ (Fig.~\subref{figure:intermolecularpotentialexample}) as function of intermolecular distance $r$ (remaining degrees of freedom are held constant). Low-density approach in Fig.~\subref{figure:ldappofc2} models more correctly the volume which is forbidden for molecular centres of masses due to impenetrability of ''hard core'' of molecules. Also in comparison to mean field in Fig.~\subref{figure:mfexampleofc2} the contribution from the minimum of $V$ is stronger.}
\end{figure}

 Up to now we dealt with exact equations; besides postulating the $L=2$ model
expansion \refeq{eq:expansionofc2l2} for $c_{2}$ (justified in case of bifurcations), we have made no approximations concerning the
direct correlation functions both in the
Taylor expansion \refeq{eq:taylorexpwithdcfs} and
bifurcation equations \refeq{eq:bifurcationeq},\refeq{eq:bifeqsecondorder}.
However, in order to find the solutions for
Eqs.~\refeq{eq:exactbifisoldl2}-\refeq{eq:bifunbx}, we need a way to calculate
$c_{2}(\coord{1},\coord{2})$ for given $\coord{1}$ and $\coord{2}$.
To this end we have to postulate a model function, and use it with
orthogonality conditions for
symmetry adapted functions \refeq{eq:orthogonalityofdelta}
to calculate the coefficients $\clmn{2}{m}{n}$ in \refeq{eq:expansionofc2l2}.
We choose the so called low-density approximation, also known as
second order viral, or Onsager approximation \cite{onsager}.
In this approach we set 
\begin{equation}
 \label{eq:ldsettings}
\begin{split}
 \beta\seconddcf{\coord{1}}{\coord{2}}{\densitysymbol_{ref}} = & \exp{\lsbra-\beta V(\coord{1},\coord{2})\rsbra}-1\equiv\meyerfunc \, , \\
 \beta c_{n}(\coord{1},\dots,\coord{n},\lsbra \densitysymbol_{ref} \rsbra) = & 0 \, , \, n \geq 3 \, ,
\end{split}
\end{equation}
where $V(\coord{1},\coord{2})$ stands for pair potential and $\meyerfunc$ for
second Meyer function. In terms of the Taylor expansion \refeq{eq:taylorexpwithdcfs} it means setting $\densitysymbol_{ref}=0$, $\excessfef{\densitysymbol_{ref}}=0$, and taking $\excessfesymbol$ as 
\begin{equation}
 \label{eq:ldapproximation}
 \beta\excessfef{\densitysymbol}=-\frac{1}{2}\int d\coord{1}d\coord{2} \, \densityf{\coord{1}}\lcbra \exp\lsbra-\beta V(\coord{1},\coord{2})\rsbra-1\rcbra\densityf{\coord{2}} \, .
\end{equation}
Both choosing the state of zero density as the reference for the expansion of
the excess Helmholtz free energy and approximating the structure of 
pair correlation function by low-density model \re{ldsettings} seem 
like great simplifications (see \figreffig{ldappofc2}). We can
expect that this approach will give higher transition temperatures than,
e.g., Monte Carlo simulations. This overestimation is mainly due to the lack of
proper treatment of entropy. Taken all this into account, the phase diagrams
produced using this method for spatially uniform mesophases are surprisingly
accurate \cite{ginzburggb}.
\nlin There exists a simple limit to the \re{ldapproximation}; when
the molecules become very long and thin,
the main contribution to the excess Helmholtz free energy will come from the
two-body excluded volume. Indeed, in this case, we can assume hard interaction,
i.e., $V(\coord{1},\coord{2})=\infty$
when molecules are in contact, and $V(\coord{1},\coord{2})=0$ otherwise,
and then \re{ldapproximation} is replaced by:
\begin{equation}
 \label{eq:onsager}
 \beta\excessfef{\densitysymbol}=\frac{1}{2}\int d\coord{1}d\coord{2}\, \densityf{\coord{1}}\Theta(\xi(\coord{1},\coord{2})) \densityf{\coord{2}} \, ,
\end{equation}
where $\Theta$ is the Heaviside step function, and where $\xi$ is the 
contact function.
\ere{onsager} represents well known Onsager model \cite{onsager},
which becomes exact for pair interactions with anisotropic hard core and
attractive soft tail in the limit of large molecular elongations
(when length-to-width ratio is greater than $10:1$).
\nlin Often even a simpler model than low-density \re{ldapproximation} is used. The so-called
mean field theory can be obtained from high temperature expansion of the
Meyer function $\meyerfunc$ in \refeq{eq:ldapproximation}\footnote{For the description of the usual method of derivation of \ere{mfapproximation} please refer to the Appendix~\refap{meanfield}.}.
In this theory the excess part of the Helmholtz free energy reads
\begin{equation}
 \label{eq:mfapproximation}
 \beta\excessfef{\densitysymbol}=\frac{1}{2}\int d\coord{1}d\coord{2} \, \densityf{\coord{1}} \lsbra \beta V(\coord{1},\coord{2}) \rsbra \densityf{\coord{2}} \, .
\end{equation}
For a representation of the above approximations see \figreffig{potentialmfld}.
\nlin In present chapter we will only consider the
spatially uniform phases, mostly in the low-density approximation.
As we have mentioned, the pair potential and correlation function
depend on the relative position and orientation of the molecules: $\meyerf{\coord{1}}{\coord{2}}=\meyerf{\orientation^{-1}_{1}\orientation_{2}}{\boldv{r}_{1}-\boldv{r}_{2}}$. In that case the intrinsic Helmholtz free energy functional \re{taylorexpwithdcfs} using \re{ldsettings}-\re{ldapproximation} can be cast as
\begin{equation}
 \label{eq:freeenergyuniform}
 \beta\intrinsicfef{P}=V \rho\int d\orientation\, \opdf\ln{\opdf}
 -\frac{1}{2} V \rho^{2}\int d\orientation d\orientationin\, \opdff{\orientation} \meyerf{\orientationin}{\orientation} \opdff{\orientationin} \, ,
\end{equation}
where as before $\rho=\average{N}/V$ stands for average number density,
$\opdf$ is the \sopdf normalized to unity, and $\meyerf{\orientationin}{\orientation}=\int d^{3}\boldv{r} \meyerf{\relativeorientation}{\vec{\mathbf{r}}}$. In equilibrium $\opdf$ fulfils the self-consistent equation \re{scuniformcase}, which reads
\begin{equation}
 \label{eq:scuniformcase2}
 \opdf=Z^{-1}\exp\lsbra \rho \int d\orientationin \, \meyerfsymbol(\orientationin,\orientation) \opdff{\orientationin} \rsbra \, ,
\end{equation}
where $Z=\int d\orientation \exp\lsbra \rho \int d\orientationin \, \meyerfsymbol(\orientationin,\orientation) \opdff{\orientationin} \rsbra$ ensures the normalization. In the above equation, the temperature is present in $\meyerfsymbol(\orientationin,\orientation)$ through the formula \re{ldsettings} and $\rho$ is an
independent parameter. It is not the case when $\meyerfsymbol(\orientationin,\orientation)=-\beta\int d^{3}\boldv{r} \, V(\orientationin,\orientation,\boldv{r})$, i.e., when mean field model is in force and the dependence between $\rho$
and $\beta$ is linear. Thus, the low-density approximation allows to
obtain the more realistic (non-linear) phase diagrams in density-temperature
plane, while, on the other hand, mean field works well for
lattice models \cite{biscarinilattice}.
\nlin The formulas \refeq{eq:freeenergyuniform}-\refeq{eq:scuniformcase2} can
be used to calculate pressure $\textnormal{P}=\mu \rho - \intrinsicfesymbol/V$
and chemical potential $\mu=\frac{\partial \lnbra\intrinsicfesymbol/V\rnbra}{\partial\rho}$:
\begin{equation}
 \label{eq:pressureandchemicalpotential}
\begin{split}
 \beta \textnormal{P} = & \rho-\frac{1}{2}\rho^{2}\int d\orientation\, d\orientationin\, \opdff{\orientationin} \meyerf{\orientationin}{\orientation} \opdff{\orientation} \, , \\
 \beta \mu = & 1+\ln\rho+\int d\orientation\, \opdf\ln\opdf-\rho \int d\orientation\, d\orientationin\, \opdff{\orientationin} \meyerf{\orientationin}{\orientation} \opdff{\orientation} \, .
\end{split}
\end{equation}
Once we have chosen the approximation for the direct correlation function $c_{2}$
like \refeq{eq:ldapproximation} or \refeq{eq:mfapproximation}, we can 
calculate the coefficients $\{\clmn{2}{m}{n}\}$ in \refeq{eq:expansionofc2l2},
and then use \eqrefeq{opdfinops} and \refeq{eq:orderparamsset}
to obtain the self-consistent equations for order parameters:
\begin{equation}
 \label{eq:orderparametersselfconsistent}
 \orderparameter{2}{m}{n}=Z^{-1}\int d\orientation \, \exp\lsbra\rho\int d\orientationin\,\meyerfsymbol(\orientationin,\orientation)\opdff{\orientationin} \rsbra \symmetrydelta{2}{m}{n}{\orientation} \, .
\end{equation}
Then we can solve the bifurcation equations \refeq{eq:exactbifisoldl2}
and \refeq{eq:bifunbx}.
\nlin Another possibility to obtain the behaviour of a system close to 
phase transition is to choose a trial \sopdf $\opdffin{trial}$ with some
adjustable parameters and use the variational principle \refeq{eq:variationalforf}
for the Helmholtz free energy $\intrinsicfef{P}$ \refeq{eq:freeenergyuniform}
to locate \sopdf corresponding to equilibrium state by finding a set
of the parameters that give $\opdffin{trial}$ which minimizes $\intrinsicfef{P_{trial}}$. We choose the trial function of the following form,
inspired by \ere{scuniformcase2}, and similar to mean field \re{mfeqapp}:
\begin{equation}
 \label{eq:usedmodelopdf}
 \opdffin{trial}=Z^{-1}_{trial}\exp\lsbra \sum_{m,n}\alpha_{mn}\symmetrydelta{2}{m}{n}{\orientation} \rsbra \, ,
\end{equation}
where as usual $Z^{-1}_{trial}=\int d\orientation \exp\lsbra \sum\alpha_{mn}\symmetrydelta{2}{m}{n}{\orientation} \rsbra$ ensures normalization.
As mentioned in Appendix~\refap{opdf}, this is not the only model available.
Employing the above formula allows to find the minimum
of $\intrinsicfef{P_{trial}}$ as function of $\alpha_{mn}$, but it is easier to
rewrite \refeq{eq:scuniformcase2} as the equations for $\alpha_{mn}$, and once
those are solved, locate the minimizing set. The resulting equations have
similar form to the \eqrefeq{orderparametersselfconsistent}, namely
\begin{equation}
 \label{eq:scforalpha}
 \alpha_{mn}=\frac{5}{8\,\pi^{2}}\,\rho\int d\orientation \, d\orientationin \, d^{3}\vec{\mathbf{r}} \,\, \symmetrydelta{2}{m}{n}{\orientation} \meyerfsymbol(\orientationin,\orientation,\vec{\mathbf{r}}) \opdfffin{\orientationin}{trial} \, .
\end{equation}
When the equilibrium set of parameters $\left.\alpha_{mn}\right|_{eq}$ is found,
we can use the equilibrium \sopdf $\left.\opdffin{trial}\right|_{eq}=\left.Z^{-1}_{trial}\right|_{eq}\exp\lsbra \sum\left.\alpha_{mn}\right|_{eq}\symmetrydelta{2}{m}{n}{\orientation} \rsbra$
to calculate the order parameters along \eqrefeq{orderparamsset}.
\nlin In the following section, we present the bifurcation study of a
simple $L=2$ model in mean field approximation (MFA), then, using the
low-density approach we consider the models of biaxial Gay-Berne interaction
and \sbentcore systems. Mostly we present the solutions of the
equations \refeq{eq:exactbifisoldl2} and \refeq{eq:bifunbx} using \re{ldsettings}-\re{ldapproximation} and determining coefficients $\clmn{2}{m}{n}$ 
by calculating numerically the integrals in \re{clmncoeffs}. In one case we
apply the model \sopdf \refeq{eq:usedmodelopdf} and minimize the
Helmholtz free energy. A more technical description of the way $\clmn{2}{m}{n}$
were obtained and the methods of addressing \eqrefeq{orderparametersselfconsistent} are presented in Appendix~\refap{numericaldetails}.
\section{Mean field of $L=2$ model of biaxial nematic}
 \label{section:meanfield} 
 Present section is devoted to the detailed analysis of the generalized 
Straley's \cite{straley} intermolecular potential model. It is the simplest
interaction giving rise to the biaxial nematic. In the same way as 
in the Maier-Saupe case, where the orientational part of the potential energy
possesses the same symmetry as the uniaxial nematic \cite{maiersaupe}, present
model is \dtwh -- symmetric and describes systems of molecules of according
symmetry. It contains a complete set of four order parameters needed to
fully describe the biaxial nematic. It allows to construct a
generic bifurcation diagram in MFA, which is exact in this case, as function
of the coupling constants, and also gives general ideas
about the mechanisms of stabilisation of \nbx. Furthermore,
the model exhibits tricritical behaviour, and additional analysis of its
symmetries reveals the existence of the Landau regions, as we have seen
in \srs{exactlandaueqs} and \rs{tricritical}.
\nlin The generalization of the Maier-Saupe potential to the \dtwh symmetry
reads \cite{straley}
\begin{equation}
 \label{eq:v2pexpandeddeltas}
\begin{split}
 V(\relativeorientation) = \, & v_{0,0}\symmetrydelta{2}{0}{0}{\relativeorientation}+v_{0,2}\symmetrydelta{2}{0}{2}{\relativeorientation} \\
 + & v_{2,0}\symmetrydelta{2}{2}{0}{\relativeorientation}+v_{2,2}\symmetrydelta{2}{2}{2}{\relativeorientation} \, .
\end{split}
\end{equation}
Many special cases of this model were studied previously. We will briefly
describe those results, and indicate the relation to the present model. In the
following studies, it was {\it a priori} assumed that the interaction is of the
form \refeq{eq:v2pexpandeddeltas} with $v_{m,n}$ as constants, it is the
simplest possible approximation that can give rise to the biaxial nematic phase,
more than thirty years after its introduction it remains a subject of studies.
\nlin Both mean field and simulations were used in the past to deal with
the Straley's interaction \re{v2pexpandeddeltas} or even more simplified models.
Among them were dispersion \cite{luckhurstquad,mukherjeequad,biscariniquad,chiccoliquad,pasiniquad,romanoquad}, two-tensor \cite{sonnetquad} and amphiphilic \cite{romanoquad} models. \tablereftable{simpliestmodels} has the relation of those
to parameters used here.
\begin{table}[t]
 \caption{Parameters of models studied so far in relation to our model.}
 \begin{tabular}{cccc}
 \label{table:simpliestmodels}
 Model & $v_{0,0}$ & $v_{0,2}$ & $v_{2,2}$ \n \hline\hline
 Straley \cite{straley} & $\beta$ & $(2/sqrt{3})\gamma$ & $\delta$ \n
 Two-tensor \cite{sonnetquad} & $-1$ & $-\sqrt{3}\gamma$ & $-3\lambda$ \n
 Dispersion \cite{luckhurstquad,mukherjeequad,biscariniquad,chiccoliquad,pasiniquad} & $-1$ & $\pm\sqrt{2}\gamma$ & $-2\lambda^{2}$ \n
 Amphiphilic \cite{romanoquad} & $0$ & $0$  & $-1$ \n
\end{tabular}
\end{table}
\nlin For dispersion model the mean field and Monte Carlo
predicted a single Landau point for $\lambda=1/\sqrt{6}$, where second
order isotropic-biaxial nematic transition takes place \cite{luckhurstquad,mukherjeequad,biscariniquad,chiccoliquad,pasiniquad}. Simulations also suggest that
when $v_{0,2} \ne 0$ and $v_{2,2}=0$ \nbx\ does not emerge. Therefore
''the most'' minimal coupling capable of producing the biaxial nematic phase
is $v_{0,2}=0$ and $v_{2,2}<0$. That led to the extreme simplification
setting only $v_{2,2}=-1$, for which bot MFA and simulations predict a
direct \iso\ -- \nbx\ transition \cite{romanoquad}. Two-tensor model
was also studied in the context of stability of biaxial nematic and showed
isolated Landau points \cite{sonnetquad}. There also a special case
of $\gamma=0$ was examined by mean field. For $0<\lambda<0.2$ uniaxial-biaxial nematic transition was found to be of second order, 
while for $0.2<\lambda<0.22$ it was first order and finally for $\lambda>0.22$ a first 
order isotropic -- \nbx\ transition took place.
\begin{figure}[b]\includegraphics[scale=0.6]{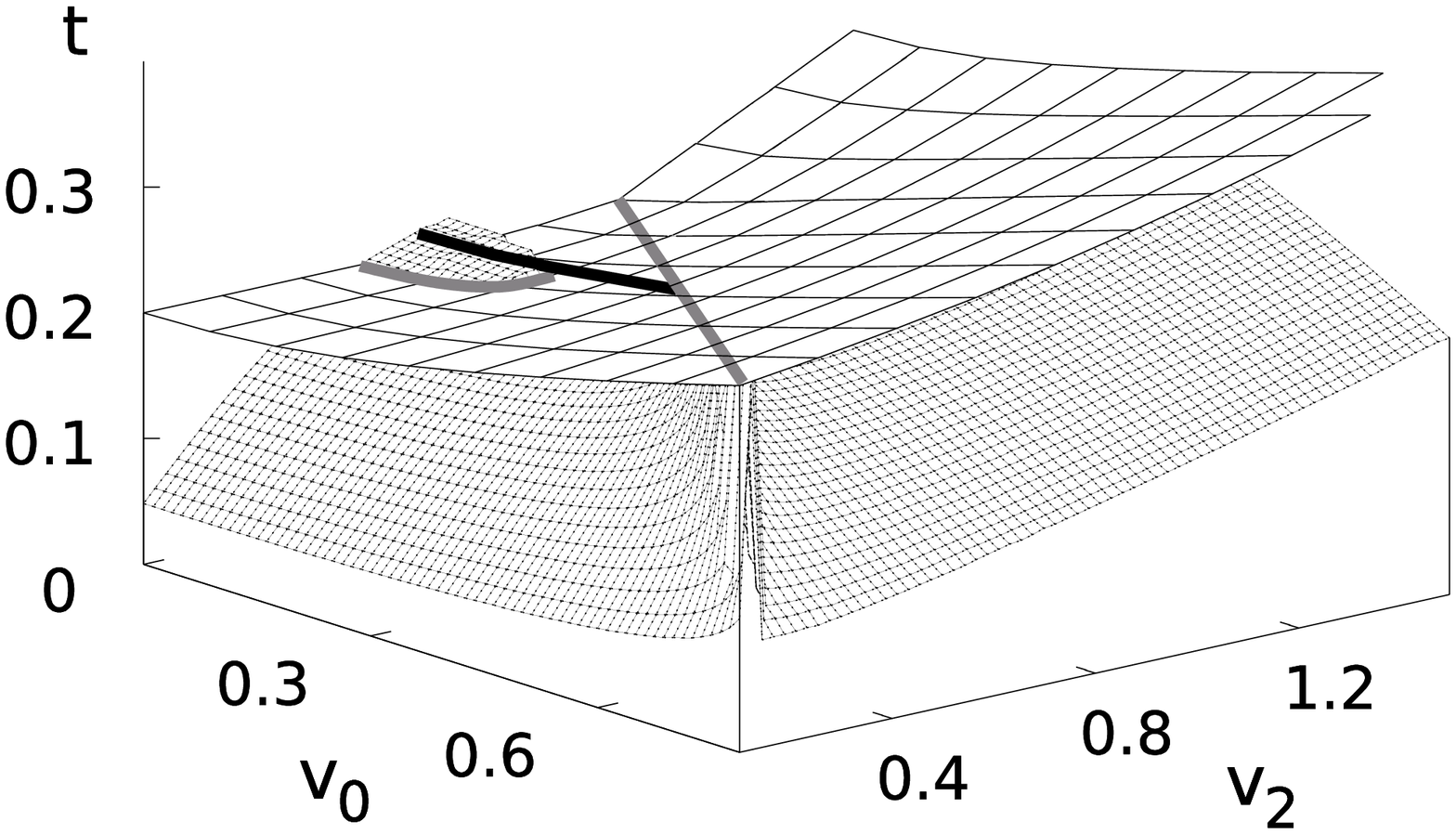}\caption[Bifurcation diagram for $L=2$ model in meanfield,]{\label{figure:l2mfdiagram}Mean field bifurcation diagram for $L=2$ model, for $sgn(v_{0,0})>0$. Tricritical points are represented by thick black line and self-dual line of Landau points is marked with grey. Upper surface stands for bifurcation from isotropic phase, the lower one to \nun\ -- \nbx\ case.}\end{figure}
\nlin The possibility of the tricritical point on the line of 
\nun\ -- \nbx\ transitions was studied for
two-tensor model \cite{sonnetquad} (where it was found) and on the line
of \iso\ -- \nbx\ transitions in \cite{matteisquad}. Also some extension to
mean field model, concerning the behaviour of order parameters and
localizing the equilibrium state, was presented \cite{bisiquad1,bisiquad2}.
\nlin In following sections we are taking into account a general model
as described by us in \cite{longa} and show the bifurcation study in MFA.
\subsection{Bifurcation for $L=2$ model}
 Having chosen a model interaction of the form of \refeq{eq:v2pexpandeddeltas},
we can apply MFA described in Appendix~\refap{meanfield}, and expressed
by \eqrefeq{mfapproximation}. In the case of $L=2$ model we can present
exact solutions for the bifurcation equations \refeq{eq:exactbifisoldl2}
and \refeq{eq:bifunbx}. The analysis which was conduced for $L=2$ model
for direct correlation function $c_{2}$ \re{expansionofc2l2} holds for the analogical
expansion for pair potential \re{v2pexpandeddeltas}. We can immediately use
the bifurcation equations \re{exactbifisoldl2}-\re{bifunbx} and conditions for
Landau (Eq.~\re{landaupointcondition}) and tricritical (Eq.~\re{tricriticaleqc2}) points,
keeping in mind that the relation between coefficients present in \re{expansionofc2l2} and \re{v2pexpandeddeltas} is: $\clmn{2}{m}{n}=-\beta v_{m,n}$.
\nlin Without loss of generality \eqrefeq{v2pexpandeddeltas}
can be rewritten as
\begin{equation}
\begin{split}
 V(\relativeorientation) = -|v_{0,0}| & \lcbra sgn(v_{0,0})\symmetrydelta{2}{0}{0}{\relativeorientation}+v_{0}\lsbra\symmetrydelta{2}{0}{2}{\relativeorientation}+\symmetrydelta{2}{2}{0}{\relativeorientation}\rsbra \right. \\
 & \left. \,\,\,\, + v_{2}\symmetrydelta{2}{2}{2}{\relativeorientation} \rcbra \, ,
\end{split}
\end{equation}
where $v_{0}\equiv v_{0,2}/\abs{v_{0,0}}$ and $v_{2}\equiv v_{2,2}/\abs{v_{0,0}}$,
and due to particle interchange symmetry we have $v_{2,0}=v_{0,2}$.
We will use the reduced temperature $t=k_{B}T/(\abs{v_{0,0}}\rho)$.
The temperature for bifurcation from isotropic phase can be calculated as
\begin{equation}
 \label{eq:mfl2bifiso}
 t=\frac{1}{10}\left[ sgn(v_{0,0}) + v_2 + \sqrt{\left(sgn(v_{0,0})-v_2\right)^2 + 4\,v_0^2} \right] \, .
\end{equation}
The bifurcation equation for the uniaxial -- biaxial nematic transition
can be cast in the following form:
\begin{equation}
 \label{eq:mfl2bifunbx}
 v_{2}=\frac{\lnbra a^{2}-b\,c \rnbra-\lsbra sgn\lnbra v_{0,0}\rnbra+2\,a\,v_{0} \rsbra\,t+t^{2}}{sgn\lnbra v_{0,0} \rnbra\lnbra a^{2}-b\,c \rnbra+b\,t} \, ,
\end{equation}
where
\begin{equation}
\begin{split}
 70\,a = & 20\,\overline{\Delta^{(2)}_{0,2}} +{\sqrt{15}}\overline{\Delta^{(4)}_{0,2}}\, , \\
 70\,b = & 14 + {{20\,\overline{\Delta^{(2)}_{0,0}} +\overline{\Delta^{(4)}_{0,0}} +{\sqrt{35}}\overline{\Delta^{(4)}_{0,4}}}}\, , \\
 70\,c = & 14  -20\,\overline{\Delta^{(2)}_{0,0}} +6\, \overline{\Delta^{(4)}_{0,0}}\, .
\end{split}
\end{equation}
\nlin The bifurcation diagram following from Eqs.~\refeq{eq:mfl2bifiso}-\refeq{eq:mfl2bifunbx} with tricritical points obtained using \ere{tricriticaleqc2} 
is presented in \figreffig{l2mfdiagram} (in the analysis we set $sgn(v_{0,0})=1$). \frf{tricriticalpoints} shows the line of tricritical points (marked with
thick black line on the upper surface in \frf{l2mfdiagram}) in $(v_{0},v_{2})$
plane in region of rod-like states. Using the duality transformation \re{dualitytransformationc2} it can be projected onto the rest of the parameter space.
\begin{figure}[t]\includegraphics[scale=0.8]{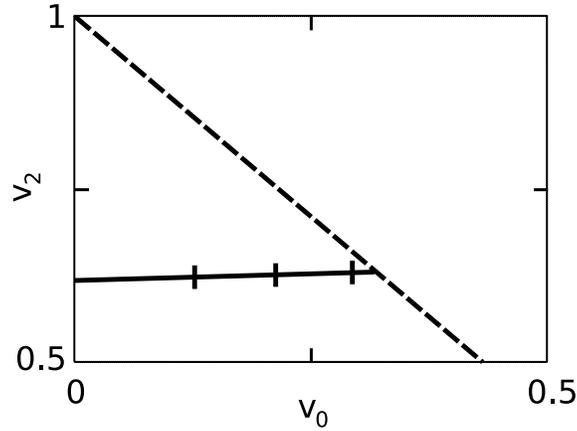}\caption[$v_2$ versus $v_0$ at tricritical point.]{\label{figure:tricriticalpoints}$v_2$ versus $v_0$ at tricritical point in the region of rod-like states. The self-dual (dashed)line $v_2=1-\frac{2}{\sqrt{3}}v_0$ is shown as well. Tricritical temperature varies from $t=0.2147$ for $v_0=0$ till $t=0.2367$ at $v_0=0.3194$, where the tricritical line meets the Landau line. Tick  marks correspond to intermediate, equally spaced temperatures.}\end{figure}
\subsection{Summary}
 The $L=2$ model \refeq{eq:v2pexpandeddeltas} developed by Straley \cite{straley}, consistent with the choice
of order parameters as averages of functions $\symmetrydelta{2}{m}{n}{\orientation}$ for $m,n=0,2$ \re{deltasl2setun}-\re{deltasl2setbx}, despite its simplicity provides a rich behaviour.
We have shown how the bifurcation temperature changes as function of the
potential parameters, and also solved a condition for the tricritical point.
Since the coefficients $v_{0,0}$, $v_{0}$, and $v_{2}$ can be related to the
model specific, microscopic parameters, the results presented here
contain a doze of generality; they are valid for any expansion of the
$L=2$ type, including \eqrefeq{expansionofc2l2}. Many potentials can be
accurately modelled in this way, some of the others can be cast on
the $L=2$ subspace and localized on the diagram in \figreffig{l2mfdiagram}.
In this way one can determine the minimal, required conditions
for \nbx\ to emerge.
In that sense the Straley's model, after over thirty years, still attains
its significance in the search for the biaxial nematic. Yet it should be 
kept in mind that terms of higher order in the expansion \refeq{eq:v2pexpandeddeltas} may play an important role and the bifurcation diagram may differ from the true
phase diagram.
\nlin In the following part of the thesis, we consider models based on the
Gay-Berne interaction: the generalized biaxial version of this potential
and the system of \sbentcore molecules constructed from
soft, uniaxial ellipsoids. Both cases are studied in the $L=2$ model
of pair direct correlation function \refeq{eq:expansionofc2l2}, which
for these potentials is a good choice. We start with the description
of the uniaxial Gay-Berne interaction since it introduces the 
methodology of soft potentials (which is also employed in \dtwh -- symmetric
case) and is used later to build \sbentcore molecules.
\section{Biaxial Gay-Berne model}
\label{section:BZ}
 In this section we investigate the biaxial Gay-Berne 
model. We begin with a definition of the generalized potential
as was done by Berardi, Fava, and Zannoni \cite{bzdevelopmentofpotential},
then, we present the results of bifurcation analysis, and where possible compare
with results published up to now. Since we are dealing with the Gay-Berne
interaction, we start with a brief description of the classic version
of this model \cite{bernepechukas,gayberne}. Next, we show the generalization
to the case of biaxial ellipsoids, as was done in \cite{bzdevelopmentofpotential}, and then turn to the results of our analysis. We present
the bifurcation diagram with the results of simulations marked, and take a
brief look at the method of minimising the Helmholtz free energy
using trial \sopdf \refeq{eq:usedmodelopdf}. In the following part of
the section, we examine more closely interaction parameter space beyond the
points studied by Monte Carlo. Subsequently, in pursuit of the dominant
factor that stabilizes the biaxial nematic phase, we analyse the influence
of shape and energy biaxiality, then the combination of those
on the bifurcation diagram, with a special focus on Landau (self-dual) points.
\nlin The model has rich parameter space, and there are many possibilities 
of investigation of their contribution to the process of stabilisation of
biaxial nematic phase. We have chosen one possible parametrization
in the form of a path in the space of constants entering the potential,
which will be described in following sections. Since we are dealing 
only with isotropic, uniaxial nematic of oblate and prolate type, and
biaxial nematic, the best candidates for most important factors 
influencing \nbx\ stability are located by investigating Landau points
positions. It is so because in those places the biaxial nematic is most
stable, in the sense that the region of stability of \nrod\ and \ndisc\ is
severely reduced in the vicinity of self-dual region while at the Landau
point isotropic phase looses stability and system stabilizes biaxial nematic.
Therefore significant part of the analysis is devoted to the localization
of Landau points as function of interaction parameters, temperature, and
density. Taken into account a large number of possibilities of investigating
of the interaction parameter space, we consider a couple of special,
representative cases.
\nlin We would like to note that we are mainly interested in the transitions to
the biaxial nematic, the behaviour of other transitions is not studied
in detail.
\subsection{Uniaxial Gay-Berne potential}
\label{section:ungb}
\begin{figure}[t]
 \centering
 \subfigure[Gay-Berne potential $V_{GB}$ as a function of length $r$ of the vector linking centres of masses $\vec{\mathbf{r}}=r\unitv{r}$, for $r>\sigma(\unitvv{u}{1},\unitvv{u}{2},\unitv{r})-\sigma_{0}$ and for fixed $\unitv{r}$. The depth of the well $\widetilde{\varepsilon}\equiv\epsilon(\unitvv{u}{1},\unitvv{u}{2},\unitv{r})$ and the position of the zero of $V_{GB}$ at $r=r_{0}\equiv\sigma(\unitvv{u}{1},\unitvv{u}{2},\unitv{r})$ in \sftf\ configuration are marked.]{\label{figure:exampleofgbpotential} \includegraphics{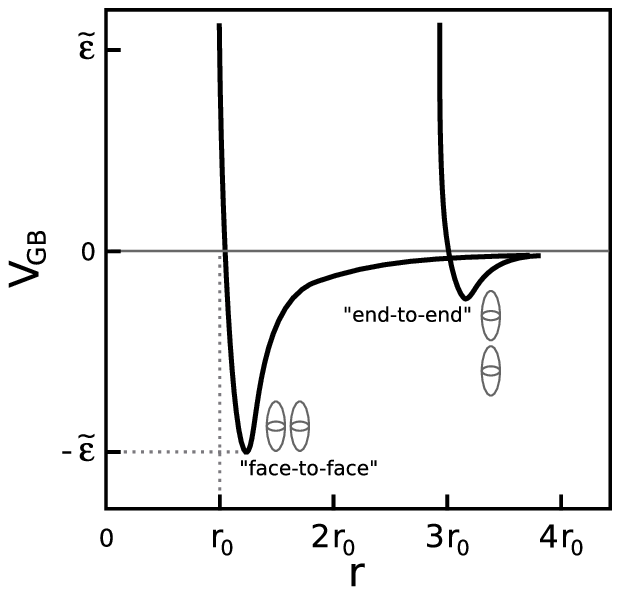}}
 \subfigure[Equipotential surfaces for molecular elongations $5:1$,
 $\tilde{\nu}=1$, $\tilde{\mu}=2$, $\epsilon_{x}/\epsilon_{z}=4$. Lines
 indicate $V_{GB}=0.0$ (most inner one), and counting from most
 outer one $V_{GB}=-0.1$, $-0.5$, $-0.9$, $-1.3$, $-1.7$, $-2.1$, $-2.5$.]{\label{figure:exampleofgbeqsurf} \includegraphics{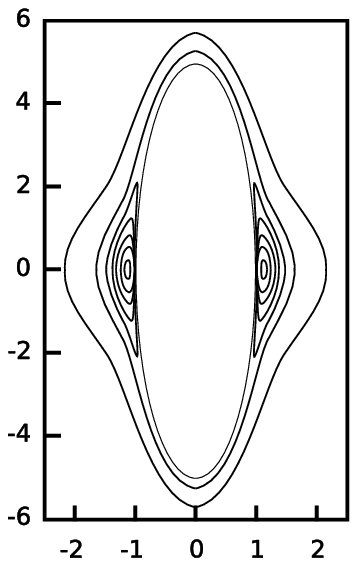}}
\caption[Uniaxial Gay-Berne interaction and surfaces of constant
potential.]{Uniaxial Gay-Berne interaction
Fig.~\subref{figure:exampleofgbpotential} and surfaces of constant potential
for parallel molecules
($\unitvv{u}{1}=\unitvv{u}{2}$) Fig.~\subref{figure:exampleofgbeqsurf}.}
\end{figure}

 The main success of the so-called Gay-Berne interaction lies in the proper,
effective inclusion of crucial for mesophase formulation
properties of the intermolecular forces. This, in connection
with mathematical tractability and low numerical cost, made it one
of the most intensively studied models in soft matter physics \cite{gbstudies0,simulationsreview,ginzburggb,velascopert1,velascopert2}. 
\nlin The uniaxial Gay-Berne interaction describes the potential energy
between two cylindrically (\dinfhns) symmetric ellipsoids of revolution. It
belongs to the class of so-called \ital{soft} potentials, which means that 
it includes attractive forces as well as short range, steric repulsions.
The key quantity, in this type of models, is a ''contact distance'' parameter
denoted by $\sigma(\orientationin,\orientation,\unitv{r})$,
which depends on the orientation of both molecules, more specifically
on relative orientation $\relativeorientation$, and orientation of the
vector linking the molecular centres of masses $\unitv{r}$. 
For length $r$ of the intermolecular vector $\boldv{r}$ equal to
$\sigma(\orientationin,\orientation,\unitv{r})$, the Gay-Berne energy
is equal to $0$; the orientation-dependent zeroth equipotential surface is
given by $\sigma(\orientationin,\orientation,\unitv{r})$. This surface,
for fixed orientations of molecules chosen so $\orientationin=\orientation$,
gives three quantities $\sigma_{x}$, $\sigma_{y}$, $\sigma_{z}$ for the
intermolecular vector direction $\unitv{r}=(1,0,0)$, $\unitv{r}=(0,1,0)$,
and $\unitv{r}=(0,0,1)$, respectively, which are identified with the
length of axes of interacting ellipsoids. For these reasons $\sigma(\orientationin,\orientation,\unitv{r})$ is called contact distance; it is 
considered to be a distance between the centres of masses of molecules
of given orientation and relative position when the ellipsoids are
in contact, in the sense described above. In case of
\dinfh symmetry, when, e.g. $\sigma_{x}=\sigma_{y}$, orientation of the molecule
is fully described by providing a unit vector along the axis of symmetry,
so we can use ${\bf{\hat{u}}}_{1}\equiv\orientationin$
and ${\bf{\hat{u}}}_{2}\equiv\orientation$, where $\unitv{u}_{1}$ and $\unitv{u}_{2}$ are unit vectors parallel to the axes of molecular symmetry.
The contact distance reads
\begin{equation}
 \label{eq:ungbsigma}
   \sigma\left( {\bf{\hat{u}}}_{1},\, {\bf{\hat{u}}}_{2},\, \bf{\hat{r}}\right) =
   \,
    \sigma_{0} \lcbra1-\frac{1}{2}\chi\lsbra
      \frac{ \left({\bf{\hat{r}}}\cdot{\bf{\hat{u}}}_{1}+{\bf{\hat{r}}}\cdot{\bf{\hat{u}}}_{2}\right)^{2} }
           { 1+\chi \left({\bf{\hat{u}}}_{1}\cdot{\bf{\hat{u}}}_{2}\right) }
     +\frac{ \left({\bf{\hat{r}}}\cdot{\bf{\hat{u}}}_{1}-{\bf{\hat{r}}}\cdot{\bf{\hat{u}}}_{2}\right)^{2} }
            { 1-\chi \left({\bf{\hat{u}}}_{1}\cdot{\bf{\hat{u}}}_{2}\right) }
      \rsbra\rcbra^{-\frac{1}{2}} \, ,
\end{equation}
where $\chi=(\kappa^{2}-1)/(\kappa^{2}+1)$, and where $\kappa=\sigma_{z}/\sigma_{x}$ describes the ''shape anisotropy'', i.e., elongation of the molecules.
\vfill
The Gay-Berne potential $V_{GB}$ is defined as
\begin{equation}
 \label{eq:ungbvgb}
\begin{split}
 V_{GB}({\bf{\hat{u}}}_{1},\, {\bf{\hat{u}}}_{2},\, \boldv{r} \,) = 
  \epsilon( {\bf{\hat{u}}}_{1},\, {\bf{\hat{u}}}_{2},\, {\bf{\hat{r}}} )\, \Bigg\{ & \lsbra \frac{\sigma_{0}}{r-\sigma \left( {\bf{\hat{u}}}_{1},\, {\bf{\hat{u}}}_{2},\, {\bf{\hat{r}}} \right)+\sigma_{0} } \rsbra^{12} \\
 - & \lsbra \frac{\sigma_{0}}{r-\sigma \left( {\bf{\hat{u}}}_{1},\, {\bf{\hat{u}}}_{2},\, {\bf{\hat{r}}} \right)+\sigma_{0} } \rsbra^{6} \Bigg\} \, ,
\end{split}
\end{equation}
where $\sigma_{0}$ is a constant with the dimension of length, and coefficient $\epsilon\left( {\bf{\hat{u}}}_{1},\, {\bf{\hat{u}}}_{2},\, {\bf{\hat{r}}} \right)$ plays a similar role to the contact distance only for the strength of
the interaction. It determines the orientation dependent depth of the minimum (as can be seen in \figreffig{exampleofgbpotential}), more precisely
\begin{equation}
 \label{eq:ungbepsilon}
\begin{split}
 \epsilon\left( {\bf{\hat{u}}}_{1},\, {\bf{\hat{u}}}_{2},\, {\bf{\hat{r}}} \right) = & \, 
  \epsilon^{\tilde{\nu}}\left( {\bf{\hat{u}}}_{1},\, {\bf{\hat{u}}}_{2}\right)
  \epsilon^{'\tilde{\mu}}\left( {\bf{\hat{u}}}_{1},\, {\bf{\hat{u}}}_{2},\, {\bf{\hat{r}}} \right) \, , \\
 \epsilon^{'}\left( {\bf{\hat{u}}}_{1},\, {\bf{\hat{u}}}_{2},\, {\bf{\hat{r}}}\right) = & \, 1 - \frac{\chi^{'}}{2}\left[ \frac{\left({\bf{\hat{r}}}\cdot{\bf{\hat{u}}}_{1}+{\bf{\hat{r}}}\cdot{\bf{\hat{u}}}_{2}\right)^{2}}{1+\chi^{'}({\bf{\hat{u}}}_{1}\cdot{\bf{\hat{u}}}_{2})} 
  + \frac{\left({\bf{\hat{r}}}\cdot{\bf{\hat{u}}_{1}}-{\bf{\hat{r}}}\cdot{\bf{\hat{u}}}_{2}\right)^{2}}{1+\chi^{'}({\bf{\hat{u}}}_{1}\cdot{\bf{\hat{u}}}_{2})}\right] \, ,
\end{split}
\end{equation}
where 
\begin{equation}
\label{eq:ungbepsilon0here}
\begin{split}
 \epsilon\left( {\bf{\hat{u}}}_{1},\, {\bf{\hat{u}}}_{2}\right)=\epsilon_{0}\left[ 1-\chi^{2} \left( {\bf{\hat{u}}}_{1}\cdot{\bf{\hat{u}}}_{2} \right)^{2}\right]^{-\frac{1}{2}} \, , \\
 \chi^{'}=\lsbra\lnbra\frac{\epsilon_{x}}{\epsilon_{z}}\rnbra^{1/\tilde{\mu}}-1\rsbra/\lsbra\lnbra\frac{\epsilon_{x}}{\epsilon_{z}}\rnbra^{1/\tilde{\mu}}+1\rsbra \, ,
\end{split}
\end{equation}
and where $\tilde{\mu}$ and $\tilde{\nu}$ are empirically chosen exponents.
The constants $\epsilon_{x}=\epsilon_{y}$ and $\epsilon_{z}$ are proportional
to the depth of the minimum of $V_{GB}(\unitvv{u}{1},\unitvv{u}{2},\unitv{r})$
when $\unitvv{u}{1}=\unitvv{u}{2}$ and when $\unitv{r}=(1,0,0)$ and $\unitv{r}=(0,0,1)$ respectively. In that sense they describe the
strength of the interactions in the directions of main axes of the molecules.
\nlin \figreffig{exampleofgbpotential} presents an example of the Gay-Berne
potential. It contains the attractive ''tail'' which behaves like $1/r^6$
and models the attractive interactions between the molecules,
like Van der Waals or dispersion forces. Those have longer range
than the repulsive part of the potential $\sim$ $1/r^{12}$ which models
the impenetrable ''core'', i.e., the excluded volume that is forbidden for the
centres of masses of molecules due to the repulsions on the shorter
distances of, e.g., Coulomb origin. The most important feature of the
Gay-Berne potential is the introduction in the consistent way of the
contact distance and analogical parameter of the interaction strength. They
both depend on the relative orientation and position of the molecules;
the former induces a shape dependent zeroth equipotential surface, while
the latter introduces anisotropy on the attractive forces. In other words,
the Gay-Berne potential opens a possibility of studies on soft ellipsoids 
of arbitrary dimensions and interaction strengths given in directions of
main axes.
\nlin The quantities in \figreffig{exampleofgbpotential} have some dimension.
In the analysis we need to operate with dimensionless objects, e.g., in
\eqrefeq{ldapproximation} the function under the exponent does not have
a dimension. It is also important from the numerical point of view.
A natural set of units in case of Gay-Berne potential is associated with
constants $\sigma_{0}$ and $\epsilon_{0}$ present in \refeq{eq:ungbsigma}
and \refeq{eq:ungbepsilon0here}. We employ $\sigma_{0}$ and $\epsilon_{0}$
as the unit of distance and energy, respectively. The set of dimensionless
quantities that will be used reads
\begin{equation}
 \label{eq:gbreducedunits}
\begin{split}
 \dimensionless{V_{GB}} = & \, V_{GB}/\epsilon_{0} \, , \,\, \dimensionless{r} = r/\sigma_{0} \, , \,\, \dimensionless{\rho}=\rho \, \sigma_{0}^{3} \, , \\
 \dimensionless{t} = & \, k_{B}T/\epsilon_{0} \, , \,\, \dimensionless{\mu}=\mu/\sqrt{\epsilon_{0}\sigma_{0}} \, ,
\end{split}
\end{equation}
where $\mu$ is the magnitude of a dipole in a dipole-dipole interaction which
will be used later. In case of uniaxial Gay-Berne $\sigma_{0}=\sqrt{2}\sigma_{x}$. For the description of the way the dimensions are treated
in the numerical integration procedure please consult Appendix~\refap{numericaldetails}.
\nlin We should note that $V_{GB}$ \refeq{eq:ungbvgb} has another zero,
apart from the one at $r_{0}=\sigma(\unitvv{u}{1},\unitvv{u}{2},\unitv{r})$
visible in \figreffig{exampleofgbpotential}. It is located
at $r=r_{0}-2\sigma_{0}$ and is considered to be ''unphysical'' together with
whole branch of $V_{GB}$ for $r<\sigma(\unitvv{u}{1},\unitvv{u}{2},\unitv{r})-\sigma_{0}$. This part of the potential is separated from the one presented
in \figreffig{exampleofgbpotential} by an infinite barrier, and has no known
physical application.

 On the historical account, the formula \refeq{eq:ungbsigma} was known as early
as in 1972, originally developed by Berne and Pechukas \cite{bernepechukas}.
They considered a "Gaussian overlap model" which could be understood as
mathematically tractable method allowing for short range, repulsive, dependent
on shape of molecule forces to be calculated. This is achieved by localizing
on each molecule a three dimensional Gauss function, which can be seen as a
mass distribution, and appropriately integrating to get $\sigma(\unitvv{u}{1},\unitvv{u}{2},\unitv{r})$. Then, in 1981, the Lennard-Jones
model \cite{lennardjones} was modified by Gay and Berne to
include $\sigma(\unitvv{u}{1},\unitvv{u}{2},\unitv{r})$
and $\epsilon(\unitvv{u}{1},\unitvv{u}{2},\unitv{r})$ as
in \refeq{eq:ungbvgb} \cite{gayberne}, and ever since was known as
the Gay-Berne potential. In mid-90s it was extended to the molecules of
different elongations \cite{generalizedungb} and generalized to the
biaxial, \dtwh -- symmetric ellipsoids \cite{bzdevelopmentofpotential}.
Now we turn to the description of the latter, the Gay-Berne potential
between two molecules with three different axes.
\subsection{Biaxial Gay-Berne interaction}
\label{section:bzmodel}
\begin{figure}[!ht]
 \centering
 \subfigure[Projection of shape biaxiality parameter $\shbx$ to the plane of axes ratios $\sigma_{x}/\sigma_{z}$ and $\sigma_{x}/\sigma_{y}$ in the rod-like regime. Darker regions indicate lower (negative) values of $\shbx$, (dashed)lines of constant $\shbx$ are drawn and labelled. The uniaxial limit at $\shbx=0$ for $\sigma_{x}/\sigma_{y}=1$ is marked with white.]{\label{figure:biaxialityrods} \includegraphics[scale=0.8]{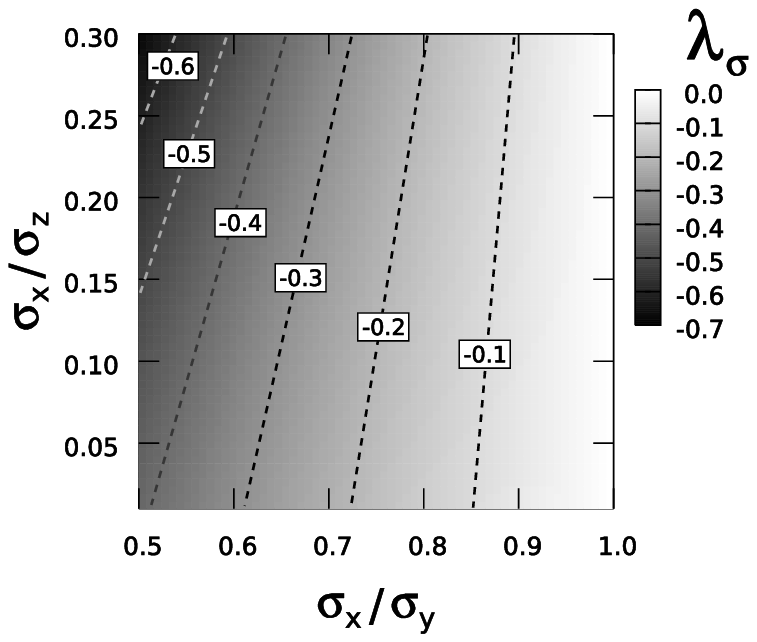}}
 \subfigure[Projection of shape biaxiality parameter $\shbx$ to the plane of axes ratios $\sigma_{x}/\sigma_{z}$ and $\sigma_{x}/\sigma_{y}$ in the disc-like regime. Whiter regions indicate larger values of $\shbx$ closer to the uniaxial shape at $\shbx=\sqrt{3/2}$, for $\sigma_{x}/\sigma_{y}=1$. Dashed lines with labels indicate the regions of constant shape biaxiality parameter.]{\label{figure:biaxialitydiscs} \includegraphics[scale=0.8]{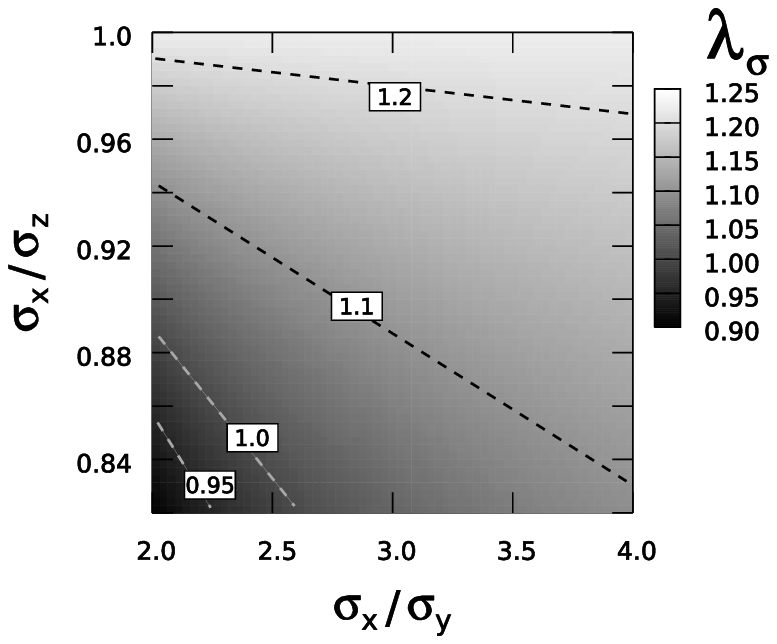}} \\
 \subfigure[Biaxiality parameter at Landau point for hard ellipsoids (at self-dual geometry) $\shbx^{sd}$ as defined by \refeq{eq:selfduallambda}.]{\label{figure:biaxialitysdlambda} \includegraphics[scale=0.8]{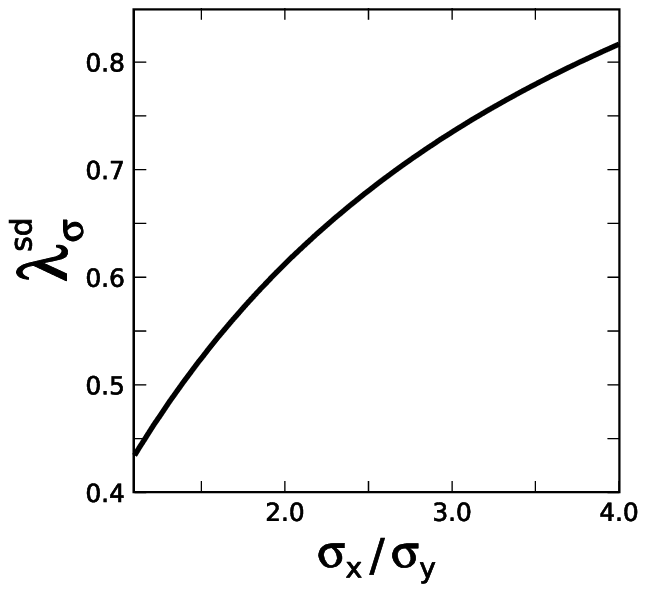}}
 \subfigure[Condition for Landau point for hard ellipsoids $\sigma_{x}=\sqrt{\sigma_{y}\sigma_{z}}$ \cite{holystponiewierski,mulderhard}.]{\label{figure:biaxialitysxszvssxsysd} \includegraphics[scale=0.8]{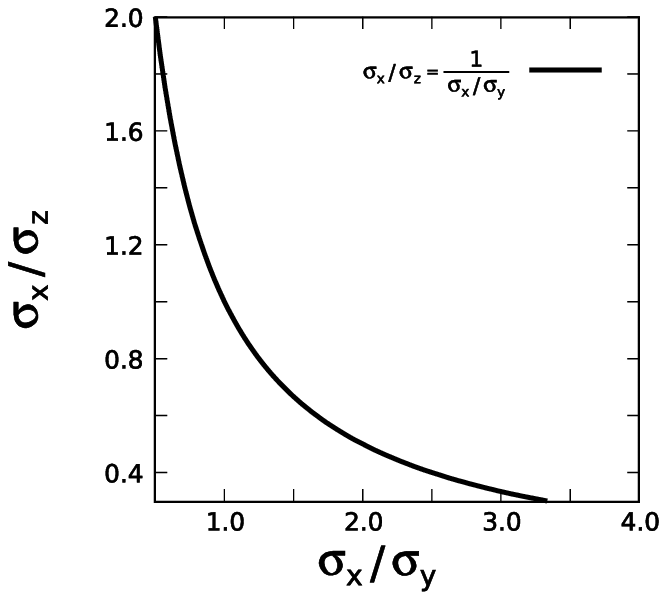}}
\caption[Closer look at the shape biaxiality parameter $\shbx$.]{\label{figure:biaxialityparams} Closer look at the shape biaxiality parameter $\shbx$ \refeq{eq:biaxialities}. On the top the density plots of $\shbx$ in oblate and prolate regimes (on the right and left, respectively) as function of axes ratios. Below on the right the dependence of axes ratios at Landau point for hard ellipsoids, on the left $\shbx$ along this curve.}
\end{figure}

 The biaxial version of the potential \refeq{eq:ungbvgb} has the same
form, namely \cite{bz2k,bzdevelopmentofpotential}
\begin{equation}
 \label{eq:bfzu}
\begin{split}
 U(\orientation_{1},\orientation_{2},\rvec\,) = 4\,\epsilon_{0}\epsilon(\orientation_{1},\orientation_{2},\unitv{r}) \Bigg\{ & \lsbra \frac{\sigma_{c}}{r-\sigma(\orientation_{1},\orientation_{2},\unitv{r})+\sigma_{c}} \rsbra^{12} \\
 - & \lsbra \frac{\sigma_{c}}{r-\sigma(\orientation_{1},\orientation_{2},\unitv{r})+\sigma_{c}} \rsbra^{6} \Bigg\} \, ,
\end{split}
\end{equation}
where $\epsilon_{0}$ and $\sigma_{c}$ are constants.
Clearly, now the interaction depends on the complete orientation of the 
frames associated with the molecules; we need to supply three
Euler angles in order to define the state of the molecule.
Parameters $\sigma(\orientation_{1},\orientation_{2},\unitv{r})$
and $\epsilon(\orientation_{1},\orientation_{2},\unitv{r})$ have exactly the
same interpretation as in the uniaxial case. In order to define
the contact distance in a general manner, we need to introduce a
so-called overlap matrix $\oper{A}$, the eigenvalues of which are proportional
to $\sigma(\orientation_{1},\orientation_{2},\unitv{r})$. It is built with
the help of shape matrix $\oper{S}$, which contains the molecular parameters,
namely $S_{ab}=\delta_{ab}\sigma_{a}$, and reads
\begin{equation}
 \label{eq:bxgboverlapmatrix}
 \oper{A}(\orientation_{1},\orientation_{2})=\oper{M}^{T}(\orientation_{1}) \oper{S}^{2} \oper{M}(\orientation_{1})+\oper{M}^{T}(\orientation_{2}) \oper{S}^{2} \oper{M}(\orientation_{2}) \, ,
\end{equation}
where $\oper{M}(\orientation)$ stands for the matrix of the operator of rotation
from laboratory to molecular frame. So, the overlap matrix $\oper{A}$
is built by rotating the squared shape matrix to the frames associated with
the molecules, in this way the dependence on the molecular relative orientations
is introduced. The dependence on the relative position $\unitv{r}$
is incorporated by calculating the eigenvalue of $\oper{A}^{-1}$ for
the vector $\unitv{r}$, which leads to the contact distance
\begin{equation}
 \label{eq:bxgbsigma}
 \sigma(\orientation_{1},\orientation_{2},\unitv{r})=\lnbra 2\unitv{r}^{T}\oper{A}^{-1}(\orientation_{1},\orientation_{2}) \unitv{r} \rnbra^{-1/2} \, \eqcolon
\end{equation}
Similarly as in the case of the uniaxial Gay-Berne, the zeroth equipotential
surface is defined by $\sigma(\orientation_{1},\orientation_{2},\unitv{r})$,
from the above equation. As before, the axes of the molecules $\sigma_{x}$, $\sigma_{y}$, and $\sigma_{z}$ are given by
\begin{equation}
 \label{eq:bxgbaxesfromequisurface}
 \sigma_{i}=\sigma(\orientation,\orientation,\unitv{r}_{i}) \, ,
\end{equation}
where $i=x,y,z$, $\unitv{r}_{x}=(1,0,0)$, $\unitv{r}_{y}=(0,1,0)$, $\unitv{r}_{z}=(0,0,1)$. When both molecules possess the same
orientation, the three relative configurations of the molecules with $\unitv{r}$ equal to $\unitv{r}_{x}=(1,0,0)$, $\unitv{r}_{y}=(0,1,0)$, and $\unitv{r}_{z}=(0,0,1)$ will be called \ssts, \sftf, and \sete, respectively. Depending on the
energy coefficient $\epsilon(\orientation_{1},\orientation_{2},\unitv{r})$
one of them gives the global minimum of the potential $U$ \re{bfzu}
(see \frf{potential}). The equipotential surfaces in planes corresponding
to these configurations are presented in \figreffig{surfaces}.
\putfigureflagslong{part2/chapter2/potential.ps}{Dimensionless biaxial Gay-Berne potential $\dimensionless{U}=U/\epsilon_{0}$ against reduced $\dimensionless{r}=r/\sigma_{0}$ for $\lnbra \sigma_{x},\,\sigma_{y},\,\sigma_{z} \rnbra=\lnbra 1.4,\,0.714,\,3.0 \rnbra$, and two sets of interaction strength parameters. On the left the \ssts\ configuration minimum is deepest; $\lnbra \epsilon_{x},\,\epsilon_{y},\,\epsilon_{z} \rnbra=\lnbra 1.7,\,1.0,\,0.2 \rnbra$, on the right \sftf\ attraction is strongest; $\lnbra \epsilon_{x},\,\epsilon_{y},\,\epsilon_{z} \rnbra=\lnbra 1.0,\,1.4,\,0.2 \rnbra$.}{potential}{!b}{Biaxial Gay-Berne potential.}
\nlin The interaction strength coefficient is defined in the analogical
way as in the uniaxial case:
\begin{equation}
\begin{split}
 \epsilon(\orientation_{1},\orientation_{2},\unitv{r}) = & \, \epsilon^{\tilde{\nu}}(\orientation_{1},\orientation_{2})\epsilon^{\tilde{\mu}}(\orientation_{1},\orientation_{2},\unitv{r}) \, , \\
 \epsilon(\orientation_{1},\orientation_{2}) = &
 \lnbra\sigma_{x}\sigma_{y}+\sigma_{z}^{2}\rnbra\sqrt{\frac{2\sigma_{x}\sigma_{y}}{\det[\oper{A}]}} \, , \\
  \epsilon(\orientation_{1},\orientation_{2},\unitv{r}) = & \, 2\unitv{r}^{T}\oper{B}^{-1}(\orientation_{1},\orientation_{2})\unitv{r} \, , 
\end{split}
\end{equation}
where the matrix $\oper{B}$ contains parameters of the potential strength
(energy is also ''biaxial'') and is defined in a manner similar to the overlap
matrix \cite{bzdevelopmentofpotential}:
\begin{equation}
 \oper{B}(\orientation_{1},\orientation_{2})=\oper{M}^{T}(\orientation_{1}) \oper{E} \, \oper{M}(\orientation_{1})+\oper{M}^{T}(\orientation_{2}) \oper{E} \, \oper{M}(\orientation_{2}) \, ,
\end{equation}
where $E_{ab}=\delta_{ab}\lnbra \epsilon_{0}/\epsilon_{a} \rnbra^{1/\tilde{\mu}}$, and $\tilde{\nu}$, $\tilde{\mu}$ are empirically chosen parameters.
\putfigureflagslong{part2/chapter2/eqsurfaces.eps}{Equipotential surfaces for the biaxial Gay-Berne potential, in three planes for interaction parameters $\lnbra \sigma_{x},\sigma_{y},\sigma_{z} \rnbra=\lnbra 1.4,0.714,3.0 \rnbra$, $\lnbra \epsilon_{x},\epsilon_{y},\epsilon_{z} \rnbra=\lnbra 1.0,1.4,0.2 \rnbra$.}{surfaces}{t}{Equipotential surfaces for the biaxial Gay-Berne potential.}
\nlin The energy and length units are set by constants $\epsilon_{0}$
and $\sigma_{0}$, respectively (see \refeq{eq:gbreducedunits}).
Also following Zannoni \cite{bz2k}, we set $\tilde{\nu}=3$, $\tilde{\mu}=1$ \cite{gbempiricalexp}, and $\sigma_{c}=\sigma_{y}$. It should be noted that the
potential \re{bfzu} for \dinfh -- symmetric molecule when, e.g., $\sigma_{x}=\sigma_{y}$, reduces to uniaxial Gay-Berne \re{ungbvgb} when $\sigma_{c}=\sigma_{0}=\sqrt{2}\sigma_{x}$.
\nlin The interaction $U$ in \refeq{eq:bfzu} has three shape-related,
and three energy-related parameters, those are respectively $(\sigma_{x},\sigma_{y},\sigma_{z})$ and $(\epsilon_{x},\epsilon_{y},\epsilon_{z})$.
They contribute to the two coefficients giving rise to two sources of biaxiality: $\epsilon(\orientation_{1},\orientation_{2},\unitv{r})$ and $\sigma(\orientation_{1},\orientation_{2},\unitv{r})$. The latter is related to the position of
the zero of the potential and is associated with the dimensions of the
molecules. The former gives the anisotropic depth of the minimum of the
potential and corresponds to the biaxiality of the attractive forces.
Less precisely we can say that the shape related parameters act along
the horizontal axis in \frf{potential}, while energy related parameters
act along vertical one. Despite the fact that $(\epsilon_{x},\epsilon_{y},\epsilon_{z})$ and $(\sigma_{x},\sigma_{y},\sigma_{z})$ are both just constants entering
the potential, we will call them energy parameters and shape parameters,
respectively. We should keep in mind that they enter the
excess Helmholtz free energy \re{ldapproximation} via
pair potential $V(\coord{1},\coord{2})$ in non-linear manner,
thus making the straight-forward interperetation of $(\epsilon_{x},\epsilon_{y},\epsilon_{z})$ and $(\sigma_{x},\sigma_{y},\sigma_{z})$ non-trivial.
\nlin It is useful to introduce a biaxiality parameter which can be treated as
a measure of how
much a given set of $\{\sigma_{i}\}$ or $\{\epsilon_{i}\}$ differs from the
uniaxial, \dinfh -- symmetric case. There are several such quantities
known in the literature. We have chosen to use the one following
from the surface tensor model \cite{surfacetensor0,surfacetensor1}
developed in \cite{ferrarini}. It is based on the assumption that the mean field
effective potential $V_{eff}(\coordsymbol)$ (see Appendix~\refap{meanfield}, \eqrefeq{mfeqapp}), which gives the equilibrium \sopdf $P_{eq}(\coordsymbol)=Z^{-1}\exp\lsbra -\beta V_{eff}(\coordsymbol) \rsbra$, can be written as a sum of
mean torque potential $V^{ext}$ and internal potential. The latter accounts for
the variations of the molecular interactions due to bond rotations,
while $V^{ext}$ describes the external interactions acting on the
given molecule in the sample, giving rise to the nematic ordering. 
Now the mean torque potential is expanded in the base of Wigner matrices $\wignerd{l}{m}{n}{\orientation}$ \cite{rose,lindner} (for brief description of these functions see Appendix~\refsec{wignerd}), following \cite{ferrarini}
\begin{equation}
 \label{eq:meantorqueexpanded}
 V^{ext}(\orientation)\sim\sum^{2}_{n=-2} T^{2,n\,*}\wignerd{2}{0}{n}{\orientation} \, ,
\end{equation}
where the $T^{2,n}$ are the components of the interaction tensor, which is
called a surface tensor. They read
\begin{equation}
 \label{eq:surfacetensorcomponents}
 T^{2,n}=-\int_{S}dS\, \wignerd{2}{0}{n}{\orientation} \, ,
\end{equation}
where $S$ denotes the surface of the molecule. Above equation is a consequence
of the assumption that each surface element has a tendency to align parallel
to the director in \nrod\ and perpendicular to it in \ndisc\ phase. This means
that $V^{ext}(\orientation)\sim \int_{S}dS\,\frac{1}{2}(3\,cos^{2}\beta-1)$, which is a base for \eqrefeq{meantorqueexpanded}. Coefficients $T^{2,n}$
in general depend on the order parameters and molecular anisotropy. The
molecular biaxiality parameter is defined as
\begin{equation}
 \label{eq:molecularbxsurfacetensor}
 \lambda=T^{2,2}/T^{2,0} \, .
\end{equation}
In case of hard parallelepiped with dimensions $a$, $b$ and $c$, along
axes $x$, $y$, and $z$, the above can be expressed as \cite{ferrarini}
\begin{equation}
 \label{eq:biaxialparametercuboid}
 \lambda=\lnbra 3/2 \rnbra^{1/2} \frac{c\lnbra a-b \rnbra}{c\lnbra a+b \rnbra - 2 ab} \, .
\end{equation}
This parameter will be used in this study (in place of $a$, $b$, $c$ we will use axes of ellipsoids).
It has some useful properties. Firstly, assuming that $a \neq 0$, $b \neq 0$, $c  \neq 0$, the only case
when $\lambda=0$ occurs for \dinfh -- symmetric shape when $a=b$, for other
uniaxial shapes when $a=c$ and $b=c$, $\lambda$ takes values of $\sqrt{3/2}$
and $-\sqrt{3/2}$, respectively. When $c$ becomes very large, $\lambda$
behaves like $\lnbra a-b \rnbra/\lnbra a+b \rnbra$, which means that 
it does not become zero in this limit. It reaches vanishingly small
value for $c$ tending to zero.
\nlin We choose the biaxiality parameters $\shbx$ and $\enbx$ respectively for shape and energy, along \refeq{eq:biaxialparametercuboid}, they read
\begin{equation}
 \label{eq:biaxialities}
\begin{split}
 \shbx = & \, \lnbra 3/2 \rnbra^{1/2} \frac{\sigma_{z}\lnbra \sigma_{x}-\sigma_{y} \rnbra}{\sigma_{z}\lnbra \sigma_{x}+\sigma_{y} \rnbra - 2 \sigma_{x}\sigma_{y}} \, , \\
 \enbx = & \, \lnbra 3/2 \rnbra^{1/2} \frac{\epsilon_{z}\lnbra \epsilon_{x}-\epsilon_{y} \rnbra}{\epsilon_{z}\lnbra \epsilon_{x}+\epsilon_{y} \rnbra - 2 \epsilon_{x}\epsilon_{y}} \, .
\end{split}
\end{equation}
\indent For hard biaxial ellipsoids there is a simple condition for self-dual
(Landau) point, it is the mentioned earlier square root rule,
obtained from \ere{landaupointcondition} \cite{holystponiewierski,mulderhard}:
\begin{equation}
 \label{eq:squarerootrule}
 \sigma_{x}=\sqrt{\sigma_{y}\sigma_{z}} \, .
\end{equation}
It defines a line in the plane of axes ratios, e.g., $(\sigma_{x}/\sigma_{z},\sigma_{x}/\sigma_{y})$ (see \figreffig{biaxialitysxszvssxsysd}), and along this
line we can represent $\enbx$ and $\shbx$ as functions
of $\ex/\ey$ and $\sigma_{x}/\sigma_{y}$, respectively (see \figreffig{biaxialitysdlambda}), namely:
\begin{align}
 \shbx^{sd} \equiv & \, \lnbra 3/2 \rnbra^{1/2} \frac{\sigma_{x}/\sigma_{y}-1}{\sigma_{x}/\sigma_{y}-2\lnbra\sigma_{x}/\sigma_{y}\rnbra^{-1}+1} \, , \label{eq:selfduallambda} \\
 \enbx^{sd} \equiv & \, \lnbra 3/2 \rnbra^{1/2} \frac{\epsilon_{x}/\epsilon_{y}-1}{\epsilon_{x}/\epsilon_{y}-2\lnbra\epsilon_{x}/\epsilon_{y}\rnbra^{-1}+1} \, . \label{eq:selfdualepsilon}
\end{align}
\nlin In order to help to visualize the way biaxiality parameters \refeq{eq:biaxialities} work, we present Figs. \reffig{biaxialityrods}-\subref{figure:biaxialitydiscs}. They show one possible method of changing $\shbx$ (the same holds for $\enbx$ and $\{\epsilon_{i}\}$) by varying the ratios of
axes $\sigma_{x}/\sigma_{z}$ and $\sigma_{x}/\sigma_{y}$.
Similar method was used in present work.
We start in the region of biaxial rod-like shapes when $\sigma_{x}<\sigma_{y}\ll\sigma_{z}$ (\frf{biaxialityrods}). Then, by increasing the ratio $\sigma_{x}/\sigma_{y}$ for given $\sigma_{x}/\sigma_{z}$ (moving to the right on \frf{biaxialityrods}) we decrease $\shbx$ to reach uniaxial, prolate molecule 
for $\sigma_{x}=\sigma_{y}$ ($\shbx=0$, white colour in \frf{biaxialityrods}).
Continuing to the disc-like region, where $\sigma_{z}>\sigma_{x}\gg\sigma_{y}$ (\frf{biaxialitydiscs}), we start with the biaxial oblate molecule (dark areas),
and by increasing the ratio $\sigma_{x}/\sigma_{z}$ for given $\sigma_{x}/\sigma_{y}$ (moving upwards on \frf{biaxialitydiscs}) we reach uniaxial, disc-like 
molecule for $\sigma_{x}=\sigma_{z}$ and $\shbx=\sqrt{3/2}$ (white areas).
As we can see, in rod-like regime, vanishing $\shbx$ indicates uniaxial molecule,
while disc-like, \dinfh -- symmetric shape entails $\shbx=\sqrt{3/2}$.

 The biaxiality parameters \re{biaxialities} for given shape and energy
parameters $\{\sigma_{i}\}$ and $\{\epsilon_{i}\}$ are not defined in a unique
way. In order to use $\enbx$ and $\shbx$ to parametrize the
bifurcation diagrams, we need to make certain assumptions
concerning $\{\sigma_{i}\}$ and $\{\epsilon_{i}\}$,
and choose the way of varying them. Assuming that for the biaxial Gay-Berne,
there exists a self-dual geometry analogical to the square root rule \re{squarerootrule} obtained for hard biaxial ellipsoids \cite{holystponiewierski,mulderhard}, we can expect
a two dimensional surface of Landau points parametrized by $\shbx^{sd}$
and $\enbx^{sd}$ \re{selfduallambda}-\re{selfdualepsilon}. The determination of
this surface is interesting, but it remains to be done in future
studies. Presently, we investigate how, by varying the biaxiality of the
molecule and forces, we intersect the self-dual plane. In other words, we 
choose a path in six dimensional space of $\{\sigma_{i}\}$
and $\{\epsilon_{i}\}$, in the manner described above,
and along it we calculate the density-temperature bifurcation diagram.
Due to chosen parametrization, we cross \sdplane\ plane in a single point.
Since the formulas \re{biaxialities}-\re{selfdualepsilon} follow from simple
models, we cannot assume \ital{a priori} that they possess any
physical meaning, especially taken into account the complicated manner 
in which \sigmaset and \epsilonset\ enter the potential in present model.
However, we can investigate if \sdplane\ is in any way related to the
surface of Landau points obtained from \re{landaupointcondition}
for biaxial Gay-Berne interaction.
\nlin The sign of both parameters $\shbx$ and $\enbx$ is of importance, for
it reflects the ratio of $\sigma_{x}/\sigma_{y}$
and $\epsilon_{x}/\epsilon_{y}$. Let's imagine two identical, rod-like, biaxial ellipsoids with $\sigma_{z}>\sigma_{x}>\sigma_{y}$ ($\shbx>0$).
It seems natural to expect that in the \sftf\ configuration the
interaction $U$ will be strongest (see \rf{potential}), which means that $\epsilon_{x}>\epsilon_{y}>\epsilon_{z}$ and in turn implies $\enbx>0$. Monte Carlo
study of the potential \refeq{eq:bfzu} did not predict a stable
biaxial nematic phase in that case \cite{bz2k}, instead the simulations 
showed \nbx\ for potential favouring the \ssts\ configuration,
namely when $\enbx<0$ while $\shbx>0$. The difference between
these two situations is depicted in \figreffig{potential},
where physical (for $r>\sigma(\orientation_{1},\orientation_{2},\unitv{r})-\sigma_{c}$) regions of $U$ are shown. In present study we refer to the case
where $\epsilon_{x}>\epsilon_{y}>\epsilon_{z}$ ($\enbx<0$) and $\sigma_{x}>\sigma_{y}$ ($\shbx>0$) by ''strong lateral interactions model''.
\putfigureflagslong{part2/chapter2/plot_withbz2k.eps}{Two bifurcation diagrams for parameter sets studied in \cite{bz2k} for $\shbx=0.58$ ($\lnbra \sigma_{x},\,\sigma_{y},\,\sigma_{z} \rnbra=\lnbra 1.4,\,0.714,\,3.0\rnbra$). On the left the model where \ssts\ configuration is preferred $\enbx=-0.06$ ($\lnbra \epsilon_{x},\,\epsilon_{y},\,\epsilon_{z} \rnbra=\lnbra 1.7,\,1.0,\,0.2 \rnbra$), on the right the \sftf\ minimum is deepest; $\enbx=0.042$ ($\lnbra \epsilon_{x},\,\epsilon_{y},\,\epsilon_{z} \rnbra=\lnbra 1.0,\,1.4,\,0.2 \rnbra$). Transition points discovered by Monte Carlo in \cite{bz2k} are marked.}{bifvsmc}{t}{Bifurcation diagrams for parameter sets studied in simulations \cite{bz2k}.}
\nlin The potential of the form of \refeq{eq:bfzu} can be immediately
put in place of $V$ into Eqs.~\re{ldsettings}-\re{ldapproximation} and used in
calculations as described in \aprefap{numericaldetails}. Then, bifurcation
equations \re{exactbifisoldl2}, \re{bifunbx} can be solved, and the bifurcation
density $\dimensionless{\rho}\equiv\dimensionless{\rho}_{0}$ and
temperature $\dimensionless{t}$ (see \re{gbreducedunits}) can be determined, provided
we had calculated coefficients $\clmn{2}{m}{n}$ and the reference state.
Unless indicated otherwise, in the following $\rho^{*}$ will stand for the
dimensionless density, following from the solutions of the
bifurcation equations.
We start with the comparison of the bifurcation results with those
obtained from Monte Carlo in \cite{bz2k}.

 The interaction \refeq{eq:bfzu} was studied by Berardi and Zannoni 
by means of Monte Carlo simulations in the isothermal isobaric ensemble
for $8192$ soft, biaxial ellipsoids. They took into account one
set of shape parameters in the rod-like regime: $\lnbra \sigma_{x},\,\sigma_{y},\,\sigma_{z} \rnbra=\lnbra 1.4,\,0.714,\,3.0\rnbra$ ($\shbx=0.58$), and
following sets of potential strength coefficients $(\epsilon_{x},\epsilon_{y},\epsilon_{z})$: $(1.0,1.2,0.2)$ ($\enbx=0.025$), $(1.0,1.4,0.2)$ ($\enbx=0.042$), $(1.4,1.0,0.2)$ ($\enbx=-0.042$), $(1.7,1.0,0.2)$ ($\enbx=-0.06$). Only for the
last one the stable biaxial nematic was found. We investigate the issue
of $\shbx$ and $\enbx$ of opposite signs in later subsections, presently we
show the comparison of our results with those obtained from simulations.
In \figreffig{bifvsmc} we plot the bifurcation diagrams for parameters studied
in simulations \cite{bz2k} and mark the transition points discovered there.
It should be noted that the Monte Carlo study \cite{bz2k} did not include
the long-range corrections, i.e., the potential involved a cut-off radius.
For this reason the comparison is only qualitative, and the 
bifurcation parameters following from low-density differ from
simulations results, especially for \iso\ -- \nun\ phase trasition.
Once the effects of cut-off are compensaed in Monte Carlo,
this comparison is improved \cite{frenkelsmit}. 
Table \reftable{mcbz2k} shows the critical densities and temperatures from
simulations in comparison with bifurcation for \iso\ -- \nrod\ and \nrod\ -- \nbx\ transitions.
\putfigureflagslong{part2/chapter2/plot_phasediagramvsbif.eps}{Phase diagram for \nun\ -- \nbx\ transition following from minimisation of the Helmholtz free energy \refeq{eq:freeenergyuniform} by solving of the equations \refeq{eq:scforalpha} (solid line), and from bifurcation analysis (dashed line) for potential parameters studied in \cite{bz2k}: $\shbx=0.58$, $\enbx=-0.06$ ($\lnbra \sigma_{x},\,\sigma_{y},\,\sigma_{z} \rnbra=\lnbra 1.4,\,0.714,\,3.0 \rnbra$, $\lnbra\epsilon_{x},\,\epsilon_{y},\,\epsilon_{z} \rnbra=\lnbra 1.7,\,1.0,\,0.2 \rnbra$).}{bifvsminof}{b}{Phase transition vs bifurcation in \nun\ -- \nbx\ case.}
\subsection{Contribution from precise reference state}
 In \srs{modelsofdcfs} we have pointed to the minimisation of
the Helmholtz free energy $\intrinsicfef{\opdfsymbol}$ \re{freeenergyuniform}
with help of the trial \sopdf \re{usedmodelopdf} as an alternative method
of obtaining a phase diagram. The above figures show how Monte Carlo results
differ from bifurcation, which, besides the mentioned long-range corrections,
may be caused by the $L=2$ model of pair direct correlation function $c_{2}$ \re{expansionofc2l2} used in the process of obtaining of the reference \sopdf
$\pref$ along \esre{opsinl2sceq}-\re{prefinun}, or by the loss of accuracy
in the bifurcation analysis.
The minimisation technique addresses these issues. In this method we do not
expand $c_{2}$ and study the minima of $\intrinsicfef{\opdfsymbol}$ directly.
Therefore it is worthwhile to present the comparison of those two approaches.
We do so in present section.
\begin{table}[t]
 \caption[Monte Carlo results vs bifurcation diagrams.]{Transition points obtained in Monte Carlo simulations (MC) by Berardi and Zannoni (model parameters as on the left plot in Fig. \reffig{bifvsmc}) compared to the results of bifurcation analysis in low-density approximation.}
 \begin{tabular}{rclccc}
 \label{table:mcbz2k}
  & & & density $\dimensionless{\rho}$ & temperature (MC) & bifurcation temperature $\dimensionless{t}$ \n \hline\hline
  \iso\ & -- & \nrod &   0.271 & 3.20 & 6.97 \n
  \nrod\ & -- & \nbx &   0.291 & 2.90 & 3.50 \n
\end{tabular}
\end{table}

 The phase diagram presented here follows from postulating
the trial \sopdf \eqrefeq{usedmodelopdf} and finding the set of parameters
$\alpha_{m,n}$ that minimises the Helmholtz free energy \refeq{eq:freeenergyuniform}, by
solving the \sceq s \refeq{eq:scforalpha}, and locating in this way
the equilibrium state. Once all such points are found,
we can easily decide which one of them is a global minimum. Having
determined the set of parameters $\alpha_{m,n}$ giving equilibrium \sopdfns,
we can calculate the order parameters using the definition \re{orderparamsset}
and derive the phase transition points. It is also possible to find 
the coexistence (or more precisely a mechanical stability) regions by looking 
for points where the pressures and chemical potentials are equal. We have done
that analysis and the results for \nun\ -- \nbx\ transition can be seen in 
\figreffig{bifvsminof}, where the phase transition as well as
bifurcation lines are shown. 
\nlin Since, by the investigation of the
order parameters, the uniaxial -- biaxial nematic transition
has been found to be second order and because in this case the bifurcation
point is a point of phase transition, the figure presents the difference
between the bifurcation point obtained with the help of reference state $\pref$
calculated using of $L=2$ model of $c_{2}$ \re{expansionofc2l2} and the exactly
obtained $\pref$, as described above. In other words,
the difference between solid and dashed line in the \frf{bifvsminof} is
the estimate of the influence of all coefficients $\{\clmn{l}{m}{n}\}$
with $l \geq 2$ in $\pref$ \re{prefinun} on the phase transition point
obtained from bifurcation equation \re{bifurcationeq}. The comparison of
bifurcation and phase diagram indicates that the critical values of density
and temperature obtained by those two methods are complementary. This proves
that the leading contribution to the transition density and temperature comes
from terms in the expansion of $c_{2}$ with angular momentum index equal $2$,
and that, at least, for transitions studied here, the bifurcation analysis
gives reasonable values of critical parameters. The numerically much more
expensive free energy minimisation shifts the diagram, but does not change
the overall phase behaviour.
\putfigureflagslong{part1/misc/ready_shbxdensity_ex1.2_ey1.2_ez0.2.ps}{Path of varying of shape biaxiality parameter $\shbx$ (thick, solid line drawn with gradient) in plane of axes ratios $(\sigma_{x}/\sigma_{y},\sigma_{x}/\sigma_{z})$, and the projection of $\shbx$. Landau point resulting from present analysis (shown in \figreffig{landaushbx}) is marked with circle at $(2.57,0.44)$, the self-dual, Landau region for hard ellipsoids ($\sigma_{x}=\sqrt{\sigma_{y}\sigma_{z}}$) is represented by dotted line. Uniaxial regimes for $\shbx=0.0$ and $\shbx=\sqrt{3/2}$ are marked with white. Thin, solid lines indicate regions of constant $\shbx$.}{shbxdensitypath}{h}{Path of varying of shape biaxiality parameter $\shbx$.}
\nlin As can be seen from Figs.~\rf{bifvsmc} and \rf{bifvsminof},
the dependence of temperature $\dimensionless{t}$ on density $\dimensionless{\rho}$ at bifurcation is almost linear; the variation of density does not change 
the topology of the phase diagram. Therefore we can choose some value of $\dimensionless{\rho}$ (keeping in mind that assumptions made in Chapter \refchap{DFT} for the Taylor expansion \re{taylorexp} may prove incorrect if it is too high) and pursuit the effects of biaxialities on bifurcation temperature
at fixed density. That will be the subject of the following section.
\subsection{Exploring the effects of biaxiality}
\label{section:bzlandaupoints}
 In the biaxial Gay-Berne potential there are two sources of biaxiality
identified with the
anisotropic forces strengths and the shape of molecules. The deviation from
uniaxial symmetry of those is expressed by parameters $\enbx$ and $\shbx$,
respectively. Currently we will show how $\enbx$ and $\shbx$ and combination
of these change the bifurcation diagram. We present the dependence of
dimensionless temperature $\dimensionless{t}$ on
dimensionless density $\dimensionless{\rho}$ at bifurcation point for
fixed $\enbx$ and $\shbx$, as well as $\dimensionless{t}$ as function
of $\shbx$ or $\enbx$ for fixed $\dimensionless{\rho}$. In the latter case the
diagrams are calculated for $\dimensionless{\rho}=0.18$, unless indicated
otherwise. It was chosen to give packing fraction, defined as $\frac{molecular\,\,volume}{average\,\,volume\,\,per\,\,molecule}$, $\eta=0.28$ (slightly lower than typical values for nematic phases: $0.4 \leq \eta \leq 0.6$).
\nlin As we mentioned before, in order to parametrize the bifurcation
diagrams by $\enbx$ and $\shbx$, we need to make additional assumptions
concerning $\{\sigma_{i}\}$ and $\{\epsilon_{i}\}$. We have decided to 
hold $\sum_{i}\sigma_{i}=const$ and $\sum_{i}\epsilon_{i}=const$
(which set the density and temperature scale), and
$\epsilon_{i}>0$, $\sigma_{i}>0$. The sets of energy
and shape parameters we refer to are listed in \trt{biaxialities},
they are denoted by \gbsete{a}-\gbsete{i} and \gbsets{a}-\gbsets{h},
respectively. As a starting point we have chosen the set studied by
Berardi and Zannoni \cite{bz2k}: \gbsete{e}, \gbsets{b}\footnote{We could have chosen the other one with $\lnbra \epsilon_{x},\,\epsilon_{y},\,\epsilon_{z} \rnbra=\lnbra \,1.7,\,1.0,\,0.2 \rnbra$, the only difference between those two is relatively stronger interaction in the latter case.}. That fixed $\sum_{i}\sigma_{i}$ and $\sum_{i}\epsilon_{i}$.
From this point, in order to decrease the shape biaxiality in a
consistent manner, we choose to 
decrease the difference between $\sx$ and $\sy$ while holding $\sz$ constant,
reaching uniaxial shape \gbsets{a} for $\shbx=0.0$. In other direction we
increase $\sz-\sy$ while keeping $\sx=const$ up to a moment where Landau point
is found at \gbsets{g}. From this set we make the molecule more disc-like,
by making $\sx-\sz$ and $\sy$ smaller, to reach uniaxial,
oblate ellipsoid \gbsets{h} for $\shbx=\sqrt{3/2}$. This path, in the plane of
axes ratios $(\sx/\sz,\sx/\sy)$, is depicted in \figreffig{shbxdensitypath}.
As we can see we are crossing the self-dual line for hard \dtwh -- symmetric
molecules once. The Landau point following from present analysis
is also depicted.
\begin{table}[t]
\caption[Biaxiality parameters $\enbx$, $\shbx$ and interaction constants sets $\{\sigma_{i}\}$,$\{\epsilon_{i}\}$.]{Biaxiality parameters $\enbx$, $\shbx$ and interaction constants sets $\{\epsilon_{i}\}$, $\{\sigma_{i}\}$.}
\begin{tabular}{clllcll}
\label{table:biaxialities}
parameter set&$\epsilon_{x}$, $\epsilon_{y}$, $\epsilon_{z}$&$\enbx$&~&parameter set&$\sigma_{x}$, $\sigma_{y}$, $\sigma_{z}$&$\shbx$\\
\hline
\hline
\gbsete{a}&$1.9,0.5,0.2$&$-0.2415$&~&\gbsets{a}&$1.057,1.057,3.0$&$0.0$\\
\gbsete{b}&$1.84,0.56,0.2$&$-0.1974$&~&\gbsets{b}&$1.4,0.714,3.0^{bz}$&$0.58$\\
\gbsete{c}&$1.4,1.0,0.2$&$-0.042$&~&\gbsets{c}&$1.4,0.637,3.077$&$0.6412^{*}$\\
\gbsete{d}&$1.2,1.2,0.2$&$0.0$&~&\gbsets{d}&$1.4,0.614,3.1$&$0.6596$\\
\gbsete{e}&$1.0,1.4,0.2^{bz}$&$0.042$&~&\gbsets{e}&$1.4,0.574,3.14$&$0.6919$\\
\gbsete{f}&$0.9,1.5,0.2$&$0.0662$&~&\gbsets{f}&$1.4,0.564,3.15$&$0.70$\\
\gbsete{g}&$0.8,1.6,0.2$&$0.0942$&~&\gbsets{g}&$1.4,0.544,3.17$&$0.7163$\\
\gbsete{h}&$0.6,1.8,0.2$&$0.1750^{**}$&~&\gbsets{h}&$2.35,0.414,2.35$&$\sqrt{3/2}$ \\
\gbsete{i}&$0.5,1.9,0.2$&$0.2415$&~&~&~&~\\
\end{tabular}
\nlin $\,\,\,\,\,\,\,\,\,\,\,\,\,\,\,\,\,\,\,$ ($*$) self-dual point for hard biaxial ellipsoids ($\sigma_{x}=\sqrt{\sigma_{y}\sigma_{z}}$).
\nlin $\,\,$ ($**$) fulfilling square root rule for energy parameters ($\epsilon_{x}=\sqrt{\epsilon_{y}\epsilon_{z}}$).
\nl ($bz$) parameters studied in Monte Carlo simulations in \cite{bz2k}.
\end{table}
\nlin In case of $\{\epsilon_{i}\}$ we do not cross the boundary between 
the ''disc-like'' and ''rod-like'' energy parameters sets. We only
follow the path linking models of strong lateral interactions and
cases where \sftf\ attractions are strongest. That is, starting
from \gbsete{d}, being the point of uniaxial energy where $\enbx=0.0$,
and holding $\ez=const$, we increase energy biaxiality by increasing
$|\ex-\ey|$ to reach \gbsete{a} and \gbsete{i}. Along this path we cross the 
point studied in simulations \cite{bz2k} \gbsete{e}, and also the model 
fulfilling square root rule \re{squarerootrule} for \epsilonset\ \gbsete{h}.
With the above sets of shape parameters, \ssts\ and \sftf\ configuration give
deepest minimum of the potential when $\ex>\ey$ and $\ex<\ey$, respectively. 
\nlin The above method of changing $\enbx$ and $\shbx$ was used in every case.
\nlin If we fix $\enbx=0$ for \gbsete{d}, we can expect that the system of
\dinfh -- symmetric, rod-like molecules with axes \gbsets{a} ($\shbx=0.0$)
stabilizes \nrod\ at the expense of isotropic state, while
for uniaxial disc-like ellipsoids, when $\shbx=\sqrt{3/2}$ for \gbsets{h},
most probably \iso\ -- \ndisc\ transition occurs first. Thus somewhere
along the path of varying molecular shapes, described
above, leading from $\shbx=0$ to $\shbx=\sqrt{3/2}$, we should cross the
boundary between oblate and prolate nematic states, and hence stabilize the
biaxial nematic. What happens at the point where we cannot tell whether the
nematic phase is of disc-like or rod-like type, and how this point is realized,
and how bifurcation temperature and density change along the chosen path 
in \epsilonset,\sigmaset\ space, will be discussed in the subsequent part
of the section.
\begin{figure}[t]
 \centering
 \subfigure[Temperature versus density at \nrod\ -- \nbx\ bifurcation, for some shape biaxialities.]{\label{figure:shapebx1} \includegraphics[scale=0.8]{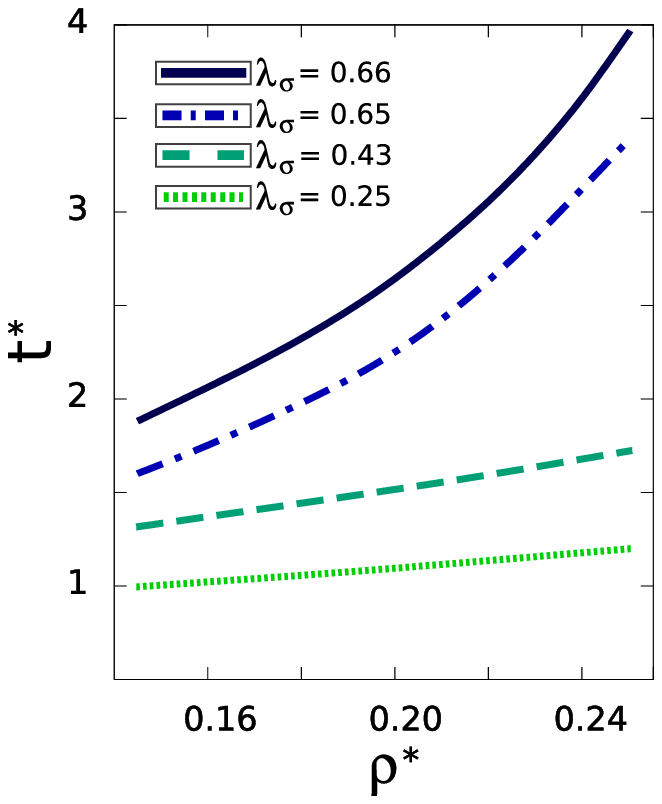}}
 \subfigure[Bifurcation temperature as function of shape biaxiality $\shbx$. \iso\ -- \nrod\ and \nrod\ -- \nbx\ branches are plotted at $\dimensionless{\rho}=0.18$.]{\label{figure:shapebx2} \includegraphics[scale=0.8]{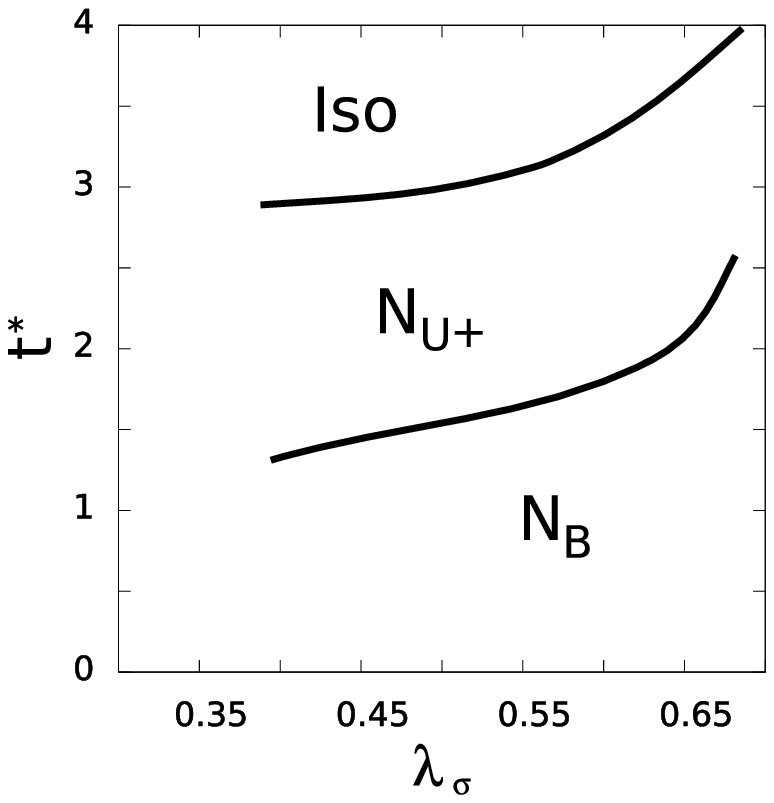}}
 \caption[Bifurcation diagrams for energy biaxiality $\enbx=0.0$.]{\label{figure:shapesbxandtvsrho} Bifurcation diagrams for energy biaxiality $\enbx=0.0$ (\gbsete{d} in \trt{biaxialities}). $\shbx$ varies along thick, gradient path in \frf{shbxdensitypath}.}
\end{figure}

 There are some matters which are of interest, and we take them systematically
into account. Firstly, there is the question of opposite signs of $\shbx$
and $\enbx$, which was pointed out by Berardi and Zannoni in \cite{bz2k}.
Next, we can wonder how the change in the shape or energy biaxiality 
separately vary the transition temperatures. Then, we can ask ourselves
how those two parameters concur in the creation of biaxial nematic, and
whether it is possible to induce a Landau point in this model by
changing $\enbx$ and $\shbx$, and if so, then how the position
of that point changes with the biaxialities. Since we know the relation between
axes of biaxial ellipsoids at self-dual point for the models of
purely repulsive forces \re{squarerootrule} (hard molecules \cite{holystponiewierski,mulderhard}), once we find this point in present model, we can estimate the influence of attractive (soft) interaction on its position, but more importantly
we can check whether a similar condition is in force for parameters \sigmaset\ and \epsilonset.
\subsubsection{Molecular shape effects}
 Firstly, we take a look at the influence of shape biaxiality on the
bifurcation temperature for the rod-like uniaxial to biaxial nematic transition.
We change $\shbx$, while keeping $\{\epsilon_{i}\}$ fixed, along the path depicted
in \frf{shbxdensitypath} (thick, gradient line).
Figure \reffig{shapebx1} represents bifurcation lines in $(\dimensionless{\rho},\dimensionless{t})$ plane for different values of $\shbx$, and since we are
interested in shape dependence alone, the energy biaxiality $\enbx=0.0$ (for \gbsete{d} in \trt{biaxialities}).
As expected, the increase in the molecular
biaxiality leads to higher temperatures for \nrod\ -- \nbx\ transition,
chances for observing the biaxial nematic phase are increasing.
\nlin \frf{shapebx2} shows the bifurcation diagram, where temperatures
are plotted as function of shape biaxiality both 
for transition from isotropic phase and for \nrod\ -- \nbx\ case.
We can see that the line of uniaxial -- biaxial nematic bifurcation
gets closer to the \iso\ -- \nrod\ one, with a prospect of meeting it eventually.
\begin{figure}[t]
 \centering
 \subfigure[Bifurcation temperature versus density for \nrod\ -- \nbx\ transition for a couple of energy biaxialities,]{\label{figure:energybx1} \includegraphics[scale=0.8]{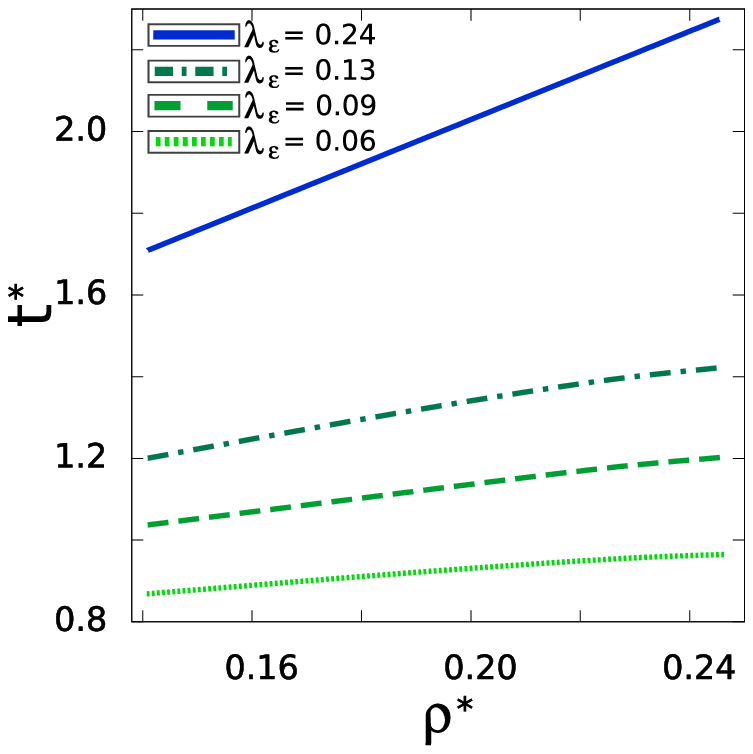}}
 \subfigure[Bifurcation temperature as function of energy biaxiality. \iso\ -- \nrod\ and \nrod\ -- \nbx\ transitions are plotted for $\dimensionless{\rho}=0.18$.]{\label{figure:energybx2} \includegraphics[scale=0.8]{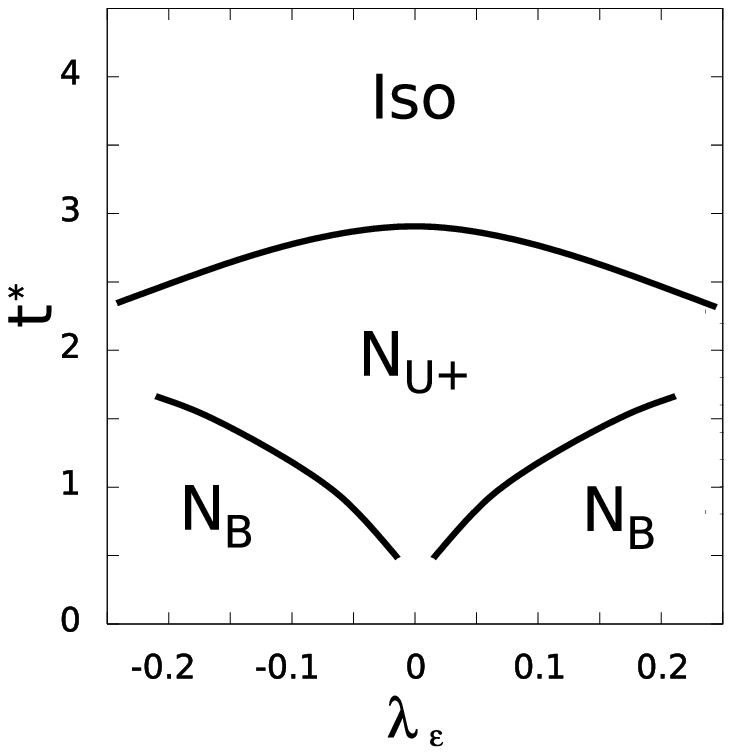}}
 \caption[Bifurcation diagrams for uniaxial molecule shape; $\shbx=0.0$.]{\label{figure:energiesbxandtvsrho} Bifurcation diagrams for uniaxial molecule shape; $\shbx=0.0$ (\gbsets{a} in \trt{biaxialities}). $\enbx$ is changed, as described before, by varying $|\ex-\ey|$, while $\ez=0.2$.}
\end{figure}
\subsubsection{Biaxiality of soft part of the potential}
 Now we show how the biaxiality stored in $(\ex,\ey,\ez)$ affects the
transition point. For fixed $\shbx$, we change $\enbx$ by
increasing $|\ex-\ey|$ while keeping $\ez=0.2$. In \frf{energybx1} we present
bifurcation lines ($\dimensionless{t}$ as function of $\dimensionless{\rho}$)
for \nrod\ -- \nbx\ transition, plotted for different values of $\enbx$ and for
uniaxial molecule ($\shbx=0.0$ for \gbsets{a} in \trt{biaxialities}).
As in the case of \figreffig{shapebx1}, the
dependence of the temperature on density in \figreffig{energybx1} is almost
trivial (linear), and is also weaker than in \frf{shapebx1}.
\nlin The Figure \reffig{energybx2} shows the bifurcation diagram with
isotropic, \nrod, and \nbx\ regions as function of $\enbx$, for
uniaxial shape $\shbx=0.0$ for \gbsets{a}. It can be seen that for low
shape biaxialities it is required to supply increasingly high energy biaxiality
in order to produce \nbx. As was mentioned earlier, $\enbx$ was changed by increasing the
difference between $\epsilon_{x}$ and $\epsilon_{y}$, and essentially we did not
cross the boundary between ''disc-like'' and ''rod-like potential'' regimes,
that is the reason why in \frf{energybx2} only prolate nematic phase is
observed. Nevertheless, the plot suggests that it is possible for the lines
of \iso\ -- \nrod\ and \nrod\ -- \nbx\ transitions to meet somewhere,
for higher energy biaxiality.
\nlin From Figures \rf{shapesbxandtvsrho} and \rf{energiesbxandtvsrho}, we can
see that the main factors affecting the diagram are the biaxiality of
molecule $\shbx$ and the biaxiality of potential $\enbx$. Thus we can
wonder how the combined shape and energy biaxialities influence
the bifurcation temperature.
\subsubsection{Concurring biaxialities}
\begin{figure}[t]
 \centering
 \subfigure[Bifurcation temperature as function of shape biaxiality, for some $\enbx$.]{\label{figure:concurringvsshbx} \includegraphics{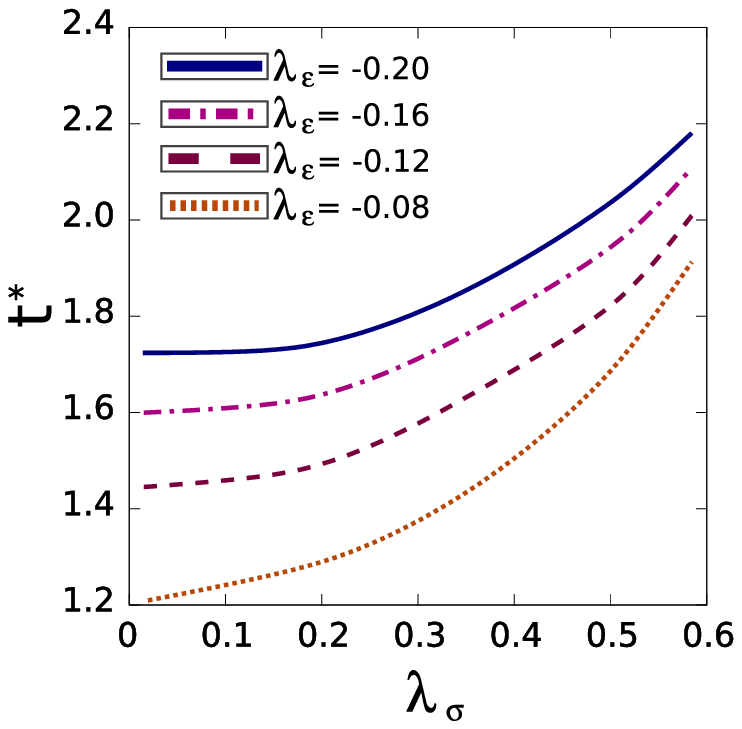}}
 \subfigure[Bifurcation temperature as function of energy biaxiality, for some $\shbx$.]{\label{figure:concurringvsenbx} \includegraphics{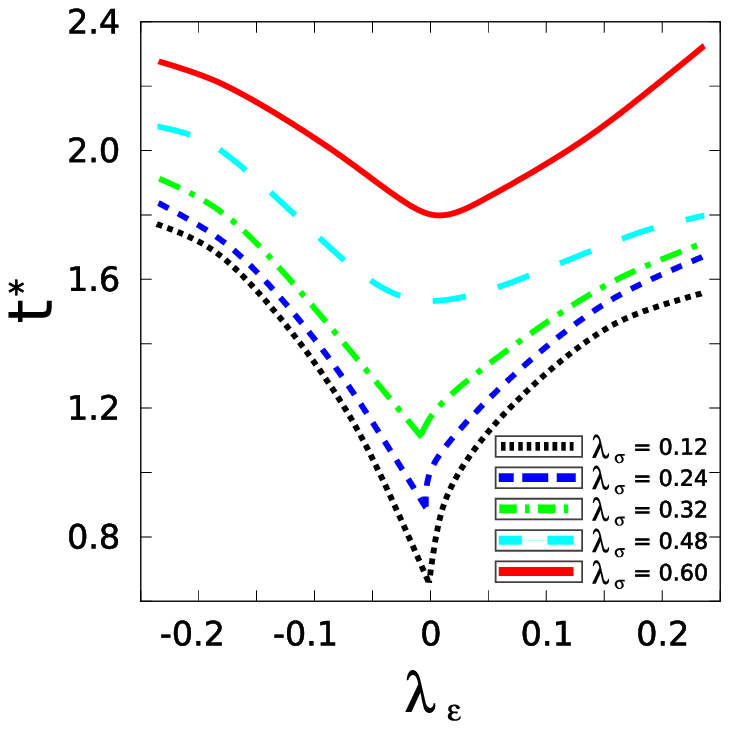}}
 \caption[Concurring $\shbx$ and $\enbx$ influence on bifurcation diagram.]{Concurring biaxialities and their influence on bifurcation diagrams for \nrod\ -- \nbx\ transition for $\dimensionless{\rho}=0.18$. $\shbx$ varies along the thick, gradient line in \frf{shbxdensitypath}, and $\enbx$ is changed by increasing $|\ex-\ey|$ while keeping $\ez=0.2$.}
\end{figure}

\putfigurescaleyoudecidelong{part2/chapter2/plot_concuringbx1.eps}{Bifurcation diagram for \nrod\ -- \nbx\ transition for negative and positive energy biaxiality $\enbx$, $(\ex,\ey,\ez)=(1.77,0.63,0.2)$ and $(\ex,\ey,\ez)=(0.63,1.77,0.2)$, respectively. $\shbx$ increases along the path in \frf{shbxdensitypath} (thick, gradient line). $\dimensionless{\rho}=0.18$.}{concurringbx0}{0.8}{\nrod\ -- \nbx\ bifurcation lines for $\enbx<0$ and $\enbx>0$.}
 Turning our attention to the question of the influence of both biaxialities,
now we allow $\enbx$ and $\shbx$ to change simultaneously. It is done as before:
shape biaxiality varies along the path in \frf{shbxdensitypath},
and $\enbx$ by increasing $|\ex-\ey|$ for $\ez=0.2$. Firstly we consider the
case of positive and negative $\enbx$, while keeping $\shbx>0$.
Figure \reffig{concurringbx0} shows the bifurcation temperatures as function
of shape biaxiality for two cases of $\enbx=-0.16$ and $\enbx=0.16$, for
\nrod\ -- \nbx\ transition. This plot shows the difference between
opposite signs of $\shbx$ and $\enbx$. It can be seen that the branch
corresponding to negative energy biaxiality is above the positive one almost
everywhere, except the point where $\shbx=0.0$ (\figreffig{energybx2} is
symmetric under the change of sign of $\enbx$), and the region where
shape biaxiality becomes (relatively to $\enbx$) high enough to take control
over the transition. A more detailed study on the concurring biaxialities is
presented in Figs.~\rf{concurringvsshbx}-\subref{figure:concurringvsenbx},
where the lines of bifurcation for \nrod\ -- \nbx\ transition are plotted as
function of $\shbx$ for different values of $\enbx$ (\frf{concurringvsshbx})
and in the opposite case (\frf{concurringvsenbx}).
\newline\indent Summary of the results can be seen in Figures
\reffig{concurringbx3} and \reffig{concurringbx4}, where the surface of 
bifurcation points is plotted against both the shape and energy biaxialities, in
the rod-like and disc-like molecular regime, respectively. The line drawn on the
upper surface (representing the transition from isotropic phase to \nun)
in \figreffig{concurringbx3} is a family of Landau points where the direct
second order transition to biaxial nematic from isotropic liquid phase
takes place. Those points will be taken into closer consideration
in the following section.
\putfigurelong{part2/chapter2/plot_bxtemp_vs_shbx_enbx.eps}{Bifurcation temperature $\dimensionless{t}$ for $\dimensionless{\rho}=0.18$ for transition from isotropic liquid to \nrod\ (upper surface), and \nrod\ -- \nbx\ transition (lower surface) as function of shape ($\shbx$) and energy biaxiality ($\enbx$). $\shbx$ changes along the path in \frf{shbxdensitypath}, $\enbx$ is varied by increasing $|\ex-\ey|$ while $\ez=0.2$. Dark line on the upper surface marks the Landau points presented in detail in \frf{landaupos}.}{concurringbx3}{Bifurcation temperature versus $\shbx$ and $\enbx$ in rod-like regime.}
\subsubsection{Landau points}
\begin{figure}[t]
 \centering
 \subfigure[$\enbx=0.0$ for \gbsete{d} (see \trt{biaxialities}). Landau point at $\shbx=0.7163$ for \gbsets{g} ($\lnbra \sigma_{x},\,\sigma_{y},\,\sigma_{z} \rnbra=\lnbra 1.4,\,0.544,\,3.17 \rnbra$) is marked. Dotted line represents the self-dual geometry for hard ellipsoids \re{squarerootrule} from \cite{holystponiewierski} at $\shbx=0.6412$ \gbsets{c} ($\lnbra \sigma_{x},\,\sigma_{y},\,\sigma_{z} \rnbra=\lnbra 1.4,0.637,3.077 \rnbra$).]{\label{figure:landaushbx}\includegraphics[scale=0.68]{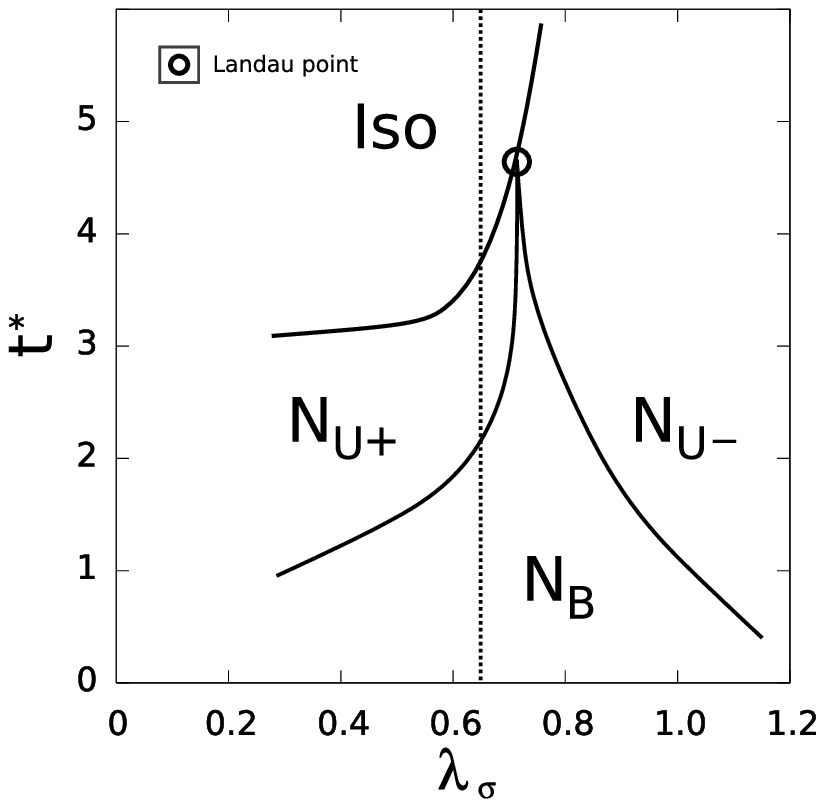}}
 \subfigure[$\shbx=0.70$ \gbsets{f}; two Landau points are visible at
 $\enbx=-0.1974$ and $\enbx=0.0662$ for(see \trt{biaxialities}) \gbsete{b} and
 \gbsete{f}, respectively. They meet at $\shbx=0.7163$ for \gbsets{g} at the Landau point in Fig.~\subref{figure:landaushbx}.]{\label{figure:landauenbx}\includegraphics[scale=0.68]{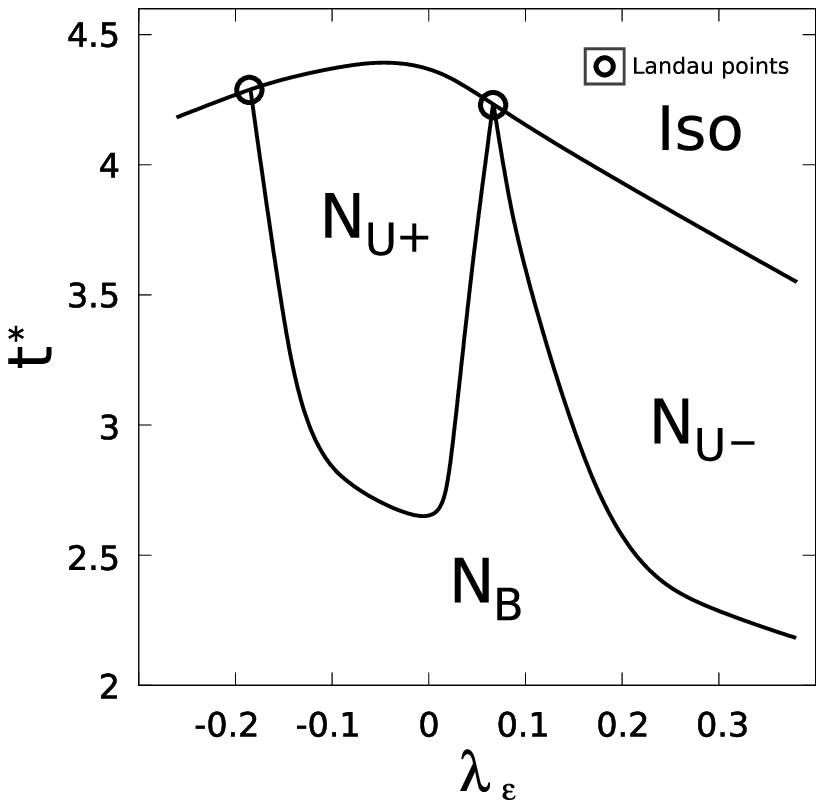}}
 \caption[Bifurcation temperature $\dimensionless{t}$ as function of $\shbx$ \subref{figure:landaushbx}, and $\enbx$ \subref{figure:landauenbx}.]{\label{figure:landaushbxlandauenbx} Bifurcation temperature $\dimensionless{t}$ for \studiedrho\ as function of $\shbx$ \subref{figure:landaushbx}, and $\enbx$ \subref{figure:landauenbx}. $\shbx$ varies along the path drawn in \frf{shbxdensitypath}, $\enbx$ is changed by increasing $|\ex-\ey|$ while $\ez=0.2$.}
\end{figure}

\indent Previously, we have seen how the parameters $\enbx$ and $\shbx$,
representing the degree of overall biaxiality of the intermolecular
potential energy \re{bfzu}, influence the bifurcation temperature and density.
In present section we will locate the Landau points
in $\lnbra \shbx,\enbx \rnbra$ space, keeping in mind that $\shbx$ is changed
along the path drawn with gradient in \frf{shbxdensitypath}, while 
in the case of $\enbx$ we hold $\ez=0.2$ and starting from $\enbx=0.0$
for $\ex=\ey$, vary the energy biaxiality by increasing $|\ex-\ey|$. We will
also comment on the meaning of the hard biaxial ellipsoids self-dual plane
$(\enbx^{sd},\shbx^{sd})$ \re{squarerootrule}-\re{selfdualepsilon},
and its relation to Landau points in present model. Firstly, we will
investigate uniaxial energy $\enbx=0.0$ (\gbsete{d} in \trt{biaxialities})
and look for shape-induced effects, then we will
turn to the case of fixed shape biaxiality and by changing $\enbx$ pursuit the
Landau point, and finally we will show how both $\enbx$ and $\shbx$
change the position of this point on the bifurcation diagram.
\newline\indent In \frf{landaushbx} we present the bifurcation line
as function of shape biaxiality for $\enbx=0.0$ for \gbsete{d}. It can be seen
that there exists a point at the boundary of disks and rods, where the 
molecule belongs to neither one of those groups and \nrod\ and \ndisc\ 
phases are indistinguishable. This is an isolated Landau point, located near 
$\shbx=0.7163$ for \gbsets{g} (see \trt{biaxialities}). It differs
from the predicted for hard biaxial ellipsoids \cite{holystponiewierski,mulderhard} self-dual geometry $\sigma_{x}=\sqrt{\sigma_{y}\sigma_{z}}$ \re{squarerootrule}, which in our
parametrization would be at $\shbx=0.6412$ for \gbsets{c}.
As we can see, the attractive forces shift the position of Landau point
towards higher values of $\shbx$.
\nlin If we can induce the transition directly to biaxial nematic phase by
changing $\shbx$, then it should be possible to do the same thing for $\enbx$.
Figure \reffig{landauenbx} shows such bifurcation diagram as function 
of energy biaxiality for $\shbx=0.70$ for \gbsets{f}. It is clear that for $\shbx \neq 0$ by
changing $\enbx$ we can induce a Landau point. We should recall, however,
that we do not cross the boundary between ''rod-like'' and ''disc-like''
energies (in terms of $(\epsilon_{x},\epsilon_{y},\epsilon_{z})$ or $\enbx$).
It turns out that for shape biaxialities close to the ones giving Landau point
by strengthening the lateral interactions we can also induce a self-dual point
(e.g. the left one in \frf{landauenbx}).
\putfigureflagslong{part2/chapter2/plot_bxtemp_vs_shbx_enbx-highshbx.eps}{Bifurcation temperature $\dimensionless{t}$ for \studiedrho\ for \ndisc\ -- \nbx\ transition as function of shape ($\shbx$) and energy biaxiality ($\enbx$) in the disc-like regime. $\shbx$ changes along the path in \frf{shbxdensitypath}, $\enbx$ is varied by increasing $|\ex-\ey|$ while $\ez=0.2$.}{concurringbx4}{b}{Bifurcation temperature versus $\shbx$ and $\enbx$ in disc-like regime.}
\newline\indent Plot presented in Figure \reffig{landauenbx} gives us also some 
insight into the issue of dependence of Landau point position on $\enbx$ and $\shbx$. There exist two such points in Fig. \reffig{landauenbx}
for $\shbx=0.70$ whereas in Fig. \reffig{landaushbx} for a bit higher
shape biaxiality $\shbx=0.7163$ there is only one Landau
point visible. Given all this we can wonder how the location of 
this point changes under the varying biaxialities. We can plot
Landau point position and divide the $\lnbra \shbx,\,\enbx \rnbra$ space 
into disc and rod-like regions, as is shown in Figure \reffig{landaupos},
while keeping in mind the way $(\sigma_{x},\sigma_{y},\sigma_{z})$
and $(\epsilon_{x},\epsilon_{y},\epsilon_{z})$ were changed and combined
into $\shbx$ and $\enbx$. It can be seen that it is ''easier'' to procure a
transition directly to biaxial nematic phase for the case of the same sign
of $\shbx$ and $\enbx$, whereas in the case of negative 
$\enbx$ it takes stronger (relatively to $\shbx$) 
energy biaxialities to induce a Landau point. The increase of $\shbx$ resulted
in meeting of the two Landau points visible on Fig. \reffig{landauenbx} 
at a single point for $\lnbra \shbx,\enbx \rnbra=\lnbra 0.7163,0.0\rnbra$ 
as presented in Fig. \reffig{landaushbx}.
\nlin More interestingly, we can make some comments on the meaning of the self-dual
condition for hard, biaxial ellipsoids \re{squarerootrule}-\re{selfdualepsilon}.
We can see that it is possible to find Landau point away from the
square root plane \sdplane\ \re{selfduallambda}-\re{selfdualepsilon}, it is
visible in \frf{landaushbx}. We have also investigated vicinity of the point
lying in this plane at $(\enbx^{sd},\shbx^{sd})=(0.175,0.6412)$ (\gbsete{h}
and \gbsets{c} in \trt{biaxialities}) and found the Landau point
at $(\enbx,\shbx)=(0.175,0.6596)$ (\gbsete{h} and \gbsets{d}).
We can conclude, although based only on the four points (lying on the upper
line in \frf{landaupos}), that
square root rule \re{squarerootrule} is a good candidate for the starting 
position in the search for Landau points; we have shown that self-dual points
can exist beyond \sdplane\ plane, but also in its vicinity. As we can see, the
upper line in \frf{landaupos} crosses the vicinity of the point on
the plane in $(\enbx,\shbx)$ space defined by square root rule.
When we consider the complicated, non-linear manner in which \epsilonset\ 
and \sigmaset\ contribute to the biaxial Gay-Berne potential, it is surprising
that the Landau points can be found close to the plane
obtained from the simple square root rule: $\ex=\sqrt{\ey\ez}$, $\sx=\sqrt{\sy\sz}$.

 The bifurcation analysis can give the position of Landau point
(using \ere{landaupointcondition}); however, in its vicinity it becomes
increasingly hard to stabilize the uniaxial nematic reference phase because it
looses stability at this point. Therefore, in order to confirm the places where
self-dual point appears to exist, we have employed the method of minimisation
of the Helmholtz free energy \re{freeenergyuniform} with
trial \sopdf \re{usedmodelopdf} by solving the equations \re{scforalpha} and
obtained the temperature and density of the transition by calculating
order parameters \re{orderparamsset}. An example of the behaviour
of the leading order parameters at the Landau point is presented
in \frf{orderparams}, and for comparison we also show the plots away from
this point in rod-like regime, both obtained from the minimisation method.
\subsection{Summary}
\indent The uniaxial Gay-Berne potential \cite{bernepechukas,gayberne} achieved
an amazing, compared to its simplicity, success in predicting liquid crystal
phase behaviour. As pointed out before, most of the mesogenic molecules are
not uniaxial. They are not biaxial either, but effectively in the
uniaxial phase they can be approximated by ellipsoids of revolution, as was
done in the original Gay-Berne interaction, and in biaxial phase
by $D_{2h}$ -- symmetric objects, as in the potential developed
by Berardi, Fava and Zannoni \cite{bzdevelopmentofpotential}.
We have investigated the latter model.
\putfigureflagslong{part2/chapter2/plot_landaupointsposition.eps}{Division of
$\lnbra \enbx,\shbx \rnbra$ space by a line of Landau points separating
disk-like and rod-like states. $\shbx$ runs along the thick, gradient path in
\frf{shbxdensitypath}, and $\enbx$ is changed by increasing $|\ex-\ey|$ while
keeping $\ez=0.2$. The line joins the marked Landau points, and the dotted
lines indicate the points where square root rule \re{squarerootrule} is
fulfilled for \epsilonset\ and \sigmaset, their intersection (point on
the \sdplane\ plane) is marked.}{landaupos}{t}{Division of $\lnbra \enbx,\shbx \rnbra$ space into oblate and prolate regions.}
\nlin We compared the bifurcation analysis of DFT in low-density approximation
with the Monte Carlo results. As expected the bifurcation temperatures were
significantly higher than those following from simulations. This tendency
was more apparent in case for \iso\ -- \nrod\ transition, while
for \nrod\ -- \nbx\ bifurcation the approach provided more accurate estimates
in relation to Monte Carlo. The comparison was only qualitative
because the simulation study neglected long-range corrections, the inclusion
of which makes the results of Monte Carlo closer to those of DFT \cite{frenkelsmit}.
\nlin Then, in order to provide some insight into the way the reference
state influences the phase diagram, we calculated the bifurcation
for \nrod\ -- \nbx\ transition using the minimisation of the
Helmholtz free energy. This method does not involve the truncation of the 
expansion of the pair direct correlation, but uses the whole function,
and therefore, compared to bifurcation for second order phase transition, 
provides the estimation of the influence of precisely calculated \sopdf
of reference phase. The results show that the inclusion of terms
with higher angular momentum index than $2$ in the expansion of
pair direct correlation function has secondary meaning.
\putfigureflagsscalelong{part2/chapter2/plot_landaupointsorderparams}{Temperature behaviour of leading order parameters at fixed density \studiedrho\ in the vicinity of Landau point (on the left) at $\shbx=0.7163$ (\gbsets{g} in \trt{biaxialities}) and, on the right, away from that point for rod-like molecule (giving \nrod) for $\shbx=0.58$, $\enbx=-0.06$ ($(\sigma_{x},\sigma_{y},\sigma_{z})=(1.4,0.714,3.0)$, $(\epsilon_{x},\epsilon_{y},\epsilon_{z})=(1.7,1.0,0.2)$). As obtained from the minimisation of the Helmholtz free energy along \ere{freeenergyuniform}, \re{orderparametersselfconsistent}, and \re{scforalpha}. }{orderparams}{t}{0.8}{Leading order parameters at Landau point and in rod-like regime.}
\nlin Next, we continued to investigate the influence of
introduced earlier shape biaxiality $\shbx$ and energy biaxiality $\enbx$ on
the bifurcation diagram, by defining a path in the six dimensional space
of potential shape and energy parameters. We presented the diagrams for
uniaxial cases of the potential ($\enbx=0.0$) for different values of the
biaxiality of shape, and uniaxial shape ($\shbx=0.0$) was studied for given energy biaxialities.
We also addressed the issue of opposite signs of $\shbx$ and $\enbx$. With
increasing positive $\shbx$, for uniaxial -- biaxial nematic bifurcation the
line of transitions corresponding to $\enbx<0$ was lying in temperatures higher
than the one associated with $\enbx>0$, apart from the point when $\shbx$ became
large enough to dominate the transition. The results showed
how the \nun\ -- \nbx\ transition is influenced by the Gay-Berne interaction
biaxiality originating from different sources. They also suggested that the
model can exhibit a Landau point; this possibility was also investigated.
\nlin We found that by increasing $\shbx$ for $\enbx=0.0$ we can cross
the boundary between the oblate and prolate nematic phases and locate the
Landau point when the distinction between them is not possible, where
\iso, \ndisc, \nrod, and \nbx\ phases ''meet'', and the second order
isotropic -- biaxial nematic phase transition occurs. This shape-induced 
self-dual point was found to be located at higher shape biaxiality
than the one obtained for hard molecules.
In terms of $\shbx$, the difference between the shape biaxiality at
Landau point following from our analysis (with $\enbx=0.0$, so the biaxiality
of the interaction originated only in shape of molecules), and the one obtained
for hard biaxial ellipsoids was found to be $0.0751$. This difference is the
estimate of the degree of the influence of attractive, anisotropic forces on
Landau point position. The self-dual points were also found to occur for fixed
shape biaxiality when $\enbx$ was changed. In this case we varied the energy 
biaxiality by going from the model of strong lateral interactions to the 
one where molecules are most attracted to their faces (we did not cross the
''rod-like'' and ''disc-like potential'' boundary in terms of $\enbx$). When
energy parameters fulfilled
the self-dual square root rule for hard ellipsoids, the Landau point 
was found for the $\shbx$ different from the one predicted for hard potentials
by $0.0184$. The same difference, far away from $\ex=\sqrt{\ey\ez}$,
for $\enbx=0.0$ was more than four times higher. The matter requires
further studies, but our results suggest that the meaning of the
square root rule maintains its importance as the qualitatively accurate
estimates for Landau region for soft biaxial Gay-Berne interaction.
We concluded our findings by the division of $(\enbx,\shbx)$ space
into regions belonging to the oblate and prolate states. It proved that
the line of Landau points crosses the vicinity of the point following from
square root rule for \epsilonset\ and \sigmaset. The existence of Landau points their position and order
of the transition were confirmed by the minimisation of the
Helmholtz free energy.
\nlin The thermotropic biaxial nematic phase discovered in simulations by
Berardi and Zannoni was reported for the model of stronger lateral interactions,
i.e., where the \ssts\ configuration is preferred (see \frf{potential}),
that is, for opposite signs of shape and energy biaxiality. Our results suggest
that while the temperature of \nrod\ -- \nbx\ transition is relatively higher
for negative $\enbx$ (while $\shbx$ is greater than zero, see \figreffig{concurringbx0}) it is necessary to provide significantly lower (negative)
energy biaxialities to induce the Landau point in that case (see \figreffig{landaupos}). Whereas for positive biaxialities the transition directly to
biaxial nematic can occur for moderate values of $\enbx$ and $\shbx$.
We shed some light on the issue of the case $\shbx>0$, $\enbx<0$ in the
last chapter, where we take into account uniaxial and biaxial smectic-A phases;
it is natural to suspect that those structures are preferred over \nbx\ when 
the interaction favours \sftf\ configuration over \ssts.

 Apparently, in case of biaxial nematic, the significance
of the Gay-Berne interaction is much smaller than for uniaxial phases.
Therefore, in order to continue the search for factors behind the stabilisation
of \nbx, we need to study a model that can be more closely related to the
real systems. We turn our attention to the \sbentcore molecules.


\section{Systems of \sbentcore molecules}
\label{section:bentcore}
 In present section we consider the models of \sbentcore (banana)
molecules built from Gay-Berne segments\footnote{Which is a simple way to do it, but at least guarantees the basic correct behaviour of the molecule.}. We study
shape-related effects and the influence of dipole-dipole interaction. Firstly,
we take into account the case of \dinfh -- symmetric arms and join two and
three prolate, soft ellipsoids, as presented in \frf{bananaconstruction},
to model banana-like molecules with bend (opening) angle $\gamma$. 
The system is studied for fixed density, and the resulting bifurcation diagrams
are presented in the plane of temperature $\dimensionless{t}$ (see \re{gbreducedunits}) and $\gamma$.
We especially focus on Landau point position, as obtained from \re{landaupointcondition}, and in order to shed more light on this issue, we also take a look at
the behaviour of Landau region when the arms deviate from \dinfh symmetry;
we incorporate two soft, biaxial ellipsoids into a \sbentcore molecule
using the Berardi-Fava-Zannoni potential \cite{bzdevelopmentofpotential},
studied in the previous section.
Inspired by recent developments \cite{mcasymmetric}, we also comment on the
asymmetric \sbentcore molecules with arms modelled by
soft, \dinfh -- symmetric ellipsoids of different elongations,
using potential from \cite{generalizedungb}.
Finally, we turn to the effects of polarity, and we include dipole-dipole
interaction between the dipole moments located along $C_{2}$ symmetry axis
of the molecule
(see \frf{bananaconstruction}). We present the phase diagrams in $(\dimensionless{t},\gamma)$ plane for fixed density at bifurcation point, again focusing on
the behaviour of Landau point.
\newline\indent We begin with a description of the model, then 
we present the bifurcation diagrams for non-polar case for uniaxial
and biaxial arms, and then we study the influence of dipole-dipole interaction.
\subsection{Models of a \sbentcore molecule}
\indent This section is devoted to the description of the pair potential that
was used as an interaction between two \sbentcore molecules. We are using
two already mentioned models, those are: the uniaxial Gay-Berne
\cite{bernepechukas,gayberne} and its biaxial version as proposed by
Berardi, Fava, and Zannoni \cite{bz2k}. \frf{bananaconstruction} shows the
construction. Each of the parts interacts with every one on the other banana,
while the interaction between segments within single molecule, giving an
irrelevant constant, is disregarded. We consider rigid bananas, i.e., we do not
allow the change of the angle $\gamma$ or dimensions of the ellipsoids
during the calculations.
\putfigureflagsscalelong{part2/chapter3/banana_construction.eps}{Construction of \sbentcore molecule and pair potential. On the left there are frames of reference associated with banana $1$: $\unitv{b}^{1}_{1},\unitv{b}^{1}_{2},\unitv{b}^{1}_{3}$, and $2$: $\unitv{b}^{2}_{1},\unitv{b}^{2}_{2},\unitv{b}^{2}_{3}$ ($\unitv{b}^{1}_{1}$ and $\unitv{b}^{2}_{1}$ are perpendicular to the picture surface).}{bananaconstruction}{b}{0.8}{Construction of \sbentcore molecule and pair potential.}
\nlin The intermolecular potential can be written as (see Fig.~\reffig{bananaconstruction}):
\begin{equation}
 \label{eq:bentcoremodelpotential}
\begin{split}
 V_{b2} = & \sum_{a,b\in \left\{1,2\right\}} V_{GB}\left(\unitv{u}_{1,a},\unitv{u}_{2,b},\boldv{r}_{ab}\right) \, , \\
 V_{b3} = & \sum_{a,b\in \left\{1,2,3\right\}} V_{GB}\left(\unitv{u}_{1,a},\unitv{u}_{2,b},\boldv{r}_{ab}\right) \, ,
\end{split}
\end{equation}
where, as before, $V_{GB}$ is either the uniaxial Gay-Berne potential, described
in \secrefsec{ungb}, or its biaxial version studied in previous section and
introduced in \secrefsec{bzmodel}, $\unitv{u}_{i,j},\, (i=1\dots 2,\, j=1\dots 3)$ are unit vectors parallel to the Gay-Berne segments symmetry axes, $\boldv{r}_{ij}$ are vectors connecting the centres of molecules. For \dtwh -- symmetric Gay-Berne,
in order to describe the orientation of the ellipsoid, it is not enough to
give the vector parallel to the longer axis, in this case we fixed the
remaining axes with respect to the plane of the angle $\gamma$, namely one axis
was set to lie in this plane and the other to be perpendicular to it. For
the uniaxial Gay-Berne the empirical exponents $\tilde{\nu}$, $\tilde{\mu}$
are chosen to be to $1$ and $2$, respectively, and the elongation of the
ellipsoids is $5:1$, and anisotropy of the potential $4:1$, in terms of constants from \re{ungbsigma} and \ere{ungbepsilon0here} it means $\kappa=5$ and $\chi^{'}=1/3$ (see \frf{exampleofgbeqsurf}). The molecular elongations are chosen
to correspond to the dimensions of the \sbentcore compounds.
\newline\indent We have also added the dipole
moment localized in the fixed position, as shown in \frf{bananaconstruction},
along molecular $C_{2}$ symmetry axis. The dipole-dipole interaction
used is of a well known form:
\begin{equation}
 V_{DD}\left(\boldv{\boldm{\mu}}_{1},\boldv{\boldm{\mu}}_{2},\boldv{r} \, \right) = 
   \frac{ \boldv{\boldm{\mu}}_{1}\cdot\boldv{\boldm{\mu}}_{2}
         - 3 \left( \boldv{\boldm{\mu}}_{1}\cdot\unitv{r} \right)
        \left( \boldv{\boldm{\mu}}_{2}\cdot\unitv{r} \right) }{ r^{3} } 
 \nonumber \, ,
\end{equation}
where $\boldv{\boldm{\mu}}_{i}=\dimensionless{\mu} \unitv{\boldm{\mu}}_{i}$, 
and where $\unitv{\boldm{\mu}}_{i}$ is a unit vector, and $\dimensionless{\mu}$
is the dimensionless (see \re{gbreducedunits}) magnitude of the dipole moment.
To provide some way of relating $\dimensionless{\mu}$ 
to the strength of $V_{GB}$, we have
calculated $\left|\frac{V_{DD}}{V_{b3}+V_{DD}}\right|$ in the ground state for
given values of $\dimensionless{\mu}$ (see Table~\reftable{muvsdd}). This
ratio gives relative weight of contribution from dipole-dipole interaction to
total potential; it will be used as a measure of dipole strength.
\nlin We employ the units of energy ($\epsilon_{0}$) and distance ($\sigma_{0}$)
as described by \re{gbreducedunits}. Results presented in this section follow
from using in place of $V$ in \eqrefeq{ldsettings}-\re{ldapproximation} $V_{b2}$
or $V_{b3}$, with $V_{DD}$ added in polar case.
\begin{table}[t]
 \caption[Dipole-dipole interaction contribution to total potential energy.]{ Dipole-dipole interaction contribution to total potential energy in the ground state. }
 \begin{tabular}{cc}
 \label{table:muvsdd}
 $\dimensionless{\mu}$ & $\left|V_{DD}/\lnbra V_{b3}+V_{DD}\rnbra\right|$ \n \hline\hline
 $2.8$ & $0.50$ \n
 $2.2$ & $0.40$ \n
 $1.6$ & $0.25$ \n
 $1.5$ & $0.22$ \n
 $1.2$ & $0.15$ \n
\end{tabular}
\end{table}
\nlin The absolute minimum of the potential between two banana-like molecules 
(the ground state) for the non-polar uniaxial model corresponds to the
configuration where ${\unitv{b}}^{1}_{i}$ and ${\unitv{b}}^{2}_{i}$ axes of
frames associated with molecule $1$ and $2$ (see \figreffig{bananaconstruction})
are parallel and the vector $\boldv{r}$ is oriented along $\unitv{b}^{1}_{1}$.
The inclusion of the dipole-dipole interaction changes that picture; in the ground state
relative orientation is the same, only $\boldv{r}$ is parallel
to $\unitv{b}^{1}_{2}$. \putfigureflagsscalelong{part2/chapter3/banana_eqsurfaces.eps}{Sample equipotential surfaces for banana composed of uniaxial Gay-Berne molecules, 3 parts (on the top) and 2 parts (on the bottom), for opening angle of $0.7\pi=126^{\circ}$. On the right, case with dipole-dipole interaction with dipole strength $\dimensionless{\mu}=2.0$. On the left, case without dipoles. Surfaces show the potential equal $0$ and $-0.2$.}{bananauneqsurfaces}{t}{1.8}{Equipotential surfaces for bananas of uniaxial Gay-Berne arms.} In numerical analysis
for two-part banana, to prevent the infinite values of potential
when dipoles are localized at the same point, we have
introduced a hard sphere between the arms with radius of $0.5\,\sigma_{0}$.
Fig.~\reffig{bananauneqsurfaces} (not containing the just-mentioned sphere)
shows the equipotential surfaces for the model where uniaxial
Gay-Berne potential was used, for four relative orientations of two molecules.
\nlin Every bifurcation diagram presented in this chapter was calculated at
fixed density chosen in the following way. If we calculate the inverse of 
the excluded volume for parallel \sbentcore molecules we will receive 
values of $0.024\,\sigma^{-3}_{0}$ and $0.016\, \sigma^{-3}_{0}$ for two,
and three -arm bananas, respectively. The densities on the diagrams were set
to be of that order, which corresponds to packing fractions
(molecular volume per average volume of molecule) of $0.2$ and $0.3$, being
slightly lower than typical values in uniaxial nematic which range 
from $0.4$ to $0.6$.
\nlin In the following section we present the bifurcation diagrams 
for \sbentcore molecules without dipole moments.
\subsection{Shape-induced effects}
 In present section we seek the answer for the most straight forward 
question of what happens to the bifurcation temperature as we change
the opening angle $\gamma$, and also whether we can find Landau points, where
direct transition from the isotropic to biaxial nematic phase takes place. Since
we pursuit only the shape related effects, we do not introduce any
dipole-dipole interactions yet. Firstly, we present the results for model
where the arms of \sbentcore molecule are uniaxial ellipsoids of 
revolution, and then we take a brief look at the case
of $D_{2h}$ -- symmetric building blocks.
\begin{figure}[t]
 \centering
 \subfigure[Diagrams for two-part, non-polar banana, for two values of bifurcation density $\dimensionless{\rho}$.]{\label{figure:diagramb2}\includegraphics{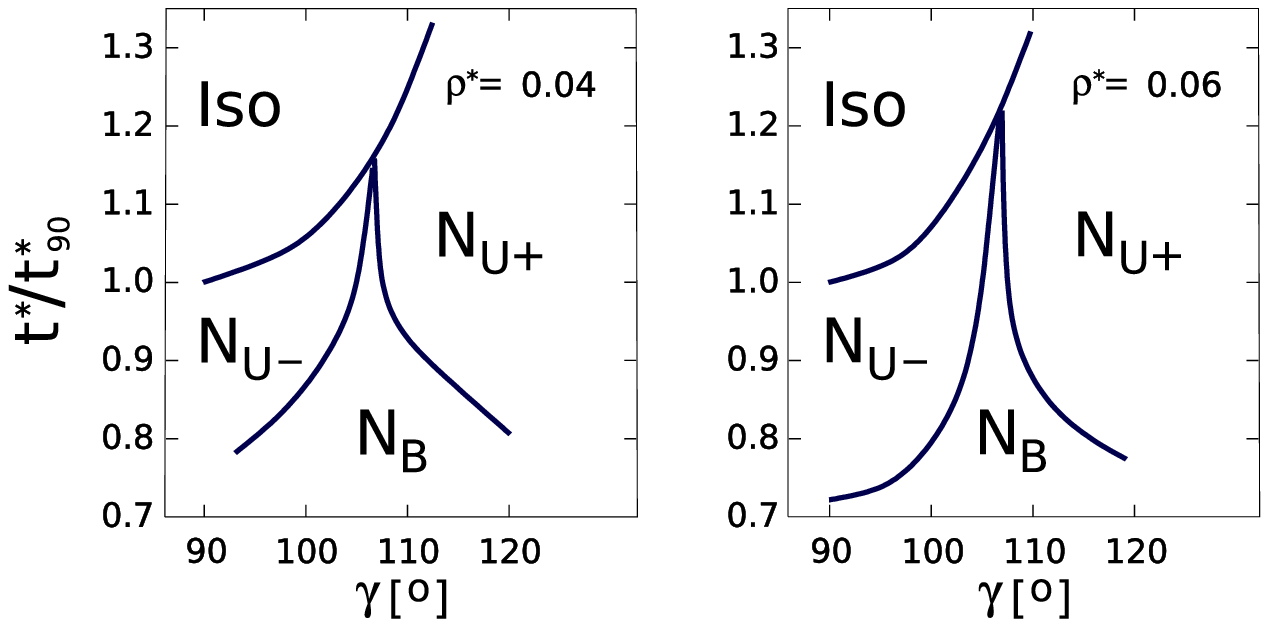}} \\
 \subfigure[Diagrams for non-polar model of three parts, for two bifurcation densities $\dimensionless{\rho}$.]{\label{figure:diagramb3}\includegraphics{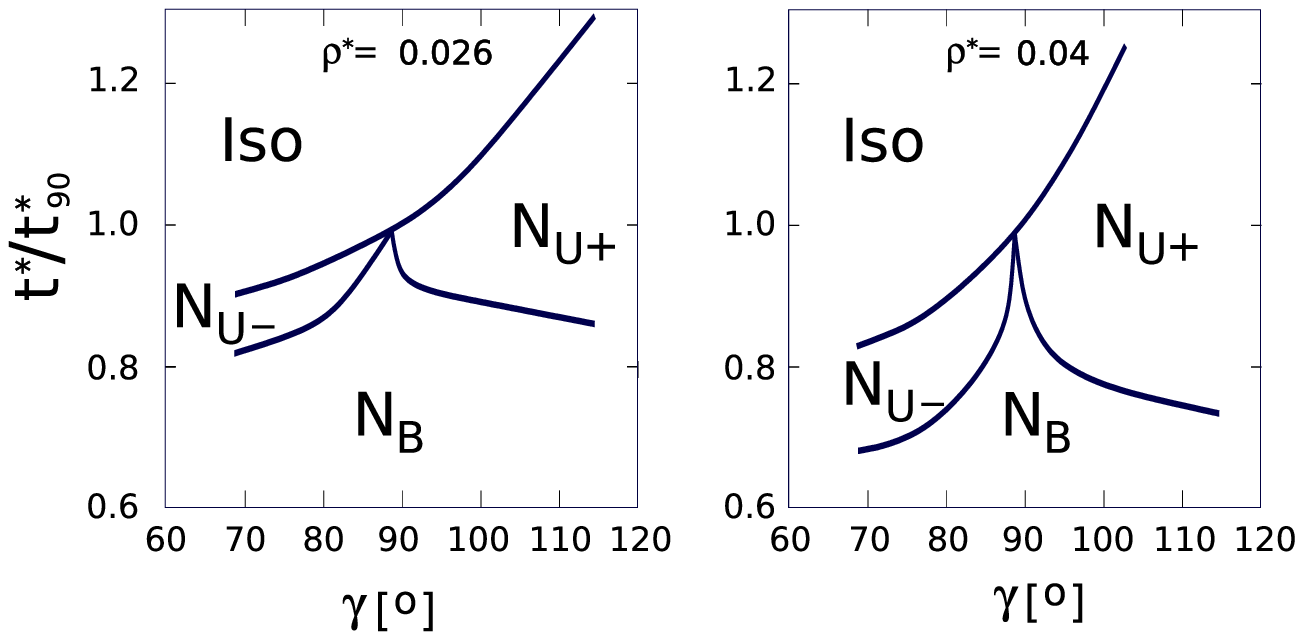}}
 \caption[Bifurcation diagrams for non-polar bananas.]{\label{figure:diagramnodd}Bifurcation diagrams for model \sbentcore molecules without dipole-dipole interaction. The temperatures are scaled by $\dimensionless{t}_{90}$, the bifurcation temperature for $\gamma=90^{\circ}$ (see Table~\reftable{rightangletemps}).}
\end{figure}
\subsubsection{Uniaxial arms}
 \label{section:bananagbun}
\indent  We can think of the arms of a banana (see Fig.~\reffig{bananasstructural}) as 
of uniaxial ellipsoids. This approximation probably will not reflect all 
properties of the \sbentcore molecule correctly, but can serve as a 
starting point for more complex models.
\begin{table}[b]
 \caption{ Temperatures of bifurcation from isotropic phase for $\gamma=90^{\circ}$. }
 \begin{tabular}{lcc}
 \label{table:rightangletemps}
 Model & $\dimensionless{\rho}$ & $\dimensionless{t}_{90}$ \n \hline\hline
 \multirow{2}{*}{two-arms \sbentcore molecule} & 
 $0.04$  & $0.84$ \n
 & $0.06$  & $0.96$ \n \hline
 \multirow{2}{*}{three-arms \sbentcore molecule} & 
 $0.026$  & $1.34$ \n
 & $0.04$  & $1.71$ \n
\end{tabular}
\end{table}
\newline\indent We have taken into account the molecules composed of two
and three Gay-Berne interacting parts. In Fig.~\reffig{diagramnodd} we show the
resulting bifurcation diagrams for non-polar banana for two densities. 
As can be seen, the opening angle at Landau point, for which
the direct transition from isotropic phase to the biaxial nematic takes place,
for two-part banana is equal to $\dg{107}$, which is in agreement with the
results for hard interactions \cite{teixeira} and close to the
mean field predictions ($\dg{109}$) \cite{luckhursttsf}. Interestingly,
for \sbentcore molecules composed of two parts the correction to the 
position of the self-dual point from attractive forces is vanishingly small.
We remember that it was not the case for convex biaxial molecules 
(see \frf{landaushbx}). The model of three arms, as is shown 
on Fig.~\reffig{diagramb3}, has the Landau point localized near the
right angle, which can be considered to be in some agreement with observations
in \cite{mlehmann}, where the mesogenic substance consisting of molecules with
bend angle of $\dg{90}$ was found to give rise to biaxial nematic phase. 
The summary of Landau points positions is presented
in Table~\reftable{landauvsmu}.
\nlin In order to pursuit the behaviour of Landau point in the two-arm
model, we have also taken into account asymmetrical bananas, i.e., we 
constructed the \sbentcore molecules from two uniaxial ellipsoids with different elongations.
We used the extended Gay-Berne potential \cite{generalizedungb} and
performed the calculations for a range of asymmetries, up to a point 
where one arm was four times longer than the other. No significant differences
in the position of the Landau point were found, it was located, as for
symmetric bananas, at $\dg{107}$.
\nlin Non-polar model of uniaxial arms is able to produce an isolated
self-dual point, given in \trt{landauvsmu}. Following sections show two
models where that behaviour is different, namely, there appears
a line of Landau points.
\subsubsection{Bent-core systems composed of biaxial parts}
 \label{section:bananagbbx}
\indent It is of interest to take into account the model where the arms of a 
banana are not uniaxial. One possibility would be to consider
the \sbentcore molecules built
of the biaxial parts. We have done so, and present the resulting phase diagram
in current section. 
\begin{table}[t]
 \caption{Landau point position versus strength of the dipole $\dimensionless{\mu}$.}
 \begin{tabular}{lcc}
 \label{table:landauvsmu}
 \sbbentcore model & $\dimensionless{\mu}$ & opening angle at Landau point \n \hline\hline
 \multirow{3}{*}{two-arms} & 
 $0.0$ & $107^{\circ}$ \n
 & $1.2$ & $104^{\circ}$ \n
 & $1.5$ & $103^{\circ}$ \n \hline
 \multirow{6}{*}{three-arms} & 
 & $\dimensionless{\rho}=0.026$ $\,\,\dimensionless{\rho}=0.04$ \n
 & $0.0$ & $\,\,\,\dg{89}$ $\,\dg{89}$ \n
 & $1.2$ & $\,\,\,\dg{86}$ $\,\dg{86}$  \n
 & $1.6$ & $\,\,\,\dg{74}$ - $\dg{86}$ $\,\dg{74}$ - $\dg{86}$ \n
 & $2.0$ & $\,\,\,\dg{63}$ - $\dg{86}$ $\,\dg{63}$ - $\dg{80}$ \n
 & $2.8$ & $\,\,\,\dg{83}$ - $\dg{97}$ $\,\dg{82}$ - $\dg{92}$ \n \hline
 two biaxial arms & 
 $0.0$  & $\dg{121}$ - $\dg{128}$ \n
\end{tabular}
\end{table}
\nlin We take the biaxial Gay-Berne ellipsoids as building blocks.
They are oriented in the following way. We fix the plane of the longer axis,
to lie in the plane of the bend angle, while the shortest axis is chosen
to be perpendicular to it. In this way we construct the \sbentcore molecule
as shown in \frf{bananaconstruction}, and build the pair interaction 
by substituting $V_{GB}$ in \re{bentcoremodelpotential} with the biaxial
Gay-Berne potential \re{bfzu}, defined in \srs{bzmodel}. The resulting 
pair interaction has the global minimum in the same configuration as
the uniaxial two-arm model described in the previous subsection.
\newline\indent We are mainly interested in the position of the Landau
point in this model. We have chosen the shape of constituting 
biaxial ellipsoids to be $\lnbra \sigma_{x},\sigma_{y},\sigma_{z} \rnbra=\lnbra 1.2,0.514,3.4\rnbra$ and the potential parameters $\lnbra \epsilon_{x},\epsilon_{y},\epsilon_{z}\rnbra=\lnbra 1.0,1.4,0.2\rnbra$. As mentioned in the previous
chapter those parameters make the attractive forces strongest in the 
face-to-face configuration, that is, in the direction of the shortest axis 
(see Sec.~\refsec{bzmodel}). This model exhibits molecular biaxiality in
the limit of $\gamma=\pi$ and $\gamma=0$; however, it is as well clear that
there should exist an angle for which Landau point appears. Indeed, the
numerical analysis shows the existence of self-dual region, in the form
of the line of Landau points, as can be seen from
Fig.~\reffig{diagramb2bz2k}. For the biaxial arms interacting by biaxial
Gay-Berne potential a line of \iso\ -- \nbx\ transitions is possible. It starts 
near $\gamma=121^{\circ}$ and ends for $\gamma=128^{\circ}$, and
reduces to a single point with a decrease in arms biaxiality. 
\putfigureflagsscalelong{part2/chapter3/diagram_b2-bz2k.eps}{Bifurcation diagram for \sbentcore molecule constructed from biaxial ellipsoids interacting by Berardi-Fava-Zannoni potential \cite{bzdevelopmentofpotential,bz2k}, for fixed density $\dimensionless{\rho}$. Landau line is in range of opening angle $\dg{121}\leq\gamma\leq\dg{128}$.}{diagramb2bz2k}{t}{0.8}{Bifurcation diagram for bananas of biaxial Gay-Berne arms.}
\newline\indent The matter should probably be investigated further, since 
the Landau region obtained for the model with biaxial arms is closer to the
experimental findings, not to mention that the topology of
the diagram is qualitatively different from what was published in the
literature so far.
\begin{figure}[h]
 \centering
 \subfigure[Diagrams for two-arm model for two densities $\dimensionless{\rho}$ and three dipole magnitudes $\dimensionless{\mu}$.]{\label{figure:diagramb2polar}\includegraphics{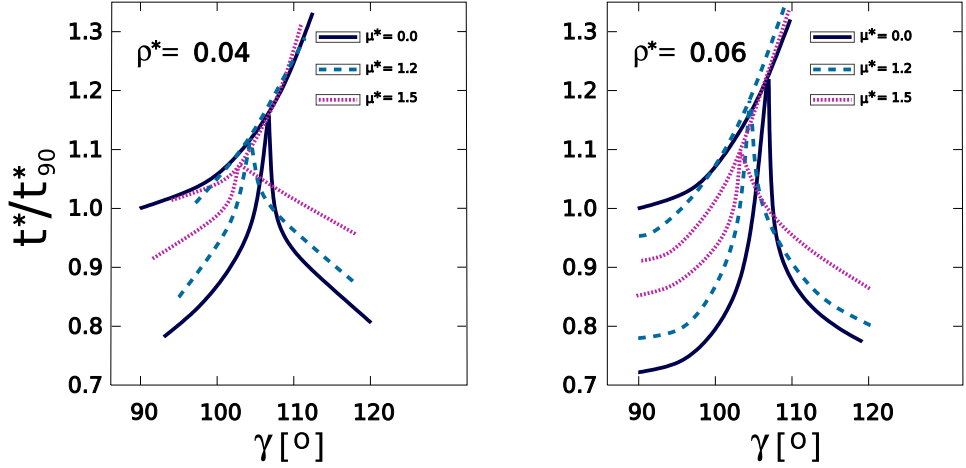}}
 \subfigure[Diagrams for three-arm banana, for two densities $\dimensionless{\rho}$ and five dipole strengths $\dimensionless{\mu}$.]{\label{figure:diagramb3polar}\includegraphics{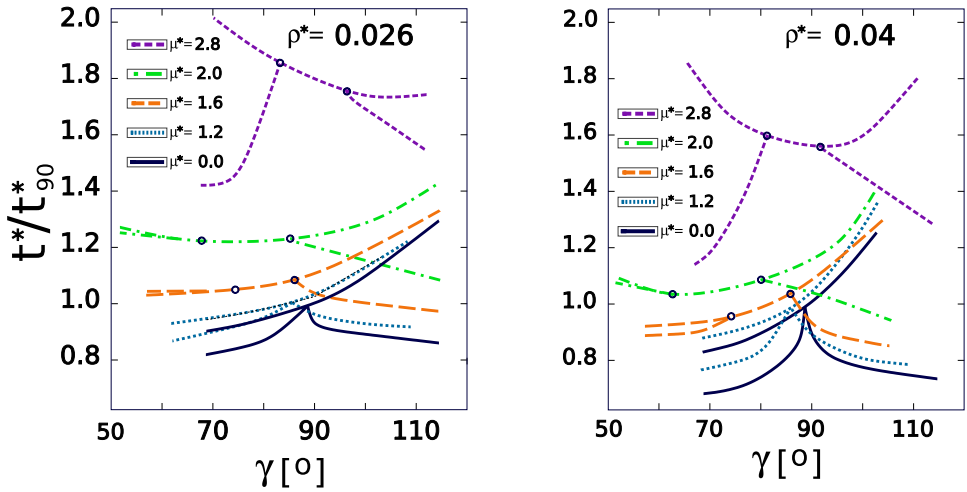}}
 \caption[Bifurcation diagrams for polar bananas.]{\label{figure:diagrampolar}Bifurcation diagrams for systems of polar \sbentcore molecules composed of uniaxial, soft ellipsoids. Temperatures are scaled by $\dimensionless{t}_{90}$, the bifurcation temperature for $\gamma=90^{\circ}$ in non-polar case (see Table~\reftable{rightangletemps}).}
\end{figure}
\subsection{Polar case}
 \label{section:bananapolar}
\indent As was mentioned before, molecules studied in \cite{madsen,acharya,mlehmann}, i.e., those for which the biaxial nematic was observed, are polar. 
Therefore, it is interesting to look at the effects of dipole-dipole 
interaction on stability of the biaxial nematic phase in model \sbentcore
systems. We have introduced the dipole moment along the molecular $C_{2}$ symmetry axis, as depicted in Fig.~\reffig{bananaconstruction}. In current section we present
the bifurcation diagrams for two and three uniaxial arm models for a couple
of dipole strengths. We again focus on the dependence of Landau point
position on the dipole magnitude (see Table~\reftable{landauvsmu}).
\newline\indent Fig.~\reffig{diagrampolar} shows the bifurcation diagrams.
As can be seen (Table~\reftable{landauvsmu}) the inclusion of the
dipole-dipole interaction shifts the Landau point for the model of
two uniaxial arms towards lower angles (Fig.~\reffig{diagramb2polar}).
In case of \sbentcore molecule modelled by three uniaxial Gay-Berne
parts (Fig.~\reffig{diagramb3polar}), for values of
dipole strength less than $20\%$ of the total potential
in the ground state (see Table~\reftable{muvsdd}), the same tendency
is maintained, i.e., Landau point is shifted towards lower angles up
to a value of $\dg{86}$. However, for dipoles above $\dimensionless{\mu}=1.4$ ($20\%$ of total potential) isolated Landau point changes into 
the second order \iso\ -- \nbx\ phase transition line.
The range of this line measured in opening angles widens;
for $\dimensionless{\mu}=1.6$ it is $\dg{12}$ and for
$\dimensionless{\mu}=2.0$ exceeds $\dg{20}$. For moderate density,
its upper limit does not change and is at $\gamma=\dg{86}$ as long as the
dipole strength $\dimensionless{\mu} \leq 2.1$ ($38\%$ of total energy).
Then, the Landau region begins to shrink and is shifted towards higher angles.
The highest dipole studied is the one of $\dimensionless{\mu}=2.8$ for
which $V_{DD}$ is $50\%$ of total potential. As can be seen
from Fig.~\reffig{diagramb3polar}, the line of \iso\ -- \nbx\ transitions
in that case is still getting shorter, and moves in the direction of higher
angles. Fig.~\reffig{landauregions} shows the evolution of Landau region for
varying dipole-dipole potential contribution to total energy. When $V_{DD}$
contribution is more than $35\%$ and less than $38\%$, the Landau
line attains its maximal width in opening angle.
\newline\indent We have presented the diagrams for two densities selected from 
lower region, as described above. It can be seen (Table~\reftable{landauvsmu})
that for strongest dipoles studied ($\dimensionless{\mu}\geq 2.0$) some
differences appear with varying density. Namely, the line of
direct isotropic -- biaxial nematic transitions gets shorter for higher density.
\newline\indent Our results suggest that there exists a certain range of 
dipole strength which makes the appearance of biaxial nematic phase
most probable. Based on the diagrams presented in \frf{diagrampolar},
and the assumed model parameters, we estimated this region to be
between $\dimensionless{\mu}=1.4$ and $2.1$, that is, when
dipole-dipole interaction contributes in range of $20$ and $38\%$ to
total potential energy.
\nlin Finally, we should mention that the only other studies in this
direction are Monte Carlo simulations of lattice Lebwohl-Lasher model presented
in \cite{mcbatesdipoles}, where the introduction of dipoles resulted in the
line of direct isotropic -- biaxial nematic transitions. \putfigureflagslong{part2/chapter3/landau_regions.eps}{Landau regions as function of dipole-dipole interaction strength contribution to total potential (see Table~\reftable{muvsdd}).}{landauregions}{t}{Landau regions as function of dipole-dipole interaction strength.}
\subsection{Conclusions}
\indent We studied the \sbentcore molecule model. Firstly, we took 
into account the bananas composed of two and three uniaxial Gay-Berne 
ellipsoids. We discussed briefly the case where arms of the molecules interact
via biaxial Gay-Berne potential, and also where one arm is a longer
uniaxial ellipsoid than the other. The dipole-dipole interactions were
introduced next, and the influence of the dipole strength on position of
Landau point was studied. Our results qualitatively support the previous
findings \cite{teixeira,luckhursttsf}, and show similar behaviour
as the less realistic polar, lattice model \cite{mcbatesdipoles}.
\newline\indent The \sbentcore Gay-Berne model of two uniaxial arms exhibits
the Landau point in agreement with previous studies for hard bananas
in Onsager approximation and mean field \cite{teixeira,luckhursttsf}. Its position
was not changed significantly by the introduction of attractive forces, nor
by the inclusion of anisotropy of the molecule. In case of the
three \dinfh -- symmetric arms, opening angle near the right angle was
distinguished as the one where Landau point appears. Recent experimental
studies \cite{mlehmann} are in line with these findings.
\nlin We tried to address the problem of disagreement between the experimental
estimates of the opening angle in the biaxial nematic phase (predicted to be 
near $\dg{140}$) and the theoretical results by
switching to biaxial arms interacting via biaxial Gay-Berne potential. In that
model with the increase of arms biaxiality, Landau point evolved to the
line of direct \iso\ -- \nbx\ second order phase transitions in range of opening angle $\gamma$ between $\dg{121}$
and $\dg{128}$. Probably, inclusion of dipoles or increase in the
arms biaxiality or taking into account molecular flexibility could widen
that region and possibly shift it towards higher angles.
\nlin That behaviour was not possible to acquire in non-polar case of
uniaxial arms. The line of direct \iso\ -- \nbx\ transitions appeared in the
model of three-part banana with dipole localized
along the $C_{2}$ symmetry axis.
It appeared that it is possible to provide dipoles strong
enough to generate the line of Landau points, which widened with the increase of
the dipole strength up to its maximal range in the opening angle
($\dg{63}<\gamma<\dg{86}$), when the dipole-dipole interaction constituted 
between $35$ and $38\%$ of total potential. Stronger dipoles began to 
shrink that line and move it towards higher angles. That should provide some hints for future research.
Possibly a study on the orientation of the dipole as well as on the 
shape of arms and shape of potential is in order,
for instance, we could wonder what would happen when shape of arms is
close to the self-dual geometry (found in \secrefsec{bzlandaupoints}).

 Our studies clearly indicate that strong dipoles were a necessary feature for
experimental discovery of the biaxial nematic phase in \sbentcore systems. In
polar case we saw that the dipole-dipole interactions not only were able to 
shift the position of the Landau point (move it towards lower angles)
but also change the topology of the bifurcation diagram to include
the self-dual line of direct isotropic -- \nbx\ transitions.
We can speculate that if the similar mechanism is at work in the 
substances where the biaxial nematic was observed, then the dipoles indeed 
became crucial, not only increasing the biaxiality of a molecule but also 
shifting the \nbx\ region to lower angles where the realization
and observation of the elusive phase becomes possible.
\nlin We were unable to account for the experimentally predicted value
of the opening angle of $\dg{140}$, as observed in the first class of
the \sbentcore biaxial materials. The stability of the biaxial nematic 
for the angles of this magnitude, in comparison to \iso\ and \nun\ phases,
was very low. The introduction of dipole-dipole interaction shifted the
Landau point in the opposite direction, so probably it is safe to conclude
that the polarity alone cannot explain the disagreement between theoretical
and experimental results. Only the inclusion of biaxial arms interacting
via biaxial Gay-Berne potential gave the values of opening angle
somewhat closer to $\dg{140}$.

 \chapter{Preliminary results}
 \label{chapter:elastic}

{ \chapabsbegin The knowledge of phase diagram does not give the
complete information about behaviour of a system, e.g.,
from the knowledge of the density-temperature region
of stability of a given structure we cannot extract
the information about response of a system to a given
deformation of the director field. For many applications
this kind of information is crucial. This chapter is devoted to
the two issues that were disregarded in the preceding part
of this thesis. One of them is the investigation of a phenomena
related to small distortions of the homeotropic biaxial nematic phase
and description of such state in terms of the biaxial elastic constants.
The other is the matter of spatially non-uniform liquid crystal states.
Presently we show the preliminary results concerning
those issues. \chapabsend }
\section{Biaxial elastic constants}
\indent Response of the biaxial nematic state to a small perturbations is
usually described by the contribution to the free energy coming from the
distortion of the directors, expanded in terms of their gradients. The
coefficients of this expansion are called \ital{elastic constants}, 
and are of great importance to both theoretical and
experimental studies. They influence most of the observed phenomena in liquid
crystals. To name only a few, we can mention their role in
light scattering effects, response to external fields \cite{degennesprost},
shape of disclination defects \cite{frank}, and
nematodynamic flow \cite{leslie}. The elastic constants are
crucial quantities in applications of liquid crystals in displays \cite{shanks},
and also are of importance in polymer-dispersed systems \cite{PRA.43.2943.1991}.
They are as well significant for the stability of the \stbxn\ phase.
The low relative magnitude of elastic constants corresponding to the
biaxial distortions in comparison to uniaxial ones can be one of the reasons
for difficulties in stabilizing \nbx.
\nlin The usual way of approaching the issues of elasticity is to study
the elastic free energy density dependence on the local 
deformations. The first derivation is due to Oseen \cite{oseen},
later phenomenologically acquired by Frank \cite{frank}, and also by
Zocher \cite{zocher}. In those classic papers the liquid is studied 
by means of the director field $\unitv{n}(\boldv{r})$ and the deformations of 
this field give rise to the increase of free energy\footnote{At the level of the molecular interpretation the free energy density of the small elastic deformations is always positive.}, typically described by expansion in gradients
of $\unitv{n}$. In the case of biaxial nematic phase, we have two additional
directors denoted by $\unitv{m}$ and $\unitv{l}$ which are not independent
on each other and on the uniaxial director. The fact that the directors form
an orthogonal tripod means that we cannot perform an independent distortion
of one director field without affecting other two. That, in turn, leads to the 
impossibility of the expression of the elastic free energy density 
as a sum of independent deformations, as in the uniaxial case.

 Current section by no means is a complete study on elasticity issues;
it serves merely as an introduction to further research. We are using the 
biaxial Gay-Berne potential, studied in the previous chapter, and 
present the set of elastic constants in the case of biaxial prolate
molecular shape, and in the vicinity of Landau point. We follow and employ
the approach from \cite{longaelastic}. We should note that there are other known distortions of the
director field where it looses the continuity, these are called
topological defects \cite{trebindefects,mermindefects} and we are disregarding
them altogether.
\subsection{Elastic constants in the biaxial Gay-Berne model}
The results presented in this section follow from the application of the 
formulas from \cite{longaelastic}. We will not repeat the full
derivation, since it can be found in the literature. Instead we briefly
write down the basic formulas.

 The elastic free energy density for biaxial nematic, in absence of polar and
chiral ordering, can be written in many ways. It is not possible to derive a
formula where the coefficients of the expansion in gradients
of $\unitv{n}$, $\unitv{l}$, and $\unitv{m}$ are independent, due to the
fact that the three directors form the orthogonal tripod. We use the form
developed in \cite{trebinelastic}:
\begin{equation}
 \label{eq:bulkelasticfreeenergy}
\begin{split}
 2\felastic = & K_{l1} ( \textnormal{div}\, \unitv{l} )^{2} + K_{l2} ( \unitv{l}\cdot\textnormal{rot}\,\unitv{l} )^{2} + K_{l3} ( \unitv{l}\times\textnormal{rot}\,\unitv{l} )^{2} + \\
 + & K_{m1} \lnbra \textnormal{div}\, \unitv{m}  \rnbra^{2} + K_{m2} \lnbra \unitv{m}\cdot\textnormal{rot}\,\unitv{m} \rnbra^{2} + K_{m3} \lnbra \unitv{m}\times\textnormal{rot}\,\unitv{m} \rnbra^{2} + \\
 + & K_{n1} \lnbra \textnormal{div}\, \unitv{n}  \rnbra^{2} + K_{n2} \lnbra \unitv{n}\cdot\textnormal{rot}\,\unitv{n} \rnbra^{2} + K_{n3} \lnbra \unitv{n}\times\textnormal{rot}\,\unitv{n} \rnbra^{2} + \\
 + & K_{mn} \lnbra \unitv{m}\cdot\textnormal{rot}\,\unitv{n} \rnbra^{2} + K_{nl} ( \unitv{n}\cdot\textnormal{rot}\,\unitv{l} )^{2}+ K_{lm} ( \unitv{l}\cdot\textnormal{rot}\,\unitv{m} ) + \\
 + & K_{l4} \textnormal{div} \lsbra (\unitv{l}\cdot\nabla)\unitv{l}-\unitv{l}\,\textnormal{div}\,\unitv{l} \rsbra + K_{m4}\textnormal{div} \Big[ (\unitv{m}\cdot\nabla)\unitv{m}-\unitv{m}\,\textnormal{div}\,\unitv{m} \Big] + \\
 + & K_{n4}\textnormal{div} \Big[ (\unitv{n}\cdot\nabla)\unitv{n}-\unitv{n}\,\textnormal{div}\,\unitv{n} \Big] \, ,
\end{split}
\end{equation}
where $K_{i1}$, $K_{i2}$, $K_{i3}$ for $i=n,l,m$ denote the elastic constants
corresponding to the deformations of director field $\unitv{n}$, $\unitv{l}$, $\unitv{m}$ called \ital{splay}, \ital{twist}, and \ital{bend}, respectively.
$K_{mn}$, $K_{nl}$, and $K_{lm}$ denote the constants associated with mixed
distortions, and the last three terms
are proportional to full divergences, and therefore play a role only
close to intrinsic surfaces (defects). Thus analysing the bulk effects we are concerned 
only with $12$ elastic constants. They are not independent,
and some of them may be negative, but the overall $\felastic>0$.
There are other methods of expressing the elastic free energy density, see
e.g., \cite{kapanowskielastic}.
\nlin In order to obtain the molecular expressions of elastic constants,
we recall the expansion \refeq{eq:taylorexpwithdcfs}
of the free energy around a given reference state $\densitysymbol_{ref}$
now representing the undeformed structure.
In order to evaluate \refeq{eq:taylorexpwithdcfs} so it contains
the local terms, which will allow to relate the coefficients $K_{ij}$
in \refeq{eq:bulkelasticfreeenergy} to microscopic quantities, we 
perform the following gradient expansion of the \sopdf
of the deformed state $\densityf{\boldv{r},\orientation}$ \cite{longaelastic}:
\begin{equation}
 \label{eq:gradientsexpforopdf}
 \densityf{\coords{2}{2}}=\densityf{\coords{1}{2}}+r^{\alpha}_{12}\partial_{\alpha}\,\densityf{\coords{1}{2}}+\frac{1}{2}r^{\alpha}_{12}r^{\beta}_{12}\partial_{\alpha}\partial_{\beta}\,\densityf{\coords{1}{2}}+\dots \, ,
\end{equation}
where $\boldv{r}_{12}$ stands for separation vector, $r^{\alpha}_{12}$
denote its Cartesian components, and summation goes over the repeated indices.
Now we can insert the above to
\refeq{eq:taylorexpwithdcfs} and obtain \cite{longaelastic}: 
\begin{equation}
 \label{eq:melasticf}
 \beta{\cal{F}}_{elastic} = \int d^{3}\boldv{r} \, \beta\felastic = \int d\boldv{r}\, M_{k^{'}k^{''}\alpha\beta\gamma^{'}\gamma^{''}}(\partial_{\alpha}n^{(k^{'})}_{\gamma^{'}})(\partial_{\beta}n^{(k^{''})}_{\gamma^{''}}) \, ,
\end{equation}
where we have set $\densityf{\boldv{r},\orientation}=\densitysymbol(\orientation^{(j)}\cdot{\unitv{n}}^{(k)}(\boldv{r}))$ with ${\unitv{n}}^{(k)}$ denoting the $k$th
director of the distorted set, i.e., we have assumed deformed director fields
to be slowly varying (see \cite{lipkinelastic}), and again the summation
convention is used. In the above
\begin{equation}
 \label{eq:mdefiniton}
 M_{k^{'}k^{''}\alpha\beta\gamma^{'}\gamma^{''}}=\frac{1}{2}\int d\boldv{r}_{12}d\orientation_{1}d\orientation_{2}\seconddcf{\orientation_{1},\orientation_{2}}{\boldv{r}_{12}}{\densitysymbol}r^{\alpha}_{12}r^{\beta}_{12}\Delta_{k^{'}\gamma^{'}}(\orientation_{1})\Delta_{k^{''}\gamma^{''}}(\orientation_{2}) \, ,
\end{equation}
where
\begin{equation}
 \label{eq:derivofopdf}
 \Delta_{kl}(\orientation)\equiv\frac{\partial\densitysymbol}{\partial\orientation_{jk}}\orientation_{jl} \, .
\end{equation}
Now by comparing \refeq{eq:bulkelasticfreeenergy} and \refeq{eq:melasticf}
we can express $K_{ij}$ in terms of
$M_{k^{'}k^{''}\alpha\beta\gamma^{'}\gamma^{''}}$. For exact derivation
please refer to \cite{longaelastic}. The results presented
in current chapter were obtained by calculating the integrals
in \eqrefeq{mdefiniton} in low-density approximation \re{ldsettings}, i.e.,
by setting 
\begin{equation}
 \seconddcf{\orientation_{1},\orientation_{2}}{\boldv{r}_{12}}{\densitysymbol}=\exp[-\beta\, U(\orientation_{1},\orientation_{2},\boldv{r}_{12})]-1 \, ,
\end{equation}
where $U$ is defined in \srs{bzmodel}, \ere{bfzu}, and then using Eqs.~(B1)-(B15)
from \cite{longaelastic} to obtain the elastic constants. We have also used
the expansion of the \sopdf in order parameters \refeq{eq:opdfinops}
remembering that $\densityf{\boldv{r},\orientation}=\rho\opdf$. To calculate
the derivatives of \sopdf in \refeq{eq:derivofopdf} we employed Eqs.~(A1)-(A4)
and Eq.~(A9) from \cite{longaelastic}.

 In \cite{longaelastic} set of elastic constants was calculated for a
lattice model of biaxial molecules \cite{biscarinilattice} interacting via a
simplified version of two-point potential \refeq{eq:v2pexpandeddeltas}
(with $v_{0,0}=-1$, $v_{0,2}=v_{2,0}=2\,\lambda$, and $v_{2,2}=-2\,\lambda^{2}$, which is a version of the dispersion model studied in \cite{luckhurstquad}).
The splay-bend degeneracy was proven and observed for all directors.
The authors also found one of the biaxial elastic constants to be negative for
every parametrization studied, but with the smallest absolute value\footnote{That is: the negative contribution to the free energy density from the deformation of director $\unitv{l}$ was overcame by the effects of remaining, positive elastic constants associated with $\unitv{m}$ and $\unitv{n}$.}. The relative values of the
elastic constants associated with uniaxial director were found to be greater than
biaxial ones. The two became relatively closer at the Landau point,
which is believed to exists at $\lambda=1/\sqrt{6}$ in that
model \cite{biscariniquad}, but still $K_{n1}$, $K_{n2}$, and $K_{n3}$ where
higher than their biaxial counterparts.
\nlin We present the plots of the set of elastic constants 
for a biaxial case calculated within low-density approximation for the
biaxial Gay-Berne model. As the results here are considered preliminary and a 
more detailed study is planned, we focus on
the $9$ bulk elastic constants $K_{i1}$, $K_{i2}$, $K_{i3}$, for $i=n,l,m$.
For comparison with the previous studies and to see how the biaxial constants
behave close to second order \iso\ -- \nbx\ transition at Landau point,
we choose to study the geometry where we found the Landau point (visible
in \frf{landaushbx}). Also to shed more light on the case of opposite sign of
energy and shape biaxiality, we study two cases for rod-like biaxial molecule
with $\enbx>0$ and $\enbx<0$. We carried out the calculations for three sets
of shape and energy biaxialities \re{biaxialities}. Firstly, we take into account
the shape studied in Monte Carlo simulations \cite{bz2k}: $\lnbra\sigma_{x},\sigma_{y},\sigma_{z}\rnbra=\lnbra 1.4,0.714,3.0\rnbra$ ($\shbx=0.58$) for
two values of $\enbx=-0.04$ ($\lnbra\epsilon_{x},\epsilon_{y},\epsilon_{z}\rnbra=\lnbra 1.4,1.0,0.2 \rnbra$) and $\enbx=0.04$ ($\lnbra\epsilon_{x},\epsilon_{y},\epsilon_{z}\rnbra=\lnbra 1.0,1.4,0.2 \rnbra$) (the diagram for that model is presented in \frf{bifvsmc}). Then we turned to the self-dual shape discovered
earlier (see \frf{landaushbx}) where $\lnbra\shbx,\enbx\rnbra=\lnbra 0.7163,0.0\rnbra$ for $\lnbra\sigma_{x},\sigma_{y},\sigma_{z}\rnbra=\lnbra 1.4,0.544,3.170\rnbra$ and $\lnbra\epsilon_{x},\epsilon_{y},\epsilon_{z}\rnbra=\lnbra 1.2,1.2,0.2 \rnbra$.
%
The calculations were performed for fixed dimensionless density $\dimensionless{\rho}=0.18$.
\putfigureflagslong{part2/chapter2/plot_elastic_three.eps}{Set of elastic
constants as function of temperature, for biaxial Gay-Berne interaction. On the
left the plot for model of positive $\enbx$ for
$\lnbra\epsilon_{x},\epsilon_{y},\epsilon_{z}\rnbra=\lnbra 1.0,1.4,0.2 \rnbra$
($\enbx=0.04$) in the middle the negative case $\enbx=-0.04$ for
$\lnbra\epsilon_{x},\epsilon_{y},\epsilon_{z}\rnbra=\lnbra 1.4,1.0,0.2 \rnbra$, both for the shape studied by Berardi and Zannoni in \cite{bz2k} for $\shbx=0.58$. On the right behaviour of the elastic constants in the vicinity of the Landau point for $\enbx=0.0$ and $\shbx=0.7163$ (as seen on \frf{landaushbx}).}{elasticthree}{t}{Set of elastic constants for biaxial Gay-Berne interaction.}
\newline\indent The results are presented in two Figures: \reffig{elasticthree}
and \reffig{elasticcompare}, where the reduced elastic constants
$\dimensionless{K}_{ij}=\beta\,K_{ij}\rho^{-2}\epsilon^{-1}_{0}\sigma^{-6}_{0}$ are plotted against dimensionless temperature $\dimensionless{t}$, where
the units are set according to \re{gbreducedunits}.
The first one shows all the elastic constants calculated,
and the second one presents a comparison of the coefficients associated with
the deformation of biaxial directors for $\enbx>0$ and $\enbx<0$.
\newline\indent The well known fact of equality of $K_{i1}$ and $K_{i3}$ 
for $i=m,n,l$ (called a splay-bend degeneracy for uniaxial
director) was found to be in force for all of the directors, which follows from
the lack of director dependence in low-density approximation for pair direct correlation function $c_{2}$ \cite{Eur.Phys.J.E.4.51}, and is due to keeping only
leading order parameters in formula \re{opdfinops}. It can be immediately seen from 
\figreffig{elasticthree} that one of the elastic constants is always
negative, this is a reflection of the impossibility of 
performing the distortion of only one director from the orthogonal set without
affecting the other. The values of the constants corresponding to the
biaxial directors are relatively low as compared with the uniaxial ones,
both for opposite and same sign of the energy and shape biaxiality.
Close to the uniaxial -- biaxial nematic transition the 
biaxial constants became equal, as can be seen in Fig. \reffig{elasticcompare}.
The situation changes qualitatively when we approach
the Landau point. There the splay-bend degeneracy evolves to the equality
of all the constants for biaxial directors $\unitv{l}$ and $\unitv{m}$:
$K_{l1}=K_{l2}=K_{l3}$ and $K_{m1}=K_{m2}=K_{m3}$, and the latter become
dominant.
\newline\indent \figreffig{elasticcompare} shows the biaxial constants for
two cases of the biaxiality of energy: $\enbx=-0.04$ and $\enbx=0.04$. It can
be seen that although for $\enbx < 0$ the constants are higher there is no
new quality revealed for the case of opposite signs of $\shbx$ and $\enbx$.
\putfigureflagslong{part2/chapter2/plot_elastic_compare.eps}{Comparison of biaxial elastic constants for positive and negative energy biaxialities.}{elasticcompare}{b}{Comparison of biaxial elastic constants for $\enbx>0$ and $\enbx<0$.}
\subsection{Summary}
 We have calculated the set of elastic constants for the biaxial Gay-Berne model
in the low-density approximation. Taken into account the importance of the
elastic issues, it seemed only fair to present the above study
to give some insight into the effects of elastic deformations in the
biaxial nematic phase formed by the molecules interacting via 
the biaxial Gay-Berne potential \cite{bzdevelopmentofpotential}.
\nlin As we can see from \figreffig{elasticcompare}, the models of
opposite and same sign of shape and energy biaxiality do not give
significantly different elastic constants. The previous analysis of
bifurcation diagrams was also unable to account for the 
apparent diversity of those cases, which was discovered in
Monte Carlo simulations \cite{bz2k}, its sources, therefore, must lie
elsewhere; we address this issue in the next section.
\nlin It is possible that the constants associated with biaxial
directors are, comparing to uniaxial ones, too small and
thermal fluctuations cause the lack of biaxial nematic ordering in experiment.
The above results show that in the vicinity of Landau point 
at $(\shbx,\enbx)=(0.7163,0.0)$ the ratios of the elastic constants
significantly change. For self-dual $\shbx$, at temperature $10$\% lower than the $\dimensionless{t}$
of \iso\ -- \nbx\ transition they are $K_{l1}:K_{m1}:K_{n1}=-0.71:2.46:1.0$ (at $\dimensionless{t}=4.14$) as compared to $K_{l1}:K_{m1}:K_{n1}=-0.18:0.33:1.0$ (at $\dimensionless{t}=1.65$; $10$\% lower than $\dimensionless{t}$ for \nrod\ -- \nbx\ transition) in rod-like regime for $\shbx=0.58$ and $\enbx=-0.04$. As we can see one of the biaxial constants associated with biaxial director
is dominant at self-dual point. However, a more detailed study, possibly involving \sbentcore molecules is needed.
\nlin In the pursuit of the factors playing a role in stabilisation
of the biaxial nematic, we continue to the next section, where we
make some comments on the simplest smectic phases.
\section{Orthogonal smectic phases}
\indent As we have seen, both theoretical and experimental
studies predict a stable biaxial nematic for \sbentcore molecules.
However, in those systems the phase diagrams are dominated by
smectic structures. Also one possible threat for the experimental observation
of the spatially uniform biaxial phase comes from the strong possibility that
the spatially non-uniform liquid crystal state can gain stability instead.
Therefore it is important to study the competition between smectic
and nematic phases. It seems, therefore, important to include those
in the analysis. This section is intended as an introduction to a
more complete study since the equations presented here need
to be solved and analysis of whether the smectic phases
win over biaxial nematic needs to be conduced in more detail.
\nlin There is a great variety of known smectic structures,
and accordingly great interest is given to the investigation of 
materials that exhibit this kind of behaviour. Without trying to be exhaustive,
we can mention the problem of the tricritical point on the line of
uniaxial nematic-smectic-A transition, existence of which is still a matter of discussion,
after over 32 years of research (see e.g. \cite{longatri,meyerlubensky}), and
a rich number of higher ordered smectics, known as $B_{n}$ phases ($n=1,\dots,7$), discovered in \sbentcore systems (see, e.g. \cite{takezoe}).
\nlin In present section we briefly show the extension of the previous
formalism to the case of the simplest spatially non-uniform liquid crystal states:
the orthogonal uniaxial and biaxial smectics. We start with an example of the
order parameters calculated in $L=2$ model including the smectic phases.
Keeping in mind that the biaxial nematic phase range can be severely limited
by the spatially non-uniform states, we again address the issue of 
opposite signs of energy and shape biaxiality (stronger
molecular lateral attractions). As we mentioned, only for this model the 
simulations \cite{bz2k} discovered stable \nbx. We show how the range of
biaxial nematic phase depends on the relative sign of $\enbx$
and $\shbx$ (defined in \re{biaxialities}). Then,
as an introduction to a more complete study, we show the additional 
bifurcation formulas for 5 transitions involving the layered phases,
these are nematic biaxial -- smectic biaxial, nematic uniaxial -- smectic uniaxial, nematic uniaxial -- smectic biaxial, isotropic -- smectic biaxial, and isotropic -- smectic uniaxial. The numerical evaluation of these equations is yet to come.
Firstly, we describe the basic smectic phases, and then show the usual methods
used in the description of those structures.
\putfigureflagslong{part1/misc/smectica.eps}{Picture of uniaxial smectic-A phase. The direction of density modulation ${\boldv{z}}$ is parallel to the nematic director ${\unitv{n}}$, and perpendicular to the layers of thickness $d$.}{smectica}{t}{Uniaxial smectic-A phase.}

 The simplest smectic phase is the so-called smectic-A (\sma).
It is characterized by a long-range ordering of the orientational degrees of
freedom, in the same way the nematic phase is; however, the positions of
molecular centres of masses are not distributed randomly; there exists
a tendency for them to on average align in layers perpendicular to the
nematic director $\unitv{n}$, while in each layer there is no long range
translational order. Therefore a modulation of the density is visible.
The inversion symmetry $\unitv{n} \rightarrow -\unitv{n}$ and invariance
under the rotation around $\unitv{n}$ are upheld.
Snapshot of such structure is pictured in \figreffig{smectica}.
Since the layers are perpendicular to the director, \sma\ belongs to the 
class of \ital{orthogonal} smectics; and due to the
rotational symmetry around $\unitv{n}$ this state is considered
to be of uniaxial symmetry, and will be referred to by \sun. It is possible
that the biaxial nematic phase acquires the tendency to form orthogonal smectic layers.
In that case the state becomes smectic with three directors, i.e., 
a biaxial smectic-A (\sbx). Interestingly, the first historically
discovered biaxially ordered phase was smectic \cite{firstbxphaseeversmectic}.
Presently, we will only consider biaxial and uniaxial \sma, for a description of
layered structures with higher degree of order see, e.g., \cite{degennesprost}.
\nlin In the smectic regime the \sopdf cannot depend only on 
orientational degrees of freedom. To account for the average tendency
of centres of mass to align in layers, it has to take, at least, one
spatial direction as an argument. Traditionally it is denoted by ${\boldv{z}}$,
and because it is chosen to be parallel to the uniaxial director $\director$: ${\boldv{z}}=z\unitv{n}$ we can conclude that $\opdff{\orientation,\boldv{z}}=\opdff{\orientation,z}$. Since $\opdfs$ has to be periodic: $\opdfs=\opdff{\orientation,z+d}$, where $d$ stands for layer thickness, usually close to the
full length of the constituent molecules, we employ the usual expansion in
the Fourier series:
\begin{equation}
 \label{eq:smecticopdfexpansion}
 \opdfs=\sum_{l,p,q} \frac{2\,l+1}{8\,\pi^{2}d}\orderparameter{l}{p}{q}\symmetrydelta{l}{p}{q}{\orientation} + \sum_{l,p,q,n}\frac{2l+1}{4\pi^{2}d}\smecticbasef{l,n}{p}{q}(z,\orientation)\average{\smecticbasef{l,n}{p}{q}} \, ,
\end{equation}
where the summation as usual goes over $0 \leq p,q \leq l$, $l>0$ even,
and $n>0$, and where
\begin{equation}
 \label{eq:smecticorderparams}
 \average{\smecticbasef{l,n}{m}{k}} = \int d\orientation \int^{d}_{0} dz\, \opdff{\orientation,z} \smecticbasef{l,n}{m}{k}(\orientation,z) \, .
\end{equation}
and where the new base functions are defined as
\begin{equation}
 \label{eq:smecticbasefunctions}
 \smecticbasef{l,n}{m}{k}(\orientation,z)=\symmetrydelta{l}{m}{k}{\orientation}cos\lnbra\frac{2\pi\,n\,z}{d}\rnbra \, .
\end{equation}
They form an orthogonal set;
\begin{equation}
 \label{eq:smortho}
 \int d\orientation \int^{d}_{0} dz\, \smecticbaseftopalign{l,n}{m}{k}(\orientation,z) \smecticbasef{l^{\prime},n^{\prime}}{m^{\prime}}{k^{\prime}}(\orientation,z)=\frac{4\pi^{2}d}{2\,l+1} \delta_{l,l^{\prime}}\delta_{m,m^{\prime}}\delta_{k,k^{\prime}}\delta_{n,n^{\prime}} \, .
\end{equation}
Similarly to the previous analysis we take $\average{\smecticbasef{l,n}{m}{k}}$ with $l=0,2$ and $n=1$
as leading order parameters; non-zero value of $\average{\smecticbasef{0,1}{0}{0}}$ and $\average{\smecticbasef{2,1}{0}{0}}$
indicates the uniaxial smectic-A, while $\average{\smecticbasef{2,1}{2}{2}}$ vanishes in \sun and becomes non-zero in biaxial smectic \sbx. In the
isotropic phase all $\average{\smecticbasef{l,n}{m}{k}}=0$ for $l,n > 0$.
\nlin In the expansion of pair direct correlation function \refeq{eq:expansionofc2l2} new terms are introduced, namely
\begin{equation}
\label{eq:smc2expansionl2}
 c_{2}(\relativeorientation,z)=c_{2}(\relativeorientation) + w_{s}\smecticbasef{0}{0}{0}(z)+\sum_{m,n}w_{m,n}\smecticbasef{2,1}{m}{n}(\relativeorientation,z) \, ,
\end{equation}
where $w_{s}$, $w_{m,n}$ need to be calculated numerically, using, e.g.
low-density approximation \re{ldsettings}-\re{ldapproximation}. They are
integrals of $c_{2}(\relativeorientation,z)$ with $\smecticbasef{2,1}{m}{n}(\relativeorientation,z)$, as can be easily obtained with help
of orthogonality relations \re{smortho}.
Since we know that the terms with angular momentum index $l=2$ will give the
leading contribution to the bifurcation, we already truncated the above
expansion accordingly; the same argument works for the Fourier series index $n$.
Since the leading contribution to bifurcation comes from $n=1$, we can fix it,
and simplify the notation by using in further calculations $\smecticbasef{2}{m}{n}(\relativeorientation,z)\equiv\smecticbasef{2,1}{m}{n}(\relativeorientation,z)$ and $\average{\smecticbasef{2}{m}{n}}\equiv\average{\smecticbasef{2,1}{m}{n}}$. Now we can derive the bifurcation equations corresponding to the
additional phase transitions from isotropic, uniaxial, or biaxial nematic to 
uniaxial or biaxial smectics. However, before we present these,
we show order parameters calculated using
the \sceq\ \re{selfconsistentequniform} with the above formulas.
\subsection{Order parameters in $L=2$ model}
\indent In present section we show an alternative method of obtaining the phase diagram following from the 
solutions of the \sceq\ \re{selfconsistentequniform} in $L=2$ model, i.e., we
take into account  the expansion of $c_{2}$ \re{smc2expansionl2}, calculate the
coefficients $\clmn{2}{m}{n}$, $w_{s}$, $w_{m,n}$ by numerical methods (see \ara{numericaldetails}), and work out the non-linear integral equations for
order parameters. In the calculations we used low-density approximation,
setting $c_{2}(\relativeorientation,z-z^{\prime})=\exp[-\beta\,U]-1$, where $U$ is the biaxial Gay-Berne potential \re{bfzu}. In the case of $w_{s}$
and $w_{m,n}$ the calculations were significantly longer, due to the fact
that the integral over $z-z^{\prime}$ has to be calculated in the similar
way as integral over $r$ in $\clmn{2}{m}{n}$, as mentioned
in \ara{numericaldetails}. 
The equation \re{selfconsistentequniform} now becomes:
\begin{equation}
 \label{eq:scnonuniform}
 \opdff{\orientation,z}=Z^{-1}_{s} \exp \lsbra \rho \int d\orientationin \int^{d}_{0} dz^{\prime} \, c^{(2)}_{2}(\relativeorientation,z-z^{\prime}) \opdff{\orientationin,z^{\prime}} \rsbra \, ,
\end{equation}
where $Z_{s}=\int d\orientation \, dz \, \exp \lsbra \rho \int d\orientationin dz^{\prime} \, c^{(2)}_{2}(\relativeorientation,z-z^{\prime}) \opdff{\orientationin,z^{\prime}} \rsbra $, and where
\begin{equation}
 \label{eq:c2nonuniformexpandedl2}
\begin{split} 
 c^{(2)}_{2}(\relativeorientation,z-z^{\prime}) = & \sum_{m,n \in \{ 0,2 \}} \clmn{2}{m}{n} \symmetrydelta{2}{m}{n}{\relativeorientation} \\ & + w_{s} \smecticbasef{0}{0}{0}(z-z^{\prime}) + \sum_{m,n \in \{0,2\}} w_{m,n} \smecticbasef{2}{m}{n}(\relativeorientation,z-z^{\prime}) \, .
\end{split}
\end{equation}
In terms of order parameters the \ere{scnonuniform} can be written (similarly to \ere{opsinl2sceq}) as:
\begin{equation}
 \label{eq:opsnonuniforml2sceq}
\begin{split}
 \average{\smecticbasef{2}{m}{n}} = Z^{-1}_{s} \int & d\orientation \int^{d}_{0} dz \, \smecticbasef{2}{m}{n}(\orientation,z) \\
 & \exp\lsbra \rho \int d\orientationin \int^{d}_{0} dz^{\prime} \, c^{(2)}_{2}(\relativeorientation,z-z^{\prime}) \opdff{\orientationin,z^{\prime}} \rsbra \, .
\end{split}
\end{equation}
\nlin Solving \esre{opsnonuniforml2sceq} and \esre{scuniformcase}
for $\orderparameter{2}{m}{n}$ in
the iterative manner we can calculate the order parameters for given density and
temperature and estimate the location of phase transitions.
\nlin An example of the results obtained in the way described above is
presented in \frf{nonuniformopsl2}. It shows dominant
order parameters for the biaxial Gay-Berne model (\srs{bzmodel}) as function
of temperature for dimensionless density $\dimensionless{\rho}=0.18$ (see \ere{gbreducedunits}). There are $9$ relevant $\average{\smecticbasef{2}{m}{n}}$,
all were calculated, but we only plot the largest ones. The range of biaxial nematic phase is limited by
biaxial smectic, and, as can be seen by comparing \frf{opsopposite}
with \frf{opssame}, depends on the sign of energy biaxiality $\enbx$.
More precisely, the region of stable \nbx\ is wider in case of 
stronger lateral interactions (where molecules are stronger attracted to their
sides, i.e., \ssts\ configuration gives the deepest minimum, see \frf{potential})
for $\shbx>0$ and $\enbx<0$ than for same signs of the biaxialities.
In that respect the present approach can explain why Monte Carlo simulations
discovered stable \nbx\ only for opposite signs of $\enbx$
and $\shbx$ \cite{bz2k}. Probably the spatially non-uniform structures gained
stability earlier. As we can see, the transition from biaxial nematic leads
directly to biaxial smectic, which is in agreement with \cite{bz2k},
also $L=2$ model gives $\average{\smecticbasef{0}{0}{0}}=\average{\smecticbasef{2}{0}{0}}$.
The results obtained from the solution of the \sceq\ for order parameters 
for \iso\ -- \nrod\ and \nrod\ -- \nbx\ transitions 
are equivalent to those obtained by bifurcation analysis, the main difference
being the calculation time, which is significantly shorter in case
of \ere{exactbifisoldl2} and \ere{bifunbx}. Thus it is useful to have
analogical equations for the bifurcations involving \sbx\ and \sun.
We present them in the following section.
\begin{figure}[b]
 \centering
 \subfigure[Case of stronger lateral interactions, for this model Monte Carlo simulations showed stable \nbx\ \cite{bz2k} $(\epsilon_{x},\epsilon_{y},\epsilon_{z})=(1.7,1.0,0.2)$ ($\enbx=-0.06$).]{\label{figure:opsopposite} \includegraphics[scale=0.7]{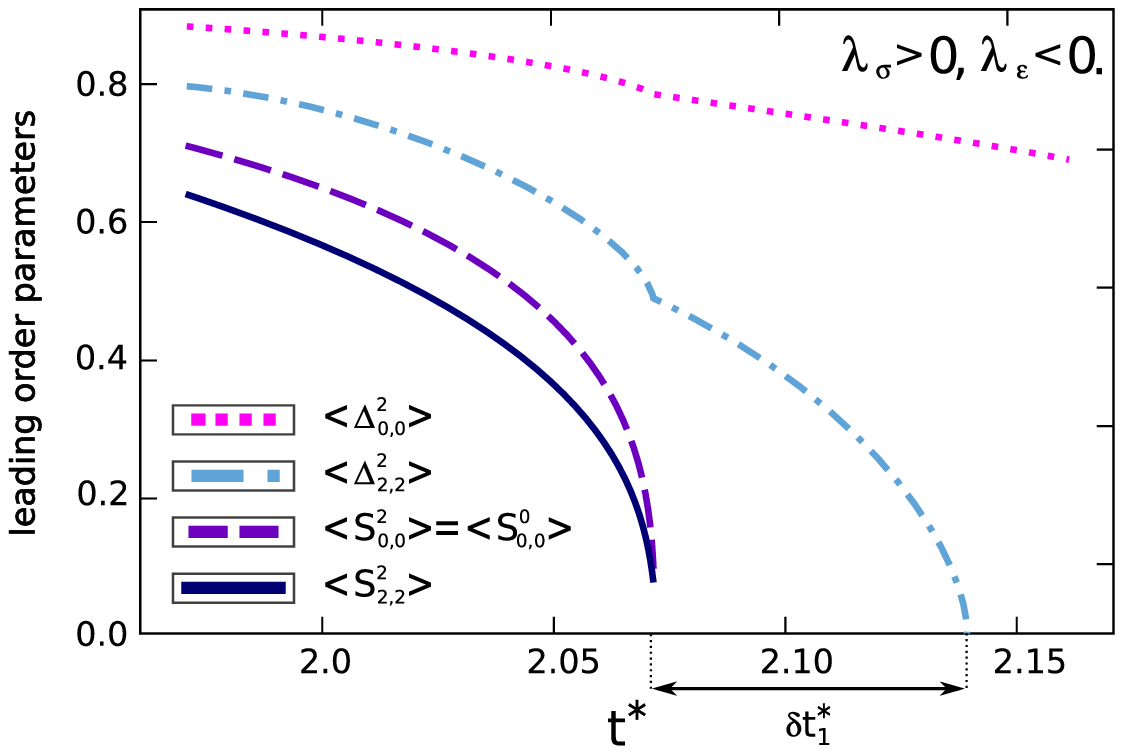}}
 \subfigure[Case of same sign of $\shbx$ and $\enbx$; minimum in the \sftf\ configuration is deepest, $(\epsilon_{x},\epsilon_{y},\epsilon_{z})=(1.0,1.4,0.2)$ ($\enbx=0.04$).]{\label{figure:opssame} \includegraphics[scale=0.7]{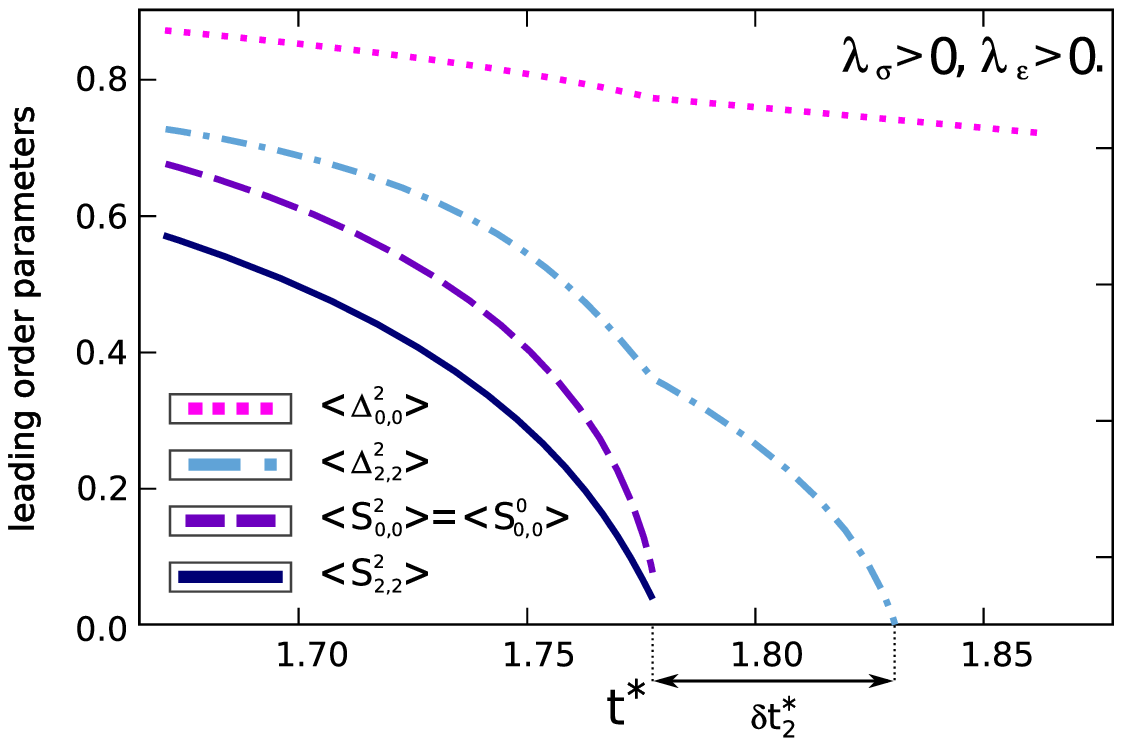}}
\caption[Order parameters in $L=2$ model including smectics.]{\label{figure:nonuniformopsl2} \nbx\ range for positive and negative energy biaxiality $\enbx$; leading, dominant order parameters are plotted: uniaxial nematic $\orderparameter{2}{0}{0}$, biaxial nematic $\orderparameter{2}{2}{2}$, uniaxial smectic $\average{\smecticbasef{0}{0}{0}}$, $\average{\smecticbasef{2}{0}{0}}$, and biaxial smectic $\average{\smecticbasef{2}{2}{2}}$, as calculated from $L=2$ model for biaxial Gay-Berne interaction using equation \re{opsnonuniforml2sceq}, for $(\sigma_{x},\sigma_{y},\sigma_{z})=(1.4,0.714,3.0)$ ($\shbx=0.58$) and two values of $\enbx$.}
\end{figure}
\subsection{Bifurcation equations}
 Given the above it is obvious that we have a larger number of possible
bifurcation scenarios. Disregarding the smectic phase as a reference,
we can expect five additional bifurcations; those are:
\nbx\ -- \sbx, \nun\ -- \sun, \nun\ -- \sbx, \iso\ -- \sun, and \iso\ -- \sbx. 
We list the bifurcation equations \re{bifurcationeq} in the form
of \ere{bifurcationeqcompiled}. They follow, as in the previous cases, from
the expansion
\begin{equation}
 \label{eq:p1expansionnonspatial}
 \bcorf{1}{\orientation,z}=\sum_{l,m,n} \slmn{l}{m}{n} \smecticbasef{l}{m}{n}(\orientation,z) \, ,
\end{equation}
which is equivalent to \ere{isotosomep1expanded}.
The derivation is analogical to the one presented for the spatially uniform
states. We show the equations with the help of the bifurcation matrix $\bifmat{2}$,
and as in case of \ere{bifurcationeqcompiled} the bifurcation point $\rho_{0}$
is determined from the roots of the characteristic polynomial, \ere{eigenequation},
which for the present normalizations can be written as 
\begin{equation}
 \label{eq:characteristic}
 \det\lsbra \bifmat{2}-2\,\rho^{-1}_{0}\oper{\mathbbm{1}} \rsbra=0 \, ,
\end{equation}
where $\oper{\mathbbm{1}}$ stands for the unit matrix. From the solutions of the
above equation the bifurcation point is chosen as the one with the
lowest $\rho_{0}$, as was described previously in \srs{bifeqsspatiallyuniform}.
In the following all the averages $\orderparameter{l}{m}{n}$ are calculated 
in the reference state.
\nlin {\bf{1. Nematic uniaxial -- smectic uniaxial.}}
\nlin In this case the bifurcation matrix is $3x3$, and the
bifurcation equation reads
\begin{equation}
\label{eq:smnunsun}
\lnbra \begin{array}{c}
 \slmn{0}{0}{0} \\
 \slmn{2}{0}{0} \\
 \slmn{2}{0}{2} \\
\end{array} \rnbra = \frac{\rho_{0}}{2} \bifmat{2}_{1} \lnbra \begin{array}{c}
 \slmn{0}{0}{0} \\
 \slmn{2}{0}{0} \\
 \slmn{2}{0}{2} \\
\end{array} \rnbra \, ,
\end{equation}
where
\begin{equation}
\bifmat{2}_{1}\equiv\lnbra \begin{array}{ccc}
 w_{s} & w_{0,0} \orderparameter{2}{0}{0}+w_{0,2} \orderparameter{2}{0}{2} & w_{2,0} \orderparameter{2}{0}{0}+w_{2,2} \orderparameter{2}{0}{2} \\
 w_{s} \orderparameter{2}{0}{0} & A_{0} w_{0,0}+B_{0} w_{0,2} & A_{0} w_{2,0}+B_{0} w_{2,2} \\
 w_{s} \orderparameter{2}{0}{2} & B_{0} w_{0,0}+C_{0} w_{0,2} & B_{0} w_{2,0}+C_{0} w_{2,2}
\end{array} \rnbra \, ,
\end{equation}
and where
\begin{equation}
\begin{split}
 35 \, A_{0} = & \, 7+10 \orderparameter{2}{0}{0}+18 \orderparameter{4}{0}{0} \, , \\ 
 35 \, B_{0} = & \, -10 \orderparameter{2}{0}{2}+3 \sqrt{15} \orderparameter{4}{0}{2} \, , \\ 
 35 \, C_{0} = & \, 7-10 \orderparameter{2}{0}{0}+3 \orderparameter{4}{0}{0}+3 \sqrt{35} \orderparameter{4}{0}{4} \, .
\end{split}
\end{equation}
\indent {\bf{2. Nematic uniaxial -- smectic biaxial.}}
\nlin  Here the uniaxial smectic parameters $\slmn{0}{0}{0}$, $\slmn{2}{0}{0}$,
and $\slmn{2}{0}{2}$ bifurcate independently, along \ere{smnunsun},
and the equation for remaining coefficients is as follows:
\begin{equation}
\label{eq:smnunsbx}
\lnbra \begin{array}{c}
 \slmn{2}{2}{0} \\
 \slmn{2}{2}{2} \\
\end{array} \rnbra = \frac{\rho_{0}}{2} \bifmat{2}_{2} \lnbra \begin{array}{c}
 \slmn{2}{2}{0} \\
 \slmn{2}{2}{2} \\
\end{array} \rnbra \, ,
\end{equation}
where
\begin{equation}
\bifmat{2}_{2}=\lnbra \begin{array}{ll}
 A_{0} w_{0,0}+B_{0} w_{0,2} & A_{0} w_{2,0}+B_{0} w_{2,2} \\
 B_{0} w_{0,0}+B_{2} w_{0,2} & B_{0} w_{2,0}+B_{2} w_{2,2}
\end{array} \rnbra \, ,
\end{equation}
and where
\begin{equation}
\begin{split}
 35 \, A_{0} = & \, 7-10 \orderparameter{2}{0}{0}+3 \orderparameter{4}{0}{0} \, , \\
 70 \, B_{0} = & \, 20 \orderparameter{2}{0}{2}+\sqrt{15} \orderparameter{4}{0}{2} \, , \\ 
 70 \, B_{2} = & \, 14+20 \orderparameter{2}{0}{0}+\orderparameter{4}{0}{0}+\sqrt{35} \orderparameter{4}{0}{4} \, .
\end{split}
\end{equation}
\indent {\bf{3. Nematic biaxial -- smectic biaxial.}}
\nlin The bifurcation eigenproblem, due to the non-trivial structure of 
the reference state, involves $5x5$ matrix. The equation reads
\begin{equation}
\label{eq:smnbxsbx}
\lnbra \begin{array}{c}
 \slmn{0}{0}{0} \\
 \slmn{2}{0}{0} \\
 \slmn{2}{0}{2} \\
 \slmn{2}{2}{0} \\
 \slmn{2}{2}{2} \\
\end{array} \rnbra = \frac{\rho_{0}}{2} \bifmat{2}_{3} \lnbra \begin{array}{c}
 \slmn{0}{0}{0} \\
 \slmn{2}{0}{0} \\
 \slmn{2}{0}{2} \\
 \slmn{2}{2}{0} \\
 \slmn{2}{2}{2} \\
\end{array} \rnbra \, ,
\end{equation}
where
\begin{equation}
\bifmat{2}_{3} = \lnbra  
\begin{array}{ccccc}
 (\bifmat{2}_{1})_{0,0} & (\bifmat{2}_{1})_{0,1} & (\bifmat{2}_{1})_{0,2} & A_{0} & A_{2} \\ 
 (\bifmat{2}_{1})_{1,0} & (\bifmat{2}_{1})_{1,1} & (\bifmat{2}_{1})_{1,2} & D w_{0,2}-C w_{0,0} & D w_{2,2}-C w_{2,0} \\
 (\bifmat{2}_{1})_{2,0} & (\bifmat{2}_{1})_{2,1} & (\bifmat{2}_{1})_{2,2} & D w_{0,0}+G w_{0,2} & D w_{2,0}+G w_{2,2} \\
 B_{0} & D w_{0,2}-C w_{0,0} & D w_{2,2}-C w_{2,0} & I w_{0,2}-H w_{0,0} & I w_{2,2}-H w_{2,0} \\
 B_{2} & D w_{0,0}+G w_{0,2} & D w_{2,0}+G w_{2,2} & I w_{0,0}+J w_{0,2} & I w_{2,0}+J w_{2,2}
\end{array}
\rnbra \, , \nn
\end{equation}
and where
\begin{equation}
\begin{split}
A_{n} = & \, \sum_{m\in\{0,2\}} w_{n,m} \orderparameter{2}{2}{m} \, , \,\,
B_{n} = w_{s} \orderparameter{2}{2}{n} \, , \\
70\,C = & \, 20 \orderparameter{2}{2}{0}-6 \sqrt{15} \orderparameter{4}{2}{0} \, , \\
14\,D = & \, 4 \orderparameter{2}{2}{2}+3 \orderparameter{4}{2}{2} \, , \\
70\,F = & \, -7+10 \orderparameter{2}{0}{0}-3 \orderparameter{4}{0}{0}-3 \sqrt{35} \orderparameter{4}{0}{4} \, , \\
70\,G = & \, 20 \orderparameter{2}{2}{0}+\sqrt{15} \orderparameter{4}{2}{0}+5 \sqrt{21} \orderparameter{4}{2}{4} \, , \\
35\,H = & \, -7+10 \orderparameter{2}{0}{0}-3 \orderparameter{4}{0}{0}-3 \sqrt{35} \orderparameter{4}{4}{0} \, , \\
70\,I = & \, 20 \orderparameter{2}{0}{2}+\sqrt{15} \orderparameter{4}{0}{2}+5 \sqrt{21} \orderparameter{4}{4}{2} \, , \\
70\,J = & \, 14+20 \orderparameter{2}{0}{0}+\orderparameter{4}{0}{0}+\sqrt{35} \orderparameter{4}{0}{4}+\sqrt{35} \orderparameter{4}{4}{0}+35 \orderparameter{4}{4}{4} \, .
\end{split}
\end{equation}
\nlin {\bf{4. Isotropic -- biaxial smectic transition.}}
\nlin In this case we have the simplest bifurcation matrix, since the
reference state has trivial, isotropic structure. The equation is as follows:
\begin{equation}
\label{eq:smisosbx}
\lnbra \begin{array}{c}
 \slmn{0}{0}{0} \\
 \slmn{2}{0}{0} \\
 \slmn{2}{0}{2} \\
 \slmn{2}{2}{0} \\
 \slmn{2}{2}{2} \\
\end{array} \rnbra = \frac{\rho_{0}}{2}\bifmat{2}_{4} \lnbra \begin{array}{c}
 \slmn{0}{0}{0} \\
 \slmn{2}{0}{0} \\
 \slmn{2}{0}{2} \\
 \slmn{2}{2}{0} \\
 \slmn{2}{2}{2} \\
\end{array} \rnbra \, ,
\end{equation}
where
\begin{equation}
\bifmat{2}_{4} \equiv 2\lnbra \begin{array}{ccccc}
 w_{s}-\frac{1}{\rho_{0}} & 0 & 0 & 0 & 0 \\
 0 & w_{0,0}-\frac{9}{\rho_{0}} & w_{2,0}  & 0 & 0 \\
 0 & w_{0,2} & w_{2,2}-\frac{9}{\rho_{0}} & 0 & 0 \\
 0 & 0 & 0 & w_{0,0}-\frac{9}{\rho_{0}} & w_{2,0} \\
 0 & 0 & 0 & w_{0,2} & w_{2,2}-\frac{9}{\rho_{0}}
\end{array} \rnbra \, .
\end{equation}
As we can see only the order parameters $\average{\smecticbasef{l,n}{q}{p}}$
for $l=2,4$ and $n=1$ are needed to localize the bifurcation point. 
In case of \nun\ -- \sbx\ transition (2), the uniaxial and biaxial parameters
bifurcate independently, and the equations involving uniaxial ones 
are the same as in case of \nun\ -- \sun\ (1). It was also the case for
the bifurcations where isotropic phase was taken as the reference
for spatially uniform states (\ere{isotosomebifmsimple}), only this time the bifurcation equations involving biaxial and
uniaxial smectic are different, due to the non-trivial structure of 
the reference state. This decoupling is removed for \nbx\ -- \sbx\ 
case (3), where the bifurcation matrix (\ere{smnbxsbx}) contains the
elements of matrix for \nun\ -- \sun\ from \ere{smnunsun}, but the biaxial
and uniaxial bifurcations do not occur separately, the coefficients are mixed.
It follows from the fact that the \dtwh symmetry is not broken during
this transition.
In case of the isotropic reference phase, as for spatially uniform states,
the coefficients associated with pure smectic order parameter $\average{\smecticbasef{0,1}{0}{0}}$, with biaxial and uniaxial smectic, bifurcate 
independently.
\subsection{Further studies}
 We have presented additional bifurcation equations for the transitions
involving uniaxial and biaxial smectic-A phases. They extend the previously
derived bifurcation equations and, with Eqs.~\re{exactbifisoldl2}, \re{bifunbx},
present a way to study the stability of the biaxial nematic against
simplest spatially non-uniform phases. The solution of those, most probably,
gives similar diagrams, as those obtained by working out
the \sceq\ \re{scnonuniform}. As an example we presented the
temperature dependence of order parameters obtained by iteratively solving
the \sceq. The phase sequence discovered with increase of the temperature
was \sbx\ -- \nbx\ -- \nun\ -- \iso. We have shown that the range of
biaxial nematic phase is influenced by the opposite signs of shape and
energy biaxiality. For stronger lateral forces (\frf{opsopposite})
the temperature range $\delta t_{1}$ of \nbx\ limited by \sbx\ was wider than
the interval $\delta t_{2}$ for same signs of $\enbx$ and $\shbx$
(\frf{opssame}), which corresponds to the model where
molecules are stronger attracted in the \sftf\ configuration. The ratio of those
ranges was found to be: $\delta t_{2}/\delta t_{1}=0.77$.
In some sense it partially explains the reasons as to why the model of
strong lateral interactions gave \nbx\ in Monte Carlo simulations \cite{bz2k}.
This method is equivalent to bifurcation, but is
numerically more demanding, and it is more worthwhile to consider using
a trial \sopdf analogical to \re{usedmodelopdf}, like
\begin{equation}
 \label{eq:modelopdfnonspatial}
 \opdfsymbol_{trial}(\orientation,z)=Z^{-1}_{trial}\exp\lsbra \sum_{m,n} \alpha_{mn}\symmetrydelta{2}{m}{n}{\orientation} + \sum_{m,n} \gamma_{m,n}\smecticbasef{2}{m}{n}(\orientation,z) \rsbra \, ,
\end{equation}
where $Z_{trial}$ ensures the normalization $\int d\orientation\,dz\,\opdfsymbol_{trial}(\orientation,z)=1$, and minimize the free energy with respect to
the parameters $\alpha_{m,n}$ and $\gamma_{m,n}$. This method will give more
accurate phase diagrams.
\nlin In the same way as we performed the analysis in search for factors that 
stabilize the biaxial nematic against \iso\ or \nun\ states, the next step
would be to localize the molecular parameters that reduce the stability
of \sun\ and \sbx\ with respect to \nbx. In that way we can obtain a more
complete picture of the region where thermotropic biaxial nematic is stable.
Yet we already know it would not be enough. As we have mentioned,
the \sbentcore systems exhibit a variety of sophisticated smectic phases,
and the inclusion of those should also be considered. 

 \chapter{Summary and future studies}
 \label{chapter:summary}

\indent We have presented a \sdft (DFT) study on the stability of the \stbxn\ (\nbx)
in comparison to isotropic liquid and uniaxial nematic phases, in three models of
intermolecular pair potential. We started with brief description of the
historical background and the research concerning biaxial nematic phase
conduced so far. Then, we introduced the theoretical tools: the \sdft
and bifurcation analysis, explored in the subsequent part of the thesis.
We derived the bifurcation equations, in both the general and specific forms,
for transitions involving spatially uniform states. We showed how the
representations numbered by angular momentum index bifurcate, and that
the contribution from pair direct correlation function to the bifurcation point
comes only from the subspace of the order equal to the one of 
bifurcating representation.
\nlin In the following part, we mainly applied DFT using the
low-density approximation for pair direct correlation function, with the
exception of the first model, where mean field theory was used. We focused
on the determination of the phase diagram at bifurcation point for
spatially uniform states, and commented on elastic constants and smectic phases.
\nlin Firstly, we analysed a generic biaxial $L=2$ model \cite{straley}
of the pair potential.
It allowed to acquire the exact bifurcation diagram in mean field,
which emphasised how the coupling constants impact the transition to
biaxial nematic, Landau points, and tricritical regions. Being the simplest
possible interaction giving rise to \nbx, it can be used to
locate the primary, effective factors influencing the stability of this
mesophase in terms of the coupling constants. 
\nlin We continued to investigate the biaxial Gay-Berne model \cite{bzdevelopmentofpotential}. It was characterized by six constants related to shape of molecules and 
interaction strengths. These were combined into two scalar quantities describing
shape and energy biaxiality, which were used to parametrize the
bifurcation diagrams. By varying these parameters we found the
Landau(self-dual) points, which mark the regions where \nbx\ is most stable
in comparison to uniaxial nematic and isotropic liquid, and, therefore,
allow to locate the microscopic parameters sets and their relative magnitudes
most important for stabilisation of biaxial nematic.
We illustrated how the shape and energy biaxiality and the inclusion of
attractive forces change Landau point position. The results suggest that the
so-called square root rule, which relates the axes of the
hard, biaxial ellipsoids at self-dual point,
is also significant for soft biaxial Gay-Berne interaction. Our analysis
rendered the plane parametrized by shape and energy biaxialities and defined
by the square root rule, as a good, but not only, candidate for the approximate
locations of the Landau points. This conclusion needs to be confirmed by more
detailed studies.
The biaxial Gay-Berne interaction was also important because Monte Carlo
study predicted for it a stable \nbx, but only for one 
set of potential parameters, namely for the model of strong lateral attractions,
where the molecules are more attracted to their sides than faces \cite{bz2k}.
In our studies \nbx\ appears stable quite easily, for many geometries,
mainly because we neglect other, lower symmetric structures. In order to 
address this issue, we presented a preliminary analysis involving orthogonal
spatially non-uniform phases. It indicated that the temperature range of
biaxial nematic between biaxial smectic-A and uniaxial nematic is wider in the
case identified in the simulations than for the model of the strongest \sftf\ 
attractions.
\nlin Biaxial Gay-Berne model served as the example for the way the
biaxiality of molecules and anisotropy of intermolecular forces influence the
transition to the biaxial nematic phase. It was also used to confirm our
findings on the contribution of the $L=2$ model of pair direct correlation function to transition points
involving isotropic and nematic phases. In pursuit for the reasons
for biaxial nematic formation in this model, we also calculated the set of
bulk biaxial elastic constants in rod-like regime and at Landau point.
Only in the vicinity of the self-dual point, relative values of
biaxial constants changed significantly, namely the ones corresponding to the
deformations of one of the biaxial directors became dominant. 
\nlin A model that can be more closely related to the experimental 
studies where biaxial nematic was observed, was the system
of \sbentcore (banana-like) molecules. We found that for banana with
two uniaxial Gay-Berne arms the Landau point is not affected by the inclusion
of attractive forces since it was found at the same opening angle $\gamma$ as
for hard \sbentcore molecules \cite{teixeira}. In case of three uniaxial arms,
Landau point was located for $\gamma$ close to the right angle, which is in
agreement with some of recent observations \cite{mlehmann}. Bent-core model was
also the only one where we introduced dipole-dipole interactions and studied
their influence on bifurcation diagram. Our findings indicate that the
Landau point was not only shifted towards lower opening angles, but also, for
three parts model a self-dual line was obtained for moderate strengths of the
dipole. In light of this, we conclude that the dipole-dipole interactions and,
therefore, the presence of electric dipoles in \sbentcore molecules are
important factors in the stabilisation of \nbx.
Using the generalized Gay-Berne potential, we also took into account
the model of two biaxial, soft ellipsoids as arms of the banana-like molecule.
Bifurcation diagram showed a line of Landau points, indicating that the
biaxiality of the arms is also important for the formation of biaxial nematic.

 A first step towards further studies is to investigate in detail the
self-dual plane for the biaxial Gay-Berne model and discuss its relation to the
one following from the square root rule. Then, one should consider the
smectic structures, 
not only orthogonal, taken into account in the last chapter, but also lower symmetric,
spatially non-uniform phases. In other words, the next(difficult) question that
should be asked is: what factors enlarge the region of stability of
biaxial nematic with respect to smectics.
\nlin Concerning the approximations for the excess Helmholtz free energy,
firstly one should take into consideration a more realistic reference state for
the Taylor expansion, e.g., a system of hard molecules. Similar strategy
was used in case of uniaxial molecules, and was proven successful in comparison
with simulation results \cite{velascopert1,velascopert2,ginzburgpert}. In those
papers, a separation (inspired by previous studies \cite{wca} and the analogy
between systems of hard spheres and hard ellipsoids \cite{somozatarazona})
of Gay-Berne potential into attractive and repulsive parts was performed.
The steric part of the free energy was calculated exactly, while the
contribution from the attractive forces was introduced in a perturbative manner.
Probably, those approaches can be generalized to the case of biaxial
ellipsoids,
and some meaning of the attractive biaxial forces as well as a better agreement
with simulations can be acquired. Also, next order terms in density in
the Taylor expansion can be included, i.e.,
higher direct correlation functions can be taken into account.
However, in this case the integrals become more challenging. One method
of addressing this issue is a so-called weighted density approximation
where the excess Helmholtz free energy depends on the density averaged
with an appropriately chosen weighting function \cite{wda}. This approach was
used for the case of hard biaxial molecules (spheroplatelets) \cite{PRL.61.2461,PRA.39.2742} and it can be extended to soft potentials.

 There are great hopes associated with the \stbxn\ phase, concerning 
not only the promising technological applications, but also a deeper
understanding of the mechanisms of mesophase formulation. Equally great
are the experimental and theoretical efforts to determine
the factors that increase the chances of observing this state.
Present work is a small contribution to this ongoing search.

\end{mainmatter}

\begin{appendices}


\chapter{Calculation of $c_{1}$}
\label{ap:firstdcf}
 Here we will show that the first direct correlation function is given by
\begin{equation}
 \label{eq:fdcfresultap}
 c_{1}\lnbra \coord{1},\lsbra\densitysymbol\rsbra \rnbra = \beta\lsbra\mu-V_{ext}\lnbra \coord{1} \rnbra \rsbra-\ln{\lsbra\Lambda\densityf{\coord{1}} \rsbra} \eqdot
\end{equation}
 To achieve this we will calculate the first derivative of excessive free 
energy functional in two ways, from \refeq{eq:idealandexcesive}, and then using
\refeq{eq:grandinpsi}. Differentiating \refeq{eq:idealandexcesive} with respect to 
$\densityf{\coord{1}}$ results in:
\begin{equation}
 \label{eq:fromfidanfdex}
 \fderivatede{\intrinsicfef{\densitysymbol}}{\coord{1}}=\beta^{-1}\ln\lsbra{\Lambda\densityf{\coord{1}}}\rsbra-\firstdcf{\coord{1}}{\densitysymbol} \eqdot
\end{equation}
Using \refeq{eq:grandinpsi} we can find the following expression for the
first derivative:
\begin{equation}
 \fderivatede{\intrinsicfef{\densitysymbol}}{\coord{1}}=\int d\coord{2} \, \fderivatepsi{\grandpotentialf{\psi}}{\coord{2}}\fderivatede{\psi\lnbra \coord{2} \rnbra}{\coord{1}}+\int d\coord{2} \, \fderivatede{\psi\lnbra \coord{2} \rnbra}{\coord{1}}\densityf{\coord{2}}+\psi\lnbra \coord{1} \rnbra \eqcolon
\end{equation}
where 
\begin{equation}
 \fderivatepsi{\grandpotentialf{\psi}}{\coord{2}}=\int d\coord{3} \, \lnbra \fderivatede{\intrinsicfef{\densitysymbol}}{\coord{3}}-\psi\lnbra \coord{3} \rnbra \rnbra \fderivatepsi{\densityf{\coord{3}}}{\coord{2}} -\densityf{\coord{2}}=-\densityf{\coord{2}} \, ,
\end{equation}
and where we have used \refeq{eq:variational} to eliminate the term under the
integral. We arrive at
\begin{equation}
 \label{eq:fromlegendre}
 \fderivatede{\intrinsicfef{\densitysymbol}}{\coord{1}}=\psi\lnbra \coord{1} \rnbra \equiv \mu - V_{ext}\lnbra \coord{1} \rnbra \, ,
\end{equation}
and from \refeq{eq:fromfidanfdex} and \refeq{eq:fromlegendre} we derive \refeq{eq:fdcfresultap}.

\chapter{Details of the calculations}
\label{ap:numericaldetails}
 Throughout the work we have employed certain technical methods, which were
not described in detail, those belong in present appendix.
Here we will describe the method of numerical integration developed
to deal with the integrals that are present in \eqrefeq{expansionofc2l2}
and in coefficients of \ere{c2nonuniformexpandedl2}, 
the methods of solving the self-consistent equations,
and comment on the magnitude of numerical errors.
\section{Integration procedure}
 Once the bifurcation equations are formulated, we are left with only
one problem: the estimation of the $\clmn{2}{m}{n}$ coefficients
in \eqrefeq{expansionofc2l2} (or $w_{s}$, $w_{m,n}$ in \ere{c2nonuniformexpandedl2}). They enter all of the equations for
branching point \refeq{eq:exactbifisoldl2}, \refeq{eq:bifunbx}.
In order to calculate them,
we need a model for pair direct correlation function. We have chosen the 
low-density approximation, as described by Eqs.~\re{ldsettings}-\re{ldapproximation}. Given the model two-point potential $V(\coord{1},\coord{2})$, we
can proceed to calculate the coefficients $\clmn{2}{m}{n}$. That task has
to be performed numerically.
\nlin We focus on the spatially uniform case; the calculations for
smectic coefficients $w_{s}$, $w_{m,n}$ are analogical. Each $\clmn{2}{m}{n}$
is, in fact, a six dimensional integral:
\begin{equation}
 \label{eq:appdetailscmn}
\begin{split}
 \clmn{2}{m}{n} = & \, \int^{2\pi}_{0} d\alpha \int^{\pi}_{0} d\beta \sin(\beta) \int^{2\pi}_{0} d\gamma \int^{2\pi}_{0} d\theta \int^{\pi}_{0} d\phi \int^{R}_{0} d\dimensionless{r} \, {\dimensionless{r}}^{2} \\ & \;\;\;\;\;\;\;\; c(\alpha,\beta,\gamma,\theta,\phi,\dimensionless{r}) \symmetrydelta{2}{m}{n}{\alpha,\beta,\gamma} \, , \\
 c(\alpha,\beta,\gamma,\theta,\phi,\dimensionless{r}) = & \, \exp\lsbra -\dimensionless{V}(\alpha,\beta,\gamma,\theta,\phi,\dimensionless{r})/\dimensionless{t} \rsbra-1 \, ,
\end{split}
\end{equation}
in the above $\epsilon_{0}$ sets the energy unit, as in \re{gbreducedunits}; $\dimensionless{V}=V/\epsilon_{0}$ and stands for the
two-point interaction, either biaxial Gay-Berne (\secrefsec{BZ}) or
model bent-core potential (\secrefsec{bentcore}), $\dimensionless{t}$ is a dimensionless temperature defined in \re{gbreducedunits}.
In \refeq{eq:appdetailscmn} $R$ is a fixed maximal radius, above which the
contribution to the integral is negligibly small. In case of Gay-Berne interaction the choice
of $R$ is fairly obvious, since there exists a distance at which most certainly
the interaction is practically zero, and the above integral 
numerically vanishes (it follows from the fact that Gay-Berne potential operates
on relatively short distances, and the long-range, attractive tail
can be at some distance considered to be negligible). However in case of
dipole-dipole interaction the choice of $R$ is not that obvious, since the
interaction is long range. In that case we have performed the integration
\refeq{eq:appdetailscmn} for $\dimensionless{r} \leq R_{GB}$, where $R_{GB}$ stands for
the radius where Gay-Berne interaction can be neglected, and for $\dimensionless{r}>R_{GB}$
we have integrated the expanded Meyer function. So the integrand
\begin{equation}
 \label{eq:appc2approxfordipoles}
 c(\alpha,\beta,\gamma,\theta,\phi,\dimensionless{r})=\left\{ \begin{array}{lr} 
  \exp\lsbra -V_{GB}/\dimensionless{t}-V_{DD}/\dimensionless{t} \rsbra-1, & \textnormal{for }\dimensionless{r} \leq R_{GB} \, , \\
  \frac{1}{2} V^{2}_{DD}/{\dimensionless{t}}^{2}, & \textnormal{for }\dimensionless{r}>R_{GB} \, . \end{array} \right. \, ,
\end{equation}
where $V_{GB}$ stands for Gay-Berne potential and $V_{DD}$
for dipole-dipole interaction.
Later the results were verified by integrating the Meyer function
symmetrized with respect to the direction of the dipole, where the 
symmetrization was imposed by the form of \sopdfns. We should keep
in mind that in present approach $c(\alpha,\beta,\gamma,\theta,\phi,\dimensionless{r})$ contributes via the \ere{ldapproximation}:
\begin{equation}
\label{eq:ldwithc}
 \frac{1}{\dimensionless{t}}\excessfef{\opdfsymbol}/\average{N}=-\frac{1}{2}\dimensionless{\rho}\int d\orientationin d\orientation d^{3}\dimensionless{\boldv{r}} \, \opdff{\orientationin}c(\relativeorientation,\boldv{r})\opdff{\orientation} \, .
\end{equation}
This equation has physical meaning, since the equilibrium state is found
as the minimum of $\intrinsicfef{\opdfsymbol}=\idealfef{\opdfsymbol}+\excessfef{\opdfsymbol}$. As we can see, the contribution to the above comes from
symmetrized $c(\relativeorientation,\boldv{r})$. The form of \ere{ldwithc} is
the base of all the above assumptions used in the calculation of $\clmn{2}{m}{n}$.
\putfigureflagslong{part1/misc/c200-ex.eps}{Sample result of the numerical integration (see \ere{appdetailscmn}).}{appexamplec222}{t}{Sample result of the numerical integration.}
\nlin In \refeq{eq:appdetailscmn} the integrand is a strongly oscillating 
function, and the straight forward approach based on the classic 
numerical methods is going to fail to give a precise result.
All of the model pair-potentials $V(\coord{1},\coord{2})$, studied here,
as functions of $\dimensionless{r}$
behave as presented in \figreffig{intermolecularpotentialexample},
which means that $c(\relativeorientation,\dimensionless{\boldv{r}})$ for fixed orientations depends on $\dimensionless{r}$ as
in \figreffig{ldappofc2}. We can see that the integral over the radius
connecting centres of mass of the molecules pose the main problems.
Therefore we have developed a ''hybrid'' integration method, where
the $5$ dimensional integral over orientations $\alpha$, $\beta$, $\gamma$,
$\theta$, and $\phi$ was calculated using Monte Carlo (MC) method, and in each
MC cycle, for fixed orientational variables, a precise adaptive\footnote{This method automatically located the areas with the largest relative error, and by subsequent divisions of the interval in these regions and integration reduced it to the given level.} $10$-point
Gauss quadrature method was employed. The number of Monte Carlo steps and
the maximum number of subdivisions of the integration interval in $\dimensionless{r}$
were tuned up so the relative error was kept to be less than $0.5$\%.
It was estimated by calculating the integral for doubled number of Monte Carlo 
steps and interval divisions, and comparing with less precise result.
That required a maximal number of Monte Carlo steps to be $4194304$,
and on average 16 subdivisions.
In \figreffig{appexamplec222} an example result of the integration 
is presented. We show the $\clmn{2}{0}{0}$ coefficient calculated for the
biaxial Gay-Berne model for biaxial, rod-like molecule as function of the 
dimensionless temperature $\dimensionless{t}$. Please recall that in order to
solve the self-consistent bifurcation equations, we need to know the values
of $\clmn{2}{m}{n}$ for given $\dimensionless{t}$.
\nlin In the case of smectic phases, the integral over $z$ in
coefficients $w_{s}$, $w_{m,n}$ has to be calculated precisely. For this reason
the calculations for spatially non-uniform structures were one order
of magnitude longer, for the integration over $z$ was performed in the 
same way as over $\dimensionless{r}$ in spatially uniform case, i.e., we are
left with a two dimensional integral for each Monte Carlo cycle.
\section{Self-consistent equations}
 Having calculated the coefficients $\clmn{2}{m}{n}$ for given
pair potential parameters, and knowing their dependence on
dimensionless temperature,
we could proceed to solve the bifurcation equations. In the case of
transitions from isotropic phase, the equations \re{exactbifisoldl2}
are ''analytical'', in the sense that once we know $\clmn{2}{m}{n}$,
we can immediately find bifurcation density for given temperature,
although not explicitly, since $\clmn{2}{m}{n}$ do not have an analytic 
dependence on $\dimensionless{t}$.
In \nun--\nbx\ case the \ere{bifunbx} are self-consistent equations for
density  depending on the order parameters in the reference
state, which have to be calculated also in the self-consistent manner
using \refeq{eq:orderparametersselfconsistent}. In order to perform the
calculations concerning the stabilisation of the reference state,
and the solution of bifurcation equations \re{exactbifisoldl2},\re{bifunbx}
we have constructed a \ital{Mathematica} notebook. 

 The integration procedure was implemented in C language, compiled using
The Portland Group Inc.~compiler. The total CPU time consumed for the
calculations, results of which are presented here, including the preliminary
studies on smectic phases was $90000$ hours.
The numerical analysis was performed at ICM under grant G27-8, using
a cluster of 110 nodes of total 292 CPUs, of which 98 machines
were IBM eServer 325 (each with two AMD Opteron 246 CPUs), and 12 were
Sun v40z (each with eight AMD Opteron 875 CPUs).
The work was financed by grant from MNiSW no N202 169 31/3455.

\chapter{Symmetry adapted functions}
\label{ap:deltas}
 In current appendix we would like to present a possible derivation of
orthogonal base in the space of real functions, which are defined on the 
orientation space, i.e., are mapping unit sphere into $\mathbb{R}$.
It is also a base in the space of solutions of bifurcation equations.
\nlin In actual calculations we have used two representations
of $\symmetrydeltana{l}{m}{n}$ functions, one with help of Euler angles
and the other with directional cosines. Also some properties of Wigner matrices
were used. We will briefly write down the formulas here. We follow the
notation and representation of Rose \cite{rose}.

\section{Wigner matrices}
\label{section:wignerd}
 Wigner matrices $\wignerdna{l}{m}{n}$ are representations of the rotation
operator in the spherical base in the same way the rotation matrix is
in Cartesian coordinates. In present work we follow the convention used
by M.~E.~Rose \cite{rose}. This section is devoted to the brief summary
of formulas which were used to obtain the symmetry adapted version
of Wigner matrices, presented in the next section.
\nlin If we consider a state $\psi_{\,l,m}$ diagonalizing the square of
angular momentum and one of its components along a given axis $\unitv{z}$,
we can wonder how this state transforms under the rotation $\oper{R}$ about 
fixed direction $\unitv{n}$. From definition that transformation is
described by Wigner matrices:
\begin{equation}
 \label{eq:rotationofm}
 {\oper{R}}\psi_{\,l,m}=\sum_{m^{'}}\wignerdna{l}{m^{'}}{m}(\alpha,\beta,\gamma)\psi_{\,l,m} \, ,
\end{equation}
where 
\begin{equation}
 \label{eq:defofwignerd}
 \wignerdna{l}{m^{'}}{m}(\alpha,\beta,\gamma)=\langle l \, m^{'} \left| e^{-i\alpha \oper{L}_{z}} e^{-i\beta \oper{L}_{y}}e^{-i\gamma \oper{L}_{z}} \right| l \, m \rangle \, , 
\end{equation}
and $\oper{L}_{x}$, $\oper{L}_{y}$, $\oper{L}_{z}$ are the angular momentum operators.
\nlin In present study we will often encounter the relative orientation
of the molecules, which means that we will need a formula for the
Wigner matrices in that case. From definition we have
\begin{equation}
 \label{eq:relativewignerdapp}
\begin{split}
 \wignerdna{l}{m^{\prime}}{m}(\relativeorientation)=\langle l \, m^{\prime} | {\oper{R}}^{\prime \, -1}{\oper{R}} | l \, m \rangle = & \, \sum_{n} \langle l \, m^{\prime} | {\oper{R}}^{\prime \, -1} | l \, n \rangle \langle l \, n \left|{\oper{R}} \right| l \, m \rangle= \\
 &\, \sum_{n}\wignerdnaconjugate{l}{n}{m^{\prime}}(\orientationin)\wignerdna{l}{n}{m^{\phantom{\prime}}}(\orientation) \, ,
\end{split}
\end{equation}
where $\wignerdnaconjugate{l}{m}{n}$ stands for complex conjugate of $\wignerdna{l}{m}{n}$.
\nlin Another important property of $\wignerdna{l}{m}{n}$ functions are the orthogonality
relations:
\begin{equation}
 \label{eq:wignerdortho}
 \int d\orientation \, \wignerdnaconjugatetopalign{l}{m}{n^{\phantom{\prime}}}(\orientation) \wignerdna{l^{\prime}}{m^{\prime}}{n^{\prime}}(\orientation) = \frac{8\pi^{2}}{2l+1}\delta_{l,l^{\prime}}\delta_{m,m^{\prime}}\delta_{n,n^{\prime}} \, .
\end{equation}
We will also need a rule for integrating three Wigner matrices, 
we can acquire it with help of the so-called Clebsch-Gordan series
(see, e.g, \cite{rose} p.57):
\begin{equation}
 \label{eq:clebschgordanseries}
 \wignerdna{l_{1}}{\mu_{1}}{m_{1}} \wignerdna{l_{2}}{\mu_{2}}{m_{2}}=\sum_{l} C(l_{1},l_{2},l;\mu_{1},\mu_{2})C(l_{1},l_{2},l;m_{1},m_{2}) \wignerdna{l}{\mu_{1}+\mu_{2}}{m_{1}+m_{2}} \, ,
\end{equation}
where $C(l_{1},l_{2},l;m_{1},m_{2})$ are Clebsch-Gordan coefficients \cite{rose},
and the summation goes over the $l$ satisfying the triangle rule with $l_{1}$
and $l_{2}$. Now using \refeq{eq:wignerdortho} we can obtain the following:
\begin{equation}
 \label{eq:threewignerorthoappp}
\begin{split}
 \int d\orientation \, \wignerdnaconjugate{l_{3}}{\nu}{n}(\orientation) \wignerdna{l_{1}}{\mu_{1}}{m_{1}}(\orientation) \wignerdna{l_{2}}{\mu_{2}}{m_{2}}(\orientation)= & \frac{8\pi^{2}}{2l_{3}+1} C(l_{1},l_{2},l_{3};\mu_{1},\mu_{2})C(l_{1},l_{2},l_{3};m_{1},m_{2}) \\
 & \delta_{\mu_{1}+\mu_{2},\nu}\delta_{m_{1}+m_{2},n} \, .
\end{split}
\end{equation}
\section{Derivation of $\symmetrydeltana{l}{m}{n}$ functions}
\label{section:basefunctions}
 \sbopdf and correlation functions,
in spatially uniform case, are defined on the orientation space. Current
section is devoted to the introduction of the base in the space of orientation
dependent, real functions, for a given model symmetry of molecules and phase.
\nlin Lets consider a set of irreducible tensors $\irrten{l}{m}$ in 
spherical base. Given some symmetry group ${\cal{G}}$, any element of the set
$\irrten{l}{m}$ can be symmetrized. Namely the ${\cal{G}}$--symmetrized
form of $\irrten{l}{m}$ can be written as
\begin{equation}
 \label{eq:symmetrizationrule}
 \irrtensymm{l}{m}{\symmg{G}} \equiv N^{l}_{m}\,\frac{1}{\left| {\cal{G}} \right|}\sum_{g\in {\cal{G}}}{\cal{D}}\lnbra g \rnbra \irrten{l}{m} \, ,
\end{equation}
where ${\cal{D}}\lnbra g \rnbra$ is the operation of symmetry element $g$,
$\left| {\cal{G}} \right|$ is the number of elements in $\symmg{G}$, and 
coefficients $N^{l}_{m}$ ensure the normalization with respect to the
following scalar product:
\begin{equation}
 \irrtensymm{l}{m}{\symmg{G}} \cdot \irrtensymm{l^{'}}{m^{'}}{\symmg{G}}=\sum_{i,j,\dots} \lsbra \irrtensymm{l}{m}{\symmg{G}} \rsbra_{i,j,\dots} \, \lsbra \irrtensymm{l^{'}}{m^{'}}{\symmg{G}} \rsbra_{i,j,\dots}=\delta_{l,l^{'}}\,\delta_{m,m^{'}} \, ,
\end{equation}
where the summation goes over Cartesian indices. $\irrten{l}{m}$ can be built
using the following relation:
\begin{equation}
 \irrten{l}{m}=\sum_{m1\,m2}\lnbra\linv 
  \begin{array}{cc}
   l-1 & 1 \\
   m_{1} & m_{2}
  \end{array} \right| 
  \begin{array}{c}
   l \\
   m
  \end{array}
  \rnbra \irrten{l-1}{m_{1}} \otimes {\bf{T}}^{1}_{m_{2}} \, ,
\end{equation}
where $\irrten{1}{0}$ and $\irrten{1}{\pm 1}$ are as follows:
\begin{equation}
\begin{split}
 \irrten{1}{0} & = {\bf{z}} \, , \\
 \irrten{1}{\pm 1} & = \frac{\pm 1}{\sqrt{2}}\lnbra {\bf{x}}\pm i \, {\bf{y}} \rnbra \, .
\end{split}
\end{equation}
 One of the issues that we are dealing with is establishing the relation between
microscopic structure and macroscopic properties of the system.
Usually we are given some shape of molecules with some model interaction,
and we are left with a problem of determining the phase sequence.
We can define the shape as invariance under some symmetry group ${\cal{G}}_{B}$,
and introduce a symmetry of the phase ${\cal{G}}_{L}$.
Then, we can build the sets of ${\cal{G}}_{B}$ and ${\cal{G}}_{L}$ -- symmetrized 
irreducible tensors \cite{jerphagon}, and choose the 
base functions as following combinations: ${\bf{L}}_{n}\cdot{\bf{B}}_{m}$ 
where ${\bf{L}}_{n}\equiv \irrtensymm{l^{*}}{n}{{\cal{G}}_{L}}$ 
and ${\bf{B}}_{m} \equiv \irrtensymm{l^{*}}{n}{{\cal{G}}_{B}}$ (the angular
momentum index $l^{*}$ is fixed\footnote{We refer to the ''lowest, weakest or minimal coupling model'' by $L=2$ having in mind $l^{*}=2$ from the formula above.}).
If we assume that both ${\cal{G}}_{m}$ and ${\cal{G}}_{l}$ are equal to \dtwh,
then the only non--vanishing, real, independent, non--trivial tensors 
remaining after symmetrization 
(performed along \refeq{eq:symmetrizationrule}) are:
\begin{equation}
 \label{eq:d2hsymmetrictensors}
\begin{split}
 \irrtensymm{2}{0}{D_{2h}} = & \frac{1}{\sqrt{6}}\,\lnbra 3\,{\bf{z}}\otimes{\bf{z}}-{\bf{1}} \rnbra \, , \\
 \irrtensymm{2}{2}{D_{2h}} = & \frac{1}{\sqrt{2}}\,\lnbra {\bf{x}}\otimes{\bf{x}}-{\bf{y}}\otimes{\bf{y}} \rnbra \, .
\end{split}
\end{equation}
We arrive at the set of \dtwh --symmetry adapted functions:
\begin{equation}
 \label{eq:symmetryadapteddeltas}
\begin{split}
 \symmetrydelta{2}{0}{0}{\orientation} = {\bf{L}}_{0}\cdot{\bf{B}}_{0} = & \, \frac{1}{4}+\frac{3}{4}\,cos\lnbra2\,\beta\rnbra \, , \\
 \symmetrydelta{2}{0}{2}{\orientation} = {\bf{L}}_{0}\cdot{\bf{B}}_{2} = & \, \frac{\sqrt{3}}{2}cos\lnbra2\,\gamma\rnbra sin\lnbra \beta \rnbra^2 \, , \\
 \symmetrydelta{2}{2}{0}{\orientation} = {\bf{L}}_{2}\cdot{\bf{B}}_{0} = & \, \frac{\sqrt{3}}{2}cos\lnbra2\,\alpha\rnbra sin\lnbra \beta \rnbra^2 \, , \\
 \symmetrydelta{2}{2}{2}{\orientation} = {\bf{L}}_{2}\cdot{\bf{B}}_{2} = & \, \frac{1}{4}\lsbra 3+cos\lnbra 2\,\beta \rnbra \rsbra\,cos\lnbra 2\,\alpha\rnbra\, cos\lnbra 2\,\gamma\rnbra \\
 & -cos\lnbra \beta \rnbra sin\lnbra 2\,\alpha \rnbra sin\lnbra 2\, \gamma \rnbra \, ,
\end{split}
\end{equation}
where:
\begin{equation}
\label{eq:symmetryadapteddircos}
\begin{split}
 {\bf{L}}_{0}\cdot{\bf{B}}_{0} & \equiv -\frac{1}{2}+\frac{3}{2} \lnbra\bodz\cdot\labz \rnbra^{2} \, , \\
 {\bf{L}}_{0}\cdot{\bf{B}}_{2} & \equiv \frac{\sqrt{3}}{2} \lcbra \lnbra \bodx\cdot\labz \rnbra^{2} - \lnbra \body\cdot\labz \rnbra^{2} \rcbra \, , \\
 {\bf{L}}_{2}\cdot{\bf{B}}_{0} & \equiv \frac{\sqrt{3}}{2} \lcbra \lnbra \bodz\cdot\labx \rnbra^{2} - \lnbra \bodz\cdot\laby \rnbra^{2} \rcbra \, , \\
 {\bf{L}}_{2}\cdot{\bf{B}}_{2} & \equiv \lnbra\bodx\cdot\labx \rnbra^{2} + \lnbra\body\cdot\laby \rnbra^{2} - \frac{1}{2}\lnbra\bodz\cdot\labz \rnbra^{2} - \frac{1}{2} \, ,
\end{split}
\end{equation}
and where $\alpha,\beta,\gamma$ are the Euler angles \cite{rose,altmann} parametrizing
the rotation transforming
the laboratory frame $\lcbra \labx,\laby,\labz \rcbra$ to the body frame $\lcbra \bodx,\body,\bodz \rcbra$.
The set of symmetry adapted base functions \refeq{eq:symmetryadapteddeltas}
is well known \cite{mulder}, in present approach we can clearly see
that it is valid for \dtwh -- symmetric phase
and molecules. Choosing the lowest possible $l^{*}$ and neglecting higher
order terms constitutes the so-called weakest coupling model.
The above allows for higher $l^{*}$ to be taken into account as well
as other symmetries. We see that the first of the lower indices
of $\symmetrydelta{l}{m}{n}{\orientation}$ is associated with the symmetry
of phase, while the other corresponds to the molecular symmetry group.
\nlin Next, we will present some properties of $\symmetrydeltana{l}{m}{n}$
functions that are useful in calculations. One of them are orthogonality relations.
Those are obvious when we realize that the set \refeq{eq:symmetryadapteddeltas}
can be represented by Wigner matrices \cite{rose,lindner} as was done 
in \cite{mulder}:
\begin{equation}
 \label{eq:deltasfromwigner}
 \symmetrydelta{l}{m}{n}{\orientation}=\dwcoeff{m}{n}\,\sum_{\sigma,\sigma^{'}\in\lcbra -1,1 \rcbra} \wignerd{l}{\sigma\,m}{\sigma^{'}\,n}{\orientation} \, ,
\end{equation}
where coefficient $\dwcoeff{m}{n}$ ensures the normalization. In case
of \dtwh symmetry $l$, $m$, and $n$ are even and positive, and
\begin{equation}
 \label{eq:deltasncoeff}
 \dwcoeff{a}{b}\equiv \lnbra \frac{\sqrt{2}}{2} \rnbra^{2+\delta_{a,0}+\delta_{b,0}}\, .
\end{equation}
Above representation is useful, since the orthogonality relations for
$\wignerdna{l}{m}{n}$ \refeq{eq:wignerdortho} are well known \cite{lindner,rose}, and we can easily see that:
\begin{equation}
 \label{eq:deltasortho}
 \int\,d\orientation\,\symmetrydeltatopalign{l}{m}{n^{\phantom{\prime}}}{\orientation}\,\symmetrydelta{l^{\prime}}{m^{\prime}}{n^{\prime}}{\orientation}=\deltasnormalization{l}\,\delta_{l,l^{\prime}}\,\delta_{m,m^{\prime}}\,\delta_{n,n^{\prime}}\, ,\,\,\deltasnormalization{l}\equiv\frac{8\,\pi^2}{2\,l+1} \, ,
\end{equation}
where $\delta_{i,j}$ stands for the Kronecker delta.
\nlin We should remember that the relation \refeq{eq:deltasfromwigner}, 
alas in present work defined only for positive and even $l,m,n$, can be easily 
extended to other molecular symmetry groups.
Following the above analysis it is possible to symmetrize the spherical tensors
to include symmetries lower than \dtwh, which for $l^{*}=2$ will mean the 
inclusion of other that even and/or positive indices $m,n$ in $\symmetrydeltana{l}{m}{n}$. In that sense the analysis conduced using the symmetry adapted functions, provided they form an orthogonal base in space of real functions,
can be considered general.
\nlin The pair \sdcf depends on relative orientation, and will be expanded
in the above base, therefore it is of interest to take 
a look at $\symmetrydelta{l}{m}{n}{\relativeorientation}$.
From the definition \refeq{eq:deltasfromwigner}, and using \refeq{eq:relativewignerdapp} we have
\begin{equation}
 \symmetrydelta{l}{m}{n}{\relativeorientation}=\dwcoeff{m}{n}\,\sum_{\sigma,\sigma^{\prime}\in\lcbra -1,1 \rcbra} \lsbra \sum_{k}\lnbra -1 \rnbra^{k-\sigma\, m}\wignerd{l}{-k}{\sigma^{\phantom{\prime}}m}{\orientationin}\wignerd{l}{k}{\sigma^{\prime}\, n}{\orientation} \rsbra \, .
\end{equation}
Using the above, and orthogonality of Wigner matrices, we can write down 
the following integration rule:
\begin{equation}
 \label{eq:relativeortho}
 \int\,d\orientationin\,\symmetrydeltatopalign{l}{m}{n^{\phantom{\prime}}}{\relativeorientation}\,\symmetrydelta{l^{\prime}}{m^{\prime}}{n^{\prime}}{\orientationin}=\deltasnormalization{l}\dwcoeff{m}{n^{\prime}}\,\symmetrydelta{l}{m^{\prime}}{n}{\orientation}\sum_{\sigma,\sigma^{\prime}\in\lcbra -1,1 \rcbra}\delta_{\sigma m,\sigma^{\prime} n^{\prime}} \, \delta_{l,l^{\prime}} \, .
\end{equation}
\nl In subsequent chapters we will also need a rule for integrating 
three symmetry adapted base functions which follows from relation
\refeq{eq:threewignerorthoappp}:
\begin{equation}
 \label{eq:threedeltasortho}
\begin{split}
 \int\,d\orientation\,\symmetrydelta{l}{m_{1}}{n_{1}}{\orientation}\symmetrydelta{2}{m_{2}}{n_{2}}{\orientation}\symmetrydelta{2}{m_{3}}{n_{3}}{\orientation}= & \deltasnorml{l}\dwcoeff{m_{1}}{n_{1}}\dwcoeff{m_{2}}{n_{2}}\dwcoeff{m_{3}}{n_{3}} \\
 & \sum_{\sigma_{i}} N^{l}_{\substack{m_{1},m_{2},m_{3}\\
         n_{1},n_{2},n_{3}}} ( \lcbra\sigma_{i}\rcbra ) \, ,\,\, l\in\lcbra 2,4 \rcbra \, ,
\end{split}
\end{equation}
where we have denoted
\begin{equation}
\begin{split}
 N^{l}_{\substack{m_{1},m_{2},m_{3}\\
          n_{1},n_{2},n_{3}}} ( \lcbra\sigma_{i}\rcbra ) \equiv & \lnbra -1 \rnbra^{\sigma_{1}\,m_{1}+\sigma_{2}\,n_{1}} \\
 & \delta_{\sigma_{3}\,m_{2}+\sigma_{5}\,m_{3},-\sigma_{1}\,m_{1}}\delta_{\sigma_{4}\,n_{2}+\sigma_{6}\,n_{3},-\sigma_{2}\,n_{1}} \\
 & \clebschgordan{2}{2}{l}{\sigma_{3}\,m_{2}}{\sigma_{5}\,m_{3}}\clebschgordan{2}{2}{l}{\sigma_{4}\,n_{2}}{\sigma_{6}\,n_{3}} \, , \\
 & i\in\lcbra 1,2,3,4,5,6 \rcbra \, ,
\end{split}
\end{equation}
and where $\clebschgordan{l_{1}}{l_{2}}{l}{m_{1}}{m_{2}}$ are Clebsch-Gordan 
coefficients \cite{rose} in decomposition of state $l,m$ on to $l_{1},m_{1}$
and $l_{2},m_{2}$, and the summation goes over all the $\sigma$s, each 
taking values of $1$ and $-1$.

 There are many ways of introducing the base functions. Most of them take into
account the analysis of the symmetries involved in the problem. We have 
presented one possible derivation which stresses out the way in which phase 
and molecular symmetry come into the picture. The other way would be to postulate 
\refeq{eq:deltasfromwigner} by requiring the proper behaviour under the 
\dtwh symmetry and realness, as was done for instance in \cite{mulder}.

 Main benefits from using the irreducible representations in present work
come from clear transformation laws under rotations, built-in
orthogonality, and completeness. The Cartesian tensors usually come in
reducible form, and therefore present a more difficult objects 
when it comes to actual calcualtions. On the other hand, since we are more 
used to the Cartesian form, some interpretation and meaning can be retrieved 
more easily in that base.

\chapter{\sbopdf and order parameters}
\label{ap:opdf}
 During phase transition a symmetry change takes place. In 
isotropic -- uniaxial nematic case the rotational invariance of the
disordered liquid  is broken, and replaced by the symmetry under the action
of elements of group containing the rotation around the director and
the reflection in the
perpendicular plane. One possible method to model that type of behaviour is to 
define some quantities that in higher symmetric phase vanish,
while in more ordered structure obtain a non--zero value. 
Those are usually defined as averages over an ensemble of the base
functions \refeq{eq:symmetryadapteddeltas}, and are called order parameters.
\nlin Of course the symmetry change happening over a phase transition has to 
influence the \sopdfns. If we employ the order parameter approach it 
seems natural that the distribution should be parametrized by those objects.
We will begin with an example of such approach.
\nlin Lets take into account two orthogonal tripods,
one associated with laboratory, usually chosen to coincide with the 
director frame: $\lcbra \labx,\laby,\labz \rcbra$,
and the other with the molecule: $\lcbra \bodx,\body,\bodz \rcbra$.
For the moment we take into account the uniaxial nematic phase, and molecules 
having the symmetry of ellipsoids of revolution, and write the \sopdf as:
\begin{equation}
 \label{eq:dircosexp}
 \opdff{\lcbra\labv{i},\bodv{i}\rcbra}=\delta\,\lnbra \bodx\cdot\labz \rnbra^{2}+\eta\,\lnbra \body\cdot\labz \rnbra^{2}+\vartheta\,\lnbra \bodz\cdot\labz \rnbra^{2} \, .
\end{equation}
The symmetry requires that in the laboratory frame there is a rotational
invariance around the director, and that the distribution is 
not changed by the reflection in the perpendicular plane. That is the
reason why only $\labz$ is present in the above formula, and that the expression
is quadratic in products of the base vectors.
We can find a transformation to the orthogonal base, and receive 
\begin{equation}
\label{eq:dircosorthogonalized}
 \opdff{\lcbra\labv{i},\bodv{i}\rcbra}=\tilde{a}\,\lsbra -\frac{1}{2}+\frac{3}{2} \lnbra\bodz\cdot\labz \rnbra^{2}\rsbra+\tilde{b}\,\lsbra \frac{\sqrt{3}}{2} \lcbra \lnbra \bodx\cdot\labz \rnbra^{2} - \lnbra \body\cdot\labz \rnbra^{2} \rcbra \rsbra \, .
\end{equation}
Now the parameters $\tilde{a}$ and $\tilde{b}$ can be calculated as 
ensemble averages of the expressions in square brackets, using the 
orthogonality of the new base functions, and identified with the order parameters.
The expansion of the form \refeq{eq:dircosexp} is inspired by the
fact that \sopdf undergoes a qualitative transformation on line leading from 
one state to another of lower symmetry.
It is of course a very simple approximation. There are several others 
available. One of them, clearly inspired by the form of the {\sceq}
\refeq{eq:selfconsistentequniform} is to assume
\begin{equation}
\label{eq:modelopdf}
 \opdf\sim\exp\lnbra\sum_{l,m,n}\alpha_{l,m,n}\symmetrydelta{l}{m}{n}{\orientation}\rnbra \, .
\end{equation}
Another is a Gaussian model, where \sopdf is taken
to be of the following form:
\begin{equation}
\label{eq:modelopdfgaussian}
 \opdff{\alpha,\beta,\gamma}\sim \exp \lsbra a\lnbra \alpha^{2}+\beta^{2}+\gamma^{2} \rnbra \rsbra \, .
\end{equation}
Expression in \eqrefeq{dircosexp} is the 
simplest possible model of \sopdfns, and should be treated with care. One
obvious flow is the fact that probability of this form is not always 
positively defined, and may become negative. Despite that problem (and others)
the approximation works surprisingly well. The Gaussian \refeq{eq:modelopdfgaussian} and exponential \refeq{eq:modelopdf} models provide a better behaviour of $\opdf$ but the relation of
coefficients $a$ and $\alpha_{l,m,n}$ to the order parameters is not as simple as in case of
\eqrefeq{dircosorthogonalized}.
\nlin Having performed the orthogonalization that led to the expansion
\refeq{eq:dircosorthogonalized}, we have changed the base to a certain class of 
functions which by construction are invariant under the operation
of \dinfh symmetry. Using some general assumptions concerning the phase and
molecules, we have received the uniaxial version of the expansion in 
the symmetry adapted base introduced 
in \refeq{eq:symmetryadapteddeltas}. In case of the molecules orientations 
of which, due to symmetry, need to be described by three vectors,
the same method can be used to obtain the whole base set in \dtwh -- symmetric
case. It should be noted that in the initial expansion
\refeq{eq:dircosexp} the higher order terms in directional cosines, which 
in the language of $\symmetrydeltana{l}{m}{n}$ would mean the higher $l$, are
not {\it a priori} forbidden. The choice of the lowest order terms
is a part of the approximation.
\nlin It is natural to choose the order parameters as averages of 
$\symmetrydeltana{2}{m}{n}$ functions, we can write
\begin{equation}
 \label{eq:opdfinopsapp}
 \opdf=\frac{1}{8\pi^{2}}+\sum_{l,m,n} \frac{2\,l+1}{8\pi^{2}}\,\orderparameter{l}{m}{n}\symmetrydelta{l}{m}{n}{\orientation} \, .
\end{equation}
Using the relations in \eqrefeq{deltasortho} we can immediately see that
\begin{equation}
 \label{eq:orderparamssetapp}
 \orderparameter{l}{m}{n} \equiv \int d\orientation\,\opdf\symmetrydelta{l}{m}{n}{\orientation} \, ,
\end{equation}
where as usual the average of quantity $A$ is defined as $\average{A}\equiv\int d\orientation \, A(\orientation)\opdf$.
From the above set $\orderparameter{2}{0}{0}$ is a well known uniaxial 
order parameter present in Maier-Saupe theory \cite{maiersaupe}. Second one, 
$\orderparameter{2}{0}{2}$, was introduced by
Freiser \cite{freiser}, $\orderparameter{2}{2}{0}$ was seen in the paper 
by Alben, McColl, and Shih \cite{ams}, and the last but not least, 
$\orderparameter{2}{2}{2}$ was for the first time defined by
Straley \cite{straley}.
\nlin The equation \refeq{eq:orderparamssetapp} can be treated
as the definition of order parameters, once a model of \sopdf has been
chosen and the equilibrium state found. 

\chapter{Mean field approximation}
\label{ap:meanfield}

There exists a class of models that treat systems 
of interacting molecules by certain approximations concerning the potential 
energy and distribution functions. Common to all of those is the presence
of a pair interaction, \sopdfns, and an effective, 
average potential that models all of the forces that a given molecule 
is subjected to. They are called \ital{mean field} theories. Presently, we 
show a way in which they are introduced.
\nlin As was mentioned in Chapter~\refchap{DFT}, the mean field approximation
can be introduced as a further expansion of two-point correlation
function modelled by $\meyerf{\coord{1}}{\coord{2}}$ (see \eqrefeq{ldapproximation} and \refeq{eq:ldsettings}).
In current appendix, we will present a different method of derivation.
\newline\indent Let's consider a system of $N$ molecules interacting by
$N\textnormal{-particle}$ potential $V\lnbra \coord{1},\dots,\coord{N}\rnbra$
and an arbitrary $N\textnormal{-particle}$ distribution function 
$\opdfsymbol_{N}\equiv P\lnbra \coord{1},\dots ,\coord{N}\rnbra$.
We can write down the free energy as a functional of $P_{N}$, namely
\begin{equation}
\label{eq:fullfeapp}
\begin{split}
 F\lsbra P_{N} \rsbra = & \, \int d\coord{1} \dots d\coord{N} \, P\lnbra \coord{1},\dots ,\coord{N}\rnbra V\lnbra \coord{1},\dots,\coord{N}\rnbra \\
 & \, + k_{B}T\,\int d\coord{1} \dots d\coord{N} \,  P\lnbra \coord{1},\dots ,\coord{N}\rnbra \ln P\lnbra \coord{1},\dots ,\coord{N}\rnbra \, .
\end{split}
\end{equation}
\newline If we assume that the $N\textnormal{-particle}$ distribution function
can be taken as a product of \sopdfs (in other words we assume the independence
of distributions, which is not necessarily true), the above changes into:
\begin{equation}
\label{eq:mffeapp}
 F\lsbra \opdfsymbol \rsbra/N=\frac{1}{2}\int d\coord{1}\,\lsbra \int d\coord{2}\,\opdff{\coord{1}}V\lnbra \coord{1},\coord{2} \rnbra \opdff{\coord{2}} \rsbra + k_{B}T \, \int d\coord{1} \, \opdff{\coord{1}} \ln{\opdff{\coord{1}}} \, .
\end{equation}
Now we can perform the minimization of the above, with normalization condition
for $\opdff{\coord{1}}$, to acquire the following equation for
equilibrium distribution
\begin{equation}
\label{eq:mfeqapp}
 \opdfsymbol_{eq}\lnbra \coord{1} \rnbra=Z^{-1} \exp{\lsbra -\beta\int d\coord{2}\, V(\coord{2},\coord{1}) \opdff{\coord{2}}\rsbra} \, ,
\end{equation}
where $V(\coord{2},\coord{1})$ is the pair potential, and where $Z$ is a
normalization constant so \linebreak $\int d\coord{1}\, \opdfsymbol_{eq}\lnbra \coord{1} \rnbra=1$.
As can be seen from the above equation, given molecule is subjected to 
an average { \it effective } potential $V_{eff}\lnbra \coord{1} \rnbra\equiv \int d\coord{2} V\lnbra \coord{2},\coord{1} \rnbra \opdff{\coord{2}}$.
\nlin \eqrefeq{mfeqapp} is well known, and widely used. Its success lies mainly
in the relative simplicity, compared to the variety of systems and behaviour
that can be modelled by it. However, clearly, the competition of the entropic
and potential energy terms in \eqrefeq{fullfeapp} is disturbed
in \eqrefeq{mffeapp}, which results in overestimated transition temperatures
obtainable from \eqrefeq{mfeqapp}.

\end{appendices}

\begin{backmatter}
 \bibliography{phd}
 \listoffigures
 \listoftables
\begin{colophon}
\indent This thesis was created using Vim text editor \href{http://www.vim.org}{http://www.vim.org} in \LaTeX{}. The style was based on the hepthesis package (available from \href{http://www.ctan.org/tex-archive/help/Catalogue/entries/hepthesis.html}{http://www.ctan.org}), heavily modified by the author. In the
process of data gathering and formatting, scripts written in awk, perl, and bash
were used.
\nlin All the plots were made with the help of gnuplot \href{http://www.gnuplot.info}{http://www.gnuplot.info}, and the drawings were designed and exported 
using inkscape \href{http://www.inkscape.org}{http://www.inkscape.org}.
\end{colophon}

\end{backmatter}

\end{document}